\documentclass[11pt,twoside]{report}
\usepackage{suthesis-2e}

\usepackage[greek,english]{babel}
\usepackage[utf8x]{inputenc}

\usepackage[pdftex,bookmarks,colorlinks]{hyperref}
\hypersetup{colorlinks=false}
\usepackage{braket}
\usepackage{amssymb}
\usepackage{epsfig}

\newcommand{\sgn}{\mathrm{sgn}}

\newcommand{\hh}{\widehat{H}}

\newcommand{\wphi}{W_{\phi}}

\newcommand{\M}{\textnormal{\tiny \textsc{m}}}
\newcommand{\W}{\textnormal{\tiny \textsc{w}}}
\newcommand{\ci}{\textnormal{\tiny \textsc{$\, \circ\, $}}}

\newcommand{\si}{\sigma}

\newcommand{\tpm}{\sqrt{2E_{m}}}

\newcommand{\Wo}{W^{0}}
\newcommand{\uh}{u^{\epsilon}}
\newcommand{\wh}{W^{\epsilon}}
\newcommand{\hb}{\epsilon}
\newcommand{\vh}{v^{\epsilon}}

\newtheorem{theorem}{Theorem}[section]
\newtheorem{lemma}[theorem]{Lemma}
\newtheorem{proposition}[theorem]{Proposition}
\newtheorem{corollary}[theorem]{Corollary}

\newenvironment{definition}[1][Definition.]{\begin{trivlist}
\item[\hskip \labelsep {\bfseries #1}]}{\end{trivlist}}
\newenvironment{example}[1][Example.]{\begin{trivlist}
\item[\hskip \labelsep {\bfseries #1}]}{\end{trivlist}}
\newenvironment{remark}[1][Remark.]{\begin{trivlist}
\item[\hskip \labelsep {\bfseries #1}]}{\end{trivlist}}

\newcommand{\qed}{\nobreak \ifvmode \relax \else
      \ifdim\lastskip<1.5em \hskip-\lastskip
      \hskip1.5em plus0em minus0.5em \fi \nobreak
      \vrule height0.75em width0.5em depth0.25em\fi}

\newcommand{\hf}{\widehat{f}}
\newcommand{\tr}{\mbox{tr}}

\usepackage{color}

\begin{document}

\title{\it{Construction of an approximate solution \\
of the Wigner equation
\\
by uniformization of WKB functions}}

\author{Konstantina-Stavroula Giannopoulou}

\submitdate{November 2015}

\dept{Mathematics \& Applied Mathematics}

\principaladviser{Prof. G.N. Makrakis}

\firstreader{Profs.\\
A. Athanassoulis, University of Leicester\\
S. Filippas, University of Crete\\
G. Karali, University of Crete\\
T. Katsaounis, King Abdullah University of Science and Technology (KAUST)\\
G. Kossioris, University of Crete\\
%G.N. Makrakis (advisor), University of Crete
P. Rozakis, University of Crete
}
%\secondreader{George Kossioris}

%\copyrightyear
%\copyrighttrue
%\copyrightfalse
\beforepreface

%\begin{flushright}
\begin{center}
\vspace{7cm}

%\huge{\bfseries
%Πτυχιακή Εργασία}\\[6mm]

\begin{center}
\large {
\selectlanguage{greek}
ΓΙΑΝΝΟΠΟΥΛΟΥ ΚΩΝΣΤΑΝΤΙΝΑ-ΣΤΑΥΡΟΥΛΑ}
\vspace{3cm}
\end{center}

\large{
\selectlanguage{greek} 
{\bf {ΚΑΤΑΣΚΕΥΗ ΜΙΑΣ ΑΣΥΜΤΩΤΙΚΗΣ ΛΥΣΗΣ\\
ΓΙΑ ΤΗΝ ΕΞΙΣΩΣΗ 
\textlatin{WIGNER}\\
ΜΕΣΩ ΟΜΟΙΟΜΟΡΦΟΠΟΙΗΣΗΣ ΣΥΝΑΡΤΗΣΕΩΝ \textlatin{WKB}
} } }\\[1mm]
\vspace{2cm}
\end{center}
%\end{flushright}
%\noindent \HRule \vspace*{\stretch{2}}

\begin{center}
\selectlanguage{greek}
\large{ΔΙΔΑΚΤΟΡΙΚΗ ΔΙΑΤΡΙΒΗ\\
\vspace{1cm}
Επιβλέπων Kαθηγητής: Γεώγιος Ν. Μακράκης }\\[3mm]
\vspace{6cm}
\end{center}

\begin{center}
\large{
\selectlanguage{greek}
ΤΜΗΜΑ ΜΑΘΗΜΑΤΙΚΩΝ \& ΕΦΑΡΜΟΣΜΕΝΩΝ ΜΑΘΗΜΑΤΙΚΩΝ\\
ΠΑΝΕΠΙΣΤΗΜΙΟ ΚΡΗΤΗΣ\\
ΝΟΕΜΒΡΙΟΣ 2015}
\end{center}

%%%%%%%%%%%%%%%%%%%%%%%%%%%%%%%%%%%%%%%%%%%%%%%%%%%%%%%%%%%%%%%%%%%
%%%%%%%%%%%%%%%%%%%%%%%%%%%%%%%%%%%%%%%%%%%%%%%%%%%%%%%%%%%%%%%%%%%%

\prefacesection{Abstract}
The Wigner equation is a non-local, evolution equation in  phase-space. It describes the evolution of the Weyl symbol of the  density operator which, in general,  is governed by the Liouville-von Neumann equation of quantum mechanics. For pure quantum states, the Wigner equation is an equivalent reformulation of the standard quantum-mechanical Schr\"odinger equation, and it could be also derived in an operational way by considering the Wigner transform of the quantum wave function, without using the Weyl calculus.

In this thesis, we construct an approximate solution of the Wigner equation in terms of Airy functions, which are semiclassically concentrated on certain Lagrangian curves in two-dimensional phase space. These curves are defined by the eigenvalues and the Hamiltonian function of the associated  one-dimensional Schr\"odinger operator, and they play a crucial role in the quantum interference mechanism in phase space. We assume that the potential of the Schr\"odinger operator is a  single potential well,  such that the spectrum is discrete. The construction starts from an eigenfunction series expansion of the solution, which  is derived here for first time in a systematic way, by combining the elementary technique of separation of variables with involved spectral results for the Moyal star exponential operator.
The eigenfunctions of the Wigner equation are the Wigner transforms of the Schr\"odinger eigenfunctions, and they are approximated in terms of Airy functions by a uniform stationary phase approximation of the Wigner transforms of the WKB expansions of the Schr\"odinger eigenfunctions. Although the WKB approximations of  Schr\"odinger eigenfunctions  have non-physical singularities  at the turning points of the classical Hamiltonian, the phase space eigenfunctions provide bounded, and correctly scaled, wave amplitudes when they are projected back onto the configuration space (uniformization). 

Therefore, the  approximation of the eigenfunction series  is an  approximated solution of the Wigner equation, which by projection onto the configuration space provides  an approximate wave amplitude, free of turning point singularities. It is generally expected that, the derived wave amplitude is bounded, and correctly scaled, even on caustics, since only finite terms of the approximate terms are significant for WKB initial wave functions with finite energy. 

The details of the calculations are presented for the simple potential of the harmonic oscillator, in order to be able to check our approximations analytically. But, the same construction can be applied to any potential well, which behaves like the harmonic oscillator near the bottom of the well. In principle, this construction could be extended to higher dimensions  using canonical forms of the Hamiltonian functions and employing the symplectic covariance inherited by the Weyl representation into the Wigner equation.
  \\
  \\
  \\
  \\
 { \huge{\bf Keywords}}
 \\
 \\
 \\
Schr\"odinger equation, Wigner equation, semiclassical limit, geometric optics,
caustics, Weyl quantization, Weyl operators, Wigner transform, uniform stationary phase method
\\
\\
\\
\\
\\
%\vspace{3cm}
This thesis has been partially supported by the “Maria Michail Manasaki” Bequest Fellowships.

\prefacesection{\textgreek{Περίληψη}}
 \selectlanguage{greek}
 {Η εξίσωση \textlatin{Wigner} (\textlatin{Wigner equation}) είναι μια μη-τοπική (\textlatin{non-local}) εξίσωση εξέλιξης στον χώρο των φάσεων (\textlatin{phase space}). Περιγράφει την εξέλιξη του \textlatin {Weyl} συμβόλου του τελεστή πυκνότητας (\textlatin{density operator}) ο οποίος, εν γένει, διέπεται από την εξίσωση \textlatin{Liouville-von Neumann} της κβαντομηχανικής.
Για καθαρές κβαντικές καταστάσεις (\textlatin{pure states}), η εξίσωση \textlatin{Wigner} είναι μια ισοδύναμη αναδιατύπωση της βασικής εξίσωσης της κβαντικής μηχανικής, της εξίσωσης \textlatin{Schr\"odinger} και θα μπορούσε επίσης να παραχθεί με έναν τελεστικό τρόπο, θεωρώντας τον μετασχηματισμό \textlatin{Wigner} (\textlatin{Wigner transform}) της κυματοσυνάρτησης, χωρίς τη χρήση του λογισμού \textlatin{Weyl} (\textlatin{Weyl calculus}).

Σε αυτήν τη διατριβή, κατασκευάζουμε μια προσεγγιστική λύση της εξίσωσης \textlatin{Wigner} εκφρασμένη σε όρους  συναρτήσεων \textlatin{Airy} (\textlatin{Airy function}), οι οποίες συγκεντρώνονται ημικλασικά πάνω σε κάποιες Λαγκραντζιανές καμπύλες (\textlatin{Lagrangian curves}) στον διδιάστατο χώρο των φάσεων. Οι καμπύλες αυτές ορίζονται από τις ιδιοτιμές και την Χαμιλτωνιανή συνάρτηση (\textlatin{Hamiltonian function}) του συσχετιζόμενου
 μονοδιάστατου τελεστή \textlatin{Schr\"odinger}, και παίζουν κρίσιμο ρόλο στον μηχανισμό της κβαντικής αλληλεπίδρασης (\textlatin{quantum interference mechanism}) στον χώρο των φάσεων. 
 Δεχόμαστε ότι το δυναμικό του τελεστή \textlatin{Schr\"odinger} είναι ένα μονό πηγάδι δυναμικού (\textlatin{single-well potential}) τέτοιο ώστε το φάσμα (\textlatin{spectrum}) να είναι διακριτό. 
 
Η κατασκευή ξεκινάει από ένα ανάπτυγμα ιδιοσυναρτήσεων (\textlatin{eigenfunction series expansion}) της λύσης, το οποίο παράγεται εδώ με έναν συστηματικό τρόπο για πρώτη φορά, συνδυάζοντας την στοιχειώδη τεχνική του χωρισμού μεταβλητών με φασματικά αποτελέσματα για τον εκθετικό τελεστή \textlatin{ Moyal} (\textlatin{Moyal star exponential operator}).
Οι ιδιοσυναρτήσεις της εξίσωσης \textlatin{Wigner} είναι οι μετασχηματισμοί \textlatin{Wigner}  των ιδιοσυναρτήσεων του τελεστή \textlatin{Schr\"odinger} και προσεγγίζονται με όρους της συνάρτησης \textlatin{Airy}, από μια προσέγγιση ομοιόμορφης στάσιμης φάσης των μετασχηματισμών \textlatin{Wigner} των αναπτυγμάτων \textlatin{WKB} των ιδιοσυναρτήσεων του τελεστή \textlatin{Schr\"odinger}. 
Μολονότι οι προσεγγίσεις \textlatin{WKB} των ιδιοσυναρτήσεων \textlatin{Schr\"odinger} έχουν μη-φυσικές ιδιομορφίες (\textlatin{non-physical singularities}) στα σημεία καμπής (\textlatin{turning points}) της κλασικής Χαμιλτωνιανής (\textlatin{classical Hamiltonian}), οι ιδιοσυναρτήσεις στον χώρο των φάσεων δίνουν φραγμένα και σε σωστή κλίμακα κυματικά πλάτη (\textlatin{wave amplitudes}) όταν αυτά προβάλλονται πίσω στον εποπτικό χώρο (\textlatin{configuration space}) (ομοιομορφοποίηση (\textlatin{uniformization)}). 

Επομένως, η προσέγγιση της σειράς ιδιοσυναρτήσεων είναι μια προσεγγιστική λύση της εξίσωσης \textlatin{Wigner}, η οποία μέσω της προβολής στον εποπτικό χώρο δίνει ένα  προοσεγγιστικό κυματικό πλάτος χωρίς ιδιομορφίες.
 Εν γένει, αναμένεται ότι, το παραγόμενο κυματικό πλάτος είναι φραγμένο και σε σωστή κλίμακα ακόμα και επάνω στις καυστικές, αφού, μόνο πεπερασμένοι όροι των προσεγγίσεων είναι σημαντικοί για αρχικές κυματοσυναρτήσεις \textlatin{WKB} \textlatin{(WKB initial wave functions)} με πεπερασμένη ενέργεια. 
 
Οι λεπτομέριες των υπολογισμών παρουσιάζονται για το απλό δυναμικό του αρμονικού ταλαντωτή , ώστε να είναι δυνατόν να ελεγχούν οι προσεγγίσεις μας αναλυτικά. Όμως, η ίδια κατασκευή μπορεί να εφαρμοστεί για οποιοδήποτε πηγάδι δυναμικού το οποίο συμπεριφέρεται όπως ο αρμονικός ταλαντωτής κοντά στον πάτο του πηγαδιού. Σε γενικές γραμμές, η κατασκευή αυτή θα μπορούσε να επεκταθεί σε υψηλότερες διαστάσεις χρησιμοποιώντας κανονικές μορφές \textlatin{(canonical forms)} των Χαμιλτωνιανών συναρτήσεων και τη συμπλεκτική συνδιακύμανση \textlatin{(symplectic covariance)}
που προκύπτει από την αναπαράσταση \textlatin{Wey} στην εξίσωση \textlatin{Wigner}.
 \\
  \\
  \\
  \\
{\huge{{\bf Λέξεις κλειδιά}}}
\\
\\
\\
Εξίσωση \textlatin{Schr\"odinger}, εξίσωση \textlatin{Wigner}, ημικλασικό όριο, γεωμετρική οπτική, καυστικές, κβάντωση \textlatin{Weyl}, τελεστές \textlatin{Weyl}, μετασχηματισμός \textlatin{Wigner}, μέθοδος ομοιόμορφης στάσιμης φάσης
81S30, 34E05, 58K55 
\\
\\
\\
%\vspace{3cm}
Η διατριβή αυτή έχει χρηματοδοτηθεί κατά ένα μέρος από τις υποτροφίες του  κληροδοτήματος <<Μαρίας Μιχαήλ Μανασάκη>>.

\selectlanguage{english}{}

%\prefacesection{Acknowledgments}
%\newpage
%\begin{flushright}
%$\Sigma\tau o u\varsigma$  $\gamma o \nu\varepsilon \acute{\iota}\varsigma$ $\mu o u$,
%\\
%Μαρία και Δημήτρη
%\end{flushright}
\afterpreface

\chapter*{Notation}
\begin{center}
\begin{tabular}[!hbp]{|c|c|}\hline
$\star_{\M}$ & the Moyal product, $(\ref{starproduct_1})$
\\
\hline 
$\uh(x,t)$ & the solution of Schr\"odinger's I.V.P. $(\ref{schr_eq_1})-(\ref{initialdata})$ \\
\hline
$\widehat{H}^{\epsilon}$ & the Schr\"odinger operator, $(\ref{schrod_op})$
\\
\hline 
$E^{\epsilon}_{n}$ & the eigenvalue of the harmonic oscillator, $n=0,1,\ldots$, $(\ref{energy_harm})$ \\
\hline 
$v^{\epsilon}_{n}(x)$ &  the eigenfunction  of the harmonic oscillator, $n=0,1,\ldots$, $(\ref{eigenf_harm})$ \\
\hline  
$W^{\epsilon}[f](x,p,t)$ & the semiclassical Wigner function of $f(x)$, $(\ref{semWigner})$\\
\hline
%$\tilde{v}_{n}^{\epsilon}(x)$ & the asymptotic approximation of the exact eigenfunction $\vh_{n}(x)$ \\
%& in the oscillatory region \\
%\hline
$W_{nm}^\epsilon(x,p)$ & the Wigner transform of $v_{n}^{\epsilon}(x)$ and $v_{m}^{\epsilon}(x)$, $(\ref{cross_wigner_v})$\\
\hline
$\psi^{\epsilon}_{n}(x)$  & WKB-approximation of eigenfunction $v^{\epsilon}_{n}(x)$, $(\ref{wkbout1})-(\ref{wkbsolution_1})-(\ref{wkbout2})$\\
\hline
$\mathcal{W}_{nm}^\epsilon(x,p)$ & the Wigner transform of $\psi_{n}^{\epsilon}(x)$ and $\psi_{m}^{\epsilon}(x)$, $(\ref{wigner_approx_nm})$\\
\hline
$W^0[f](x,p,t)$  & limit Wigner distribution,  $(\ref{diracwig})$  \\
\hline
\end{tabular}
\end{center}
\begin{table}[h]
\caption{\it{Notation} }
\end{table}

%%%%%%%%%%%%%%%%%%%%%%%%%%%%%%%%%%%%%%%%%%%%%%%%%%%%%%%%%%%%%%%%%%%%%%%%%%%%%%%%%%%%%%%%%%%%%%%%%%%%
%%%%%%%%%%%%%%%%%%%%%%%%%%%%%%%%%%%%%%%%%%%%%%%%%%%%%%%%%%%%%%%%%%%%%%%%%%%%%%%%%%%%%%%%%%%%%%%%%%%%
%%%%%%%%%%%%%%%%%%%%%%%%%%%%%%%%%%%%%%%%%%%%%%
%%%%%%%%%%%%%%%%%%%%%%%%%%%%%%%%%%%%%%
\newpage
\chapter{Introduction}

\section{Schr\"odinger equation and high-frequency waves}
High-frequency wave propagation is a fundamental problem which arises in quantum mechanics, and in classical wave theories such as acoustics, seismology, optics and electromagnetism. Wave equations modelling energy propagation including diffraction and scattering effects, in many interesting cases, as, for example, the propagation in infinite domains with non-compact boundaries, or in media with complex inhomogeneous structure, are difficult to be treated either analytically or numerically.
For this reason in several important applications,
as in underwater acoustics, or in  propagation of laser beams in the atmosphere and  the propagation of radio waves near  Earth's surface , it has been proposed, and it has been successfully implemented for practical purposes, the {\it parabolic approximation} method  \cite{Flat, Tap1, Foc1, Tap2}. The main idea is that, under certain assumptions, instead of solving the wave equations, someone approximates the slow varying component of the wave field  by a wave function which solves  an initial value problem for the quantum mechanical {\it Schr\"odinger equation}
\begin{eqnarray*}
i\epsilon\partial_{t}u^{\epsilon}(x,t)=\left[-\frac{\epsilon^2}{2}\partial_{xx}+V(x)\right]u^{\epsilon}(x,t)\quad ,\quad (x,t)\in R\times [0,\infty) \ ,
\end{eqnarray*}
$u^{\epsilon}(x,t=0)=u^{\epsilon}_{0}(x)$ being an appropriately modeled initial wavefunction. Note that in classical wave propagation, the variable $t$ is also spatial variable, usually representing the direction of long-distant one-way propagation, and not the physical time as in quantum mechanics.

The parameter $\epsilon$ is connected with Planck's constant in quantum mechanics. In classical waves it is connected with the frequency of the waves, and it is small in many interesting case where the wavelength is small compared  with the scale of spatial variations of the properties of, and the size of the domain occupied by, the propagating medium. 

The potential $V$  encodes the properties of the propagating medium in classical waves, while it describes the forces acting on moving particles in quantum mechanics. There is, however,  a certain relation between the classical waves and the quantum particles, which is described by the {\it particle-wave duality} and the {\it correspondence principle}. This duality becomes mathematically apparent in the {\it classical limit} (or, {\it high-frequency limit}) $\epsilon \to 0 \,$ where classical mechanics and geometrical optics emerge from quantum mechanics and wave theory, respectively,  in the form of the Hamilton-Jacobi and transport equations \cite{Rob, MF, NSS}.

\subsubsection{Geometrical optics, caustics and phase-space methods}

The most interesting and difficult problems concern the propagation of highly oscillatory initial data of the form 
$$ u^{\epsilon}_{0}(x)=A_{0}(x)\exp(i S_{0}(x)/\epsilon) \ , \ \ \ \ \epsilon<<1 \ . $$ 
Mathematically these problems  have traditionally been treated using the   {\it WKB  method}, which is also known as the  {\it Geometrical Optics technique}, \cite{BLP, BB}. According to this method one seeks for an approximating solution for the problem of the form  
$$u^{\epsilon}({ x},t)\approx \psi^{\epsilon}({ x},t)
=A(x,t)\exp(iS(x,t)/\epsilon) \ ,$$
where the amplitude  $A(x,t)$ and the phase $S(x,t)$  satisfy the transport  and Hamilton-Jacobi  equations, respectively. 

However, the method fails on caustics and focal points where the solution of the Hamilton-Jacobi equation becomes multi-valued and the solution of the transport diverges and it predicts physically meaningless infinite wave amplitudes. 
Also, the method fails in shadow regions (i.e. regions devoid of rays), where it yields erroneous zero fields. Formation of caustics and shadows is a typical situation in optics, underwater acoustics and seismology, as a result of multipath propagation from localized sources, \cite{TC, CMP}. In quantum mechanics the formation of such singularities are connected with the classically forbidden regions, where classical particles cannot penetrate, and the tunneling effects \cite{Raz}.

Uniform asymptotic expansions near simple caustics have been constructed , assuming that the multivalued function $S$ is known away from caustics and using {\it boundary
layer techniques} and {\it matched asymptotic expansions} \cite{BB, BaKi}. However, these analytical techniques are very complicated to apply in specific problems, since the matching procedure depends on the form of the particular caustic and it requires delicate geometrical constructions.

A different category of methods for the construction of uniform wave fields near caustics, is based on  representations of the solutions in terms of phase-space integrals. The main and most known methods in this category are {\it Maslov's
canonical operator} \cite{MF, MSS}, and the {\it Lagrangian integrals (Kravtsov-Ludwig method)} \cite{Kra, Lu, Dui1, Dui2, KO}. Both representations are special cases of Fourier integral operators \cite{Dui3, Tr}.

All the above described techniques {\it assume an ansatz for the wave field}, which for the final determination requires the knowledge of the multivalued phase
functions, or, geometrically, of the Lagrangian manifold generated by the bicharacteristics
of the underlying Hamiltonian system in phase space.

\section{Wigner transform and Wigner equation}

An alternative approach is to  {\it reformulate the evolution equations in configuration space to kinetic-type equations in phase space} by using {\it phase space transforms} of the wave functions (see, e.g.,  \cite{MCD} for a review of this idea in several wave problems of classical physics).

The most popular phase space transform appears to be the Wigner transform. This is a function
defined on phase space as the Fourier transform of the two-point correlation of the wave
function,
\begin{eqnarray*}
W^\hb[u^{\hb}](x,p,t) :=
(2\pi\hb)^{-1}\int_{R}^{} e^{-\frac{i}{\epsilon}p y} u^{\hb}\left(x+\frac{y}{2},t\right)\overline{u^{\hb}}\left(x-\frac{y}{2},t\right)\, dy \ .
\end{eqnarray*}
This object was introduced by E.Wigner \cite{Wig} for modeling purposes in quantum thermodynamics,
and recently, it has been successfully used in semiclassical analysis for the reformulation
of wave equations as non local equations in phase space and the study of homogenization
problems in high-frequency waves \cite{GMMP}. 

In the case of the Schr\"odinger equation, the corresponding Wigner function $W^{\hb}[u^{\hb}]$ satisfies the \emph{semiclassical Wigner equation},
\begin{eqnarray*}
&&i\epsilon {\partial_t} W^\hb[u^\epsilon](x,p,t)=H(x,p)\star_{\M} W^\hb[u^\epsilon](x,p,t)-W^\hb[u^\epsilon](x,p,t)\star_{\M} H(x,p) \ , 
\end{eqnarray*}
where $H(x,p)$ is the classical Hamiltonian, and the Moyal product $\star_{\M}$ is defined by
\begin{eqnarray*}
\star_{\M}:=
\exp\left[\frac{i\epsilon}{2}\left(
\overleftarrow{\partial _x}
\overrightarrow{\partial _p}-
\overleftarrow{\partial _p}
\overrightarrow{\partial _x}\right)\right] \ .
\end{eqnarray*}

This is a linear evolution equation in phase space, and non locality stems from the Moyal star product coupling the Hamiltonian with the Wigner function. This non-commutative product, which is the Weyl image of operator composition, encodes all the  important features of quantum interferences, and it is interpreted  in a physically plausible and mathematically elegant way in the framework of deformation quantization (Bopp shift) of the classical Hamiltonian mechanics \cite{Gro, BaFFLS1}.

The basic and most interesting derivation of the Wigner equation in quantum mechanics comes from the Weyl representation of the Liouville-von Neumann equation for the evolution of quantum
density operator, and, for pure quantum states, it is equivalent to the Schr\"odinger equation \cite{Gro, CFZ2}.
There are a few basic theoretical for the Wigner equation, mainly the works of P.Markowich, et al. \cite{MA, SMM} for the equivalence of Wigner and   Schr\"odinger equations, and the related asymptotic analysis, and also some results on scattering theory by  H.Rogeon \& P.Emamirad \cite{EmRo1, EmRo2}. However, there are not efficient constructive techniques for solving the Wigner equation, probably because of the complicated and rather unusual characters of this equation, that are described in the sequel. 

For smooth
potentials, the non local operations in the equation can be reformulated as an infinite order singular perturbation (with dispersion terms with respect to the momentum of the phase space) of the Liouville equation of classical mechanics.
In particular, for the Schr\"odinger equation, which, as it will be explained later, is the quantization of standard particle Hamiltonian $H(x,p)=\frac{p^2}{2}+V(x)$ ,  the Wigner equation can be written as 
\begin{eqnarray*}
\left({\partial_t} -V'(x)\partial_{p}+p\partial_{x}\right)W^\hb[u^\hb](x,p,t)=\sum_{n=1}^{\infty}\frac{(-1)^n}{(2n+1)!}\left(\frac{\hb}{2}\right)^{2n}
\frac{d^{2n+1}}{dx^{2n+1}} V(x) {\partial_p}^{2n+1}W^\hb[u^\hb](x,p,t) \ .
\end{eqnarray*}

It becomes now apparent that the Wigner equation has non-constant coefficients, and, at least formally, it combines two different features, namely those of transport  and  of dispersive equations. The first one arises the Hamiltonian system of the Liouville equation in the left hand side  it is related the underlying classical mechanics of the problem. The second one arises from odd derivatives with respect to momentum in the right hand side. It is related to the quantum energy dispersion  inside a boundary layer around the Lagrangian manifold of the Hamiltonian system and the modification of local scales according to the geometry of the evolving manifold.

At the classical limit $\epsilon \rightarrow 0$ formally  the series in the right hand side of the equations disappears  and the equation becomes the classical Liouville equation of classical mechanics. In a fundamental work, P.L.Lions \& T.Paul \cite{LP} have shown that in fact the solution of the Wigner equation has a weak solution, to the so called Wigner measure, and this measure solves the Liouville equation. For relatively smooth initial phase functions $S_0$, the weak limit is equivalent to the single phase geometrical optics, in the sense that it produces the same results with the WKB method for the energy density.

In the case of multi-phase optics and caustic formation,
the Wigner measure is  not the appropriate tool for the study of the
semiclassical limit of waveamplitudes, that we are interested in. It has been shown by S.Filippas \&
G.N.Makrakis   \cite{FM1, FM2}, by solving analytically  certain simple porblems {\it for the
case of time-dependent Schr\"odinger equation} that the Wigner
measure ($a$) cannot be expressed as a distribution with respect
to the momentum for a fixed space-time point,  and thus it cannot
produce the amplitude of the wavefunction, and ($b$) is unable to
``recognize'' the correct frequency scales of the wavefield
near caustics. It was however explained  that the solutions of the
integro-differential Wigner equation are able to
capture the correct frequency scales, and therefore it became promising to look for asymptotic approximations of the solution of the Wigner equation.

Therefore, the deeper study of asymptotic solutions of the Wigner equation for small $\epsilon$ seems to be promising
both for understanding the structure of the solutions of the equation, and for computing energy densities and probability densities, in multiphase geometrical optics, shadow zones and classically forbidden regions.

 Before proceeding to the review and discussion of asymptotic solutions we would like to mention some interesting and useful numerical approaches for the Wigner equation. Such methods have been mainly proposed in quantum optics and quantum chemistry. First of all, it has been proposed  to construct solutions of classical Liouville equation,
as an alternative to apply the WKB method,  by attempting to capture numerically, in a kinetic way,
the multivalued solutions far from the caustic (see, e.g.,
\cite{JL, Ru1, Ru2}). In order to apply this technique, it is necessary to introduce an priori  {\it closure assumption} for a system of equations for
the moments of the Wigner measure, which essentially fixes a
finite number of rays passing through a particular point. Such a closure condition leads to systems which have many similarities with incompressible hydrodynamics and then several popular methods of computational fluid mechanics can be used.
Among many numerical solutions of the Wigner equation which have been developed for specific applications in physics, chemistry and quantum chemistry, we would like to mentions some of them which we believe that give also some mathematical insight to the problem. First, the spectral method of M.Hug, C.Menke \& W.P.Schleich \cite{HMS1, HMS2}, which is based on the approximation of Wigner functions by Chebyshev polynomials in phase space. Second, the particle technique which was proposed by A.Arnold \& F.Nier \cite{AN}. This technique was applied for the numerical investigation of simple problems with the presence of caustics  by  E.Kalligiannaki \cite{Kal1}, where the results for the wave amplitude are compared with those obtained by a finite element code for the Schr\"odinger equation and  E.Fergadakis \cite{Ferg} who applied a particle in cell method for the same problem in the case of the  propagation of plane waves in a linear layered medium (where caustics coincide with the turning points of rays). Third, and most important, is the numerical solution of the smoothed Wigner equation proposed by A.Athanassoulis \cite{Ath1, Ath2} for the accurate and efficient numerical treatment of wave propagation problems. The key point of his approach is that the smoothed Wigner transform can be used to compute the solution at a chosen spatio-spectral resolution. Thus leading to some degree of averaging. The novel idea, which is not common in more traditional techniques, is that the separation of
the averaging operator from the error, leads to an approximate coarse-scale solution.

\section{Asymptotic solutions of the Wigner equation}

Asymptotic solutions  of the Wigner equation have a relatively short history, and most of them are based on perturbation methods, guided partly from the mathematics of the equations and partly from the physics of the wave problem. The techniques that have been proposed so far, could be classified as follows \footnote{This classification reflects somehow our point of view for the problem, and it is guided by the idea of deformation quantization mentioned above.}.

\medskip
\noindent
(a) \underline{\it Distributional expansions:} H.Steinr\"uck \cite{Ste} \& M.Pulvirenti \cite{P} have constructed perturbation distributional  expansions near the solution of the classical Liouville equation,
by expanding the initial data in a distributional series with respect to the small parameter. The first work seems to be the first formal attempt in this direction, while the second one initiated the rigorous investigation of such expansions.

\medskip
\noindent
(b)  \underline{\it Semiclassical expansions using modified characteristics:}
E.J.Heller  \cite{He}, guided by his deep insight from quantum chemistry, noted that the distributional expansions are not physically appropriate for  studying the evolution of singular (with respect to the semiclassical parameter $\epsilon$) initial conditions. Instead, he proposed a different expansion where the first order term is the solution of a classical Liouville equation but with an {\it effective potential}. The use of {\it modified characteristics} resulting from the  effective potential,  aims to include indirectly some quantum phenomena. This is a general philosophy  for the treatment of the quantum Liouville equation by physicists (method of Wigner trajectories, see H.W.Lee \cite{L}), and it has been motivated by  quantum hydrodynamics (Bohm equations) and the technique of Gaussian beams. For similar reasons,  F.Narcowich \cite{N} has proposed a different expansion near the classical Liouville equation, somehow smoothing the problem by imposing  where $\epsilon$-dependent initial data, instead of distributional ones, to the Liouville equation, thus avoiding  the distributional expansions of H.Steinr\"uck and M.Pulvirenti.

%\bibitem[N]{N}F. Narkowich, \emph{On the quantum Liouville equation}, Physica, {\bf 134A},  193-208, 1985

\medskip
\noindent
 (c)  \underline{\it Airy-type semiclassical expansions:} More recently, S.Filippas \& G.N.Makrakis 
\\ 
 \cite{FM1, FM2}, used  asymptotic expansions of Airy type  for multivalued WKB functions as asymptotic solutions of the Wigner equation and they showed that such solutions are the correct approximations at least near  simple caustics. The basic tool for the construction of the asymptotic solution is Berry's semiclassical Wigner function \cite{Ber}. This function is an Airy type approximation of the Wigner transform of a single-phase WKB function, which has been derived by the Chester-Friedman-Ursell technique  of uniform  stationary phase for one-dimensional Fourier integrals \cite{CFU}.
The key observation for the construction of this asymptotic solution  is that when the semiclassical Wigner function is transported by the Hamiltonian flow, it remains a good approximation of the Wigner equation at least until critical time when the Lagrangian manifold develops singularities and the first caustics appears. However, it turned out a posteriori, that the asymptotic formula remains meaningful even after the critical time. In the simplest case of fold and cusp caustics the asymptotic formula reproduces the already known asymptotic approximations of the wave field and it captures the correct dependence of the wave function on the frequency, near and on the caustics.  

\medskip
\noindent
 (d) \underline{\it Semiclassical expansions near harmonic oscillator:} In a different direction, 
\\E.K.Kalligiannaki \cite{Kal} (see also \cite{KalMak}), has proposed in her doctoral dissertation a new strategy for the construction of asymptotic approximations of the Wigner equation, departing  from the eigenfunction expansion of Schr\"odinger equation's  solution, in the case of single potential well.
The interrelation between the solutions of the Wigner equation in phase space and those of the Schr\"odinger equation in the configuration space has bee very important for understanding the structure of the phase space solutions. This idea goes back to the pioneering work of J.E.Moyal \cite{Mo} for statistical interpretation of quantum mechanics, and it has been rigorously exploited  for first time by P.Markowich \cite{MA} using the functional analysis of Hilbert-Schmidt operators.  The interrelation of the spectra of the Wigner and Schr\"odinger equations has been considered first  by H.Spohn \cite{SP}, and it has been further clarified by I.Antoniou et al. \cite{ASS}. At least in the case when the Schr\"odinger equation has purely discrete spectrum it is clear that the Wigner equation has discrete spectrum too, and we can construct the phase space eigenfunctions of the Wigner equation as the Wigner transforms of the eigenfunctions of the Schr\"odinger equation in the configuration space.
Kalligiannaki's construction relies on the combination of  the Wigner transform with the  perturbations expansion of the eigenfunctions of the Schr\"odinger equation about the harmonic oscillator derived by B.Simon \cite{Si}. It turned out that the expansions of the Wigner eigenfunctions satisfy approximately the pair of equations which govern the phase space eigenfunctions. On the basis of these expansions she proposed an asymptotic ansatz of the solution of Wigner equation. The time-dependent coefficients of the expansion were computed from a hierarchy of equations arising through a regular perturbation scheme. The asymptotic nearness of the expansion to the true solution is studied for small times by using the technique developed by A.Bouzouina \& D.Robert \cite{BR} for the uniform approximation of quantum observables. Although she did not proved estimates for large time, she explained that the application of the expansion to caustic problems for the harmonic and quartic quantum oscillator leads to reasonable numerical approximations of the wave function near and on some caustics.  
 An approximation of this type  naturally produces expansions which are compatible and come from the ``exact'' solution of Schr\"odinger equation, which compared with the WKB solution does not reveal caustic problems (which anyway appear  exactly when we use Geometrical Optics technique). Under certain conditions on the potential function the produced expansion (``harmonic approximation''--``harmonic expansion'') is in accordance with the expansion near classical Liouville equation's solution (``classical approximation''--``classical expansion''), even at caustics in the appropriate semiclassical regime for the parameters of the problem.

\section{Approximation by  uniformization of  WKB functions}

Inspired by the general idea of deformation quantization \footnote{This idea, although not mentioned explicitly,  naturally underlies the construction of semiclassical Wigner function \cite{Ber}, and the approximation of the Wigner equation in \cite{FM1}.}, we have explained in the works  \cite{G, GM}, how the Wigner transform can be used to {\it uniformize two-phase WKB functions near turning points (caustics)}. A two-phase WKB function is the sum of two WKB functions, corresponding to the two geometric phases near the turning point. The Wigner transform of this function consists of  four Fourier integrals, which are approximated by Airy functions either by the uniform formula  of Chester-Friedman-Ursell technique \cite{CFU}, or by the standard stationary phase formula, depending on the phase involved in each integral and the position in phase space. The approximations were combined and matched appropriately  ({\it asymptotic surgery}) in order to derive a {\it uniform Airy expression of the Wigner function the  over the whole phase space} \footnote{We occasionally use the term Wigner function instead of solution of the Wigner equation, when this is clear from the context.}. Then, by projecting the Wigner function onto the configuration we derive wave amplitudes that are bounded at the turning points and they have the correct scale with respect to the semiclassical parameter $\epsilon$.

The details of the calculations were presented for the particular example of two-phase WKB approximation of the Green’s function for the Airy equation with an appropriate radiation condition at infinity. This is a  simple scattering problem with continuous spectrum, which models the propagation of acoustic waves emitted from a point source in a linearly stratified medium. For this model the WKB solution has a simple fold caustic. The applicability of the uniformization  technique for a general fold caustic was asserted by exploiting the geometrical similarity of  a folded Lagrangian manifold  with the simple parabolic manifold of  our model problem. 

It was also shown that the derived Wigner function is an approximate solution of the semiclassical Wigner equation corresponding to the Airy equation. Note that Airy equation is the simplest  stationary Schr\"odinger equation, with linear potential and continuous spectrum.  

Therefore, we have naturally raised the following question:
 {\it Can we construct approximate solutions of the semiclassical Wigner equation by uniformization of WKB functions, when the Schr\"odinger operator has discrete spectrum as, for example, happens in the case of potential wells?} This problem is much more complicated than the above described model one, and we expect to face a very complicated structure of the Wigner function. The  reason is that instead of a single Lagrangian manifold, now we have a {\it fan of closed Lagrangian manifolds} corresponding to the energy levels (bound states) of the anharmonic oscillator. Accordingly, it is expected that the Wigner function  must contain {\it new terms due to the interaction between the manifolds}, in addition to the interactions between the branches of each individual manifold. 

\section{The scope and the contents of the thesis}

The present thesis aims to the construction of an approximate solution of the Wigner equation by uniformization of WKB approximations of the Schr\"odinger eigenfunctions in the case of the quantum harmonic oscillator. We have chosen to work with this simple potential, in order to minimise the geometrical details, and also in order to be able to check some of our approximation results analytically.
The same construction can be applied to any potential well, which behaves like the harmonic oscillator near the bottom of the well, by employing  perturbations of  the Schr\"odinger eigenfunctions  as was done in \cite{Kal}.

The construction of the approximation of the Wigner function departs from an eigenfunction expansion of the solution of the Wigner equation and it relies on Airy type approximations of the  eigenfunctions of the equation (from now on referred as the  {\it Wigner eigenfunction}). The approximate Wigner eigenfunctions are constructed by uniformization of the two-phase WKB eigenfunctions of the corresponding Schr\"odinger operator.
It turns out that  approximation series of the Wigner function is an approximation of the solution of the Wigner equation, and it is well behaved everywhere in the phase space. The projection of the approximation series of the Wigner function onto the configuration space, provides an approximate wave amplitude. This amplitude is  uniformized in the sense that it is bounded on the turning point of the WKB eigenfunctions of the corresponding Schr\"odinger operator.

%%%%%%%%%%%%%%%%%%%%%%%%%%%%%%%%%%%%%%%%%%%%%%%%
%%%%%%%%%%%%%%%%%%%%%%%%%%%%%%%%%%%%%%%%%%%%%%%%
The \underline{contents of the thesis}  are as follows. In Chapters 2-4 and in Chapter 6,, we collect some background or known advance material, which, however, has been  considerably elaborated for use in the developments of  Chapters 5, 7 \& 8.

In \underline{Chapter \ref{chapter2}} we  give a short description of classical and quantum mechanics. We introduce the Schr\"odinger equation as the basic dynamic law governing the evolution of quantum states (wavefunctions), and the more general formulation based on the Liouville-von Neumann equation governing the density operator \cite{NSS}. We explain how E.Schr\"odinger \cite{Schr} invented the Schr\"odinger equation  by quantizing the classical Hamiltonian function. This fundamental idea had far-reaching consequences in quantum physics but also in functional analysis where it has offered crucial motivation for the development of pseudodifferential operators \cite{BaBMo}. Also we explain how the semiclassical parameter $\epsilon$ follows from physical Planck's constant when we use dimensionless time and coordinates, which gives meaning to the notion of semiclassical regime.

In \underline{Chapter \ref{chapter3}} we collect some basic results about the Cauchy problem for the Schr\"odinger equation. First, we describe the construction of the asymptotic solution when $\epsilon$ is small
by the WKB method, we introduce the ideas of geometrical optics  and we explain how the  method fails when caustics, shadow zones or other singularities of the ray congruence develop.
Second, we describe the eigenfunction series expansion of the solution, which in principle is valid for any value of $\epsilon$ and free of caustics. This solution is derived by standard method of separation of variables. Assuming that the potential $V(x)$ is 
 a single well, growing without bound at infinity, so that the spectrum of the Schr\"odinger operator to be purely discrete. For $\epsilon \ll 1$  we present  asymptotic approximations  of  the higher eigenvalues as derived from the Bohr-Sommerfeld rule, and the WKB expansions of eigenfunctions \cite{Fed}. Also, for later use, we present explicit formulas for the eigenvalues and WKB eigenfunctions of  the quantum harmonic oscillator.
 
In \underline{Chapter \ref{chapter4}}  we present some ideas from the theory of pseudodifferential operators, mainly the Weyl method of quantization of symbols (classical observables). We present in details the construction of the Weyl operator following the excellent exposition of F.A.Berezin \& M.A.Shubin \cite{BS}, and we introduce the notion of star product utilizing the Weyl symbol of the composition of operators. Then, we introduce the {\it semiclassical Wigner function} as the Weyl symbol of the pure state density operator, and we derive the Wigner equation,  by applying the Weyl correspondence rule on the Liouville-von Neumann equation governing the evolution of the density operator. In the case of pure quantum states, it turns out that the Wigner equation can be derived also by applying the Wigner transform in a operational way directly onto the  Schr\"odinger equation.

In \underline{Chapter \ref{ch5}}, by exploiting the linearity of the Wigner equation we construct  {\it directly in phase space} an eigenfunction series expansion of the solution of the Cauchy problem for the Wigner equation. The separation procedure leads to  {\it  a pair of phase-space eigenvalue equations}, one corresponding to the Moyal (sine) bracket and the other to the Baker (cosine) bracket. The emergence of these phase space eigenvalue equations is natural when we derive the Wigner equation from the Heisenberg equation by Weyl calculus, in contrary to the most familiar derivation, when one starts from configuration space and transforms the Cauchy problem for the Schr\"odinger equation by introducing the Wigner transform of the wavefunction in an operational way.   

In the study of the eigenvalues equations we used known fundamental results for the spectra of quantum Liouvillian. More precisely, for the eigenvalues of the first (Moyal (sine) bracket) equation we used the results of  H.Spohn \cite{SP}  and I.Antoniou et al. \cite{ASS}, while for the eigenvalues of second (Baker (cosine) bracket) equation we used the results derived in the thesis by E.Kalligiannaki \cite{Kal} (see also \cite{KalMak}), for the special case {\it when the Hamiltonian operator has discrete spectrum}. Then, the {\it representation of the Wigner eigenfunctions as Wigner transforms of the Schr\"odinger eigenfunctions} was proved by using the recent spectral results by M.A.de Gosson \& F.Luef \cite{GL} for the $\star$-genvalue equation.

Our construction of the eigenfunction series expansion of the solution of the Wigner equation by the elementary method of separation of variables, directly in phase space, seems to be a new one. It is a folk statement of physicists that phase-space formulation of quantum mechanics is an independent approach. However,  they always work with Wigner equation by using formal star calculus \cite{CFZ1}, and transferring  by Wigner transform spectral results for the Schr\"odinger equation from configuration space into phase space, whenever they need to complete the phase picture.

In \underline{Chapter \ref{chapter6} } we present some results for the  Wigner transform of WKB functions, and we explain why it is necessary to work with the semiclassical Wigner transform in order to be consistent with geometrical optics. It is very interesting to note that the semiclassical Wigner transform has already bounced independently from Weyl calculus. The main purpose of this chapter is to present the details of Berry's construction of the semiclassical Wigner function of a single-phase WKB function \cite{Ber}. Also we present the Wigner transform of multi-phase WKB functions and its limit Wigner distribution. 

%%%%%%%%%%%%%%%%%%%%%%%%%%%%%%%%%%%%%%%%%%%%%%%%

In \underline{Chapter \ref{ch7}} we construct  Airy-type asymptotic approximations of the Wigner eigenfunctions from the uniform asymptotic approximation of the Wigner transform of the two-phase WKB eigenfunctions  of the harmonic oscillator. This is the main achievement of the thesis, since these approximations are the basic ingredients for the construction of the the approximate solution of the Wigner equation. 

The technique of the construction is somehow an extension of the {\it uniformization procedure} developed in \cite{G, GM} for the semiclassical Airy equation.
 However, there are certain fundamental differences between the two problems, which cause many new technical difficulties. 
 These differences basically arise from the fact that the spectrum of the semiclassical Airy function is continuous, while the spectrum of the harmonic oscillator is discrete. They crucially affect the geometry of the Lagrangian manifolds (actually Lagrangian curves, since the phase space is two dimensional). The Lagrangian curve of the semiclassical Airy equation is an open curve extending to infinity (parabola).  The Lagrangian curves of the  harmonic oscillator  form an infinite fan of closed curves (circles), corresponding to the eigenvalues of the oscillator.
Although the local behaviour at the turning points is qualitatively the same in both cases (namely the behaviour of a fold singularity),  in the case of the harmonic oscillator the presence of a second turning point creates new interactions between the upper and lower branches of the Lagrangian curves. Moreover, there are additional new interactions between the Lagrangian curves of the fan.

Thus, we have two groups of Wigner eigenfunctions. The diagonal Wigner eigenfunctions, associated with a particular Lagrangian curve,  are semiclassically concentrated as Airy functions on this curve. The off-diagonal Wigner eigenfunctions, associated with a pair of Lagrangian curves are again semiclassically concentrated as Airy functions, but on an effective curve between the Lagrangian curves of the pair, and they are modulated with an oscillatory factor depending on the angular direction of phase space. This factor is responsible for the fact that the net contribution to the wave energy of the off-diagonal Wigner eigenfunctions is zero, and therefore the interactions between different Lagrangian curves of the oscillator, are responsible only for the energy exchange between different modes in the solution of the Cauchy problem for the Wigner equation.

The Airy approximations of the Wigner eigenfunctions  have been compared in certain regimes of the parameters with the exact Wigner eigenfunctions of the harmonic oscillator, which are expressed through Laguerre polynomials. This comparison partially confirms the validity of our approximation technique. Moreover, the classical limits of the Airy approximations of the Wigner eigenfunctions are solutions of the classical limits of the Moyal and Baker eigenvalue equations, and this observation provides a further  confirmation of  the validity of our approximation technique.

In \underline{Chapter \ref{ch8}} we combine the eigenfunction series expansion of the solution of the Wigner equation, the approximate Wigner eigenfunctions and Berry's semiclassical function for the WKB initial datum, to derive an approximate solution of the Cauchy problem for the Wigner equation. This is the second main achievement of the thesis. The expansion is the sum of a coherent (time-independent) part spanned by the diagonal Wigner eigenfunctions, and an incoherent (time-dependent) part, spanned by the off-diagonal Wigner eigenfunctions.
The coefficients of the expansion of the coherent part (and in special case also of the incoherent part) of the solution are evaluated either analytically for quadratic initial phases, or approximately by combining a novel asymptotic decomposition of the Airy function with analytic integration, for more general phases. 
The  calculations show that the dependence of the coefficients on the semiclassical parameter $\epsilon$, is crucially dependent  on the initial phase. The derived approximate Wigner function is well-behaved everywhere in the phase space. Moreover, its integration with respect to the momentum results in a wave amplitude which is meaningful even on the turning points  of the WKB approximations of the Schr\"odinger eigenfunctions. Therefore, the  approximate Wigner function derived by the uniformization procedure is  a promising tool to smooth out the caustic singularities in time-dependent problems.

Finally, in \underline{Chapter \ref{ch9}} we give  a short discussion of the main results and the achievements of the thesis, and we discuss about a new question raised by our investigation. This concerns the asymptotic nearness between two approximate solutions of the Wigner equation, that is the time-dependent semiclassical Wigner function constructed in \cite{FM1} and the series approximation constructed in Chapter \ref{ch8} of this thesis.

%%%%%%%%%%%%%%%%%%%%%%%
%%%%%%%%%%%%%%%%%%%%%%%%%%%%%%%%%%%%%%%%%%%%%%%%%%%%%%%%%%%%%%%%%%%%
\chapter{Classical and Quantum mechanics}\label{chapter2}

A classical or quantum mechanical system $\bf{S}$ is described  by defining certain basic mathematical objects which necessarily include: {\it states, observables, dynamical law and transformations}. At each instant of time the system resides in a particular state $\bf{s}$ which is an element of an appropriate linear space (state space). The nature of the states depends essentially on the adopted physical modelling of the system, and they can be elects of finite-dimensional vector spaces, function space or even more general linear spaces.  The experimental observations of the system provide measurements whose interpretation leads to the notion of observable (observable quantity). The evolution of the system is described by the dynamical law which defines the state of the system at any particular time. Finally, both for physical and mathematical reasons, we need to define certain transformations acting on states, which provide equivalent descriptions of the system.

%%%%%%%%%%%%%%%%%%%%%%%%%%%%%%%%%%%%%%%%%%%%%%%%%%%%%%%%%%%%%%%%%%%%%%%%%%%%%%%%%%%%%%%%%%%%%
%%%%%%%%%%%%%%%%%%%%%%%%%%%%%%%%%%%%%%%%%%%%%%%%%%%%%%%%%%%%%%%%%%%%%%%%%%%%%%%%%%%%%%%%%

\section{Classical mechanics}

The simplest system  $\bf{S}$ in classical Newtonian mechanics has $d$ degrees of freedom which form the $d-$dimensional {\it configuration space} $R^d$. For example, if the system consists of $k$ particles moving in three-dimensional space under the action of certain forces, and without any kinematical constraints, we have $d=3k$. The state of such a system is determined uniquely by specifying the (generalised) position and momentum vectors of each particle. Therefore, a state of $\bf{S}$ is a point ${\bf{s}=(x,p)}\in R_{{\bf {xp}}}^{2d}$ , where ${\bf{x}}= (x_1, ..., x_d)$ is the vector of coordinates and  ${\bf{p}}=(p_1, ...,p_d)$ is the vector of momentums. The even-dimensional space $R^{2d}=R_{{\bf xp}}^{2d}$ is called the {\it phase space} of the mechanical system  $\bf{S}$. For a more complicated mechanical systems the phase space is a symplectic manifold, namely a smooth manifold $M$ with a non-degenerate closed two-form $\omega$ defined on it. In this case the phase space is the tangent space $T^{*}M$ \cite{ARN1}. 
 
An observable $\bf{f}$ of the system in a state $\bf{s}$, is a function $f:R^{2d}\rightarrow R$, and the measurement of  $\bf{f}$ gives invariably the same value $f({\bf x},{\bf p})$. The simplest observables are the position and the momentum, i.e. $f({\bf x},{\bf p})=x_i$ and $f({\bf x},{\bf p})=p_i$, respectively.
The most important observable is the total energy $\bf{E}$ of the system. In the state $\bf{s}=(x,p)$ the energy has the value  $E=H({\bf x},{\bf p})$, where $H({\bf x},{\bf p})$ is the Hamiltonian of the system \cite{ARN1, GOLD}. 

In general, scalar functions on the phase space can be multiplied pointwise, i.e. $(fg)({\bf x})=f({\bf x})g({\bf x})$ 
where the multiplication fulfills the following properties: commutativity, linearity and associativity.
So that the observables equipped with the addition and the pointwise multiplication form a commutative algebra of observables.

The dynamics of the system  is uniquely determined by the Hamiltonian function. More precisely, it is postulated that  the evolution of the state $\bf{s}=(x,p)$ (phase point) is governed by the {\it Hamiltonian system} of ordinary differential equations
\begin{eqnarray}\label{hamiltoniansystem_1}
\frac{{d\bf x}(t)}{dt}&=&\nabla_{{\bf p}}H({\bf x,p})\\
&=&(\partial_{p_1}H({\bf x,p}),\ldots\partial_{p_d}H({\bf x,p})) \nonumber
\end{eqnarray}
\begin{eqnarray}\label{hamiltoniansystem_2}
\frac{{d\bf p}(t)}{dt}&=&-\nabla_{{\bf x}}H({\bf x,p})\\
&=&-(\partial_{x_1}H({\bf x,p}),\ldots\partial_{x_d}H({\bf x,p})) \ , \nonumber
\end{eqnarray}
for $i=1,\ldots,d$ .

The Hamiltonian system $(\ref{hamiltoniansystem_1})$, $(\ref{hamiltoniansystem_2})$ under certain smoothness assumptions a map $g_{H}^{t}:R^{2d}\rightarrow R^{2d}$, $t\in R$,  of the phase space into itself,
The map $g_{H}^{t}$, which is known as the {Hamiltonian flow}, preserves the symplectic 2-form $\omega^2=dp\wedge dq$, i.e. $(g_{H}^{t})^{*}\omega ^2=\omega ^2 $, and the Poisson bracket, and therefore it is an admissible canonical transformation of the phase space \cite{GOLD}.

By using the Hamiltonian system, we can compute the time derivative of a time-independent observable $f({\bf x},{\bf p})$ along the Hamiltonian flow, as follows

\begin{eqnarray*}
\dot{f}= \frac{d}{dt} f({\bf x}(t),{\bf p}(t))&=&\dot{{\bf p}}\cdot\nabla_{\bf p} f+\dot{\bf x}\cdot \nabla_{\bf x} f\\
&=&\nabla_{{\bf p}}H\cdot\nabla_{\bf x} f-
\nabla_{{\bf x}}H\cdot\nabla_{\bf p}f \ .
\end{eqnarray*}

Therefore, the evolution of $f$ is governed by the {Liouville equation}
\begin{eqnarray}\label{liouville}
\dot{f}=  {\lbrace H,f\rbrace}_{PB} \ ,
\end{eqnarray}
where ${\lbrace H,f\rbrace}_{PB}$ is the Poisson bracket of the Hamiltonian and the observable $f$ .

The {\it Poisson bracket} of the functions $f, g$, is defined by
\begin{eqnarray}\label{poissonbracket}
\lbrace{f,g}\rbrace_{PB}({\bf x,p})=\sum_{i=1}^{d}\left( \frac{\partial f}{\partial x_{i}}\frac{\partial g}{\partial p_{i}}-\frac{\partial f}{\partial p_{i}}
\frac{\partial g}{\partial x_{i}}\right)\Bigg|_{({\bf x,p})} \ .
\end{eqnarray}
Sometimes it is written symbolically in the form
\begin{eqnarray}
\lbrace{f,g}\rbrace_{PB}({\bf x,p})=\sum_{i=1}^{d}f\left(\overleftarrow{\partial_{x_i}}\overrightarrow{\partial_{p_i}}-
\overleftarrow{\partial_{p_i}}\overrightarrow{\partial_{x_i}}\right)g \Big |_{({\bf x,p})}\quad , 
\end{eqnarray}
where the vector arrows indicate in which direction the differentiation acts. For example, $f\overleftarrow{\partial_{x_i}}g=\frac{\partial f}{\partial x_{i}} g$ and $f\overrightarrow{\partial_{p_i}}g=f\frac{\partial g}{\partial p_{i}} $ .

The Liouville equation $(\ref{liouville})$ is a fundamental equation for the description of general mechanical systems, and it lies in the heart of more general theories, as for example the classical statistical mechanics \cite{B}. Most important, it has an analogue in the phase space description of quantum mechanics, the Wigner equation (see, Chapter  $\ref{chapter4}$, below), and, in this connection, it lies also in the heart of quantization and deformation theories \cite{BaFFLS1, BaFFLS2}.

\section{Quantum Mechanics}\label{section22}

\subsection{Quantum states and  observables}

{\bf States.} A quantum mechanical system $\bf{S}$ is described by postulating that the states $\bf{s}$ are  elements $u\in\mathcal{H}$ of an infinite-dimensional, separable, complex  Hilbert space $\mathcal{H}$ with inner product  $ (\cdot, \cdot)_{\mathcal H}$. We assume that the states $u$ are normalized $||u||_{\mathcal H}=1$.

In Dirac's notation \cite{D}, which is very popular in quantum mechanics \cite{B}          \footnote{although someone can find many  books on quantum mechanics which avoid this notation. See, e.g., \cite{S, Gr} and \cite{Tak}}, we denote the vectors $u\in\mathcal{H}$ by the symbols $\ket{u}$ which are called \emph{kets}. The linear functionals $V$ in the dual space $\mathcal{H'}$, are called \emph{bras} and they are denoted by $\bra{v}$. The numerical value of the functional is denoted as 
$V(u)= \bra{v}\ket{u} \ .$
This notation is justified by the
\\
{\bf Riesz theorem.} There is a one-to-one correspondence between linear functionals $V$, in the dual space $\mathcal{H'}$, and vectors $v$  in  $\mathcal{H}$, such that $V(u)=(v,u)_{\mathcal H} \ .$

Indeed, this theorem implies that there is a one-to-one correspondence between bras and kets \footnote{In his original presentation, Dirac had assumed this correspondence between bras and kets, but it is not at all clear if this was a mathematical or physical assumption. But by Riesz theorem, there is no need, and indeed no room, for such an assumption.}, and therefore we can write the equation
$$V(u)= \bra{v}\ket{u}=(v,u)_{\mathcal H} \ ,$$
which says that we can think of Dirac's "bracket" $\bra{\cdot}\ket{\cdot}$ simply as the inner product $(v,u)_{\mathcal H}$.
Moreover, by Riesz theorem, the correspondence between bras and kets is antilinear, therfore
$$\overline{(c\ket{u})}=\overline {c}\bra{u}\,$$
for any complex number $c$, $\overline {c}$ being its complex conjugate. Finally, by the normalization condition $\sqrt{\braket{u|u}}=1$.

%The outer product between a ket $\ket{u}$ and a bra $\bra{\phi}$ is written as $\ket{u}\bra{\phi}$ and is an operator. If we apply $\ket{u}\bra{\phi}$ on a ket $\ket{\chi}$ then  $(\ket{u}\bra{\phi})\ket{\chi}=\ket{u}\braket{\phi|\chi}$ , this is a new ket and is proportional to $\ket{u}$ since the inner product $\braket{\phi|\chi}$ is a number (complex). It defines a projector operator $\widehat P_{j}=\ket{u_j}\bra{u_j}$ , if ${\lbrace\ket{u_{i}}\rbrace}_{i=1}^{\infty}$ is an orthonormal basis of Hilbert space $\mathcal{H}$, then any ket $\ket{u}$ can be expressed as $\ket{u}=\sum_{i}^{\infty}c_{i}\ket{u_i}$ and the action of $\widehat P_{j}$ on $\ket{u}$ has the property $\widehat P_{j}\ket{u}=\widehat P_{j}\sum_{i=1}^{\infty}c_{i}\ket{u_i}=c_{j}\ket{u_j}$. Also $\widehat P_j$ is linear, self-adjoint operator and finally $\widehat P^{2}_j=\widehat P_j$.

\noindent
{\bf Observables.} The most striking feature of quantum mechanics, as opposed to classical mechanics where the observables are functions of the phase point, is that the quantum observables $\bf{f}$ are represented by linear self-adjoint operators 
$$\widehat{f}: \mathcal{H} \rightarrow \mathcal{H} \ ,$$
which are, in general, unbounded.

This choice of the observables is intimately related with the fundamental  fact that the measurement of an observable, while the quantum system resides in a particular state, is not uniquely determined by the state, but it provides various values, that are modelled as a random variable obeying a certain distribution law \cite{NEUM, BS}.

More precisely it is postulated that any measurement of the observable $\bf{f}$ yields a number $\lambda$ belonging to the spectrum $\sigma(\widehat{f})$ of $\widehat{f}$.
For simplicity, we assume that the spectrum is discrete with eigenvalues  $\sigma(\widehat{f})={\lbrace \lambda_{i} \rbrace}_{i=1}^{\infty}$ and  let ${\lbrace\ket{u_{i}}\rbrace}_{i=1}^{\infty}$ be the corresponding orthonormal basis (ONB) of eigenfunctions of $\widehat{f}$ . In other words $\lambda_{i}$ and $\ket {u_{i}}$ satisfy the eigenvalue problem
\begin{eqnarray}\label{eigenvaluepr}
\widehat{f}\ket{u_{i}}=\lambda_{i}\ket{u_{i}} \ .
\end{eqnarray}
The expansion of the state  $\ket{u}$ in the ONB ${\lbrace\ket{u_{i}}\rbrace}_{i=1}^{\infty}$ of the eigenfunctions is
$\ket{u}=\sum_{i=1}^{\infty} c_{i}\ket{u_i}$.
Then the probability of obtaining the value $\lambda=\lambda_{i}$ in the  measurement of  the observable of $\widehat{f}$ when the system resides in the state $u$, is given by
\begin{eqnarray}\label{prob}
\mathbf{P}_{u}(\lambda=\lambda_{i})={\vert (u_{i},u)_{\mathcal{H}}\vert}^2={|\braket{u_{i}|u}|}^2=|c_i|^2\quad .
\end{eqnarray}
Since ${||\ket{u}||}_{\mathcal{H}}=1$ , it follows that these probabilities sum to the unit.

%Moreover, using the property 
%\begin{eqnarray*}\label{conj_rel}
%\overline{(c\ket{u})}=\overline {c}\bra{u}\, 
%\end{eqnarray*}
%for a complex number $c$ and the bar $^{-}$ indicates the complex conjugate,
%and $\overline{\braket{u_{i}|u}}=\braket{u|u_{i}}$ for the inner product on a complex linear space we write the probability in terms of a projection operator $\widehat P_i=\ket{u_i}\bra{u_i}$,
%\begin{eqnarray}
%\mathbf{P}_{u}(\lambda=\lambda_{i})={|\braket{u_{i}|u}|}^2&=&\braket{u_{i}|u}\overline{\braket{u_{i}|u}}\\
%&=&\braket{u_{i}|u}\braket{u|u_{i}}\\
%&=&\braket{u|u_{i}}\braket{u_{i}|u}\\
%&=&\braket{u|\widehat P_{i}|u}\, .
%\end{eqnarray}

The expectation value (mean value) of the random variable provided by the measurement of the observable $\widehat{f}$ when the system is  in the state $\ket{u}$ , is given by
\begin{eqnarray}\label{expectvalue}
\overline{{\widehat{f}}} &=&\sum_{i=1}^{\infty}\lambda_{i}\mathbf{P}_{u}(\lambda=\lambda_{i}) \nonumber \\
&=&\sum_{i=1}^{\infty}\lambda_{i}{|\braket{u_{i}|u}|}^2\nonumber\\
&=&\sum_{i=1}^{\infty}\lambda_{i} \braket{u_{i}|u}\braket{u|u_i} \nonumber\\
&=& \braket{\widehat{f}u | u }=  \braket{\widehat{f}}_{u } \ ,
\end{eqnarray}
since $\widehat{f}\ket{u_i}=\lambda_{i}\ket{u_i}$.
By this result, it is also referred as the mean value $\braket{\widehat{f}}_{u }$ of the observable $\widehat{f}$.
 
By standard results from functional analysis, the expectation satisfies the following conditions  
$$
\braket{\widehat{f}}_{u}(I)=1 \quad , \qquad
\braket{\widehat{f}{\widehat{f}}^{*}}_{u}\geq 0 \quad , \qquad
\braket{{\widehat{f}}^{*}}_{u}= \overline{\braket{\widehat{f}}_{u}}\quad .
$$
Here $I$ denotes the identity operator and ${\widehat{f}}^{*}$ is the adjoint operator of $\widehat{f}$ .

We must note that when the spectrum $\sigma(\widehat{f})$ is continuous, then $\widehat{f}$ admits generalised eigenfunctions $u_{\lambda}$ indexed by the point $\lambda$ in the continuous spectrum. Then, the sums in the above formulas are substituted  by integrals and the probabilities are substituted by probability measures $\mu_{u} (\lambda)=\braket{u_{\lambda}| u}d\lambda$ .  Also, when the spectrum is mixed, someone must further modify the formulas by combining sums and spectral integrals.

We must also note that because  we are interested for the mean values of the observables, and not for the values of the particular realisation of the random variable provided by the measurement process, and the expectation is the same for any two states $u$ and $u'$ differing by a unimodular factor, it becomes clear why it is enough to work with the projective space $\cal{P}(H)$ .

\subsection{Evolution of quantum states and observables}
The evolution of a quantum system, just in classical mechanics, is determined by the distinguished observable of energy $\bf{E}$ which is represented by the energy operator $\widehat{H}^{\hbar}$. It will be explained later that by the process of quantization,  this operator corresponds to the Hamiltonian $H({\bf x},{\bf p})$ (total energy) of the corresponding classical system, and for this reason we will call it the {\it Hamiltonian} operator. Due to the quantization process the operator $\widehat{H}^{\hbar}$ depends on Planck's constant $\hbar$. This is a physical constant \footnote{Actually $\hbar:=h/2\pi$ is the re-scaled physical Planck's constant which has dimensions of action and its numerical value is $h=6.626176 \times 10^{-27} erg\cdot sec$.}
which has its roots in the fundamental physics of quantum theory \cite{Foc}.
There are several different, in some sense, equivalent pictures of quantum mechanics. The most known is  the 
Schr\"odinger picture which looks for the evolution of the wavefunction, and the Heisenberg picture, which looks for the evolution of the observables.

%In the Schr\"odinger picture, the time dependence is in the wavefunction or the state vector, whereas in the Heisenberg picture, the time dependence is in the operator that represents an observable. Mathematically, in quantum mechanics an observable is a self-adjoint operator that acts on state vectors in the Hilbert space. 
%The eigenvectors of self-adjoint operators form an orthonormal basis of the state space for the system.

%In the Heisenberg picture, it is the
%operators which change in time while
%the basis of the space remains fixed.

In the  \emph{Schr\"odinger picture}  the  state vector (a wavefunction) changes in time according to the  Schr\"{o}dinger equation %$(\ref{SchrodingerH})$ , 
\begin{eqnarray}\label{schrep}
i\hbar{\partial_{ t}}\ket{u_{{t}}}=\widehat{H}^{\hbar}\ket{u_{ t}} \quad ,
\end{eqnarray}
where $u_{ t}$ is considered as a vector of $\mathcal{H}$ which varies with time.

If the Hamiltonian function is not explicitly dependent
on time, the solution of this equation with initial value $u_{ t}\mid_{t=0} = u_{0}$ is given by 
\begin{eqnarray}
\ket{u_{ t}}=U_{ t}\ket{u_{0}}
\end{eqnarray}
where $U_{t}=e^{-\frac{i}{\hbar}\widehat{H}^{\hbar}{ t}}=\sum_{n=0}^{\infty}\frac{1}{n!}{\left(-\frac{i}{\hbar}\widehat{H}^{\hbar}{ t}\right)}^{n}$ is the one-parameter group of unitary operators, which generated by the self-adjoint operator $\widehat{H}^{\hbar}$ . 

%Although the time evolution of the system is naturally described by  the evolution of  state. 
However, we can think in terms of the initial data $u_{0}$ and a time-dependent version of the observable, when we compute mean values of observables.
The mean value of $\widehat{f}$ at time $t$ is given by
\begin{eqnarray*}
\braket{ \widehat{f}}_{u_{ t}}&=&\braket{u_{ t}|\widehat{f}|u_{ t}}\\
&=&\braket{U_{ t}u_{0}|\widehat{f}|U_{ t}u_{0}}\\
&=&\braket{u_{0}|U_{t}^{*}\ci\widehat{f}\ci U_{ t}|u_{0}}  \ .
\end{eqnarray*}
Thus,  we are led to define the observable
$$\widehat{f}_{ t}^{H}:=U_{ t}^{*}\ci \widehat{f}\ci U_{ t}=e^{\frac{i}{\epsilon}\widehat{H}^{\hbar}{ t}}\ci \widehat{f}\ci e^{-\frac{i}{\hbar}\widehat{H}^{\hbar}{t}} \ , $$
such that $\widehat{f}_{ t}^{H}\mid_{t=0} = \widehat{f}$.
Then, the mean value is expressed by
\begin{eqnarray*}
\braket{ \widehat{f}}_{u_{ t}}=\braket{u_{0}|\widehat{f}_{ t}^{H}|u_{0}} \ ,
\end{eqnarray*}
as the expectation of  expectation of $\widehat{f}_{ t}^{H}$, when the system is in state $\ket{u_0}$.
This approach to the description of quantum systems is referred as the \emph{Heisenberg picture}. 

By direct computation, using $(\ref{schrep})$, we find that $\widehat{f}_{ t}^{H}$ satisfies the operator equation
\begin{eqnarray}\label{heisrep}
-i\hbar\frac{d\widehat{f}_{t}^{H}}{dt}=[\widehat{H}^{\hbar},\widehat{f}_{t}^{H}]
\end{eqnarray}
where $[\widehat{H}^{\hbar},\widehat{f}_{t}^H]=\widehat{H}^{\hbar}\ci\widehat{f}_{t}^{H}-\widehat{f}_{t}^{H}\ci\widehat{H}^{\hbar}$ ,  denotes the commutator of $\widehat{H}^{\hbar}$ and $\widehat{f}_{ t}^{H}$.

\subsection{Density operator}
So far we have considered that the  quantum system is described by a single state. More general systems have been considered by J.von Neumann \cite{NEUM} in his statistical theory of measurements on quantum systems.  The basic idea is the following. Let an ensemble of  quantum states (wavefunctions) $\ket{u_1},\ket{u_2},\ldots$ 
and let $p_{i} \ , \ i=1,2, \dots$  be the probability of finding the system in the  state $\ket{u_{i}}$. Then, we consider that the system resides in a \emph{mixed state}  (somehow in a mixture of states) which is described by the  self-adjoint \emph{density operator} 
\begin{eqnarray}\label{density}
\widehat{\rho}=\sum_{i=1}^{\infty}p_{i}\ket{u_i}\bra{u_i} \ .
\end{eqnarray}
It turns out that $\ket{u_i}$ are eigenfunctions and $p_{i}$ eigenvalues of $\widehat{\rho}$.
\begin{remark}
Mixed states arise in the description of quantum systems when they are considered as subsystems of larger systems with more degrees of freedom as it is necessary to do in scattering problems and open systems. In all these cases the mixing of the sates comes from the averaging with respect to the additional degrees of freedom of the larger system with respect to the smaller one \cite{NSS}, Ch.1, and \cite{BP}.
\end{remark}

Contrary to mixed states, when a state of a system is accurately known, we say that it resides in  \emph{pure state}. This means that  the probability $p_{i} =1$ for some $i$,  and all others are zero. For this case the density operator takes the simple form 
\begin{eqnarray}\label{densitypure1}
{\widehat{\rho}}_{u}=\ket{u}\bra{u}\quad .
\end{eqnarray}
The expression $(\ref{densitypure1})$ shows that the density operator ${\widehat{\rho}}_{u}$ for a pure state is a projection operator. We refer to it as the \emph{density matrix corresponding to the state $u$}. In this sense, we say that a \emph{mixed state is a collection of different pure states}.

\begin{definition}
Let $\widehat{f}$ be an operator on $\mathcal{H}$, and an  ONB ${\lbrace\ket{u_{i}}\rbrace}_{i=1}^{\infty}$ of $\mathcal{H}$. Then,  its \emph{trace} $\tr(\widehat f)$  is defined by the formula
 \begin{eqnarray*}\label{traceform}
\tr(\widehat f):=\sum_{i=1}^{\infty}\braket{u_i|\widehat f|u_i}=\sum_{i=1}^{\infty}\braket{u_{i}|\widehat fu_{i}}=\sum_{i=1}^{\infty}(u_{i},\widehat fu_{i}) \ ,
\end{eqnarray*}
assuming that the series converge. If  $\tr(\widehat f)$ is finite, we say that the operator  $\widehat{f}$ is of \emph{trace class}.
\end{definition}

Then, by the definition $(\ref{densitypure1})$ of pure state density operator, we have
${\widehat{\rho}}_{u}\ket{u_{i}}=\ket{u}\braket{u|u_{i}}$,
and therefore
\begin{eqnarray}\label{trace_exp}
\tr(\widehat{\rho}\widehat{f})&=& \sum_{i=1}^{\infty}\braket{u_{i}|{\widehat{\rho}}_{u} \widehat{f}u_{i}} \nonumber\\
&=& \sum_{i=1}^{\infty}\braket{u_{i}| \lambda_{i} {\widehat{\rho}}_{u}u_i} \nonumber \\
&=&
\sum_{i=1}^{\infty}\bra{u_{i}}\lambda_{i}\ket{u}\braket{u|u_{i}}\nonumber\\
&=&\sum_{i=1}^{\infty}\lambda_{i}\braket{u_{i}|u}\braket{u|u_{i}}\nonumber \\
&=&\sum_{i=1}^{\infty}\lambda_{i}|\braket{u_{i}|u}|^2 \ .
\end{eqnarray}

Thus, by $(\ref{expectvalue})$, we can write the expectation  in the form 
\begin{eqnarray}\label{trace_exp}
\overline{{\widehat{f}}}=\braket{\widehat{f}}_{u}=\tr({\widehat{\rho}}_{u}\widehat{f}) \ .
\end{eqnarray}

The formula $(\ref{trace_exp})$ provides the reason why we want to introduce the density operator in the description of quantum systems. Density operator is the tool for computing the expectation values, which are the only source of information about the system.

The density operator as introduced above has the following properties \footnote{On the basis of these properties, we can generally define as density operator $\widehat\rho$, any selfadjoint, positive semi-definite operator with  $\tr(\widehat\rho)=1$.}:
\begin{enumerate}
\item
The density operator is self-adjoint , $\widehat\rho=\widehat{\rho}^{*}$
\item
The trace of the density operator is equal to 1, $\tr(\widehat\rho)=1$
\item
The eigenvalues of a density operator satisfy $0\leq \lambda_i\leq 1$
\item
For a pure state $\ket{u}$, ${\widehat{\rho}}_{u}^{2}={\widehat{\rho}}_{u}$ and $\tr({\widehat{\rho}}_{u})=1$
\item
For a mixed state $\tr(\widehat{\rho}^2)<1 \ .$
\end{enumerate}

The simplest (mathematical) example of mixed state is constructed as follows.
\begin{example} Let ${u_j}$ a sequence of states in $\mathcal{H}$ and ${\alpha_j}$ a sequence of nonnegative numbers suche that $\Sigma_{j=1}^{\infty}{\alpha_j}=1$. The operator $\widehat{\rho}= \Sigma_{j=1}^{\infty}{\alpha_j}\widehat{\rho}_{u_j}$ is the density matrix of a mixed state. This simple mixed state is the mixture of pure states and, roughly speaking, it has the physical interpretation that the system resides in state ${u_j}$ with probability ${\alpha_j}$. Then, it is consistent with the interpretation of  $tr({\widehat{\rho}}\widehat{f})$ as the expectation of $\widehat{f}$ in the state represented by $\widehat{\rho}$.
\end{example}

We close our brief discussion on density operator, by clarifying its connection with coherence properties in quantum systems.
\begin{remark}
Let us consider a pure state $\ket{u}$ expanded in the form 
\begin{eqnarray}\label{densitypure}
\ket{u}=\sum_{i=1}^{\infty}c_{i}\ket{u_i}
\end{eqnarray}
with respect to an ONB  $\ket{u_i}$ for all $i=1,\ldots,\infty$ of the Hilbert space $\mathcal{H}$. By 
$(\ref{densitypure})$ we write the pure state density operator as follows
$\widehat{\rho}=\ket{u}\bra{u}$ as follows 
\begin{eqnarray}\label{proj_op}
 \widehat{\rho}=\sum_{i=1}^{\infty}|c_{i}|^2\ket{u_i}\bra{u_i}+\sum_{i\neq j=1}^{\infty}c_{i}\overline{c_{j}}\ket{u_i}\bra{u_j}
\end{eqnarray}
The first term in $(\ref{proj_op})$ relates to the probability of the system being in the state $\ket{u_i}$, since $\braket{u_i|\widehat{\rho}|u_i}=|c_i|^2$. For the second term, we write the complex number $c_i$ in polar form, $c_{i}=|c_i|e^{i\varphi_i}$ thus $\braket{u_{i}|\widehat\rho|u_{j}}=c_{i}\overline{c_{j}}=|c_{i}||c_{j}|e^{i(\varphi_{i}-\varphi_{j})}$ . The phase difference in the exponential expresses the coherence of terms in the state to interfere with each other. 
%{\color{red} This is a characteristic of a pure state.something is missing here...}
\end{remark}

\subsection{The time-evolution of density operator}\label{density_evol}

Let the density operator  
\begin{eqnarray}\label{densityexpansion}
\widehat{\rho}( t)=\sum_{i=1}^{\infty}p_{i}\ket{u_{i}}\bra{u_{i}}
\end{eqnarray}
Since each state $\ket{u_{i}}$ satisfies the Schr\"odinger equation, 
\begin{eqnarray*}
i\hbar\frac{\partial}{\partial t}\ket{u_{i}}=\widehat{H}^{\hbar} \ket{u_{i}}
\end{eqnarray*}
and also
\begin{eqnarray*}
-i\hbar\frac{\partial}{\partial t}\bra{u_{i}}=\bra{u_{i}}\widehat{H}^{\hbar}  \ .
\end{eqnarray*}
By differentiating $(\ref{densityexpansion})$, we have
\begin{eqnarray*}
\frac{d}{d t}\widehat{\rho}(t)&=&\sum_{i=1}^{\infty}p_{i}\frac{\partial}{\partial t}\left(\ket{u_{i}}\bra{u_{i}}\right)\\
&=&\sum_{i=1}^{\infty}p_{i}\left(\frac{\partial}{\partial t}\ket{u_{i}}\right)\bra{u_{i}}+\sum_{i=1}^{\infty}p_{i}\ket{u_{i}}\left(\frac{\partial}{\partial  t}\bra{u_{i}}\right)\\
&=&\frac{1}{i\hbar}\sum_{i=1}^{\infty}p_{i}\left(\widehat{H}^{\hbar}\ket{u_{i}}\right)\bra{u_{i}}+\frac{1}{-i\hb}\sum_{i=1}^{\infty}p_{i}\ket{u_{i}}\left(\bra{u_{i}}\widehat{H}^{\hbar} \right)\\
&=&\frac{1}{i\hbar}\left( \widehat{H}^{\hbar} \widehat{\rho}({ t})-\widehat{\rho}({t}) \widehat{H}^{\hbar}\right) \ ,
\end{eqnarray*}
Therefore, we conclude that $\widehat{\rho}$ satisfies the {\it Liouville-von Neumann equation}
\begin{eqnarray}\label{vonneumann}
{i\hbar} \frac{d}{d t}\widehat{\rho}({t})=[\widehat{H}^{\hbar},\widehat{\rho}(t)]\, .
\end{eqnarray}

\subsection{Schr\"odinger equation in coordinate representation}

\subsubsection{Schr\"odinger's derivation}

In the simplest case of a quantum mechanical particle (e.g. atom),  we deal with the quantum analogue of a classical particle. By classical particle we mean a point mass moving in Euclidean configuration (position) space $R^{d}_{\bf x}$ according to Newton's second law of classical mechanics under the action of  a potential $V({\bf x})$. This point moves  according to the Hamiltonian system $(\ref{hamiltoniansystem_1})$, with Hamiltonian function
\begin{equation}\label{particle_hamiltonian}
H({\bf x} \ , {\bf p})=\frac{{\bf p}^2}{2{\bf m}}+V({\bf x}) \ ,
\end{equation}
which is the total energy $E$, i.e. the sum of kinetic and potential energy, of the particle.

In dealing with this simple quantum system, it is convenient to work with the \emph{standard coordinate representation}, in which the state space ${\mathcal H}$ is just the space $L^2(R^d_{{\bf x}})$. This is the space of functions which are  continuous, infinitely differentiable,  and square integrable, i.e. $\int_{R^{d}_{\bf x}}|\phi({\bf x})|^{2}\, d\bf x<\infty$, with inner product $(\psi({\bf x}),\phi({\bf x}))_{L^2(R^d_{{\bf x}})}=\int_{R^{d}_{{\bf x}}}\psi({\bf x})\overline{\phi({\bf x})}\, d\bf x$ . In this representation, a ket $\ket{u}$ is  a function $u({\bf x}) \in L^2(R^d_{{\bf x}})$. 
The correspondence between a ket and its associated wavefunction is $u(\bf x) \equiv \bra{ \bf x}$ $ \ket{u}= \braket{{\bf x} |u}$ , where $\bra{\bf x}$ is the {\it bra} corresponding to ${\bf x} \in R^{d}_{\bf x}$ .

Then, it turns out that the fundamental observables are the position and momentum operators
$\widehat{q}=(\widehat{q}_1 \ ,\dots \, \widehat{q}_d)$ and $\widehat{p}=(\widehat{p}_1 \ ,\dots \, \widehat{p}_d)$, respectively. The operators  $\widehat{q}_j u({\bf x})=x_j u({\bf x})$ act as multiplication by position, while the operators  
$\widehat{p}_k u({\bf x})= -i\hbar\frac{\partial}{\partial_{x_k}} u({\bf x})$ act by differentiation,  $\hbar$ being the Planck's constant. 
These observables satisfy the relations 
\begin{equation}\label{coord_commut}
[\widehat{p}_{j} \ , \widehat{p}_{k}]= [\widehat{q}_{j} \ , \widehat{q}_{k}] =0 \ , \ \  [\widehat{p}_{j} \ , \widehat{q}_{k}]= -i\hbar \delta_{jk} \ .
\end{equation}
Then, the energy operator 
\begin{equation}\label{hamiltonian_op}
\widehat{H}^{\hbar}=-\frac{\hbar^{2}}{2{\bf m}}\nabla_{{\bf x}}^{2} +V({\bf x}) \ ,
\end{equation}
is obtained from the classical particle Hamiltonian $(\ref{particle_hamiltonian})$ by the formal substitution
\begin{equation}\label{Schrodinger_quant}
\widehat{H}^{\hbar}=H(\widehat{q} \ , \widehat{p})= H({\bf x}, -i\hbar\nabla_{{\bf x}})=\frac{1}{2{\bf m}}(-i\hbar\nabla_{{\bf x}})^{2} +V({\bf x}) \ .
\end{equation}

The substitution $(\ref{Schrodinger_quant})$, which associates to a scalar function defined on classical phase space $R^{2d}_{{\bf x}{\bf p}}$, a partial differential operator acting on the Hilbert space ${\mathcal H}=L^2(R^d_{{\bf x}})$, is referred in the literature as the  \emph{Schr\"odinger quantization}. It has been introduced in 1926 by E.Schr\"odinger \cite{Schr}, who also assumed that the classical particle energy $E$ must be associated with the time derivative $i\hbar\partial_{t}$, in order to derive form the energy conservation $E=H({\bf x} \ , {\bf p})$ 
the \emph{Schr\"odinger equation} governs the evolution of the quantum-particle wavefunction (see \cite{Z}, Ch. 5.11.4 , for a concise description of the derivation).  This equation reads as
\begin{eqnarray}\label{schr_p}
i\hbar{\partial_t}u^{\hbar}({\bf x},{ t})=\left[-\frac{\hbar ^2}{2\bf m}\Delta_{\bf x} +V({\bf x})\right]u^{\hbar}({\bf x},{ t})\quad , \quad {\bf x}\in R^{d}\quad , \quad t\in [0,\infty)
\end{eqnarray}
where $\Delta_{\bf x}=\sum_{i=1}^{n}\frac{\partial^2}{\partial {{ x_i}}^2}$ is the Laplacian operator. 
Note that $(\ref{schr_p})$ is the equation $(\ref{schrep})$ written in the coordinate representation.

\subsubsection{Dimensionless form of the Schr\"odinger equation}

We now transform the Schr\"odinger equation $(\ref{schr_p})$ into a dimensionless equation. In order to do this we define dimensionless coordinates 
$ {\bf x}\leftarrow {\bf x}/L$, where $L$ is an arbitrary length and $t\leftarrow{  t}/T$ with $T$ is a unit of time, that we will choose to make $(\ref{schr_p})$ simpler. This change of variables gives $\partial_{{ t}}u\leftarrow\partial_{ t}u/T$ , $\partial_{{\bf x}}u\leftarrow\partial_ {\bf x}u/L$ and $\Delta_{{\bf x}}\leftarrow\Delta_{ \bf x}/L^2$ then, the Schr\"odinger equation for the dimensionless potential $V({\bf x}) T^{2}/{\bf m}L^2$ takes the form 
\begin{eqnarray}\label{schr_param}
i{\partial_t}u^{\hbar}({\bf x},t)=\left[-\frac{\epsilon T}{2{\bf m} L^2}\Delta_{\bf x} +\frac{{\bf m}L^2}{\hbar T}  V({\bf x})\right]u^{\hbar}( {\bf x},t)\quad .
\end{eqnarray}

The dimensionless parameter $2\pi\cdot \hbar T/{\bf m}L^2$ into $(\ref{schr_param})$  can be written as 
$$2\pi\cdot \frac{\hbar}{\frac{{\bf m}L}{T}L}=\frac{h}{\frac{{\bf m }L}{T}L}$$ 
where $L/T$ is the speed of a classical particle of mass $\bf m$ and ${\bf m}L/{T}$ expresses its momentum. In quantum mechanics we assume that $2\pi\hbar\ll{\bf m}L^{2}/T$, this means that the Broglie wavelength of the particle $2\pi\hbar T/{\bf m}L$ is small compared to the length scale $L$. Thus we can say that the parameter $\epsilon:=\hbar T/{\bf m} L^2$ in  $(\ref{schr_param})$ controls the quantum effects. This means that as $\epsilon$ becomes small, the quantum effects become negligible, and classical mechanics dominates the particle motion. 

The regime of physical quantities where the dimensionless parameter $\epsilon$ is nonzero, but it can be arbitrarily small: $0<\epsilon\ll 1$, is referred as the \emph{semiclassical regime} and $\epsilon$ is referred as  the \emph{semiclassical parameter}. The limit $\epsilon\rightarrow 0$ is designated as the  \emph{classical limit} and it is associated with the transition from quantum mechanics to classical mechanics.

In the sequel we use the Schr\"odinger equation in the dimensionless form
\begin{eqnarray}\label{Schrodinger}
i\epsilon\partial_{t}u^{\epsilon}({\bf x},t)=\left[-\frac{\epsilon^2}{2}\Delta_{{\bf x}}+V({\bf x})\right]u^{\epsilon}({\bf x},t)\quad ,\quad ( {\bf x},t)\in R^{d}\times [0,\infty) \ ,
\end{eqnarray}
where $u^{\epsilon}({\bf x},t)=u^{\hbar}({\bf x}/L,{ t}/T) \ , \hbar= \epsilon {\bf m} L^{2}/T$.

Introducing the {\it Schr\"{o}dinger operator}
\begin{equation}\label{schrod_op}
\widehat{H}^\epsilon=-\frac{\epsilon^2}{2}\Delta_{{\bf x}}+V({\bf x}) \ ,
\end{equation}
we sometimes write $(\ref{Schrodinger})$ in the symbolic form 
\begin{eqnarray}\label{SchrodingerH}
i\epsilon{\partial_t}u^\epsilon({\bf x}, t)=\widehat{H}^\epsilon u^\epsilon({\bf x},t)\quad , \quad {\bf x}\in R^{d}\quad , \quad  t\in[0,\infty)\  .
\end{eqnarray}

\begin{remark}
Apart from quantum mechanics, the Schr\"odinger equation $(\ref{Schrodinger})$ arises in many contexts in classical
wave propagation problems, as the paraxial approximation of forward propagating waves
\cite{Flat, LF}. Thus, it is of practical importance for computing wave intensities in many
applied fields such as radioengineering \cite{Foc1, Foc2},  laser optics \cite{Tap1}, underwater acoustics
\cite{Tap2}, the investigation of light and sound propagation in turbulent atmosphere
\cite{Tat1}, and seismic wave propagation in the earth’s crust \cite{SF}, to mention but a
few. In these cases, the potential $V$ is explicitly related to the refraction index of the
propagating medium.
\end{remark}
%%%%%%%%%%%%%%%%%%%%%%%%%%%%%%%%%%%%%%%%%%%%%%%%
%%%%%%%%%%%%%%%%%%%%%%%%%%%%%%%%%%%%%%%%%%%%%%%%
%%%%%%%%%%%%%%%%%%%%%%%%%%%%%%%%%%%%%%%%%%%%%%%%

\chapter{QM in configuration space: Schr\"odinger equation}\label{chapter3}

As we have seen  in Section $\ref{section22}$ the evolution of the wave function in quantum mechanics is governed by  the Schr\"{o}dinger equation $(\ref{Schrodinger})$. From now on we will consider only the one-dimensional case $d=1$.

A problem of primary interest, both for quantum mechanics and for the classical wave problems that we mentioned at the end of previous chapter, is the evolution of highly oscillatory  wavefunctions.
Therefore, we consider the Cauchy problem with WKB initial data
\begin{eqnarray}
i\epsilon{\partial_t}u^{\epsilon}(x,t)&=&\left[-\frac{\epsilon ^2}{2}\partial_{xx}+V(x)\right]u^{\epsilon}(x,t)\quad , \quad x\in R_{x} \quad ,\, \, \, t\in [0,T) \ , \label{schr_eq_1} \\
u^{\epsilon}(x,t=0)&=&u_{0}^{\epsilon}(x)=A_{0}(x)\, e^{\frac{i}{\epsilon}S_{0}(x)} \ . \label{initialdata}
\end{eqnarray}
We assume the potential $V(x) \in C^{\infty}(R_x)$ is real valued, and $T$ is some positive constant.
Moreover, we assume that $A_{0}(x)\in C_{0}^{\infty}(R_x)$ and $S_{0}(x) \in C^{\infty}(R_x)$.

We are interested for  solutions $u^{\epsilon}(x,t)$ in $L^{2}(R_x)$, for any fixed $t \in [0,T)$. This choice is motivated by the conservation of quantum energy (probability), that is, the conservation of $L^2$-norm of the wavefunction  $u^{\hb}(x,t)$. Indeed, by direct calculation, using the Schr\"odinger equation, it folows that $\frac{d}{dt}\Vert u^{\hb}(x,t)\Vert_{L^{2}(R_{x})}=0$, which implies that $\Vert u^{\hb}(x,t)\Vert_{L^{2}(R_{x})}=
\Vert u_{0}^{\epsilon}(x)\Vert_{L^{2}(R_{x})} \ , \mathrm{for \ any} \  t>0$.

In the sequel we present two different constructions of the wavefunction. The first one is the  WKB method (geometrical optics) which deals with the construction of asymptotic solution when $\epsilon$ is small. The second one leads to an eigenfunction series expansion of the solution, and, in principle, is valid for any value of $\epsilon$. However, such series are, in general, very slowly convergent when $\epsilon$ is small, and therefore are not efficient for solving the Schr\"odinger equation in the semiclassical regime. We will use it as an intermediate tool for the investigation of the Wigner equation in phase space (see Section $\ref{sec52}$ ).
 
\section{The time-dependent WKB method}\label{section31}

When the semiclassical parameter $\epsilon$ is  small, highly oscillatory solutions of the problem 
 $(\ref{schr_eq_1})$-$(\ref{initialdata})$ have been traditionally studied by WKB method  (Wentzel-Kramers-Brillouin), known also as geometrical optics (see, e.g., \cite{BB, BLP, KO}).

The WKB method seeks for an approximate solution of the form ,
\begin{eqnarray}\label{wkb}
u^{\epsilon}(x,t)\approx \psi^{\epsilon}(x,t)=A^{\epsilon}(x,t)\, e^{\frac{i}{\epsilon} S(x,t)}\, ,
\end{eqnarray}
where $S(x,t)$ , $A^\epsilon(x,t)$ are real-valued functions \footnote{extensions of the method for complex-valued phases $S$  have been also developed, but we do not consider them in this work}. 

For the moment we assume that the amplitude is sufficiently smooth and we expand it as a power series
$$A^\epsilon=A^{(0)}+\epsilon A^{(1)}+\epsilon^2 A^{(2)}+\ldots \  $$
for small $\epsilon$.

\subsection{Hamilton-Jacobi and transport equation}

Substituting $\psi^{\epsilon}$ into $(\ref{schr_eq_1})$, and retaining the terms of order $O(1)$ and $O(\epsilon)$ ,  we obtain  that the phase function   $S(x,t)$ satisfies the Hamilton-Jacobi equation (also called the  {\it eikonal} equation)
\begin{equation}\label{eikonal}
\partial_{t}S(x,t) + \frac{1}{2}\left({\partial_x}S(x,t)\right)^{2} +V(x) =0 \quad , \quad S(x, t=0)= S_{0}(x) \ ,
\end{equation}
and  that the (zeroth order) amplitude  $A:=A^{(0)}$ satisfies the {\it transport} equation %\begin{equation}\label{transport}
%2\partial_{t}A +2\partial_{x}A\partial_{x}S +A\partial^{2}_{x}S =0
%\ , \ \ \ \ A(x,t=0)=A_{0}(x) \ .
%\end{equation}
\begin{equation}\label{transport}
{\partial_t}A(x,t) +{\partial_x}A(x,t)\,\partial_{x} S(x,t)+\frac{1}{2}A(x,t)\,\partial_{xx}S(x,t)  =0
\quad , \quad A(x,t=0)=A_{0}(x) \ .
\end{equation}

\subsection{Bicharacteristics, rays and caustics}

The Hamilton-Jacobi equation $(\ref{eikonal})$ is a  first-order nonlinear partial differential equation, and   a standard way for solving it is based on the method of bicharacteristics (see, e.g.,  \cite{Ho}, Vol. I, Chap. VIII, and \cite{Jo}, Chap. 2).
 
Let $H(x,p)$ be the Hamiltonian function \footnote{Note that this is the Hamiltonian of a classical particle of unit mass, exhibiting one-dimensional motion under the action of the potential $V(x)$.} $H(x,p)=\frac{1}{2}p^{2}+V(x)$. 
Let the trajectories
$$\{x=x(x_0,t) \ , \ p=p(x_0,t)\}$$ 
be obtained by solving the  Hamiltonian system
\begin{eqnarray}
 \frac{dx}{dt}&=&\partial_{p}
H(x,p)=p \quad , \label{hamsys1} \\
\frac{dp}{dt}&=&-\partial_{x}
H(x,p)=-V'(x)\label{hamsys2}
\end{eqnarray}
with initial conditions
\begin{equation}\label{initbich}
 x(x_0,t=0)=x_0\quad ,\quad
p(x_0,t=0)=p_{0}(x_0)=S_{0}'(x_0) \ .
\end{equation}

The trajectories $\left\{x=x(x_0,t)\ , \
p=p(x_0,t)\right\}$ which solve the initial value problem $(\ref{hamsys1})$-$(\ref{hamsys2})$,
$(\ref{initbich})$,  in the phase space $R_{xp}^{2}$
are called {\it bicharacteristics}, and their projection
$\{x=x(x_0,t)\}$ onto $R_{x}$ are called {\it rays}.

The phase $S(x(x_0,t),t)$ is obtained by integrating the eikonal equation $(\ref{eikonal})$ along the rays, that is to integrate the ordinary differential equation
\begin{eqnarray}\label{int_phase}
\frac{dS}{dt}=\partial_{t}S+\partial_{x}S\, \frac{dx}{dt}=-H(x(x_0,t),p(x_0,t))+ p^2=\frac{1}{2}\, (p(x_0,t))^2-V(x(x_0,t))
\end{eqnarray}
with initial condition $S(x_0,t=0)=S_{0}(x_0)$. Note that along the bicharacteristics $p=p(x_0,t)=\partial_{x}S(x(x_0,t),t)$.

%%%%%%%%%%%%%%%%%%%%%%%%%%%%%%%%%%%%%%%%%%%%%
%%%%%%%%%%%%%%%%%%%%%%%%%%%%%%%%%%%%%%%%%%%%%%

On the other hand, the solution of the  transport equation $(\ref{transport})$ for
the amplitude $A$ along the rays, is obtained by
applying divergence theorem in a ray tube,

\begin{equation} \label{amplitude}
A(x(x_0,t),t)=\frac{A_0(x_0)}{\sqrt{J(x_0,t)}}
\end{equation}
where
\begin{equation}\label{jacobian}
 J(x_0,t )= \frac{\partial x(x_0,t)}{\partial x_0 }
\end{equation}
is the Jacobian of the ray transformation $x_0\mapsto
x(x_0,t)$ (see, e.g., \cite{BB, Zau}).

Since $(\ref{eikonal})$ is a non-linear equation, it has, in general, a smooth solution only up to
some finite time. The points $x=x(x_0,t)$ at which $J(x_0,t)=0$ , are called {\it focal points}, and the line formatted from these points is called {\it caustic}.

Whenever $J=0$, it can happen that $x_0=x_0(x,t)$ to be non-smooth or multi-valued
functions of $x$, the rays may intersect or have an envelope or touch each other, and
in general have singularities.  Then, the phase function
$S=S(x(x_0,t))$ may be a multi-valued or even a
non-smooth function. 

It must be emphasized that, however, the bicharacteristics
never develop singularities in the phase space, and the singularities of the rays develop due to the projection from the phase space to the physical space.

Since at a focal point the Jacobian $J(x_0, t)$ is zero, by $(\ref{amplitude})$ the amplitude becomes infinite. 
Therefore, on the caustics  the WKB method predicts infinite wave amplitudes. 

It has been shown by other methods that this phenomenon is a mathematical deficiency of the WKB ansatz $(\ref{wkb})$ which cannot capture  local scales of the wavefunction which are of order different than $O(\epsilon)$.
For example,
assuming that the multivalued phase function can be constructed, uniform asymptotic formulas for the wave field near the caustics have been developed using boundary layer techniques in physical space \cite{BaKi, BuKe}, as well as phase-space techniques, like Lagrangian integrals \cite{Lu, Kra, Dui1, GS, KO}, and the method of canonical operator \cite{MF, Va1}. Given the practical importance of the problem, a number of numerical techniques have been also proposed in order to compute the multivalued phase functions, see,
e.g., \cite{Ben, FEO, ER, Ru2} and the references therein.

\begin{example}
We give a simple example of caustic. Consider the Schr\"odinger equation $(\ref{schr_eq_1})$, $(\ref{initialdata})$, with potential $V(x)\equiv 0$ ,  
\begin{eqnarray}
i\epsilon{\partial_t}u^{\epsilon}(x,t)=-\frac{\epsilon ^2}{2}\partial_{xx} u^{\epsilon}(x,t) \ , \ (x,t)\in R\times [0,\infty)
\end{eqnarray}
and  initial data   
\begin{eqnarray}
u_{0}^{\epsilon}(x)=e^{-x^2}\, e^{-ix^{2}/2} \ .
\end{eqnarray}
Then, $A_{0}(x)=e^{-x^2} \ , \ \ S_{0}(x)=-x^{2}/2$.

Solving  the Hamiltonian system  $(\ref{hamsys1})$-$(\ref{hamsys2})$ with initial conditions $(x_0 \ , p_0=S'_{0}(x_0)=-x_0)$, we obtain the bicharacteristics 
$$\{x(x_0,t)=x_0(1-t) \ , \ \ p(x_0,t)=-x_0\} \ .$$
Therefore, the rays are represented by the lines $x=x_0(1-t)$ in configuration space $R^{2}_{xt}$ .

By integrating the equation $(\ref{int_phase})$, that is 
$$ \frac{dS}{dt}=-H(x(x_0,t),p(x_0,t))+ p^2=\frac{1}{2}\, (p(x_0,t))^2= \frac{1}{2}x_0^2 \ ,$$
we derive the phase
$$S(x,t)={x}^2/2(t-1) \ .$$

The Jacobian $(\ref{jacobian})$ of the ray transformation is $J(x_0,t)=1-t$, and it vanishes at the critical time $t_c=1$ .
Thus, we get a focal point (caustic)  $(x,t)=(0,1)$ in $R_{xt}^2$ (see Fig. 3.1). 
Then, by $(\ref{amplitude})$ the amplitude 
\begin{eqnarray*}
A(x,t)=\frac{A_0(x)}{\sqrt{1-t}}\quad  ,
\end{eqnarray*}
is infinite at the focal point. We observe that  the amplitude is a real-valued function when $t<t_c= 1$ and  it becomes complex-valued for $t>t_c=1$ .
\newpage
\begin{figure}[h] \label{caustic_ex}
\centering
\includegraphics[width=0.7\textwidth]{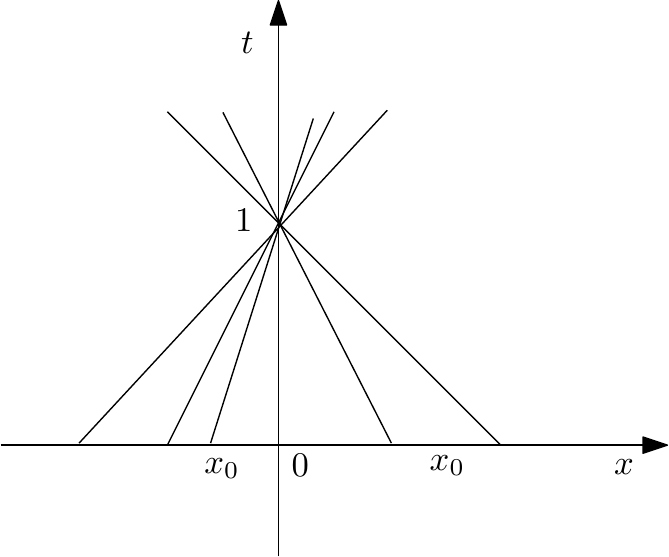}
\caption{{\it Focal point for the free case $V \equiv 0.$ }}
\end{figure}
\end{example}

%{\color{red} since focal points are degenerate caustics, it would be better for explanation to include the example with the fold; that in the paper with Filippas, I had showed you---you must decide if you will do} 

%%%%%%%%%%%%%%%%%%%%%%%%%%%%%%%%%%%%%%%%%%%%%%%%%%%%
%%%%%%%%%%%%%%%%%%%%%%%%%%%%%%%%%%%%%%%%%%%%%%%%%%%%%

\section{Eigenfunction series expansion of the wave function}\label{32}

Since the potential $V$  is assumed time-independent, the Schr\"odinger equation $(\ref{schr_eq_1})$ can be solved by separation of variables. Thus, we seek a solution in the form
\begin{eqnarray}\label{sepvar}
u^{\epsilon}(x,t)=v^{\epsilon}(x)\,\tau^{\epsilon}(t) \quad  \ .
\end{eqnarray}
Substituting $(\ref{sepvar})$ into $(\ref{schr_eq_1})$ we obtain
\begin{eqnarray}\label{sep}
\frac{1}{v^{\epsilon}(x)}
\left[-\frac{\epsilon^{2}}{2}{v^{\epsilon}}''(x)+V(x)v^{\epsilon}(x)\right]=i\epsilon\frac{1}{\tau^{\epsilon}(t)}{\tau^{\epsilon}}'(t)=E^{\epsilon}  \ ,
\end{eqnarray}
where $E^{\epsilon}$ is the separation constant (which must depend on $\epsilon$). 

For $\tau^{\epsilon}(t)$ we have,
\begin{eqnarray}\label{timedep}
i\epsilon {\tau^{\epsilon}}'(t)=E^{\epsilon} \tau^{\epsilon}(t)
\end{eqnarray}
and the solution is $\tau^{\epsilon}(t)=e^{-\frac{i}{\epsilon}E^{\epsilon}t}$, up to a constant.

Then, the left hand side of $(\ref{sep})$ gives the equation
\begin{eqnarray}\label{statschr}
\widehat{H}^{\epsilon}v^{\epsilon}(x)=\left[-\frac{\epsilon^{2}}{2}\frac{d^{2}}{dx^{2}}+V(x)\right]v^{\epsilon}(x)=E^{\epsilon}v^{\epsilon}(x) \quad ,
\end{eqnarray}
which is known as the {\it time-independent} or {\it stationary Schr\"{o}dinger equation}.

In addition to the assumption that the potential $V(x) \in C^{\infty}(R_x)$ is real valued, we also assume that it is positive, and  $\lim_{| x| \rightarrow \infty}V(x)=\infty$. Then, for each fixed $\epsilon \in(0,\epsilon_0)$ with arbitrary $\epsilon_0>0$,  the operator $\widehat{H}^{\epsilon}$ has purely discrete spectrum  and the eigenvalues can be arranged in an increasing sequence (see e.g. \cite{BS, HS})
\begin{eqnarray}\label{energy_seq}
0<E_{0}^{\epsilon}< E_{1}^{\epsilon}\leq\ldots\leq E_{n}^{\epsilon}\leq\ldots, \quad \lim _{n\rightarrow \infty}E_{n}^{\epsilon}=+\infty \quad  \ .
\end{eqnarray}
The corresponding eigenfunctions are $v_{n}^{\epsilon}(x) \in L^2(R_x)$.

Therefore, the solution $u^{\epsilon}(x,t)$ of the Cauchy problem $(\ref{schr_eq_1})$-$(\ref{initialdata})$ is expanded in the eigenfunction series 
\begin{eqnarray}\label{serexpsch}
u^{\epsilon}(x,t)=\sum_{n=0}^{\infty}c^{\epsilon}_{n}(0)v_{n}^{\epsilon}(x)e^{-\frac{i}{\epsilon}E_{n}^{\epsilon} t} \ .
\end{eqnarray}
The coefficients  
$c^{\epsilon}_{n}(0)$ are the $L^{2}-$projections 
$$c^{\epsilon}_{n}(0)=(u^{\epsilon}_{0},v_{n}^{\epsilon})_{L^{2}(R)}\quad \mathrm{for \,\,\,  all}\,\, \, n=0,1,2,\ldots \ \ ,$$
of the initial data $(\ref{initialdata})$ onto the eigenfunctions.

The eigenfunction expansion of the wavefunction is an exact 
solution of the Cauchy problem $(\ref{schr_eq_1})$-$(\ref{initialdata})$, provided that we can compute the eigenfunctions $v_{n}^{\epsilon}(x)$. This is not, in general, an easy task, but fortunately it is possible to construct asymptotic expansions of the eigenvalues and eigenfunctions for large orders $n$ and small values of the parameter $\epsilon$. These asymptotic approximations are based on the WKB method and they  will be described in next section.

In principle, the eigenfunction expansion, in contrary to the WKB solution which we presented in previous Section $\ref{section31}$, is a solution that by construction does not face caustic problems, although, for small  $\epsilon$, its pointwise convergence is dramatically slow  at the points $(x,t)$ lying on caustics of the WKB solution.

%%%%%%%%%%%%%%%%%%%%%%%%%%%%%%%%%%%%%%%%%%%%%%%%%%%%%%%%%%%%%%%%%%%%%%%%%%%%%%%%%%%%%%%%%%%%%%%%%%%%%%%%%%%%%%%%%%
%%%%%%%%%%%%%%%%%%%%%%%%%%%%%%%%%%%%%%%%%%%%%%%%%%%%%%%%%%%%%%%%
\section{WKB asymptotic expansion of eigenfunctions}

We look for the eigenvalues $E^{\epsilon}$ and the eigenfunctions 
$v^{\epsilon}(x) \in L^2(R_x)$
of  the operator $\widehat{H}^{\hb}=-\frac{\epsilon ^2}{2}\frac{d^2}{dx^2}+V(x)$ for small  $\epsilon$. The eigenvalues and eigenfunctions are constructed using WKB asymptotic solutions of the problem
\begin{eqnarray}\label{eigenval_eq}
\left[-\frac{\epsilon ^2}{2}\frac{d^2}{dx^2}+V(x)\right]v^{\epsilon}(x)=E^{\epsilon}v^{\epsilon}(x)\quad , \quad x\in R \ .
\end{eqnarray}

We define $q^{\epsilon}(x):=2(V(x)-E^\epsilon)$ and we rewrite $(\ref{eigenval_eq})$ in the form
\begin{eqnarray}\label{eigenval_eq_2}
\frac{d^2}{dx^2}v^{\epsilon}(x)=\frac{1}{\epsilon^2}\, q^{\epsilon}(x)v^{\epsilon}(x)\quad  \ .
\end{eqnarray}

Then, by setting 
$$\lambda=1/\epsilon \ , \ \ \ E^{\epsilon}=E \ , \  \ \ q(x):=2(V(x)-E) \ ,$$
the problem $(\ref{eigenval_eq_2})$ becomes
\begin{eqnarray}\label{eq_fedoriouk}
\frac{d^2}{dx^2}v^{\lambda}(x)={\lambda^2} q(x)v^{\lambda}(x) \ .
\end{eqnarray}
This is the standard form of eigenvalue problems for second-order differential equations.
In this problem we think of $E$ as fixed and  the large parameter $\lambda=\lambda(E)$ plays the role of the eigenvalue. This consideration allows for the construction of the so called Bohr-Sommerfeld quantization rule for the eigenvalues.
 
Therefore, we must consider the spectrum of the operator $\widehat{L}^{\lambda}:=\frac{d^{2}}{dx^{2}}-\lambda^2 q(x)$. Assuming that the  function $q(x)$ is real-valued and continuous with $\lim_{ x \rightarrow \pm\infty}q(x)=+\infty$ , we replace the condition $v^{\lambda}(x)\in L^2(R_x)$ by the boundary conditions at infinity $\lim_{x \rightarrow \pm\infty}v^{\lambda}(x)=0$ .
Under these conditions for the function $q(x)$ , the spectrum of the operator $\widehat{L}^{\lambda}$ , is purely discrete and consists of a countable set of eigenvalues \cite{Fed},
$$0<\lambda_0<\lambda_1\leq\ldots\leq\lambda_n\leq\ldots ,  \quad \lim _{n\rightarrow \infty}\lambda_{n}=+\infty\quad ,$$ and the corresponding eigenfunctions are $v_{n}(x)=v^{\lambda_{n}}(x)$ .

The details of the WKB construction of the eigenvalues and eigenfunctions for large  orders $n$  can be found, for example, in  the books of M.Fedoriuk \cite{Fed} and K.Yosida \cite{Y}. 

%{\color{red} I removed the appendix. I have moved this part to an independent supplement. I will make it if there is time after finishing everything in the PhD draft---}

The asymptotics of the eigenvalues $\lambda_n$, for large $n$,  are given by
\begin{eqnarray}\label{lambda}
\lambda_n=\left(\int_{x_{1}(E)}^{x_{2}(E)}|\sqrt{q(x)}|\, dx\right)^{-1}\left(n\pi+\frac{\pi}{2}\right)+O(1/n) \ ,
\end{eqnarray}
 where $q(x):=2(V(x)-E)$ with some fixed $E$ . 
If we restore the actual value of  $\lambda=\lambda_n=1/\epsilon$ (eq. $(\ref{eq_fedoriouk})$ into $(\ref{lambda})$), it follows that it must be $E=E^{\hb}_{n}$, where $E^{\hb}_{n}$ are the solutions
\begin{eqnarray}\label{bohrsom}
f(E_{n}^{\hb})\equiv\int_{x_{1}(E_{n}^{\hb})}^{x_{2}(E_{n}^{\hb})}\sqrt{2(E_{n}^{\hb}-V(x))}\, dx\approx\pi\left(n+\frac{1}{2}\right)\epsilon \ ,
\end{eqnarray}
for large $n$. It is clear that $E^{\hb}_{n}$ are the eigenvalues of $\widehat{H}^{\epsilon}$ (see eq. $(\ref{statschr}))$.

The formula $(\ref{bohrsom})$ is known as the {\it Bohr-Sommerfeld quantization rule} (\cite{MF, Fed}). It provides the only allowed  values of energy of the quantum particle, in contrary to classical mechanics where any value of energy can be prescribed.

Recall that the potential $V(x)$ is positive, continuous function and  $\lim_{| x| \rightarrow \infty}V(x)=\infty$. In addition, we  assume that the equation $(\ref{eq_fedoriouk})$ has exactly two simple turning points $x_1(E)$, $x_2(E)$ with $x_1(E)<x_2(E)$ both simple. A turning point of $(\ref{eigenval_eq})$  is defined as a point $x$ at which $V(x)=E$, that is $q(x)=0$ . For example, simple turning points appear in the case of a single-potential well  (see Fig. \ref{well} ), and we have a couple of turning points for each eigenvalue $E=E_{n}^{\epsilon}$. The  turning points divide the real axis into regions where either $V(x)>E_{n}^{\epsilon}$ or  $V(x)<E_{n}^{\epsilon}$.

The asymptotic approximations of the eigenfunctions, for small $\epsilon$ and large $n$,  in the different regions, read as follows.

\noindent
(1) $x>x_{2}(E_{n}^{\hb})$ 
\begin{eqnarray}\label{wkbout1}
v_{n}^{\epsilon}(x)\approx \psi_{n}^{\epsilon}(x)={[2(V(x)-E_{n}^{\hb})]}^{-1/4}e^{-\frac{1}{\epsilon} \int_{x_{2}(E_{n}^{\hb})}^{x}\sqrt{2(V(t)-E_{n}^{\hb})} \, dt}\quad  \ ,\nonumber\\
&&
\end{eqnarray}
(2) $x_{1}(E_{n}^{\hb})<x<x_{2}(E_{n}^{\hb})$ 
\begin{eqnarray}\label{wkbsolution_1}
v_{n}^{\epsilon}(x)\approx\psi_{n}^{\epsilon}(x)={[2(E_{n}^{\hb}-V(x))]}^{-1/4}\cos\left({\frac{1}{\epsilon}\int_{x_{2}(E_{n}^{\hb})}^{x}\sqrt{2(E_{n}^{\hb}-V(t))} \, dt+\frac{\pi}{4}}\right)\quad \ , \nonumber \\
&& \,
\end{eqnarray}
(3) $x<x_{1}(E_{n}^{\hb})$ 
\begin{eqnarray}\label{wkbout2}
v_{n}^{\epsilon}(x)\approx\psi_{n}^{\epsilon}(x)= (-1)^n{[2(V(x)-E_{n}^{\hb})]}^{-1/4}e^{\frac{1}{\epsilon}\int_{x_{2}(E_{n}^{\hb})}^{x}\sqrt{2(V(t)-E_{n}^{\hb})} \, dt}\quad \ . 
\end{eqnarray}

%\newpage
\begin{figure}[h!]
\centering
\includegraphics[width=0.7\textwidth]{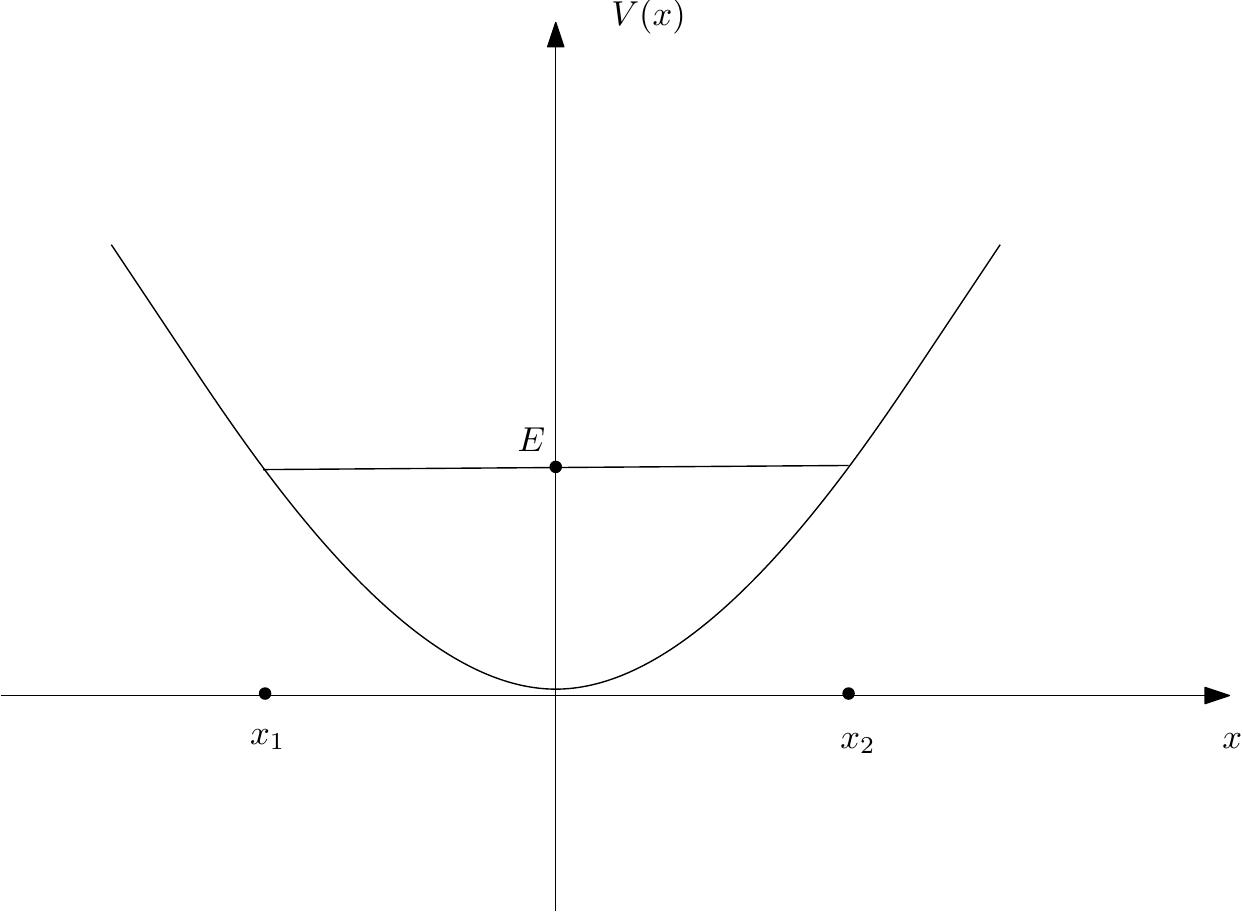}
\caption{{\it Single potential well}}
\label{well}
\end{figure}

These WKB approximations break down at the turning points where $V(x)=E_{n}^{\hb}$, since the amplitudes $[2(V(x)-E_{n}^{\hb})]^{-1/4}$ diverge at these points.

In the region $\{x\in R : V(x)>E^{\hb}_{n}\}$, the eigenfunctions decay exponentially as $|x|\rightarrow \infty$, while in the  region $\{x\in R : V(x)<E^{\hb}_{n}\}$ the eigenfunctions are rapidly oscillatory functions. The mechanical interpretation of this behaviour is the following. In classical mechanics, a particle with energy $E$ remains trapped in the region $V(x)<E$ (classically allowed region, or illuminated region in the terminology of optics), and it can never go out to the region $V(x)>E$ (classically forbidden region  or shadow region in the terminology of optics). In quantum mechanics, the wavefunction  is not restricted to the classically allowed region  $V(x)<E$ only, but in the   classically forbidden region  it is exponentially small. This means that the probability of finding the quantum particle in this region is negligible for small $\epsilon$.

%%%%%%%%%%%%%%%%%%%%%%%%%%%%%%%%%%%%%%%%%%%%%%%%%%%%%%%%%%%%%%%%%%%%%%%%%%%%%%%%%%%%%%%%%%%%%%%%%%%%%%%%%%%%%%%%%
%%%%%%%%%%%%%%%%%%%%%%%%%%%%%%%%%%%%%%%%%%%%%%%%%%%%%%%%%%%%%

\subsection{Example: The harmonic oscillator}\label{314}

We consider as our basic example the harmonic oscillator $V(x)=x^{2}/2$ . 
The integral in $(\ref{bohrsom})$ is computed analytically, and we get the approximations
\begin{eqnarray}\label{energy_appr}
E_{n}^\epsilon\approx\left(n+\frac{1}{2}\right)\epsilon \quad ,
\end{eqnarray}
for small $\epsilon$ and large $n$.

On the other hand, the eigenvalues and the eigenfunctions  of the harmonic oscillator can be computed analytically (see, e.g. \cite{Foc, Tak}).  The exact eigenvalues are
\begin{eqnarray}\label{energy_harm}
E_{n}^{\hb} = \left(n + {1\over 2}\right)\hb \, , \quad n=0,1,\ldots \ ,
\end{eqnarray}
and they coincide with the first-order approximation provided by the Bohr-Sommerfeld formula.
The exact eigenfunctions, normalized to one,  are
\begin{eqnarray}\label{eigenf_harm}
\vh_{n}(x)=\frac{e^{-x^2/2\epsilon}}{(\pi\epsilon)^{1/4}\sqrt{2^{n}n!}}\, H_{n}\left(\frac{x}{\sqrt{\epsilon}}\right)\, ,\quad \mathrm{for}\quad n=0,1,\ldots 
\end{eqnarray}
where $H_n$ are the Hermite polynomials
$H_n(x):=(-1)^n e^{x^2}\frac{d^n}{dx^n}\left(e^{-x^2}\right)\ .$ 

It can be showed by direct computations that the set $\{\vh_{n}(x)\}$ for $n=0,1,\ldots$ , forms an orthonormal basis of $L^{2}(R)$ \cite{WoM}.

By the relation $(\ref{wkbsolution_1})$, the WKB approximation of $\vh_{n}(x)$ in the oscillatory region $x^2/2<E_{n}^{\hb}$, is given by
\begin{eqnarray}\label{wkbcos}
v_{n}^{\epsilon}(x)&\approx &\left(\frac{2}{\pi}\right)^{1/2}\psi_{n}^{\epsilon}(x)\nonumber\\
&=&\left(\frac{2}{\pi}\right)^{1/2}\left(2E_{n}^{\hb}-x^2\right)^{-1/4}\cos\left({\frac{1}{\epsilon}\int_{\sqrt{2E_{n}^{\hb}}}^{x}\sqrt{2E_{n}^{\hb}-t^2} \, dt+\frac{\pi}{4}}\right)\ ,\nonumber \\
&& \,
\end{eqnarray}

%{\color{red} we must explain why the $\left(\frac{2}{\pi}\right)^{1/2}$ normalizing constant }

In Appendix \ref{comparison_sol} by exploiting  appropriate asymptotic expansions of the Hermite polynomials (\cite{Do}), we check that  the asymptotic expansion of $(\ref{eigenf_harm})$ 
coincides  with the WKB approximation  $(\ref{wkbcos})$  in the  oscillatory region.

Finally, in order to emphasize the {\it two-phase structure of the WKB approximations near the turning points}, and for using in the construction of asymptotics of the Wigner functions, we rewrite $(\ref{wkbcos})$  in terms of complex exponentials
\begin{equation}\label{wkbhosc}
v_{n}^{\hb}(x)\approx\left(\frac{2}{\pi}\right)^{1/2}\psi^{\epsilon}_{n}(x)={A_{n}^{\hb}}^{+}(x)\,
e^{\frac{i}{\epsilon}{S_{n}^{\hb}}^{+}(x)}+{A_{n}^{\hb}}^{-}(x)\,
e^{\frac{i}{\epsilon}{S_{n}^\hb}^{-}(x)} \ .
\end{equation}

The phases ${S_{n}^\hb}^{\pm}$ and the amplitudes ${A_{n}^{\hb}}^{\pm}$ are given by the formulae

\begin{eqnarray}
{A_{n}^{\hb}}^{+}(x)&:=&\frac{1}{2}\left(\frac{2}{\pi}\right)^{1/2}\left(2E_{n}^{\hb}-x^2\right)^{-1/4}e^{i\pi /4}\quad ,\label{wkbampl_1}\\
{A_{n}^{\hb}}^{-}(x)&:=&\frac{1}{2}\left(\frac{2}{\pi}\right)^{1/2}\left(2E_{n}^{\hb}-x^2\right)^{-1/4}e^{-i\pi /4}\label{wkbampl_2}
\end{eqnarray}
and
\begin{eqnarray}
{S_{n}^{\hb}}^{+}(x)&:=&\int_{\sqrt{2E_{n}^{\hb}}}^{x}\sqrt{2E_{n}^\epsilon - t^2}dt\quad ,\label{wkbph_1}\\
{S_{n}^{\hb}}^{-}(x)&:=&-\int_{\sqrt{2E_{n}^{\hb}}}^{x}\sqrt{2E_{n}^\epsilon - t^2}dt=-{S_{n}^{\hb}}^{+}(x)\quad ,\label{wkbph_2}
\end{eqnarray}
where $E_{n}^\epsilon=\left(n+1/2\right)\epsilon$ are the eigenvalues. The amplitudes $A_{n}^{\pm}(x)$ diverge at the turning points $x=\pm\sqrt{2E_{n}^{\hb}}$ .

%%%%%%%%%%%%%%%%%%%%%%%%%%%%%%%%%%%%%%%%%%%%%%%%%%%%%%%%%%%%%%%%%%%%%%%%%%%%%%%%%%%%%
%%%%%%%%%%%%%%%%%%%%%%%%%%%%%%%%%%%%%%%%%%%%%%%%%%%%%%%%%%%%%%%%%%%%%%%%%%%%%%%%%%%%%%

%%%%%%%%%%%%%%%%%%%%%%%%%%%%%%%%%%%%%%%%%%%%%%%%%%%%%%%%%%%%%%%%%%%%%%%%%%%%%%%%%%%%%
%%%%%%%%%%%%%%%%%%%%%%%%%%%%%%%%%%%%%%%%%%%%%%%%%%%%%%%%%%%%%%%%%%%%%%%%%%%%%%%%%%%%%%

%%%%%%%%%%%%%%%%%%%%%%%%%%%%%%%%%%%%%%%%%%%%%%%%%%%%%%%%%%%%%%%%%%%%%%%%%%%%%%%%%%%%%%%%%%%%%%%%%%%%%%%%%%%%%%%%%%%%%%%%%%%%%%%%%%%%%%%%%%%%%%%%%%%%%%%%%%%%%%%%%%%%%%%%%%%%%
\chapter{QM in Phase space: Moyal \& Wigner equation}\label{chapter4}

In classical mechanics the observables are ordinary functions $f(x,p)$ on phase space \footnote{Recall that we work in configuration space of dimension $d=1$, and therefore the phase space is $2-$dimensional.}, which naturally commute under multiplication. In quantum physics the observables become operators $\widehat{f^{\hb}}$ in Hilbert space. These observables depend, in general, on the small parameter $\hb$, and they usually do not commute under composition. This is a radical difference in the mathematical nature between the classical and quantum observables. It is a folk statement, which is in fact a theorem under certain conditions, that classical mechanics is derived by taking the classical limit $\hb\rightarrow 0$ of quantum mechanics. 
On the other hand,
quantum mechanics results from  classical mechanics by a procedure called {\it {quantization}}. We have already shown in Section $\ref{section22}$ how the Schr\"odinger operator results from the classical particle Hamiltonian (see eq. $(\ref{Schrodinger_quant})$).

For these reasons, we say that, there is a correspondence between a quantum operator $\widehat{f^{\hb}}$ and  its {\it {symbol}}  $f(x,p)$ . 

A {\it quantization rule}, is a mapping  $f\longmapsto \hf^{\hb}$ which  satisfies the following conditions:
\begin{itemize}
\item
it is a linear and one-to-one  correspondence that depends on parameter $\hb$
\item
the function $f$ is, in a sense, the ``limit" of the operator $\widehat{f^{\hb}}$ as $\hb\rightarrow 0$ .
\end{itemize}

%In the sequel, in order to simplify the notation,  we omit the superscript $\hb$ when we write the quantum operators, but someone must always keeps in mind that they depend on this small parameter.

\section{Operators and Symbols}

As we have already explained a quantization procedure is a rule to associate an operator to a classical observable.
In passing from classical to quantum mechanics, the phase space coordinates are replaced by coordinate and momentum operators, $\widehat{q}=x$ (the operator of multiplication) and $\widehat{p}=-i\epsilon\frac{\partial}{\partial x}$ (the operator of differentiation), respectively, which obviously satisfy the commutation relation $[\widehat{p},\widehat{q}]=-i\hb$. 

If we try to extend this correspondence rule for an arbitrary function of $(x,p)\in R^{2}$ we meet a serious and fundamental difficulty. For instance we may associate the product $px$ with the operators $\widehat{p}\widehat{q}$ or  $\widehat{q}\widehat{p}$ or $\left(\widehat{p}\widehat{q}+\widehat{q}\widehat{p}\right)/2$.
This simple example shows that quantization is not uniquely defined. 

There are  many ways to quantize a symbol (classical observable) but, the symbol is uniquely determined by the operator, given the quantization rule. The most common quantization rules are the {\it{qp-quantization}}, {\it{pq-quantization}} and the {\it{Weyl quantization}} (or symmetric quantization) (see, e.g.,  the books by A.Martinez \cite{Mar}, M.Zworski \cite{Zw}, V.E.Nazaikinskii et al. \cite{NSS}, for other quantization schemes). Since the quantization process makes essential use of ideas \footnote{many ideas in this theory have been motivated by quantum mechanics} from the theory of pseudodifferential operators (see e.g. the books by F.Treves \cite{Tr}, and M.Taylor \cite{Tayl}, for a general description of this theory), in this section we present some necessary ideas  from this theory.

Let us consider an $m-$th order differential operator with non-constant coefficients
\begin{eqnarray}
\hf=\sum_{\alpha\leq m}^{}a_{\alpha}(x)D_{x}^{\alpha} \ ,
\end{eqnarray}
where 
$D_{x}=\frac{1}{i}\frac{\partial}{\partial x}$.
Formally we can derive the  differential operator by the substitution $\xi=D_{x}$ in the polynomial
\begin{eqnarray}
f(x,\xi):=\sum_{|\alpha|\leq m}^{}a_{\alpha}(x)\xi ^{\alpha} \ .
\end{eqnarray}
It is important to note that since the coefficients depend on $x$, the ordering (i.e. the operators are kept in the right of the coefficients) is crucial . Let us assume that the operator acts on functions in $L^{2}(R)$. Then, by the properties of Fourier transform we have 
\begin{eqnarray*}
\hf[u(x)]=\sum_{\alpha\leq m}^{}{(2\pi)}^{-1}\int_{R}^{} e^{ix\cdot\xi}a_{\alpha}(x) \xi^{\alpha}
 \widehat{u}(\xi)\, d\xi \ ,
\end{eqnarray*}
that is
\begin{eqnarray*}
\hf[u(x)]={(2\pi)}^{-1}\int_{R}^{}e^{ix\cdot\xi}f(x,\xi)\widehat{u}(\xi)\, d\xi \ ,
\end{eqnarray*}
where $\widehat{u}(\xi)$ is the Fourier transform of $u(x)$.
By substituting the Fourier transform  $\widehat{u}(\xi)$ by its definition in the last equation, we get 
\begin{eqnarray}\label{oscdo}
\hf[u(x)]=\frac{1}{2\pi}\int_{R}^{} \int_{R}^{}e^{i(x-y)\cdot\xi}f(x,\xi)u(y)\, dyd\xi \ .
\end{eqnarray}

The formula $(\ref{oscdo})$ motivates the definition \footnote{the definition adopting the formula 
$(\ref{oscdo})$ corresponds to the so called standard representation of $\Psi DO$. There are other representations... } of  a pseudodifferential operator ($\Psi DO$) as a generalization of the differential operator,  since it makes possible to define the operator $\hf$ for  a broad class of symbols,  which are not necessarily polynomials with respect to $\xi$.
For doing this, one must gives meaning to the oscillatory integral $(\ref{oscdo})$ as an operator acting on certain classes of functions which of course depend on the class of symbols that are used. For instance, it can be proved 
(see, e.g., \cite{Zw}, Sec. 4.1.2), that if  $f \in \mathcal{S}(R^{2})$, then $\hf$ can be defined as continuous mapping from $\mathcal{S}'$ to $\mathcal{S}$, while for distributional symbol $f \in \mathcal{S}'(R^{2})$, $\hf$ can be defined as continuous mapping from $\mathcal{S}$ to $\mathcal{S}'$. Here $\mathcal{S}$ is the Schwartz space and $\mathcal{S}'$ the space of tempered distributions. 

For later use we introduce the semiclassical $\Psi DO$ corresponding to $(\ref{oscdo})$, by the formula 
\begin{eqnarray}\label{scloscdo}
\hf^{\epsilon}[u(x)]=\frac{1}{2\pi{\epsilon}}\int_{R}^{} \int_{R}^{}e^{\frac{i}{\epsilon}(x-y)\cdot\xi}f(x,\xi)u(y)\, dyd\xi \ .
\end{eqnarray}
Formally, this operator corresponds to the use of semiclassical Fourier transform defined with $\frac{i}{\epsilon}$ in place of $i$ in the usual transform, but as we will see in the next section semiclassical operators naturally appear in quantum mechanics\footnote{For this reason, often the semiclassical Fourier transform is referred as quantum Fourier transform in the literature.}.

\subsection{Weyl operators \& Wigner transform}

In quantum mechanics, and also in the high frequency analysis of many wave problems, the preferred representation of $\Psi DOs$ is the Weyl representation, although the operator calculus in this representation  is much more complicated  than the one   for other representations (like, e.g., the standard $(\ref{oscdo})$). Besides many other reasons related to the formulation of quantum mechanics itself, Weyl operators have the important feature that when the symbol is real, the operator is selfadjoint \footnote{This is not true for other representations.}. 
The construction of the Weyl operator which we present in this section is based on the books by  F.A.Berezin \& M.A.Shubin \cite{BS}. See also \cite{Mar, Zw, HO} for a complete mathematical exposition, and \cite{FOL, G1} for the emergence of Weyl calculus in quantum mechanics and harmonic analysis in phase space.

%{\color{red} Weyl ops bounded in $L^2$, Zworski, p.82}

By the Fourier transform \footnote{The  Fourier transforms  in the sequel are assumed to act on $L^2(R)$ functions, see e.g. \cite{Z}.}  we decompose the symbol 
\begin{eqnarray}\label{fourier}
f(x,p)=\int_{R}^{}\int_{R}^{} e^{i(r{x}+s{p})} \phi(r,s)\, dr ds\, ,
\end{eqnarray}
in terms of its inverse Fourier transform
\begin{eqnarray}\label{invfourier}
\phi (r,s)= (2\pi)^{-2}\int_{R}^{} \int_{R}^{} e^{-i(r q+s p)}
f(q,p) \, dq dp \ .
\end{eqnarray}

Then, we associate $f$ with the operator
\begin{eqnarray}\label{fourier_op}
\hf_{\W}^\epsilon=\int_{R}^{}\int_{R}^{} e^{i(r \widehat{q}+s \widehat{p})} \phi(r,s)\, dr ds \ ,
\end{eqnarray}
where the exponential operator $e^{i(r \widehat{q}+s \widehat{p})}$ acts as follows 
\begin{eqnarray}\label{cbh_form}
e^{i(r \widehat{q}+s\widehat{p})}[u(x)]=e^{i(r x+\epsilon r s/2)}u(x+\epsilon s) \ .
\end{eqnarray}
The rule $(\ref{cbh_form})$ follows, for example, from the Campbell-Baker-Hausdorff formula (see,  e.g. \cite{Za2} , Zassenhaus expansion)
%{\color{red} we have better references for that}), 
$$e^{t(\widehat{A}+\widehat{B})}= e^{t\widehat{A}}e^{t\widehat{B}}e^{-\frac{t^2}{2}[\widehat{A},\widehat{B}]}\ .$$ 
Indeed, by putting $\widehat{A}= ir\widehat{q}$, $\widehat{B}=is \widehat{p}$, $t=i$ and $[\widehat{A},\widehat{B}]=ir s[\widehat{q},\widehat{p}]=\hb r s$, $[\widehat{A},[\widehat{A},\widehat{B}]]=[\widehat{B},[\widehat{A},\widehat{B}]]=0$   we obtain
\begin{eqnarray*}
e^{i(r \widehat{q}+s \widehat{p})}[u(x)]=e^{ir\widehat{q}}e^{i\frac{\hb}{2}r s}e^{is \widehat{p}}[u(x)]=e^{ir x}e^{i\frac{\hb}{2}r s}u(x+\hb s) \ .
\end{eqnarray*}
In the last equation, we use the fact that
the operator $\widehat{T}_{\alpha}:=e^{a\frac{d}{d x}}$ ($a$: parameter) acts as the shift operator \cite{FOL}.
%{\color{red} Nazaikinskii:Non commutative operators, Maslov: Operational methods and some papers we have read ...may be better references}
This is formally proved by expanding the exponential in Taylor series, as follows
\begin{eqnarray*}
\widehat{T}_{\alpha}[g(x)]=e^{a\frac{d}{d x}}[g(x)]=\sum_{n=0}^{\infty}\frac{\alpha^n}{n!}\frac{d^n}{dx^n}g(x)=
%\sum_{n=0}^{\infty}\frac{\alpha^n}{n!}g^{(n)}(x)=
g(x+\alpha)\, .
\end{eqnarray*}

Combining $(\ref{fourier_op})$ with $(\ref{cbh_form})$, and substituting $(\ref{invfourier})$, we have
\begin{eqnarray*}
\hf_{\W}^\epsilon[u(x)]&=&(2\pi)^{-2}\int_{R}^{}\int_{R}^{}\int_{R}^{}\int_{R}^{} e^{i[{r(x-q)-s(p-\epsilon r/2)}]} f(q,p) u(x+\epsilon s) \, drdsdqdp\\
&=& (2\pi)^{-2} \int_{R}^{}\int_{R}^{}\int_{R}^{}\int_{R}^{} e^{-i s p}e^{-ir[q-(x+\epsilon s/2)]} f(q,p) u(x+\epsilon s)\, drdsdqdp \\
&=& (2\pi)^{-1} \int_{R}^{}\int_{R}^{}\int_{R}^{} e^{-i s p} \delta \left(q-(x+\epsilon s/2)\right) f(q,p) u(x+\epsilon s)\, dqdpds \\
&=& (2\pi)^{-1}\int_{R}^{} \int_{R}^{} e^{-i s p} f(x+\epsilon s/2,p)u(x+\epsilon s) \, dpds \, .
\end{eqnarray*}
By the change $y=x+\epsilon s$ of the integration variable $s$, we finally derive the formula
\begin{eqnarray}\label{wel_op}
\hf_{\W}^\epsilon[u(x)]=(2\pi\hb)^{-1}\int_{R}^{}\int_{R}^{} e^{\frac{i}{\hb}(x-y) p} f\left(\frac{x+y}{2},p\right) u(y) \, dydp \ .
\end{eqnarray}
The formula $(\ref{wel_op})$ is the definition of \emph{the Weyl operator $\hf_{\W}^\epsilon$ with symbol $f$. }

It follows from $(\ref{wel_op})$ that the  Schwartz's kernel of $\hf_{\W}^\epsilon$ is given by
\begin{eqnarray} \label{weylkernel}
K^\epsilon(x,y)=(2\pi\hb)^{-1}\int_{R}^{} e^{\frac{i}{\hb}(x-y) p} f\left(\frac{x+y}{2},p\right) \, dp \, .
\end{eqnarray}
Hence we have
\begin{eqnarray*}
K^\epsilon(x+\hb\zeta/2,x-\hb\zeta/2)=(2\pi\hb)^{-1} \int_{R}^{}e^{i\zeta p} f(x,p) \, dp 
\end{eqnarray*}
and by the Fourier inversion formula we express the symbol in terms of the Schwartz's kernel, as follows
\begin{eqnarray}\label{weylsymbol}
f(x,p)= \int_{R}^{} e^{-\frac{i}{\hb}y p} K^\epsilon(x+y /2,x-y /2) \, dy\, \  . 
\end{eqnarray}

The definition $(\ref{wel_op})$ of the Weyl operator and the interrelations $(\ref{weylkernel})$, $(\ref{weylsymbol})$ between the symbol and the Schwartz's kernel imply the following important properties:

\begin{enumerate}
\item
{\it The symbol $f$ is real iff ${\widehat{f_{\W}^\epsilon}}^* = \widehat{f}_\W^\epsilon$.}

This follows easily from $(\ref{weylkernel})$, $(\ref{weylsymbol})$ and the fact that  if ${\widehat{f_{\W}^\epsilon}}^*$ is the adjoint operator of $\widehat{f}_\W^\epsilon$,  then the Schwartz's kernels are complex conjugate.  

\item
{\it  If $f \in \mathcal{S}(R^{2})$ then $\hf_{\W}^\epsilon: L^2(R) \to  L^2(R)$ is bounded independently of $\hb$}

For the proof see the book by M.Zworski \cite{Zw}, Sec. 4.5.1.

\end{enumerate}

The interrelation $(\ref{weylsymbol})$ between the symbol and the Schwartz's kernel when applied to the density operator of quantum mechanics (see, eq. $(\ref{densitypure1})$) leads to the definition of the semiclassical Wigner transform (see eq. $(\ref{semWigner})$ below).
Indeed,  if $\ket{\psi^\hb}$ is a pure state, the density operator has the form $\widehat{\rho}_{\psi^\hb}=\ket{\psi^\hb}\bra{\psi^\hb}$ and it acts as a projection operator. Thus, for a function $\phi(x)\in L^2(R)$, we have
\begin{eqnarray*}
\widehat{\rho}_{\psi^\hb}\phi(x)&=&(\phi,\psi^\hb)_{L^2(R)} \psi^\hb(x)=\int_{R}^{} \overline{\psi^\hb}(y)\psi^\hb(x)\phi(y)\, dy \ ,
\end{eqnarray*}
which implies that the Schwartz's kernel of $\widehat{\rho}_{\psi^\hb}$ is $K^\hb(x,y)=\psi^\hb(x) \overline{\psi^\hb}(y)$. Here $\overline{\psi^\hb}$  is the complex conjugate of $\psi^\hb$.
If we denote by  $\rho_{\W}(x,p)$ the symbol of operator $\widehat{\rho}_{\psi^\hb}$ considered in the Weyl representation,
then by $(\ref{weylsymbol})$ we obtain
\begin{eqnarray*}\label{wig_ro}
\rho_{\W}^\epsilon(x,p)=\int_{R}^{} e^{-\frac{i}{\epsilon}y p} \psi^\hb\left(x+\frac{y}{2}\right)\overline{\psi^\hb}\left(x-\frac{y}{2}\right) \, dy \, .
\end{eqnarray*}

The Weyl symbol  $\rho_{\W}^\epsilon$ of the density operator $\widehat{\rho}_{\psi^\hb}=\ket{\psi^\hb}\bra{\psi^\hb}$ has appeared independently in E.Wigner's study of quantum thermodynamics \cite{Wig}. For this reason it is usually referred in the literature as the {\it semiclassical Wigner transform  of $\psi^\hb(x)$}, and it is denoted by 
 \begin{eqnarray}\label{semWigner}
W^{\hb}[\psi^\hb](x,p)=(2\pi\hb)^{-1}\int_{R}^{} e^{-\frac{i}{\epsilon}y p} \psi^\hb\left(x+\frac{y}{2}\right)\overline{\psi^\hb}\left(x-\frac{y}{2}\right) \, dy \ .
\end{eqnarray}

Furthermore, we consider the functions  $\psi(x),\phi(x)\in L^2(R)$ and we define the operator $\widehat{\rho}_{\phi\psi}$, with  Schwartz kernel $K_{\psi\phi}(x,y):=\psi(x)\overline{\phi}(y)$ , by 
 \begin{eqnarray}\label{cross_den}
 \widehat{\rho}_{\phi\psi}u(x):=\int_{R}^{}K_{\phi\psi}(x,y)u(y)\, dy=\psi(x)(u,\phi)_{L^2(R)}
 \end{eqnarray}
By $(\ref{weylsymbol})$, the Weyl symbol $ \widehat{\rho}_{\phi\psi}$ of  is given by
\begin{eqnarray}\label{cross_wig}
W_{\phi}^{\hb}[\psi](x,p)(x,p)=(2\pi\hb)^{-1}\int_{R}^{} e^{-\frac{i}{\epsilon}p y} \psi\left(x+\frac{y}{2}\right)\overline{\phi}\left(x-\frac{y}{2}\right) \, dy \ .
\end{eqnarray}
The symbol $(\ref{cross_wig})$ is  referred in the literature as the {\it cross Wigner transform of $\psi$ and $\phi$}.

\subsection{Composition of operators and the Moyal product}

%We are interested  to find a deformation of the algebra of classical observables. For this, we present a  construction of  a larger algebra of formal power series in the deformation parameter $\epsilon$, with coefficients in the classical algebra on which deformed (non-commutative) product, a so called {\it star product} is defined.

In this section we define the Moyal star product $f_{1}\star_{\M}f_2$ as the symbol of composition of operators $\hf_{\W _1}^{\hb}\ci \widehat{f}_{\W _2}^{\hb}$ in the Weyl representation.

Let $\hf_{\W _1}^{\hb}, \widehat{f}_{\W _2}^{\hb}$ be Weyl operators with  $f_1,f_2\in L^2(R^{2}_{xp})$ symbols, respectively. 
We want to calculate  the Weyl symbol  $f$ of the operator
$\hf_{\W}^{\hb}=\hf_{\W_1}^{\hb}\ci \hf_{\W_2}^{\hb}$, where $\ci$ denotes the composition of the operators.  If $K_{1}^{\hb}(x,y)$, $K_{2}^{\hb}(x,y)$ are the Schwartz  kernels of the operators $\hf_{\W_1}^{\hb}$ and $\hf_{\W_2}^{\hb}$, respectively, then, the  kernel $K^{\hb}(x,y)$ of $\hf_{\W}^{\hb}$ is given by 
\begin{eqnarray}
K^{\hb}(x,y)=\int_{R}^{}K_{1}^{\hb}(x,z)K_{2}^{\hb}(z,y)\, dz \ .
\end{eqnarray}
Indeed,  for $u(x)\in L^2(R)$ and $v^{\hb}(x)=\hf_{\W_2}^{\hb}[u(x)]$, we have
\begin{eqnarray*}
(\hf_{\W_1}^{\hb}\ci \hf_{\W_2}^{\hb})[u(x)]&=&\hf_{\W_1}^{\hb}[\hf_{\W_2}^{\hb}[u(x)]]\\
&=&\hf_{\W _1}^{\hb}[v^{\hb}(x)]\\
&=&\int_{R}^{}K_{1}^{\hb}(x,z)v^{\hb}(z)\, dz\\
&=&\int_{R}^{}\int_{R}^{}K_{1}^{\hb}(x,z)K_{2}^{\hb}(z,y)u(y)\, dz dy\, .
\end{eqnarray*}

By $(\ref{weylsymbol})$, the Weyl symbol of the composition is given by 
\begin{eqnarray*}
f(x,p)&=&\int_{R}^{}e^{\frac{i}{\epsilon} p\xi}K^{\hb}(x-\xi /2,x+\xi /2)\, d\xi\\ 
&=&
2 \int_{R}^{}\int_{R}^{}e^{2\frac{i}{\epsilon}p\eta} K_{1}^{\hb}(x-\eta,z)K_{2}^{\hb}(z,x+\eta) \,dz d\eta
\end{eqnarray*}
Using $(\ref{weylkernel})$, we rewrite the last equation  in terms of Weyl symbols $f_{1} \ , f_{2}$,
\begin{eqnarray*}
f(x,p)&=&
2{(2\pi\epsilon)}^{-2}
\int_{R}{}\int_{R}^{}\int_{R}^{}\int_{R}^{} e^{\frac{i}{\epsilon}
[(x-\eta-z)p_{1}+
(z-x-\eta) p_{2}+2p\eta]}\\
&&\cdot
f_{1}\left(\frac{x-\eta+z}{2},p_{1}\right)
f_{2}\left(\frac{x+\eta+z}{2},p_{2}\right) dp_{1}dp_{2}dz d\eta \ ,
\end{eqnarray*}
and by the change of variables 
$$x_1=(x-\eta+z)/2 \ , \ \ x_{2}=(x+\eta+z)/2$$ 
we get 
$$f(x,p)=(f_{1}\star_{\M} f_{2})(x,p)\ ,$$
where 

\begin{eqnarray}\label{starproduct_1}
(f_{1}\star_{\M} f_{2})(x,p)&:=&
(2\pi\epsilon)^{-2}\int_{R}^{}\int_{R}^{}\int_{R}^{}\int_{R}^{}e^{2
\frac{i}{\epsilon}[(x-x_2) p_{1}+(x_1-x)p_{2}+(x_{2}-x_{1})p]} \\ \nonumber
&&\cdot
f_{1}(x_1,p_1)f_{2}(x_2,p_2)\, dp_{1}dp_{2}dx_{1}dx_{2}  
\end{eqnarray}

The  RHS of  $(\ref{starproduct_1})$ is an integral representation of the Weyl symbol of the composition
$\hf_{\W}^{\hb}=\hf_{\W_1}^{\hb}\ci \hf_{\W_2}^{\hb}$, and it provides the definition of the {\it Moyal product } $f _{1}\star_{\M} f_{2}$ of the symbols $f_1$ and $f_2$. 

%This product appeared for first time in \cite{Mo} and, then, helped Groenewold \cite{Gro} to demonstrate how Poisson brackets contrast crucially  to quantum commutators ("Groenewold theorem").

The argument of the exponential function in the above expression has an interesting geometrical interpretation. Let 
$\Delta=AA_{1}A_{2}$ be the triangle with vertices
$A=(x,p)$, $A_1=(x_1,p_1)$ and $A_2=(x_2,p_2)$ in the phase space $R^{2}_{xp}$ . Then the length of the side $AA_{1}$ of the  triangle is given by 
\begin{eqnarray*}
\int_{A}^{A_1} p \, dq&=& \int_{0}^{1}[p+(p_{1}-p)t](x_{1}-x)\, dt\\
&=& \frac{1}{2}(p_{1}+p)(x_{1}-x)\, ,
\end{eqnarray*}
and the area of triangle $\Delta$ is given by 
\begin{eqnarray*}
\int_{\Delta}^{}p \, dq&=&\frac{1}{2}\, \left[(p_{1}+p)(x_{1}-x)+(p_{2}+p_{1})(x_{2}-x_{1})+(p+p_{2})(x-x_2)\right]\\
&=&-\frac{1}{2}\, \left[(x-x_2) p_{1}+(x_1-x) p_{2}+(x_{2}-x_{1}) p\right ]\, .
\end{eqnarray*}
Thus, the phase in the integral representation of the Moyal product can be translated as the geometrical area  \footnote{In higher dimensions by similar calculations this area appears through the symplectic product   \cite{G1}.}  $-2\int_{\Delta}^{}p \, dq$ .
Departing from this observation, C.Zachos \cite{Za1} has given a survey of the geometry related to the Moyal product. 

A symbolic representation of the Moyal product which reveals its connection with the deformation quantization process is derived as follows.

The formula $(\ref{starproduct_1})$, by the change of varianles $x_1=x-\eta\epsilon/2$ and $p_1=p+\eta\epsilon/2$ , is written as follows.
\begin{eqnarray*}\label{starproduct_2}
f(x,p)=(f_{1}\star_{\M} f_{2})(x,p)&=&
(2\pi)^{-2}\int_{R}^{}\int_{R}^{}\int_{R}^{}\int_{R}^{}e^{i[(x-x_2)\xi +(p-p_2)\eta]} \nonumber \\
&&\cdot f_{1}(x-\eta\epsilon / 2,p+\xi\epsilon /2)f_2(x_2,p_2)\, d\eta d\xi dx_{2}dp_{2} \ .
\end{eqnarray*}
Then, we expand the term $ f_{1}(x-\eta\epsilon / 2,p+\xi\epsilon /2)$ in Taylor series we integrate the Fourier integrals involving the powers of $\xi$, $\eta$ in terms of derivatives of the Dirac function and then we integrate the remaining Fourier integrals w.r.t. the variables $x_{2}$, $p_{2}$ . In this way we get a series involving derivatives of $f_{2}$ , which is summed by the use of the shift rule
$$e^{a\frac{d}{d x}}[g(x)]=\sum_{n=0}^{\infty}\frac{\alpha^n}{n!}\frac{d^n}{dx^n}g(x)=g(x+\alpha) \ .$$
The calculation results to the symbolic formula
\begin{eqnarray}\label{starproduct_2}
(f_{1}\star_{\M} f_{2})(x,p)=f_{1}\left(x+\frac{i\epsilon}{2}\partial_{p_2},p-\frac{i\epsilon}{2}\partial_{x_2}\right)f_{2}(x_2,p_2)|_{x_2=x,p_2=p} \ ,
\end{eqnarray}
which, according to its derivation, has to be interpreted as
\begin{eqnarray}\label{taylorprod}
f(x,p)&=&
(f_{1}\star_{\M} f_{2})(x,p)\nonumber\\
&=&
\sum_{\alpha,\beta=1}^{\infty}\frac{(-1)^{\beta}}{\alpha !\beta!}{\left(\frac{i\epsilon}{2}\right)}^{\alpha+\beta}[\partial^{\alpha}_{x}\partial^{\beta}_{p}f_{1}(x,p)]
[\partial^{\beta}_{x}\partial^{\alpha}_{p}f_{2}(x,p)] \ .
\end{eqnarray}
The symbolic formula $(\ref{taylorprod})$ can be further written in the form
\begin{eqnarray}
(f_{1}\star_{\M} f_{2})(x,p)&=&
f_{1}\left(x+\frac{i\epsilon}{2}\overrightarrow{\partial_ p},p-\frac{i\epsilon}{2}\overrightarrow{\partial _x}\right)f_{2}(x,p)\label{starformula_1}\\
&=&f_{1}(x,p)f_{2}\left(x-\frac{i\epsilon}{2}\overleftarrow{\partial_ p},p+\frac{i\epsilon}{2}\overleftarrow{\partial_ x}\right)\label{starformula_2}\\
&=& f_{1}\left(x+\frac{i\epsilon}{2}\overrightarrow{\partial_ p},p\right)f_{2}\left(x-\frac{i\epsilon}{2}\overleftarrow{\partial _p},p\right)\\
&=&f_{1}\left(x,p-\frac{i\epsilon}{2}\overrightarrow{\partial _x}\right)f_{2}\left(x,p+\frac{i\epsilon}{2}\overleftarrow{\partial_ x}\right)\label{starformula1} \, .
\end{eqnarray}
where the vector arrows over the derivatives indicate in which direction the differentiation acts.

Finally, $(\ref{taylorprod})$ can also be rewritten in terms of the so called star-exponential operator as follows (\cite{Gro}; see also \cite{Tak}, Sec. 3.4)
\begin{eqnarray}\label{starexp}
(f_{1}\star_{\M} f_{2})(x,p)=
f_{1}(x,p)\exp\left[\frac{i\epsilon}{2}\left(
\overleftarrow{\partial _x}
\overrightarrow{\partial _p}-
\overleftarrow{\partial _p}
\overrightarrow{\partial _x}\right)\right]
f_{2}(x,p)\ .
\end{eqnarray}
Expanding the exponential  in $(\ref{starexp})$ into power series, we obtain
\begin{eqnarray}\label{moyalexp}
f_{1}\star_{\M} f_{2}=f_{1}f_{2}+
\sum_{n=1}^{\infty}{\left(\frac{i\epsilon}{2}\right)}^{n} C_{n}(f_{1}, f_{2})
\end{eqnarray}
where 
$C_{n}(f_{1}, f_{2})=\frac{1}{n!}f_{1}
(\overleftarrow{\partial _x}
\overrightarrow{\partial_ p}-
\overleftarrow{\partial_ p}
\overrightarrow{\partial_ x})^{n}
f_{2}\, .$
This formula shows that the Moyal product is a  deformation of the usual  product  of functions (symbols). The small parameter $\epsilon$ plays the role of deformation parameter. A detailed analysis of the deformation theory and its application to quantization is given by F.Bayen, et al. \cite{BaFFLS1, BaFFLS2} see also the book by B.Fedosov \cite{Fedos}).

From $(\ref{moyalexp})$ it follows that 
\begin{eqnarray}
f_{1}\star_{\M} f_{2}=f_{1}f_{2}+O(\epsilon) \ ,
\end{eqnarray}
and also
\begin{eqnarray}\label{moyal_com}
f_{1}\star_{\M} f_{2}-f_{2}\star_{\M} f_{1}=i\epsilon\lbrace f_{1} ,f_{2}\rbrace_{PB}+O(\epsilon^2) \ .
\end{eqnarray}

The quantity $f_{1}\star_{\M} f_{2}-f_{2}\star_{\M} f_{1}$ is known as the {\it Moyal} or {\it quantum commutator} \cite{BaFFLS2}. It has the important feature that in the classical limit $\epsilon =0$ converges to the Poisson bracket (see eq. $(\ref{poissonbracket})$), a fact which can be thought of as the phase space picture of the  {\it correspondence principle}. On other words this convergence is as  mathematical statement of the folk theorem saying that classical mechanics is inferred for the quantum mechanics in the classical limit (recall the introductory comments on quantisation at the beginning of this chapter).

Finally, it is not difficult to check that the Moyal product is associative 
\begin{eqnarray}
(f_{1}\star_{\M} f_{2})\star_{\M}f_{3}=f_{1}\star_{\M}(f_{2}\star_{\M} f_{3}) \ ,
\end{eqnarray}
and  noncommutative
\begin{eqnarray}
f_{1}\star_{\M} f_{2}\neq f_{2}\star_{\M} f_{1} \ .
\end{eqnarray}

For later use in Section \ref{sec43new}, we introduce the cosine and sine bracket of  $f_{1}$, $f_{2}$ which are derived by antisymmetrizing and symmetrizing the Moyal bracket, respectively
\begin{itemize}
\item 
 the Moyal (sine) bracket (\cite{Mo, Gro})
\begin{eqnarray}\label{moyalbr}
\lbrace f_{1} ,f_{2}\rbrace_{MB}=\frac{f_{1}\star_{\M} f_{2}-f_{2}\star_{\M} f_{1}}{2i}=f_{1}(x,p)\sin\left[\frac{\hb}{2}\left({\overleftarrow{\partial_
x}}\overrightarrow{\partial_p}-\overleftarrow{\partial_p}\overrightarrow{\partial_
x}\right)\right]f_{2}(x,p)\nonumber\\
& &
\end{eqnarray}
\item
 the Baker's cosine bracket (\cite{Ba, F, FMan})
\begin{eqnarray}\label{bakerbr}
\lbrace f_{1} ,f_{2}\rbrace_{BB}=\frac{f_{1}\star_{\M} f_{2}+f_{2}\star_{\M} f_{1}}{2}=f_{1}(x,p)\cos\left[\frac{\hb}{2}\left({\overleftarrow{\partial_
x}}\overrightarrow{\partial_p}-\overleftarrow{\partial_p}\overrightarrow{\partial_
x}\right)\right]f_{2}(x,p)\, .\nonumber\\
& &
\end{eqnarray}
\end{itemize}

Then, we can write the Moyal star product $(\ref{starexp})$, in terms of the Baker and Moyal brackets $(\ref{bakerbr})$, $(\ref{moyalbr})$ as follows 
\begin{eqnarray}\label{starexpdecomp}
(f_{1}\star_{\M} f_{2})(x,p)&=&
f_{1}(x,p)\exp\left[\frac{i\epsilon}{2}\left(
\overleftarrow{\partial _x}
\overrightarrow{\partial _p}-
\overleftarrow{\partial _p}
\overrightarrow{\partial _x}\right)\right]f_{2}(x,p)  \nonumber \\
&=&f_{1}(x,p)\cos\left[\frac{\hb}{2}\left({\overleftarrow{\partial_
x}}\overrightarrow{\partial_p}-\overleftarrow{\partial_p}\overrightarrow{\partial_
x}\right)\right]f_{2}(x,p) \nonumber \\
&&+ i f_{1}(x,p)\sin\left[\frac{\hb}{2}\left({\overleftarrow{\partial_
x}}\overrightarrow{\partial_p}-\overleftarrow{\partial_p}\overrightarrow{\partial_
x}\right)\right]f_{2}(x,p) \nonumber\\
&=&\lbrace f_{1} ,f_{2}\rbrace_{BB}(x,p)+i\lbrace f_{1} ,f_{2}\rbrace_{MB}(x,p) \ .
\end{eqnarray}

\begin{remark}
Similar start products $f_{1}\star f_{2}$ and analogous calculus can be defined when any other (than the Weyl) representation of the operators is used \cite{BaFFLS1}, and no matter which calculus we use, it holds that  as $\epsilon \rightarrow 0 $, 
$$(f_{1}\star f_{2})(x,p)\rightarrow
f_{1}(x,p) f_{2}(x,p) \ , \ \ \mathrm{and} \ \
\frac{1}{i\epsilon}(f_{1}\star f_{2}-f_{2}\star f_{1})\rightarrow\lbrace f_{1} ,f_{2}\rbrace_{PB} \ ,$$
that is, all quantisations are asymptotically equivalent.
\end{remark}

\section{The Moyal equation}\label{section42}
\subsection{Derivation}

Let us consider the Liouville-von Neumann equation (see $(\ref{vonneumann})$) which governs  the evolution  of the density operator 
\begin{eqnarray}\label{vonneumann2}
i\epsilon\frac{d}{dt}\widehat{\rho}=[\widehat{H}^\hb ,\widehat{\rho}]=\widehat{H}^\hb \ci\widehat{\rho}-\widehat{\rho}\ci\widehat{H}^\hb \ .
\end{eqnarray}
Note that here $\widehat{H}^\hb$ is the operator $(\ref{schrod_op})$, and that in the definition of the density operator must be used 
the wavefunction $u^{\epsilon}$ which is the solution of the dimensionless Schr\"odinger equation   $(\ref{SchrodingerH})$.

In the Weyl correspondence (recall eq.  $(\ref{wel_op})$), the operators $\widehat{H}^\hb \ci\widehat{\rho}$ and $\widehat{\rho}\ci\widehat{H}^\hb$ have Weyl symbols $H\star_{\M}\rho_{\W}$  and $\rho_{\W}\star_{\M}H$ respectively, where $H$ is the Weyl symbol of $\widehat{H}^\hb$ and $\rho_\W$ is the Weyl symbol of $\widehat{\rho}$.
Therefore, in the  Weyl representation, from the equation  $(\ref{vonneumann2})$ we derive the{\it{ Moyal equation}} (\cite{EW, Mo, BaFFLS2})
\begin{eqnarray}\label{moyaleq}
i\epsilon{\partial_t}\rho_{\W}(x,p,t)=H(x,p)\star_{\M}\rho_{\W}(x,p,t)-\rho_{\W}(x,p,t)\star_{\M}H(x,p) \ .
\end{eqnarray}
This is a non-local linear equation in phase space which governs the evolution of the Weyl symbol $\rho_\W$ of the density operator $\widehat{\rho}$. Within the framework of Weyl correspondence it is equivalent to the Liouville-von Neumann equation, and it thus provides a reformulation of quantum problem in phase space.

\subsection{The $\star$-solution of the Moyal equation}

The Moyal equation $(\ref{moyaleq})$ is formally solved through a $\star_{\M}$-unitary evolution operator (see \cite{BaFFLS2} for the definition of a general $\star$-unitary evolution). We define 
\begin{eqnarray}\label{ustar}
U_{\star_{\M}}(x,p;t)&=&e^{\frac{i}{\epsilon}tH}_{\star_{\M}} \nonumber \\
&:=&
1+\frac{i}{\epsilon}tH(x,p)+{\left(\frac{i}{\epsilon}t\right)}^{2} \frac{1}{2!} H(x,p)\star_{\M} H(x,p) \nonumber \\
&&+
{\left(\frac{i}{\epsilon}t\right)}^{3} \frac{1}{3!} H(x,p)\star_{\M} H(x,p)\star_{\M} H(x,p)+\ldots \ .
\end{eqnarray}
Then, $\rho_{\W}(x,p,t)$ is derived by the time evolution of $\rho_{\W}(x,p,0)$ according to (see \cite{CFZ2})
\begin{eqnarray}\label{solution_moyal}
\rho_{\W}(x,p,t)=U_{\star_{\M}}^{-1}(x,p,t)\star_{\M}\rho_{\W}(x,p,0)\star_{\M} U_{{\star}_{\M}}(x,p,t) \ ,
\end{eqnarray}
which is quite analogous to the evolution by Schr\"odinger semigroup in configuration space.

Now, we prove that $(\ref{solution_moyal})$ satisfies the Moyal equation $(\ref{moyaleq})$. 
Indeed, using the local notation $f(t)=\rho_{\W}(x,p,t)$, we have
\begin{eqnarray*}
i\hb\frac{d}{dt}f=i\hb\partial_{t}\left\{\sum_{n=0}^{\infty}\frac{1}{n!}\left(-\frac{i}{\hb}t\right)^{n}(H\star_{\M})^{n}\star_{\M} f(0)\star_{\M} \sum_{m=0}^{\infty}\frac{1}{m!}\left(\frac{i}{\hb}t\right)^{m}(H\star_{\M})^{m}\right\}
\end{eqnarray*}
and by the general property  $$\frac{d}{dt}\left(f_1 \star f_2\right)=\left(\frac{d}{dt}f_1\right) \star f_2 +f_1 \star \left(\frac{d}{dt} f_2\right) $$ for smooth time-dependent symbols $f_1$, $f_2$, and for every $\star$-product (\cite{BaFFLS2}, eq. (2.9)),
we get
\begin{eqnarray*}
i\hb\frac{d}{dt}f=&i\hb&\frac{d}{dt}\left\{\sum_{n=0}^{\infty}\frac{1}{n!}\left(-\frac{i}{\hb}t\right)^{n}(H\star_{\M})^{n}\right\}\star_{\M} f(0)\star_{\M} \sum_{m=0}^{\infty}\frac{1}{m!}\left(\frac{i}{\hb}t\right)^{m}(H{\star}_{\M})^{m}\\
&+&\sum_{n=0}^{\infty}\frac{1}{n!}\left(-\frac{i}{\hb}t\right)^{n}(H\star_{\M})^{n}\star_{\M} f(0)\star_{\M} i\hb\frac{d}{dt}\left\{\sum_{m=0}^{\infty}\frac{1}{m!}\left(\frac{i}{\hb}t\right)^{m}(H{\star}_{\M})^{m}\right\}\quad .
\end{eqnarray*}
Moreover,  for every $\star$-product, the $\star$-exponentials satisfy  the following identities (\cite{BaFFLS2}, eq. 4.3)
\begin{eqnarray*}
i\hb\frac{d}{dt}U_{\star}^{-1}(x,p,t)=H\star U_{\star}^{-1}(x,p,t) \  ,
\end{eqnarray*}
\begin{eqnarray*}
-i\hb\frac{d}{dt}U_{\star}(x,p,t)=H\star U_{\star}(x,p,t) \ ,
\end{eqnarray*}
and hence, by using the series representation $(\ref{ustar})$, we obtain
$$H\star U_{\star}(x,p,t)=U_{\star}(x,p,t) \star H$$  
 
Thus,  we  obtain
\begin{eqnarray*}
i\hb\frac{d}{dt}f(t)=H(x,p)\star_{\M} f(t)- f(t)\star_{\M} H(x,p) \ ,
\end{eqnarray*}
that is the solution $(\ref{solution_moyal})$.
%%%%%%%%%%%%%%%%%%%%%%%%%%%%%%%%%%%%%%%%%%%%%%%
%%%%%%%%%%%%%%%%%%%%%%%%%%%%%%%%%%%%%%%%%%%%%%%
%%%%%%%%%%%%%%%%%%%%%%%%%%%%%%%%%%%%%%%%%%%%%%%%
%%%%%%%%%%%%%%%%%%%%%%%%%%%%%%%%%%%%%%%%%%%%%%%%%

\section{The Wigner equation}\label{sec43new}
\subsection{From the Moyal to the Wigner equation}\label{431}

Recall the Cauchy problem of the Schr\"{o}dinger equation (Section $\ref{section22}$)
\begin{eqnarray}\label{schr_eq_5}
i\epsilon{\partial_t}u^{\epsilon}(x,t)=\widehat{H}^{\epsilon} u^{\epsilon}(x,t)
\end{eqnarray}
with initial data
\begin{eqnarray}\label{initialdata5}
u^{\epsilon}(x,t=0)=u_{0}^{\epsilon}(x)\in L^2(R)\quad ,
\end{eqnarray}
where $\widehat{H}^{\epsilon}:=-\frac{\epsilon ^2}{2}\partial_{xx}+V(x)$ is the Schr\"odinger operator.

For a pure state that satisfies the Schr\"odinger equation $(\ref{schr_eq_5})$, the density operator has the form $\widehat{\rho}_{u^\hb}=\ket{u^\hb}\bra{u^\hb}$, and its Schwartz kernel is $K^{\hb}(x,y,t)=u^{\hb}(x,t)\overline{u^\hb}(y,t)$. Then,  the Weyl symbol $\rho_{\W}$ of $\widehat{\rho}_{u^\hb}$ is given by the semiclassical Wigner transform $W^\hb[u^\hb](x,p,t)$ defined by $(\ref{semWigner})$, and it is real-valued function.
From $(\ref{moyaleq})$ it follows that $W^\hb[u^\hb]$   satisfies the  Moyal equation
\begin{eqnarray}\label {wignereq_moy}
i\epsilon {\partial_t} W^\hb[u^\epsilon](x,p,t)=H(x,p)\star_{\M} W^\hb[u^\epsilon](x,p,t)-W^\hb[u^\epsilon](x,p,t)\star_{\M} H(x,p) \ ,
\end{eqnarray}
with initial data
\begin{eqnarray}\label{wignereq_in}
W^\hb_{0}[u^\epsilon](x,p)=W^\hb[u^\epsilon](x,p,t=0)=W^\hb[u^{\hb}_{0}](x,p) \ ,
\end{eqnarray}
where $W^\hb[u^{\hb}_{0}](x,p)$ is the semiclassical Wigner transform of initial wavefunction $u_{0}^{\epsilon}$, defined by $(\ref{semWigner})$.

Equation $(\ref{wignereq_moy})$    is usually referred as the quantum Liouville equation 
in quantum mechanics \cite{MA}, but here we adopt the term {\it Wigner equation} which is preferred by the wave propagation community \cite{PR}.

The Wigner equation $(\ref{wignereq_moy})$ is formally solved by $(\ref{solution_moyal})$, i.e., by conjugating the initial data with the $\star$-unitary evolution operator
$U_{\star_{\M}}(x,p;t)=e^{\frac{i}{\epsilon}tH}_{\star_{\M}}$. The solution is given by
\begin{eqnarray}\label{wignersolution}
W^\hb[u^\epsilon](x,p,t)=U_{\star_{\M}}^{-1}(x,p,t)\star_{\M} W^\hb_{0}[u^\epsilon](x,p)\star_{\M} U_{\star_{\M}}(x,p,t) \ .
\end{eqnarray}

Using $(\ref{moyalbr})$ we can rewrite the Wigner equation $(\ref {wignereq_moy})$ in the form
\begin{equation}\label{wignereqi}
{\partial_t} W^\hb[u^\hb](x,p,t)+\mathcal{L}^{\hb} W^\hb[u^\hb](x,p,t)=0 \ ,
\end{equation}
where the operator  $\mathcal{L}^{\hb}$ is defined in terms of the Hamiltonian  $H(x,p)$ and the Moyal bracket
$(\ref{moyalbr})$ as follows
\begin{eqnarray}\label{opl}
\mathcal{L}^{\hb}\, \bullet:=-\frac{2}{\hb}\, H(x,p)\sin\left[\frac{\hb}{2}\left({\overleftarrow{\partial_
x}}\overrightarrow{\partial_p}-\overleftarrow{\partial_p}\overrightarrow{\partial_
x}\right)\right]\, \bullet= -\frac{2}{\hb}\,\{H(x,p),\bullet\}_{MB} \ .
\end{eqnarray}
For  later use in the study of the spectrum of $\mathcal{L}^{\hb}$, it is also necessary to consider the operator
\begin{eqnarray}\label{op_m}
\mathcal{M}^{\hb}\, \bullet:= H(x,p)\cos\left[\frac{\hb}{2}\left({\overleftarrow{\partial_
x}}\overrightarrow{\partial_p}-\overleftarrow{\partial_p}\overrightarrow{\partial_
x}\right)\right]\, \bullet=\{H(x,p),\bullet\}_{BB} \,
\end{eqnarray}
which corresponds to the Baker bracket.

Now, by using the Taylor series of  sine into $\mathcal{L}^{\hb}$ we formally obtain
\begin{eqnarray*}
&&\sin\left[\frac{\hb}{2}\left({\overleftarrow{\partial_
x}}\overrightarrow{\partial_p}-\overleftarrow{\partial_p}\overrightarrow{\partial_
x}\right)\right]\\
&=&
\sum_{n=0}^{\infty}\frac{(-1)^n}{(2n+1)!}\left(\frac{\hb}{2}\right)^{2n+1}\left({\overleftarrow{\partial_
x}}\overrightarrow{\partial_p}-\overleftarrow{\partial_p}\overrightarrow{\partial_
x}\right)^{2n+1}\\
&=&\frac{\hb}{2}\left({\overleftarrow{\partial_
x}}\overrightarrow{\partial_p}-\overleftarrow{\partial_p}\overrightarrow{\partial_
x}\right)+\sum_{n=1}^{\infty}\frac{(-1)^n}{(2n+1)!}\left(\frac{\hb}{2}\right)^{2n+1}\left({\overleftarrow{\partial_
x}}\overrightarrow{\partial_p}-\overleftarrow{\partial_p}\overrightarrow{\partial_
x}\right)^{2n+1} \ 
\end{eqnarray*}
and thus the operator $\mathcal{L}^{\hb}$ becomes 
\begin{eqnarray*}
\mathcal{L}^{\hb}\bullet =\left[-H(x,p)\left({\overleftarrow{\partial_
x}}\overrightarrow{\partial_p}-\overleftarrow{\partial_p}\overrightarrow{\partial_
x}\right)-\frac{2}{\hb}H(x,p)
\sum_{n=1}^{\infty}\frac{(-1)^n}{(2n+1)!}\left(\frac{\hb}{2}\right)^{2n+1}\left({\overleftarrow{\partial_
x}}\overrightarrow{\partial_p}-\overleftarrow{\partial_p}\overrightarrow{\partial_
x}\right)^{2n+1}\right]\bullet \ .
\end{eqnarray*}
Then, the Wigner equation $(\ref{wignereqi})$  takes the form  
\begin{eqnarray*}
&&\left({\partial_t} -\partial_{x}H(x,p)\partial_{p}+\partial_{p}H(x,p)\partial_{x}\right)W^\hb[u^\hb](x,p,t) \\
&=&H(x,p)\sum_{n=1}^{\infty}\frac{(-1)^n}{(2n+1)!}\left(\frac{\hb}{2}\right)^{2n}\left({\overleftarrow{\partial_
x}}\overrightarrow{\partial_p}-\overleftarrow{\partial_p}\overrightarrow{\partial_
x}\right)^{2n+1}W^\hb[u^\hb](x,p,t) \ .
\end{eqnarray*}
For the standard Hamiltonian $H(x,p)=\frac{p^2}{2}+V(x)$, by taking into account that it is quadratic in $p$,the above formula can be successfully written as 
\begin{eqnarray}\label{wigner_ser}
&&\left({\partial_t} -V'(x)\partial_{p}+p\partial_{x}\right)W^\hb[u^\hb](x,p,t)\nonumber \\
\nonumber \\
&=&\sum_{n=1}^{\infty}\frac{(-1)^n}{(2n+1)!}\left(\frac{\hb}{2}\right)^{2n}H(x,p)\left({\overleftarrow{\partial_
x}}\overrightarrow{\partial_p}-\overleftarrow{\partial_p}\overrightarrow{\partial_
x}\right)^{2n+1}W^\hb[u^\hb](x,p,t) \nonumber  \\
&=&\sum_{n=1}^{\infty}\frac{(-1)^n}{(2n+1)!}\left(\frac{\hb}{2}\right)^{2n}H(x,p)\left({\overleftarrow{\partial_
x}}\overrightarrow{\partial_p}\right)^{2n+1}W^\hb[u^\hb](x,p,t) \nonumber  \\ 
&=&\sum_{n=1}^{\infty}\frac{(-1)^n}{(2n+1)!}\left(\frac{\hb}{2}\right)^{2n}
{\partial_x}^{2n+1} H(x,p) {\partial_p}^{2n+1}W^\hb[u^\hb](x,p,t) \nonumber \\
&=&\sum_{n=1}^{\infty}\frac{(-1)^n}{(2n+1)!}\left(\frac{\hb}{2}\right)^{2n}
\frac{d^{2n+1}}{dx^{2n+1}} V(x) {\partial_p}^{2n+1}W^\hb[u^\hb](x,p,t) \ .
\end{eqnarray}

The  form $(\ref{wigner_ser})$ of the Wigner equation reveals the interesting structure of this equation. It is a combination of the classical transport (Liouville) operator in the left hand side, with a dispersion operator of infinite order in the right hand side. Formally speaking, this combination suggests that the phase space evolution of $W^\hb[u^\hb]$ results from the interaction between the classical transport of the Lagrangian manifold generated by the Hamiltonian and a non-local dispersion of energy out from the manifold towards the surrounding region of phase space. This picture is consistent with the fact that in the classical limit $\hb \rightarrow 0$ the dispersion mechanism formally disappears, the evolution of the Wigner measure is governed by classical mechanics and all quantum effects disappear.

In fact, as the small parameter $\hb$ tends to zero, the Wigner function $W^\hb[u^\hb](x,p,t)$ tends weakly to a positive measure $W^0(x,p,t)$ called  limit {\it Wigner measure}  \cite{LP}. The  Wigner measure satisfies the Liouville equation of classical mechanics,
$$\left({\partial_t} -V'(x)\partial_{p}+p\partial_{x}\right)W^0(x,p,t)=0\quad ,$$
with initial datum $W_{0}^{0}(x,p)$ being the limit Wigner measure of $W^\hb_0[u^\hb](x,p)=W^{\hb}[u^{\hb}](x,p,t=0)=W^{\hb}[u^{\hb}_{0}](x,p)$ (see also eq. $(\ref{diracwig})$ below, and the related discussion there).

\subsection{Spectrum of the Moyal and Baker brackets}\label{432}

Let $v^{\hb}_{n}(x)$, $v^{\hb}_{m}(x)$ be eigenfunctions of the  Schr\"odinger operator. Then, the density operator $(\ref{proj_op})$ is given by
\begin{eqnarray}\label{def_dens}
\widehat{\rho}_{nm}v(x):=\widehat{\rho}_{v^{\hb}_{n}v^{\hb}_{m}}v(x)=v^{\hb}_{n}(x)(v,v^{\hb}_{m})_{L^2(R_x)} \quad \mathrm{for\, \,  every}\, \,  v\in {L^2(R_x)} \ .
\end{eqnarray}
See eq. $(\ref{cross_den})$ for the definition of $\widehat{\rho}_{v^{\hb}_{n}v^{\hb}_{m}}$.

%the eigenvalue equations
% $\widehat{H}^{\hb} v^{\hb}_n (x)=E^{\hb}_n v^{\hb}_n(x)$ and %$ \widehat{H}^{\hb}\overline{v^{\hb}_m}(x)=E^{\hb}_m  \overline{v^{\hb}_m}(x)$, 

The operators $\widehat{\rho}_{nm}$ satisfy the eigen-equations
\begin{eqnarray}\label{eq_1}
\widehat{H}^{\hb}\ci \widehat{\rho}_{nm}\bullet=E^{\hb}_n \widehat{\rho}_{nm}\bullet 
\end{eqnarray}
and
\begin{eqnarray}\label{eq_2}
 \widehat{\rho}_{nm}\ci\widehat{H}^{\hb}\bullet=E^{\hb}_m  \widehat{\rho}_{nm}\bullet  \ , \ \ \mathrm{for \ all}
\ \ n,m=0,1,\ldots \ .
 \end{eqnarray}

Indeed, we have
\begin{eqnarray*}
\widehat{H}^{\hb} \ci\widehat{\rho}_{nm}v(x)=\widehat{H}^{\hb} v^{\hb}_n(x)(v,v^{\hb}_m)_{L^2(R_x)} =E^{\hb}_n v ^{\hb}_n(x)(v,v^{\hb}_m)_{L^2(R_x)} =E^{\hb}_n \widehat{\rho}_{nm}v(x)  \ ,
\end{eqnarray*}
and 
\begin{eqnarray*}
\widehat{\rho}_{nm}\ci \widehat{H}^{\hb} v(x)= v^{\hb}_n(x)(\widehat{H}^{\hb} v,v^{\hb}_m
)_{L^2(R_x)}&=&v^{\hb}_n(x)( v,\widehat{H}^{\hb}v^{\hb}_m )_{L^2(R_x)} \\
&=&E^{\hb}_m v^{\hb}_n(x)(v,v^{\hb}_m)_{L^2(R_x)}\\
& =& E^{\hb}_m\widehat{\rho}_{nm}v(x) \ ,
\end{eqnarray*}
by the self-adjointness of $\widehat{H}^{\hb}$.

Subtracting and adding  the equations $(\ref{eq_1})$ and $(\ref{eq_2})$ we get the equations
\begin{equation}\label{op2}
\left[\widehat{H}^{\hb},\widehat{\rho}_{nm}\right]=\widehat{H}^{\hb}\ci 
\widehat{\rho}_{nm}-\widehat{\rho}_{nm}\ci\widehat{H}^{\hb}=(E^{\hb}_n-E^{\hb}_m)
\widehat{\rho}_{nm} \qquad \textrm{  (commutator) }
\end{equation}
and 
\begin{equation}\label{op1}
\widehat{H}^{\hb} \ci\widehat{\rho}_{nm}+\widehat{\rho}_{nm}\ci\widehat{H}^{\hb}=(E^{\hb}_n+E^{\hb}_m)
\widehat{\rho}_{nm} \qquad \textrm{ (anti-commutator) }
\end{equation}
respectively, where $[\cdot\,  ,\cdot]$ denotes the commutator of operators.

Applying the  Weyl correspondence rule on $(\ref{op2})$ and $(\ref{op1})$ for the operators $\widehat{\rho}_{nm}$ and $\widehat{H}^{\hb}$, we obtain the following eigenvalue equations in the phase space,
\begin{equation}\label{eig1}
\mathcal{L}^{\hb}\wh_{nm}(x,p)=\frac{i}{\hb}(E^{\hb}_n-E^{\hb}_m)\wh_{nm}(x,p) \ , 
\end{equation}
\begin{equation}\label{eig2}
\mathcal{M}^{\hb}\wh_{nm}(x,p)=\frac12(E^{\hb}_n+E^{\hb}_m)\wh_{nm}(x,p) \ ,
\end{equation}
for the cross Wigner functions $\wh_{nm}:=W^\hb_{v_{m}^{\hb}}[v_{n}^{\hb}]$ (see, eq. $(\ref{cross_wig})$)
\begin{eqnarray}\label{cross_wigner_v}
\wh_{nm}(x,p)&:=&W^\hb_{v_{m}^{\hb}}[v_{n}^{\hb}](x,p) \nonumber \\
&=&(2\pi\hb)^{-1}\int_{R}^{} e^{-\frac{i}{\epsilon}p y} v_{m}^{\hb}\left(x+\frac{y}{2}\right)\overline{v_{n}^{\hb}}\left(x-\frac{y}{2}\right)\, dy \ .
\end{eqnarray}

The operators $\mathcal{L}^{\hb}$, $\mathcal{M}^{\hb}$  are defined by $(\ref{opl})$ and $(\ref{op_m})$ 
\begin{eqnarray*}%\label{op_l}
\mathcal{L}^{\hb}\, \bullet:=-\frac{2}{\hb}\, H(x,p)\sin\left[\frac{\hb}{2}\left({\overleftarrow{\partial_
x}}\overrightarrow{\partial_p}-\overleftarrow{\partial_p}\overrightarrow{\partial_
x}\right)\right]\, \bullet =-\frac{2}{\hb}\,\{H(x,p),\bullet\}_{MB}
\end{eqnarray*}
and 
\begin{eqnarray*}%\label{op_m}
\mathcal{M}^{\hb}\, \bullet:= H(x,p)\cos\left[\frac{\hb}{2}\left({\overleftarrow{\partial_
x}}\overrightarrow{\partial_p}-\overleftarrow{\partial_p}\overrightarrow{\partial_
x}\right)\right]\, \bullet=\{H(x,p),\bullet\}_{BB} \,
\end{eqnarray*}
Recall that the Moyal and the Baker bracket are given by $(\ref{moyalbr})$ and $(\ref{bakerbr})$, respectively.

It is known that the spectrum of the Wigner equation (quantum Liouville equation) can be determined from the spectrum of the corresponding Schr\"odinger operator $\widehat{H}^\hb$ (see, e.g., \cite{MA, SP}). In general, someone anticipates the formula 
$$\sigma(\mathcal{L}^\hb)=\left\{\frac{i}{\hb}(E^{\hb}-E^{\hb '})\, ,\quad E, \, E^{\hb '}\in \sigma(\widehat {H}^{\hb})\right\}
$$
to hold. In fact, this relation holds {\it for the discrete spectrum} 
\begin{equation}
\si_{p}(\mathcal{L}^\hb)=\left\{\frac{i}{\hb}(E_{n}^{\hb}-E_{m}^{\hb})\, ,\quad E_{n}^{\hb},\,  E_{m}^{\hb}\in \sigma_p(\widehat {H}^{\hb})\right\} \ . 
\end{equation}
A similar formula holds for the point spectrum of the cosine braket operator $\mathcal{M}^\hb$ , that is 
\begin{equation}
\si_{p}(\mathcal{M}^\hb)=\left\{\frac{1}{2}(E_{n}^{\hb}+E_{m}^{\hb})\, ,\quad E_{n}^{\hb},\,  E_{m}^{\hb}\in \sigma_p(\widehat {H}^{\hb})\right\} \ . 
\end{equation}

However, these formulae are not in general true for the absolutely and singular continuous spectrum. These spectral questions have been studied first by H.Spohn \cite{SP} and later by I.Antoniou et al. \cite{ASS}, who have proved the negative result
$$
\si_{sc,ac}(\mathcal {L}^{\hb})\neq \left\{ \frac{i}{\hb}(E^{\hb}-E^{\hb '})\, ,\quad E^{\hb},\,  E^{\hb '}\in \sigma_{sc,ac}(\widehat {H}^{\hb})\right\}\quad ,
$$
where $\si_{sc,ac}$ denote the singular and absolutely continuous spectrum, respectively.

Here we avoid the complications arising from the continuous spectrum (although this arises in the most interesting cases of scattering problems), since we consider operators $\widehat{H}^\hb$ with purely discrete spectrum $\si(\widehat{H}^{\hb})=\si_{p}(\widehat{H}^{\hb})$ , therefore $\si(\mathcal{L}^{\hb})=\si_{p}(\mathcal{L}^{\hb})$ and  $\si(\mathcal{M}^{\hb})=\si_{p}(\mathcal{M}^{\hb})$ . When the potential $V(x)\in L_{loc}^{1}(R_x)$ is bounded below and $\lim _{|x|\rightarrow \infty}V(x)=\infty$ it is known that $\widehat{H}^{\hb}$ has purely discrete spectrum and therefore the operators $\mathcal{L}^{\hb}$, 
$\mathcal{M}^{\hb}$ have also purely discrete spectrum. 

The functions $\{W_{nm}^{\hb}\}_{n,m=0,1,\ldots}$ (eq. $(\ref{cross_wigner_v})$) form a complete orthonormal basis in $L^2(R_{xp}^{2})$ and they are common eigenfunctions of operators 
$\mathcal{L}^{\hb}$ and $\mathcal{M}^{\hb}$, i.e., they satisfy the eigenvalue problems (\ref{eig1}) and (\ref{eig2}), respectively. 

As it has been shown by certain examples in  \cite{CFZ1, KP}, for the direct computation of Wigner functions in the phase space, we need both eigenvalue problems, (\ref{eig1}) and (\ref{eig2}). However, it seems that there is no evolution equation in phase space which  corresponds to the eigenvalue equation  (\ref{eig2}), which could be deduced from  configuration space, as it happens for the quantum Liouville equation. This means that (\ref{eig2}) cannot be derived naturally from some initial value problem for the Wigner function. In an attempt to derive  the second eigenvalue equation directly from the phase space, D.B.Fairlie \& C.A.Manogue  \cite{FMan} augmented the variables of the Wigner function by introducing an imaginary time  $s$. This leads to a second initial value problem, with time derivative $i\partial_{s}$ and space operator  $\mathcal{M}^{\hb}$, for the extended Wigner function. The study of this equation is still open.

\subsection{From the Schr\"odinger to the Wigner equation}\label{433}

%The Wigner function can be constructed from a time dependent wave function through the density operator. Then it should be possible to derive a differential equation for the Wigner function from the time dependent Schr\"odinger equation.

The evolution equation of the Wigner function can derived directly form the Schr\"odinger equation using the Wigner transform formalism in a operational way.

If $u^{\epsilon}(x,t)$ is the solution of the  initial-value problem $(\ref{schr_eq_5})$, $(\ref{initialdata5})$ for the Schr\"{o}dinger equation,   then, $\uh\left(x+ \epsilon y/2,t\right) \ , \overline{\uh}\left(x-\epsilon y/2,t\right)$
obviously satisfy the equations

\begin{eqnarray*}
&&\Biggl( i\epsilon\partial_t u^{\epsilon}\left(x+\frac{\hb y}{2},t\right)\Biggr) \overline{\uh}\left(x-\frac{\hb y}{2},t\right)\\
&=& -\frac{\epsilon^2}{2}\partial_{xx}u^{\epsilon}\left(x+\frac{\hb y}{2},t\right)
 \overline{\uh}\left(x-\frac{\hb y}{2},t\right)
+V^{\hb}\left(x+\frac{\hb y}{2}\right)u^{\epsilon}\left(x+\frac{\hb y}{2},t\right)\overline{\uh}\left(x-\frac{\hb y}{2},t\right)\ ,
\end{eqnarray*}
and
\begin{eqnarray*}
&&\Biggl(-i\epsilon\partial_t \overline{\uh}\left(x-\frac{\hb y}{2},t\right)\Biggr)u^{\epsilon}\left(x+\frac{\hb y}{2},t\right)\\
&=&\Biggl(-\frac{\epsilon^2}{2}\partial_{xx}
 \overline{\uh}\left(x-\frac{\hb y}{2},t\right)\Biggr)u^{\epsilon}\left(x+\frac{\hb y}{2},t\right)
 +V^{\hb}\left(x-\frac{\hb y}{2}\right)u^{\epsilon}\left(x+\frac{\hb y}{2},t\right)\overline{\uh}\left(x-\frac{\hb y}{2},t\right) \ .
\end{eqnarray*}

Subtracting the last two equations we take
\begin{eqnarray}\label{wignertr_schr}
&&i\hb\left[\partial_t u^{\epsilon}\left(x+\frac{\hb y}{2},t\right)\overline{\uh}\left(x-\frac{\hb y}{2},t\right)+
\partial_t \overline{\uh}\left(x-\frac{\hb y}{2},t\right)u^{\epsilon}\left(x+\frac{\hb y}{2},t\right)
\right]=\nonumber \\ 
&& -\frac{\hb ^2}{2}\left[ \partial_{xx}u^{\epsilon}\left(x+\frac{\hb y}{2},t\right)
 \overline{\uh}\left(x-\frac{\hb y}{2},t\right)-
\partial_{xx}
 \overline{\uh}\left(x-\frac{\hb y}{2},t\right)u^{\epsilon}\left(x+\frac{\hb y}{2},t\right)\right] \nonumber \\
& & +\left[V^{\hb}\left(x+\frac{\hb y}{2}\right)-V^{\hb}\left(x-\frac{\hb y}{2}\right)\right]u^{\epsilon}\left(x+\frac{\hb y}{2},t\right)\overline{\uh}\left(x-\frac{\hb y}{2},t\right)\ .
 \end{eqnarray}
 
Let  $w^{\hb}$ be  the two-point correlation function
\begin{equation}
w^{\hb}(x,y,t)=u^{\epsilon}\left(x+\frac{\hb y}{2},t\right)\overline{\uh}\left(x-\frac{\hb y}{2},t\right)  \ .
\end{equation}
Then, we have
$$   
\partial_x   
w^{\hb}=\partial_x   u^{\epsilon}\left(x+\frac{\hb y}{2},t\right)\overline{\uh}\left(x-\frac{\hb y}{2},t\right)+u^{\epsilon}\left(x+\frac{\hb y}{2},t\right)\partial_ x\overline{\uh}\left(x-\frac{\hb y}{2},t\right)
$$   
and  
\begin{eqnarray*}   
\partial_{yx}w^{\hb}=\frac{\hb}{2}\left[\partial_{xx} u^{\epsilon}\left(x+\frac{\hb y}{2},t\right)\overline{\uh}\left(x-\frac{\hb y}{2},t\right)-u^{\epsilon}\left(x+\frac{\hb y}{2},t\right)\partial_ {xx}\overline{\uh}\left(x-\frac{\hb y}{2},t\right)\right]\ ,
\end{eqnarray*}  
and the equation $(\ref{wignertr_schr})$ is written as  
\begin{eqnarray} 
i\hb\partial_t   
w^{\hb}=-\hb\partial_{yx} 
w^{\hb}+\left[V^{\hb}\left(x+\frac{\hb y}{2}\right)-V^{\hb}\left(x-\frac{\hb y}{2}\right)\right]w^{\hb} \quad .   
\end{eqnarray}  
Multiplying the last equation by ${e^{-ipy}}/2\pi$ , and integrating with respect to $y$, we get 
\begin{eqnarray*}  
&&i\hb\partial_t\left[\frac{1}{2\pi}\int_{R}e^{-ipy}w^{\hb}(x,y,t)\, dy\right]=\\
&&-\hb  
\partial_x\left[\frac{1}{2\pi}\int_{R}e^{-ipy}\partial_y w^{\hb}(x,y,t)\, dy\right]\\
&&
 +\frac{1}{2\pi}\int_{R}e^{-ipy}\left[V^{\hb}\left(x+\frac{\hb y}{2}\right)-V^{\hb}\left(x-\frac{\hb y}{2}\right)\right]w^{\hb}(x,y,t)\, dy \ .
\end{eqnarray*}
Assuming now that $u^{\hb}(x,t)$ decays fast enough as $|x| \rightarrow \infty$, we have   
$$   
\lim_{|y|\rightarrow+\infty}(e^{-ipy}w^{\hb})=0 \ ,   
$$   
and using the definition $(\ref{semWigner})$ of the Wigner transform 
$$   
W^\hb[u^\hb](x,p,t)=\frac{1}{2\pi}\int_{R}e^{-ipy}w^{\epsilon}(x,y,t)\, dy \ ,   
$$   
we derive that the  $W^\hb[u^\hb](x,p,t)$ satisfies the equation   
\begin{equation}  \label{wignereq} 
\partial_t W^\hb[u^\hb](x,p,t)=-p\, \partial_x W^\hb[u^\hb](x,p,t)-Z^{\hb}(x,p)*_p W^\hb[u^\hb](x,p,t) \ ,
\end{equation}   
where   
\begin{eqnarray}\label{ZETA}
Z^{\hb}(x,p):=\frac{1}{i\hb}\frac{1}{2\pi}\int_{R}e^{-ipy}\left(V^{\hb}\left(x+\frac{\hb y}{2}\right)-V^{\hb}\left(x-\frac{\hb y}{2}\right)\right) \, dy  \ .
\end{eqnarray}
The equation $(\ref{wignereq})$ is an integro-differential form of the {\it{Wigner equation}}. Indeed, by expanding the potential in $(\ref{ZETA})$ in Taylor series and integrating formally the series, we get the equation $(\ref{wigner_ser})$, that is,
\begin{eqnarray}\label{wignerschr}
\left(\partial_{t}+p{\partial_x}-V'(x){\partial_p}\right)W^{\epsilon}[u^\hb](x,p,t)=\Biggl(\sum_{n=1}^{\infty}\alpha_{n}\epsilon^{2n}V^{(2n+1)}(x){\partial_p}^{2n+1}\Biggr)W^{\epsilon}[u^\hb](x,p,t)\nonumber\\
\end{eqnarray}
where $\alpha_n = (-1)^n/(2n+1)!\, 2^{2n}$ and $V^{(2n+1)} (x) =\frac{d^{2n+1}}{dx^{2n+1}}V(x)$ , $n=1,2,\ldots$ .

\chapter{Eigenfunction series solution of the Wigner equation}\label{ch5}

In this section we construct an eigenfunction series expansion of the solution $W^{\hb}[u^\epsilon]$  of the initial value problem (eqs. $(\ref{wignereq_moy})$, $(\ref{wignereq_in})$)

\begin{eqnarray}\label{wignereq_moy_1}
&&i\epsilon {\partial_t} W^\hb[u^\epsilon](x,p,t)=H(x,p)\star_{\M} W^\hb[u^\epsilon](x,p,t)-W^\hb[u^\epsilon](x,p,t)\star_{\M} H(x,p) \ ,   \\
\nonumber \\
&&W^\hb[u^\epsilon](x,p,t=0)=W^\hb_{0}[u^\epsilon](x,p):=W^\hb[u^{\hb}_{0}](x,p)  \label{wignereq_in_1}\ .
\end{eqnarray}

Here $u^{\epsilon}(x,t)$ is the solution of the  initial-value problem $(\ref{schr_eq_5})$, $(\ref{initialdata5})$ for the Schr\"{o}dinger equation, and recall that we have assumed that the Schr\"odinger operator $\widehat{H}^{\epsilon}=-\frac{\epsilon ^2}{2}\frac{d^2}{dx^2}+V(x)$ has purely discrete spectrum.

\section{Separation of variables and eigenfunctions}\label{section52}

Since equation $(\ref{wignereq_moy_1})$ is linear and of the Schr\"odinger-type, and moreover we know some results for the eigenfunctions of $\mathcal{L}^{\hb} $ (see (\ref{opl}) ), we attempt to construct the expansion of $W^{\hb}[u^\epsilon]$ by the method of  separation of variables.

Thus,  we look for a solution of the form 
\begin{equation}\label{wigner_sep}
W^\hb[u^\epsilon](x,p,t)=T^{\hb}(t)\Psi^{\hb}(x,p) \ . 
\end{equation}
Then, since the Moyal product is time independent, the equation $(\ref{wignereq_moy_1})$ becomes 
\begin{eqnarray}
i\hb {T^{\hb '}}(t)\Psi^\hb(x,p)=T^{\hb}(t)\left(H(x,p)\star_{\M}\Psi^\hb(x,p)-\Psi^\hb(x,p)\star_{\M}H(x,p)\right) \ .
\end{eqnarray}
Dividing this equation by $T^{\hb}(t)\Psi^{\hb}(x,p)$ and assuming that  $\Psi^{\hb}$ is not zero, we get the equations 
\begin{eqnarray}
i\hb T^{\hb '}(t)={\mathcal{E}}^\hb T^{\hb}(t)\quad ,
\end{eqnarray}
and
\begin{eqnarray}\label{eq_spatial}
 H(x,p)\star_{\M} \Psi^{\epsilon}(x,p) -\Psi^{\epsilon}(x,p)\star_{\M} H(x,p)={\mathcal{E}}^{\hb}\Psi^{\hb}(x,p)
\end{eqnarray}
where ${\mathcal{E}}^{\hb}$ is the separation constant. The first equation has the solution   ${T^\hb}(t)=e^{-\frac{i}{\hb}{\mathcal{E}}^{\hb}t}$ up to a multiplicative constant. 

To deal with the second equation $(\ref{eq_spatial})$ , we use the properties $(\ref{starformula_1})$ and $(\ref{starformula_2})$ for the Moyal product, and we rewrite it in the form

\begin{eqnarray}
H\left(x+\frac{i\epsilon}{2}\overrightarrow{\partial_ p},p-\frac{i\epsilon}{2}\overrightarrow{\partial _x}\right)\Psi^{\epsilon}(x,p)-\Psi^{\epsilon}(x,p)H\left(x-\frac{i\epsilon}{2}\overleftarrow{\partial_ p},p+\frac{i\epsilon}{2}\overleftarrow{\partial_ x}\right)={\mathcal{E}}^{\hb}\Psi^{\hb}(x,p) \ ,
\end{eqnarray}
which is then obviously written in the usual form of  eigenvalue equation
\begin{eqnarray}\label{phase_sp_eq}
\left[H\left(x+\frac{i\epsilon}{2}\overrightarrow{\partial_ p},p-\frac{i\epsilon}{2}\overrightarrow{\partial _x}\right)-H\left(x-\frac{i\epsilon}{2}\overrightarrow{\partial_ p},p+\frac{i\epsilon}{2}\overrightarrow{\partial_ x}\right)\right]\Psi^{\hb}(x,p)={\mathcal{E}}^{\hb}\Psi^{\hb}(x,p) \ .
\end{eqnarray}
We introduce the operator (see Appendix $\ref{appendix_a}$)
\begin{eqnarray*}
\mathbb{H}^\hb:=H\left(x+\frac{i\epsilon}{2}\overrightarrow{\partial_ p},p-\frac{i\epsilon}{2}\overrightarrow{\partial _x}\right) \ \ \left( \mathrm{resp.} \ \ \overline{\mathbb{H}^\hb}=H\left(x-\frac{i\epsilon}{2}\overrightarrow{\partial_ p},p+\frac{i\epsilon}{2}\overrightarrow{\partial_ x}\right) \right)\ .
\end{eqnarray*}
By $(\ref{moyalbr})$, $(\ref{bakerbr})$ and $(\ref{starexpdecomp})$ we get 
\begin{eqnarray*}
\mathbb{H}^\hb=\mathcal{M}^{\hb}-i\frac{\hb}{2}\mathcal{L}^{\hb} \ \ \left( \mathrm{resp.}\ \  \overline{\mathbb{H}^\hb}=\mathcal{M}^{\hb}+i\frac{\hb}{2}\mathcal{L}^{\hb} \right)\ ,
\end{eqnarray*}
where operators $\mathcal{L}^{\hb}$, $\mathcal{M}^{\hb}$  are defined by $(\ref{opl})$ and $(\ref{op_m})$. 

Then, the eigenvalue equation $(\ref{phase_sp_eq})$ becomes
 \begin{eqnarray*}
 (\mathbb{H}^\hb-\overline{\mathbb{H}^\hb})\Psi^{\hb}(x,p)=-i\hb\mathcal{L}^{\hb}\Psi_{}^{\hb}(x,p)={\mathcal{E}}^{\hb}\Psi^{\hb}(x,p) \ .
 \end{eqnarray*}
Therefore we have transformed  $(\ref{eq_spatial})$ to the following eigenvalue equation for $\mathcal{L}^{\hb}$,
\begin{eqnarray}
-i\hb\mathcal{L}^{\hb}\Psi_{}^{\hb}(x,p)={\mathcal{E}}^{\hb}\Psi_{}^{\hb}(x,p)\ .
\end{eqnarray}
 
According to the results of Section $\ref{432}$ (eq. $(\ref{eig1})$), the eigenvalues of $\mathcal{L}^{\hb}$ are
$\frac{i}{\hb}(E_{n}^\hb-E_{m}^\hb) $, $n,m=0,1, \ldots$ , and therefore the eigenvalues ${\mathcal{E}}^{\hb}$ take the values $\mathcal{E}_{nm}^{\hb}:=E_{n}^{\hb}-E_{m}^{\hb}$ ,  and the corresponding eigenfunctions are $\Psi_{nm}^{\epsilon}(x,p)=\wh_{nm}(x,p)$ .

{\it It is now natural to ask  if there exist any other eigenfunctions for $\mathcal{L}^{\hb}$}.
The spatial equation $(\ref{eq_spatial})$  becomes 
\begin{eqnarray}\label{stat_eq_phs}
H(x,p)\star_{\M} \Psi_{nm}^{\epsilon}(x,p) -\Psi_{nm}^{\epsilon}(x,p)\star_{\M} H(x,p)=(E_{n}^{\hb}-E_{m}^{\hb})\Psi_{nm}^{\epsilon}(x,p)\ .
\end{eqnarray}
We recall from Section $\ref{432}$ that if $v^{\hb}_{n}(x)$, $v^{\hb}_{m}(x)$ are eigenfunctions of the  Schr\"odinger operator $\widehat{H}^{\hb}$, the cross Wigner functions $\wh_{nm}(x,p):=W^\hb_{v_{m}^{\hb}}[v_{n}^{\hb}](x,p)$ (see eq. $(\ref{cross_wigner_v})$), obey the 
eigenvalue equation  $(\ref{eig1})$ for the operator $\mathcal{L}^{\hb}$ .

For this, we rewrite  
$(\ref{stat_eq_phs})$ in the form
\begin{eqnarray*}
\left(H(x,p)\star_{\M} \Psi_{nm}^{\epsilon}(x,p) -E_{n}^{\hb}\Psi_{nm}^{\epsilon}(x,p)\right)-\left(\Psi_{nm}^{\epsilon}(x,p)\star_{\M} H(x,p)-E_{m}^{\hb}\Psi_{nm}^{\epsilon}(x,p)\right)=0
\end{eqnarray*}
and we make the arbitrary, although plausible, assumption that the  equations
\begin{eqnarray}\label{ass1}
H(x,p)\star_{\M} \Psi_{nm}^{\epsilon}(x,p) =E_{n}^{\hb}\Psi_{nm}^{\epsilon}(x,p)
\end{eqnarray}
and 
\begin{eqnarray}\label{ass2}
\Psi_{nm}^{\epsilon}(x,p)\star_{\M} H(x,p)=E_{m}^{\hb}\Psi_{nm}^{\epsilon}(x,p)
\end{eqnarray}
must hold simultaneously in order to $(\ref{stat_eq_phs})$  holds.

On the basis of  Theorem $(\ref{mainresult})$ and of Corollary $(\ref{corollary})$ in  the Appendix \ref{appendix_a}, equation $(\ref{ass1})$ implies that the phase space eigenfunctions are given by the cross Wigner functions 
$W_{nm}^{\epsilon}(x,p)$ . 

But on the other hand, the equation $(\ref{ass2})$ is valid for $\Psi_{nm}^{\epsilon}(x,p)=W_{nm}^{\epsilon}(x,p)$ , by the formula $(\ref{starexp})$ for the star exponential. Indeed,  we have that
$$\overline{W_{nm}^{\hb}\star_{\M}H}=\overline{W_{nm}^{\hb}\, e^{\frac{i\hb}{2}\left(
\overleftarrow{\partial _x}
\overrightarrow{\partial _p}-
\overleftarrow{\partial _p}
\overrightarrow{\partial _x}\right)}H}=H\,  e^{\frac{i\hb}{2}\left(
\overleftarrow{\partial _x}
\overrightarrow{\partial _p}-
\overleftarrow{\partial _p}
\overrightarrow{\partial _x}\right)}W_{mn}^{\hb}=H\star_{\M} W_{mn}^{\hb} \ ,$$ 
that is
\begin{eqnarray}
\overline{W_{nm}^{\hb}(x,p)\star_{\M}H(x,p)}=H(x,p)\star_{\M}W_{mn}^{\hb}(x,p)\ .
\end{eqnarray}

Since the Wigner functions $W_{mn}^{\hb}(x,p)$ satisfy the eigenvalue equation (Appendix \ref{appendix_a}; also \cite{CFZ1})
$$H(x,p)\star_{\M}W_{mn}^{\hb}(x,p)=E_{m}W_{mn}^{\hb}(x,p) \ ,$$  
we conclude that
$\overline{W_{nm}^{\hb}\star_{\M}H}=E_{m}W_{mn}^{\hb}(x,p)$ and  we finally get the equation
$$W_{nm}^{\hb}\star_{\M}H=E_{m}W_{nm}^{\hb}(x,p)\quad .$$

Then,  by $(\ref{wigner_sep})$, and that above discussion for the eigenfunctions of $(\ref{eq_spatial})$,  
the solution of the Wigner equation $(\ref{wignereq_moy_1})$ can be expanded in {\it the eigenfunction  series}
\begin{eqnarray}\label{wigner_exp}
W^{\hb}[u^\epsilon](x,p,t)=\sum_{n=0}^{\infty}\sum_{m=0}^{\infty} c_{nm}^{\hb}(0)\, e^{-\frac{i}{\hb}(E_{n}^{\hb}-E_{m}^{\hb})t} W_{nm}^{\hb}(x,p)\ .
\end{eqnarray}
The coefficients are given by 
\begin{eqnarray*}
c_{nm}^{\hb}(0)=
(W_{0}^{\hb}[u^\hb],W_{nm}^{\hb})_{L^2(R_{xp}^2)}
 \ , \ \ n,m=0,1,\ldots 
\end{eqnarray*}
where $W^\hb_{0}[u^\epsilon](x,p):=W^\hb[u^{\hb}_{0}](x,p)$ is the initial Wigner function.

 Note that by the isometry of Wigner transform, the coefficients can be expressed also in terms of the eigenfunctions of the
 Schr\"odinger operator as follows 
 \begin{eqnarray*}
c_{nm}^{\hb}(0)=(2\pi\hb)^{-1}(u_{0}^{\hb},v_{n}^{\hb})_{L^2(R_x)}\overline{(u_{0}^{\hb},v_{m}^{\hb})}_{L^2(R_x)} \ .
\end{eqnarray*}
%%%%%%%%%%%%%%%%%%%%%%%%%%%%%%%%%%%%%%%%%%%%%%%%%%%%%
%%%%%%%%%%%%%%%%%%%%%%%%%%%%%%%%%%%%%%%%%%%%%%%%%%%%%%%
%%%%%%%%%%%%%%%%%%%%%%%%%%%%%%%%%%%%%%%%%%%%%%%%%%%%%%%%
\section{Alternative derivations of eigenfunction series}\label{sec52}

\subsection{Star-product representation and eigenfunction series}

An alternative derivation of the eigenfunction series expansion $(\ref{wigner_exp})$ of the solution of the initial value problem $(\ref{wignereq_moy_1})$-$(\ref{wignereq_in_1})$ starts from the formal star representation $(\ref{wignersolution})$, that is
\begin{eqnarray*}
W^{\hb}[u^\epsilon](x,p,t)=U_{\star_{\M}}^{-1}(x,p,t)\star_{\M} W^{\hb}_0[u^\epsilon](x,p)\star_{\M} U_{\star_{\M}}(x,p,t)\quad .
\end{eqnarray*}
Indeed,  by expanding the initial datum $(\ref{wignereq_in_1})$ in terms of the eigenfunctions $W_{nm}^{\hb}$, we have
\begin{eqnarray}\label{wigner_exp_star}
W^{\hb}[u^\epsilon](x,p,t)=
U_{\star_{\M}}^{-1}(x,p,t)\star_{\M}\sum_{n=0}^{\infty}\sum_{m=0}^{\infty}c_{nm}^{\hb}(0)W_{nm}^{\hb}(x,p) \star_{\M} U_{\star_{\M}}(x,p,t) \ .
\end{eqnarray}
By using the star-resolution of identity \cite{CFZ2}, we expand $U_{\star_{\M}}(x,p,t)$ in terms of the eigenfunctions \footnote{Note that this property holds for every star product.}
\begin{eqnarray*}
&&U_{\star_{\M}}(x,p,t)=e^{\frac{it}{\epsilon}H}_{\star_{\M}}=e^{\frac{it}{\epsilon}H}_{\star_{\M}}\star_{\M} 1=
e^{\frac{it}{\epsilon}H}_{\star_{\M}}\star_{\M} 2\pi\hb\sum_{n=0}^{\infty}W_{nn}^{\hb}(x,p)=2\pi\hb\sum_{n=0}^{\infty}e^{itE_{n}/\hb}W_{nn}^{\hb}(x,p) \ ,
\end{eqnarray*}
 and we rewrite the double series $(\ref{wigner_exp_star})$ in the form
\begin{eqnarray}\label{wigner_exp_star_1}
W^{\hb}[u^\epsilon](x,p,t)&=&
(2\pi\hb)^{2}\sum_{n=0}^{\infty}\sum_{m=0}^{\infty}\sum_{k=0}^{\infty}\sum_{\ell=0}^{\infty}c_{nm}^{\hb}(0)e^{\frac{i}{\hb}t(E_{\ell}^{\hb}-E_{k}^{\hb})}  \nonumber \\
&&\cdot\ 
W_{kk}^{\hb}(x,p)\star_{\M} W_{nm}^{\hb}(x,p)\star_{\M} W_{\ell\ell}^{\hb}(x,p) \ .
\end{eqnarray}
Finally, by the projection formula \cite{CFZ2}
\begin{eqnarray*}
&&(2\pi\hb) W_{mn}^{\hb}(x,p)\star_{\M} W_{k\ell}^{\hb}(x,p)=\delta_{nk} W_{m\ell}^{\hb}(x,p) \ ,
\end{eqnarray*}
 we trivially sum a double series in  $(\ref{wigner_exp_star_1})$, and we get $(\ref{wigner_exp})$.

%%%%%%%%%%%%%%%%%%%%%%%%%%%%%%%%%%%%%%%%%%%%%%%%%%%%%%%%%%

\subsection{From configuration to phase space eigenfunction series}

A second alternative derivation of  the expansion $(\ref{wigner_exp})$, is to apply termwise the Wigner transform on the eigenfunction expansion of the Schr\"odinger equation $(\ref{schr_eq_5})$ which reads as follows (eq. $(\ref{serexpsch})$),
\begin{eqnarray}\label{serexp_sch}
u^{\epsilon}(x,t)=\sum_{n=0}^{\infty}c^{\epsilon}_{n}(0)v^{\epsilon}_{n}(x)e^{-\frac{i}{\epsilon}E^{\epsilon}_{n} t} \ .
\end{eqnarray}
The coefficients $c^{\epsilon}_{n}(0)=(u^{\epsilon}_{0},v^{\epsilon}_{n})_{L^{2}(R_x)}$  are the projections of initial wave function $u^{\epsilon}_{0}(x)$  (eq. $(\ref {initialdata5})$) onto the eigenfunctions  $v^{\epsilon}_{n}(x)$  of $\widehat{H}^\hb$.

Indeed, by operating with the semiclassical Wigner transform $(\ref{semWigner})$ onto the series $(\ref{serexp_sch})$ we obtain
\begin{eqnarray}
W^{\epsilon}[u^{\hb}](x,p,t)&=&W^{\epsilon}\left[\sum_{n=0}^{\infty}c^{\epsilon}_{n}(0)v^{\epsilon}_{n}(x)e^{-\frac{i}{\epsilon}E^{\epsilon}_{n} t}\right]\nonumber\\
&=&\sum_{n=0}^{\infty}\sum_{m=0}^{\infty}c^{\epsilon}_{nm}(t)W^{\epsilon}_{v_{m}^{\hb}}[v_{n}^{\hb}](x,p)\nonumber\\
&=&\sum_{n=0}^{\infty}\sum_{m=0}^{\infty}c^{\epsilon}_{nm}(t)W^{\epsilon}_{nm}(x,p)\quad ,\label{eigenexp}
\end{eqnarray}
where $W^{\epsilon}_{nm}(x,p):=W^{\epsilon}_{v_{m}^{\hb}}[v_{n}^{\hb}](x,p)$ is the cross Wigner function 
of eigenfunctions $v_{n}^{\hb}(x)$ and $v_{m}^{\hb}(x)$ (see eq. $(\ref{cross_wigner_v})$) and 
\begin{eqnarray}
c^{\epsilon}_{nm}(t)=c^{\epsilon}_{nm}(0)e^{-\frac{i}{\epsilon}(E^{\epsilon}_n-E^{\epsilon}_m)t}=c^{\epsilon}_{n}(0)\overline {c^{\epsilon}_{m}}(0)e^{-\frac{i}{\epsilon}(E^{\epsilon}_n-E^{\epsilon}_m)t}\ .
\end{eqnarray}

%Separating the cross Wigner functions of $(\ref{eigenexp})$, we see that the eigenfunction series expansion of the Wigner function consists of a coherent term  and an incoherent term,
%\begin{eqnarray}\label{wignersol}
%W^{\epsilon}(x,p,t)=\sum_{n=0}^{\infty} \left[ |c^{\epsilon}_{n}(0)|^2 W^{\epsilon}_{n}(x,p) + %\sum_{m=0,m\neq n}^{\infty} c^{\epsilon}_{nm}(t)W^{\epsilon}_{nm}(x,p)\right]\, . 
%\end{eqnarray}

\section{Asymptotics of the phase-space eigenfunction series}\label{sec53}

The construction of an asymptotic solution of the Wigner equation, starting form the eigenfunction series $(\ref{wigner_exp})$, requires the  following ingredients:
\begin{itemize}
\item
Asymptotic approximations of the Wigner eiegnfucntions 
\begin{eqnarray*}
W^{\hb}_{nm}(x,p)=(2\pi\hb)^{-1}\int_{R}^{} e^{-\frac{i}{\epsilon}y p} v^\hb_{n}\left(x+\frac{y}{2}\right)\overline{v^\hb_{m}}\left(x-\frac{y}{2}\right) \, dy \,
\end{eqnarray*}

Semiclassical approximations of these eigenfunctions will be constructed in Chapter \ref{ch7}.

\item
Asymptotic approximations of the coefficients $c_{nm}^{\hb}(0)$ ,
\begin{eqnarray*}
c_{nm}^{\hb}(0)=
(W_{0}^{\hb}[u^\hb],W_{nm}^{\hb})_{L^2(R_{xp}^2)}
\end{eqnarray*}
for certain type of WKB initial data $u^{\epsilon}_{0}(x)$  (eq. $(\ref {initialdata5})$).

Approximations of these coefficients will be presented in Chapter \ref{ch7}.

\end{itemize}

%%%%%%%%%%%%%%%%%%%%%%%%%%SEMICLASSICAL WIGNER FUNCTIONS%%%%%%%%%%%%%%%%%%%%%%%%%%%

\chapter{Semiclassical Wigner functions}\label{chapter6}

\section{The Wigner transform}

For any smooth complex valued function $f(x)$ rapidly decaying at infinity, say
$f
\in \mathcal{S}(R)$ , the Wigner transform of $f$ is a
quadratic transform defined by
\begin{equation} \label{wig}
W(x,p):=W[f](x,p)=\frac{1}{2\pi}\int_{R}{e}^{-ip
y}\, f\left(x+\frac{y}{2}\right)\, \overline f\left(x-\frac{y}{2}\right)\, dy
\end{equation}
where $\overline{f}$ is the complex conjugate of $f$. The Wigner transform is defined in phase space $R^{2}_{xp}$, it is real and bilinear function.

%It is interesting to note that there are Wigner functions which are  non-negative everywhere. It is proved that the Wigner function of a pure quantum state is non-negative if and only if the state is Gaussian, \cite{HUD}.

Some of the most important properties of Wigner functions which are useful for the computation of physical quantities, not only in the theory of classical wave propagation but also in quantum mechanics are the following.

\begin{itemize}
\item 
The integral of $W(x,p)$ w.r.t. $p$ gives the squared
amplitude (energy density) of $f$,
\begin{equation}\label{energy_density}
\int_{R}W(x,p)\, dp=|f(x)|^2 \quad .
\end{equation}
\item
The first moment  of $W(x,p)$ w.r.t. to $p$ gives the
(energy) flux of $f$,
\begin{equation}
\int_{R}pW(x,p)\, dp=\frac{1}{2i}\left (f(x)\, \overline{f^{'}}(x)-
\overline{f}(x)\, f^{'}(x)\right )={\mathcal F}(x) \quad .
\end{equation}
\item
The $xp$-integral of $W(x,p)$ gives the total energy 
\begin{equation}
\int_{R}^{}\int_{R}^{}W(x,p)\, dxdp=||f||_{L^2} \quad .
\end{equation}
\end{itemize}

In the same way the {\it cross Wigner transform of two functions $f$ and $g$} is defined as 
\begin{equation}\label{wigcross}
W_{g}[f](x,p)=\frac{1}{2\pi}\int_{R}{e}^{-ip
y}\, f\left(x+\frac{y}{2}\right)\, \overline g\left(x-\frac{y}{2}\right)\, dy \quad .
\end{equation}

This is a bilinear mapping $W_{g}:S(R)\rightarrow S(R^2)$ , where $f\in S(R)$ and is anti-linear in $g$ as well. Moreover, same properties for the density and flux hold true, for details you can  see here \cite{GG}.

\section{Wigner transform of single-phase WKB functions}

\subsection{Scaled Wigner transform and Wigner measures}

As we have explained in Chapter \ref{chapter3}, in the case of high
frequency wave propagation, it is useful to use WKB wave functions
of the form
\begin{equation}\label{wkbfunc}
\psi^{\epsilon}(x)= A(x)\, e^{i{S(x)}/{\epsilon}} \quad ,
\end{equation}
where $S(x)$ is a real-valued and smooth phase, and $A(x)$ is a
real-valued and smooth amplitude of compact support or at least
rapidly decaying at infinity. 
In this chapter when we write $\psi^{\epsilon}(x)$ we mean a snapshot of wavefunction $\psi^{\epsilon}(x,t)$ , i.e. the wave at a given time $t=$const.

The Wigner transform of $\psi^{\epsilon}(x)$ 
\begin{equation}
W(x,p)= W[\psi^{\epsilon}](x,p) = \frac{1}{2\pi}
\int_{R}e^{-ipy}\, e^{\frac{i}{\epsilon}S(x+\frac y2)}\, A\left(x+\frac
y2\right)\, e^{-\frac{i}{\epsilon}S(x-\frac y2)}\, \overline{A}\left(x-\frac y2\right)\, dy \quad ,
\end{equation}
does not converge  to a nontrivial
limit, as $\epsilon \rightarrow 0$ . However, the
rescaled version of $W(x,p)$,  which will be referred in the sequel as the {\it semiclassical Wigner
transform} of $\psi^{\epsilon}$ ,
\begin{equation}\label{scaled_wig}
W^{\epsilon}(x,p)=\frac{1}{\epsilon}W\left(x,\frac{p}{\epsilon}\right) \
\end{equation}
converges weakly as $\epsilon\rightarrow 0$  to the limit Wigner
distribution (see \cite{PR, LP})
\begin{equation}\label{diracwig}
W^{0}(x,p)=[A(x)]^2\frac{1}{2\pi}\int_{R}
e^{-i(p-S'(x))y}\, dy=[A(x)]^2\delta(p-S'(x)) \quad ,
\end{equation}
which is a Dirac measure concentrated on the Lagrangian manifold
$p=S'(x)$ , associated with the phase of the WKB  wavefunction, and
it is the correct weak limit of $W^{\epsilon}$ (see, e.g., P.L.Lions \& T.Paul
\cite{LP}).
%Thus, from the limit Wigner measure we can recover the modulus of the amplitude $A$ and the derivative of the phase $S$ of the WKB approximation.

This result can be derived formally as follows. We rewrite $W^{\epsilon}$ in the form
$$
W^{\epsilon}(x,p)=\frac{1}{2\pi}\int_{R}e^{-ipy}\, A\left(x+\frac{\epsilon y}{2}\right
)\, \overline{A}\left(x-\frac{\epsilon y}{2}\right)\, e^{\frac{i}{\epsilon}\left[ S(x+\frac {\epsilon
y}{2})-S(x-\frac{\epsilon y}{2} )\right]}\, dy \ ,
$$
and we expand both $A\left(x \pm \frac{\epsilon
y}{2}\right )$ and $S\left(x \pm \frac{\epsilon y}{2}\right )$ in Taylor series about $x$  . Then, we have
\begin{eqnarray*}
A\left(x+\frac{\epsilon y}{2}\right )\, \overline{A}\left(x-\frac{\epsilon y}{2}\right)
&=&\left(A(x)+\frac{\epsilon}{2}yA^{'}(x)+\dots\right)\left(\overline{A}(x)-
\frac{\epsilon}{2}y\overline{A}^{'}(x)+\dots\right)\\
&=&A(x)\, \overline{A}(x)+ O(\epsilon)\\
&=&|A(x)|^2 + O(\epsilon)\\
&=&[A(x)]^2 + O(\epsilon) \quad ,
\end{eqnarray*}
and
\begin{eqnarray*}
 S\left(x+\frac {\epsilon y}{2}\right)-S\left(x-\frac{\epsilon y}{2}
\right )&=&\left(S(x)+\frac{\epsilon}{2}yS^{'}(x)+\frac{\epsilon ^2}{8}y^2S^{''}(x)+\dots
 \right)\\
 &&-
\left(S(x)-\frac{\epsilon}{2}yS^{'}(x)+\frac{\epsilon ^2}{8}y^2S^{''}(x)-\dots
 \right)\\
 &=&\epsilon y S^{'}(x) +O(\epsilon ^3) \quad .
\end{eqnarray*}
Retaining only terms of order $O(1)$ in $A$ and $O(\epsilon)$ in $S$, and
integrating the expansion termwise we obtain that $W^{\epsilon}(x,p)$
``converges formally " to $(\ref{diracwig})$. We will see later in this chapter, that similar result arises for the Wigner transform of multiphase WKB functions.

The above observations suggest that the semiclassical Wigner transform
\begin{eqnarray}\label{wige}
W^{\epsilon}(x,p)&=&\frac{1}{\epsilon}W\left(x,\frac{p}{\epsilon}\right) \nonumber\\
&=&\frac{1}{2\pi}\int_{R}e^{-ipy}\, \psi^{\epsilon}\left(x+\frac{\epsilon y}{2}\right)\,
\overline {\psi^{\epsilon}}\left(x-\frac{\epsilon y}{2}\right)\, dy  \quad ,
\end{eqnarray}
is the correct phase-space object for analysing high frequency
waves. Recall here that the semiclassical Wigner transform has already emerged as the Weyl symbol of the density operator (See eq.  $(\ref{semWigner})$) .

The Wigner function is not in general positive, except when $\psi^\hb$ is a Gaussian function (see \cite{LP}) but, it always becomes positive in the high frequency limit. 
For this reason it is not a pure probability distribution, although it has been  introduced as a substitute of such a function in quantum statistical mechanics \cite{Wig}. 
This is exactly the property that makes the Wigner function a powerful tool  for the study of wave and quantum interference phenomena.

%The scaled version for the evolved Wigner density  (it is defined in analogy to $(\ref{wige})$) is
%\begin{equation} \label{wig_time}
%W^{\hb}(x,p,t)=W[\psi^{\hb}](x,p,t)=\frac{1}{2\pi}\int_{R}{e}^{-ip
%y}\, \psi^{\hb}\left(x+\frac{\hb y}{2},t\right)\, \overline {\psi^{\hb}} \left(x-\frac{\hb y}{2},t\right)\, dy \, .
%\end{equation}

%%%%%%%%%%%%%%%%%%%%%%%%%%%%%%%%%%%%%%%%%%%%%%%%%%%%%%%%%%%%%%%%%%%%%%%%%
%%%%%%%%%%%%%%%%%%%%%%%%%%%%%%%%%%%%%%%%%%%%%%%%%%%%%%%%%%%%%%%%%%%%%%%%%

\subsection{Berry's semiclassical Wigner function}\label{section422}

Consider now the Wigner function \footnote{From now on, we use the symbol ${\mathcal{W}}^{\epsilon}$ to denote the semiclassical Wigner transform of a WKB function.}
\begin{eqnarray}\label{rescaled_wig}
{\mathcal{W}}^{\epsilon}(x,p)=\frac{1}{\pi\epsilon}\int_{R}^{}\ \psi^{\epsilon}(x+\si)\,
\overline{{\psi}^{\epsilon}}(x-\si)\, \ e^{-\frac{i}{\epsilon}2p\si}\, d\si
\end{eqnarray}
of the WKB wave function
\begin{equation}\label{wkb}
\psi^{\epsilon}(x)=A(x)\, e^{iS(x)/\epsilon} \ .
\end{equation}
We assume that the amplitude $A$ and the phase $S$ are smooth and real-valued functions, and that $S'(x)$ is globally convex.

Function $(\ref{rescaled_wig})$ is written in the form of the Fourier integral 
\begin{equation}\label{wigwkb}
{\mathcal{W}}^{\epsilon}(x,p)=\frac{1}{\pi\epsilon}\int_{R}^{}D(\si,x)\,
e^{i\frac{1}{\epsilon}F(\si,x,p)}\, d\si \quad ,
\end{equation}
where
\begin{equation}
D(\si,x)=A(x+\si)A(x-\si)
\end{equation}
is the Wigner amplitude, and
\begin{equation}\label{wigphase}
F(\si,x,p)=S(x+\si)-S(x-\si)-2p\si
\end{equation}
is the \emph{Wigner phase}. 

A uniform  asymptotic expansion of this integral has been for first time constructed by M.V.Berry  \cite{Ber} using the phase transformation developed by C.Chester, B.Friedman and F.Ursell \cite{CFU}. The formula has been derived in \cite{FM1} by using a uniform stationary formula in the form presented in \cite{Bor} (see Appendix  \ref{app_a}  for a brief presentation of this result), in a way that emphasises  the role of the Lagrangian curve $\Lambda =\{p=S'(x)\}$ of the WKB function and it is appropriate for handling the evolution of Wigner function in the corresponding time-dependent problem. The construction of the approximation goes as follows.

\begin{figure}[h]
\centering
\includegraphics[width=0.7\textwidth]{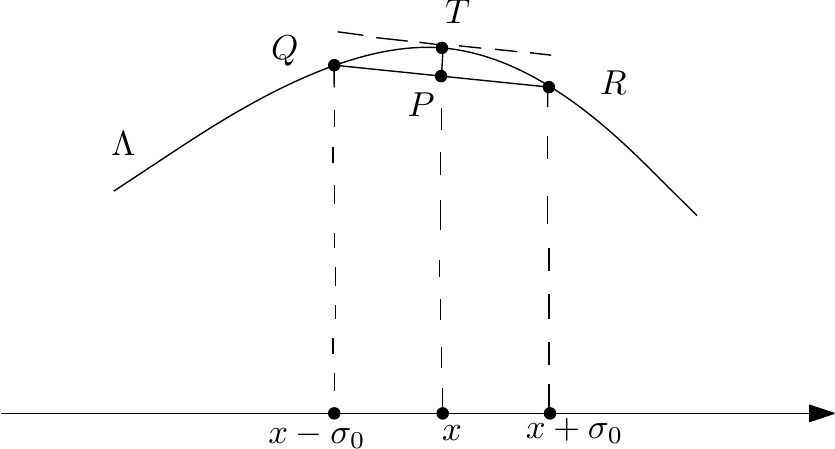}
\caption{{\it Lagrangian curve \& Berry's chord}}
\label{berry_ch}
\end{figure}

The critical  points of the
phase $F(\si,x,p)$, w.r.t. $\sigma$, are the roots of
\begin{equation}\label{alpha}
\partial_{\si}F(\si,x,p)=S'(x+\si)+S'(x-\si)-2p=0 \ .
\end{equation}

M.V.Berry has introduced  an invariant geometrical interpretation of this equation (Figure \ref{berry_ch}), by observing that 
$(\ref{alpha})$ has a pair of symmetric roots $\pm\si_{0}(x,p)$ such that the point $P=(x,p)$ be the middle of a chord $QR$ (Berry's chord) with endpoints $Q(x-\si_{0} \ , S'(x-\si_{0}))$ and $R(x-\si_{0}  \ , S'(x+\si_{0}))$
on the Lagrangian (``manifold") curve $\Lambda =\{p=S'(x)\}$ of the WKB function. 
As $P$ approaches toward $\Lambda$, the chord $QR$ becomes
to the tangent of $\Lambda$ and $\si_{0}(x,p)\rightarrow 0$ .
Therefore, the two stationary  points of $(\ref{alpha})$ coalesce to the double point $\si=0$ as $(x,p)$ moves towards $\Lambda$.

In fact, by
\begin{equation}
\partial_{\si\si}F(\si,x,p)=S''(x+\si)-S''(x-\si) \ ,
\end{equation} 
and
\begin{equation}
\partial_{\si\si\si}F(\si,x,p)=S'''(x+\si)+S'''(x-\si) \ ,
\end{equation}
we have 
\begin{equation}
\partial_{\si\si}F(\si=0,x,p)=0\, , \quad \partial_{\si\si\si}F(\si=0,x,p)=2S'''(x)\not=0 \ ,
\end{equation}
which assert that $\si=0$ is a double stationary point of $F$.

Since the ordinary stationary-phase formula (\ref{statphform}) fails when  $\partial_{\si\si}F(\si=0,x,p)=0$,  we must to apply the uniform stationary formula  (\ref{ap4}) of Appendix  \ref{app_a}.  For applying this formula, we need first  to identify the parameter $\alpha$, which controls the distance between the stationary points $\pm\si_{0}(x,p)$ of the Wigner phase. In order to do this, we expand $F$ in Taylor series about $\si=0$ ,
\begin{eqnarray*}
F(\si,x,p)&=& S(x)+\si S'(x)+\frac{\si^2}{2}S''(x)+\frac{\si^3}{6}
S'''(x)+...\\
&&- \left(S(x)-\si
S'(x)+\frac{\si^2}{2}S''(x)-\frac{\si^3}{6}S'''(x)+...\right)-2p\si\\
&=&-2(p-S'(x))\si+\frac{1}{3}S'''(x)\si^3+O(\si^5) \ .
\end{eqnarray*}
It becomes evident that for $P$ lying close enough to $\Lambda$,
the parameter $\alpha$ has to be identified as
\begin{equation}
\alpha=\alpha(x,p):=p-S'(x) \ ,
\end{equation}
since by
\begin{eqnarray*}
\partial_{\si\si}F(\si,x,p)= -2(p-S'(x)) + S'''(x)\si^2+O(\si^4) \ ,
\end{eqnarray*}
we easily see that $\si=0$ is a double stationary point for $p=S'(x)$ .

Then, for any fixed $x$, we rewrite the Wigner phase $F$ in the
form\begin{eqnarray}
F(\si,\alpha,x)&=& S(x+\si)-S(x-\si)-2\si(\alpha+S'(x))\nonumber\\
&=& \left (S(x+\si)-S(x-\si)-2\si S'(x)\right )-2\si\alpha \quad ,
\end{eqnarray}
and we have
\begin{equation}
\partial_{\si\si}F(\si=0,x,p)=0\quad , \quad  \partial_{\si\si\si}F(\si=0,x,p)=2S'''(x)\quad , \quad
F_{\si\alpha}(\si=0,\alpha,x)=-2 \neq 0\ .
\end{equation}
These are exactly the conditions  on the phase which are required
for applying the uniform asymptotic formula (\ref{ap4}).

It is important to note that for any fixed $x$ we think of $F(\si,x,p)$ as $F(\si,\alpha;x)$ where $x$ plays the role of parameter. This is necessary for applying the uniform approximation theory in Appendix  \ref{app_a}.

By the approximations (\ref{289})-(\ref{32}),   we further approximate  the various quantities entering (\ref{ap4}) as $\alpha\rightarrow 0$, as follows

\begin{eqnarray}\label{xiapprox}
\xi(x,p) \approx \left(S'''(x)\right)^{-1/3}\left(p-S'(x)\right)\label{zeta} \ ,
\end{eqnarray}

\begin{equation}\label{beta} 
\partial_{\si\si}F(\si=\si_{0},x,p)\approx 2\si_0(x,p)S'''(x)  \ ,
\end{equation}

\begin{eqnarray}
\si_{0}(x,p) \approx   \left(\ \frac{2}{S'''(x)}(p-S'(x))\ \right)^{1/2} \ ,
\end{eqnarray}

\begin{equation}
\partial_{\si\si}F(\si_{0},x,p)\approx 2\si_{0}S'''(x) =2 \left[\
\frac{2(p-S'(x))}{S'''(x)}\ \right]^{1/2}S'''(x) \ ,
\end{equation}
and 
\begin{equation}
\frac{\xi^{1/4}}{\mid \partial_{\si\si}F\mid ^{1/2}} =\left(\frac{2}{\mid
S'''(x)\mid} \right)^{1/3}\frac{1}{2^{5/6}} \quad  .
\end{equation}

Moreover, since 
$$D(\si_{o},x)=D(-\si_{0},x)\approx A^{2}(x)  \ ,$$ 
by using {(\ref{A0})}, {(\ref{B0})}
and the approximation $(\ref{zeta})$ of $\xi$, we get
\begin{equation}\label{delta}%\label{epsilon}
A_{0}\approx\frac{1}{2^{1/3}}\left(\frac{2}{\mid S'''(x)\mid}
\right)^{1/3} A^{2}(x) \ , \ \ \ B_{0} =0
\end{equation}
We also have $F(\si=0,\alpha=0,x)=0$.

Substituting the approximations $(\ref{xiapprox})$-$(\ref{delta})$ into the formula  (\ref{ap4}), and  keeping only the zero-order terms  $A_{0}$ , $B_{0}$ in the amplitudes, we finally obtain the approximation
\newpage
\begin{eqnarray}\label{sclwigairy}
{\mathcal{W}}^{\epsilon}(x,p)\approx {\widetilde{\mathcal{W}}}^\epsilon(x,p)&:=& \frac{2^{2/3}}{\epsilon^{2/3}}\left(\frac{2}{\mid
S'''(x)\mid} \right)^{1/3} A^{2}(x) \nonumber \\
&&\cdot Ai\left(-\frac{2^{2/3}}{\epsilon^{2/3}}\left(\frac{2}{S'''(x)} \right)^{1/3}(p-S'(x))\right)     \ , \ \ (x,p) \ \ \mathrm{near} \ \ \Lambda \ . \nonumber\\
\end{eqnarray}
This is Berry's semiclassical approximation of ${\mathcal{W}}^{\epsilon}$.
In the sequel we refer to the formula $(\ref{sclwigairy})$ as  {\emph{the semiclassical Wigner function (associated to the WKB function $(\ref{wkb})$)}}.

Someone should always have in mind that  that $(\ref{sclwigairy})$ is an approximation of
$(\ref{wigwkb})$ which holds simultaneously  locally near the manifold $\Lambda$ ( i.e. $\alpha=p-S'(x)$ very small), and for small $\epsilon$.

%%%%%%%%%%%%%%%%%%%%%%%%%%%%%%%%%%%%%%%%%%%%%%%%%%%%%%%%%%%%%%%%%%%%%%%
%%%%%%%%%%%%%%%%%%%%%%%%%%%%%%%%%%%%%%%%%%%%%%%%%%%%%%%%%%%%%%%%%%%%%%%

\section{Wigner transform of multi-phase WKB functions}\label{sec43}

\subsection{Limit Wigner distribution}

In Section \ref{section31}, we explained that 
 the WKB method for the Schr\"odinger equation leads to the solution of  the Hamilton-Jacobi  equation (\ref{eikonal}) for the phase  and the transport equation (\ref{transport}) for the amplitude.
Recall  that the Hamilton-Jacobi equation is a nonlinear first order equation,  and thus, it does not have a smooth solution in the large $t$ (i.e. for all $t$). Usually the solution is smooth up to some finite time $t_c$ (critical  time), then caustics appear, the rays cross each other, the phase function becomes multi-valued
(it breaks into several branches), and  $(\ref{diracwig})$ is no longer valid.
Also the amplitude $A(x)$ of the WKB-wavefunction becomes infinite on the caustics, and complex-valued for $t>t_c$.

In such cases, we naturally want to include multivalued phases, and we adopt
a WKB-ansatz of the  form
\begin{eqnarray}\label{wkb_mult}
u^{\hb}(x,t)\approx\psi^{\epsilon} (x,t) = \sum_{j=1}^{N}A_{j}(x,t)\, e^{\frac{i}{\epsilon}S_{j}(x,t)} \ .
\end{eqnarray}
Here $N=N(x,t)$ is the number of phases, which is equal to the number of crossing rays at  point $(x,t)$ near the caustics. For every fixed $(x,t)$ which does not lie on the caustic, 
the amplitudes  $A_j(x,t)$, which are now complex valued functions, and phases $S_j(x,t)$ (real valued functions)  satisfy the  transport (\ref{transport}) and Hamilton-Jacobi  (\ref{eikonal}) equation, respectively. 

Now, applying the semiclassical Wigner transform $(\ref{semWigner})$  on  $(\ref{wkb_mult})$, we have
\begin{eqnarray*}
{\mathcal{W}}^{\epsilon}(x,p,t)&=&W^{\epsilon}[\psi^{\epsilon}](x,p,t) \nonumber \\
&=&(\pi\epsilon)^{-1}\int_{R}
e^{-\frac{i}{\epsilon}2y p} \psi^\hb\left(x+y,t\right)\overline{\psi^\hb}\left(x-y,t\right) \, dy \nonumber \\
&=&\sum_{j,k=1}^{N}\left(\mathrm{I}_j +\mathrm{I}_{jk}\right)\quad ,
\end{eqnarray*}
where 
\begin{eqnarray*}
\mathrm{I}_j:=\frac{1}{\pi}\int_{R}A_{j}\left(x+y,t\right)\overline{A_{j}}\left(x-y,t\right)e^{\frac{i}{\epsilon}\left[S_{j}(x+y,t)
-S_{j}(x-y,t)\right]}e^{-i2py/\epsilon}\quad  dy\nonumber\\
&&
\end{eqnarray*}
and 
\begin{eqnarray*}
\mathrm{I}_{jk}:=\frac{1}{\pi}\int_{R}A_{j}\left(x+y,t\right)\overline{A_{k}}\left(x-y,t\right)e^{\frac{i}{\epsilon}\left[S_{j}(x+y,t)
-S_{k}(x-y,t)\right]}e^{-2ipy/\epsilon} \, dy \quad .\nonumber\\
&&
\end{eqnarray*}
Expanding in Taylor series about $x$, both amplitudes and phases, we get
\begin{eqnarray*}
A_{j}\left(x+\frac{\epsilon y}{2},t\right)\overline{A_{k}}\left(x-\frac{\epsilon y}{2},t\right)=A_j(x,t)\overline{A_{k}}(x,t)+O(\epsilon) \ ,
\end{eqnarray*}
and
\begin{eqnarray*}
&& S_{j}\left(x+\frac{\epsilon y}{2},t\right)-S_{k}\left(x+\frac{\epsilon y}{2},t\right)\\
\\
&=&\left(S_{j}(x)-S_{k}(x)\right)+\frac{\epsilon}{2}y \left(\partial_{x}S_{j}(x,t)+\partial_{x}S_{k}(x,t)\right)+O(\epsilon^2)\ .
\end{eqnarray*}

Thus, as $\epsilon \rightarrow 0$,
\begin{eqnarray*}
\mathrm{I}_j \rightarrow |A_{j}(x,t)|^2 \delta\left(p-\partial_{x}S_{j}(x,t)\right) \ .
\end{eqnarray*}

On the other hand, $\mathrm{I}_{jk}$ converges weakly to zero, since we get in front of the integral the highly oscillatory term
$$
e^{{\frac{i}{\epsilon}}\left[S_{j}(x,t)-S_{k}(x,t)\right]} \quad ,
$$
which tends weakly to zero as $\epsilon \rightarrow 0$ , when $S_{j}(x,t)\neq S_{k}(x,t)$ . Note that these phase factors describe  the interference between the waves in $(\ref{wkb_mult})$ which move along the different rays which cross at the point $(x,t)$ near the caustic. These wave effects disappear in the high frequency limit where geometrical optics dominates.

Therefore, we conclude that  $W^{\epsilon}[\psi^\epsilon]$ converges weakly as $\epsilon\rightarrow 0$ to the  limit Wigner distribution  $\Wo(x,p,t)$  which solves a Liouville equation in phase space,
\begin{eqnarray}\label{multisol}
\Wo
(x,p,t)=\sum_{j=1}^{N}|A_{j}(x,t)|^{2}\, \delta\left(p-\partial_{x}S_{j}(x,t)\right)
\quad .
\end{eqnarray}

By $(\ref{energy_density})$ , the energy density $\eta^{\epsilon}(x,t)=|\psi^{\epsilon}(x,t)|^2$ converges to
\begin{equation}
\eta^{0}(x,t)=\int_{R}W^{0}(x,p,t)\, dp=\sum_{j=1}^{N}|A_{j}(x,t)|^{2}\ .
\end{equation}

Therefore,  the limit Wigner distribution of the WKB solution predicts the geometrically expected  amplitude in the illuminated zone and away from the caustics. Formally, the limit Wigner distribution vanishes in the shadow zone where no rays penetrate, and thus it predicts zero wave intensity, as WKB function $(\ref{wkb_mult})$ does too, in this zone.

\subsection{Semiclassical Wigner functions}

It has been explained by S.Filippas \& G.N.Makrakis \cite{FM1, FM2}, by using certain simple examples of fold caustics, that in the multiphase case, the limit Wigner distribution, although when considered as semiclassical measure on phase space is well-defined, it is not a well defined distribution in $\mathcal{D'}(R_{p})$ for fixed $(x,t)$ on a caustic. For this reason, it is not the appropriate tool for the computation of energy densities at a fixed point of configuration space \footnote{except from particular degenerate singularities where an infinite number of rays converge to a focal point, \cite{SMM}}, because: (a) it cannot be expressed as a distribution with respect to the momentum for a fixed space-time point, and thus it cannot be used to compute the amplitude of the wavefunction, on caustics, and (b) it is unable to ``recognise" the correct scales of the wavefield near caustics.  
Moreover, it has been shown, by using special initial data for the free time-dependent Schr\"odinger equation, and employing Berry's semiclassical Wigner function, that is possible to construct time-dependent approximations of the solution of the Wigner equation, which capture the correct scales and amplitudes on simple fold and cusp caustics.

In a related but different direction, V.Giannopoulou \& G.N.Makrakis \cite{G, GM} have explained how the Wigner transform of a two-phase WKB function with turning points (fold caustics) can be uniformized by using semiclassical Airy approximations of the corresponding Wigner function. It has been shown that the constructed semiclassical Wigner function is a solution of the Wigner equation corresponding to a model stationary scattering problem for the Airy equation.

Therefore, there is enough evidence that Airy semiclassical approximations of the Wigner transform of multi-phase WKB functions are appropriate solutions of the Wigner equation, and they are promising  tools for computing energy densities and other wave quantities by projecting smooth phase space functions, and not geometrical objects, in order to avoid singularities of configuration space.

%%%%%%%%%%%%%%%%%%%%%%%%%%%%%%%%%%%%%%%%%%%%%%%%%%%%%%%%%%%
%%%%%%%%%%%%%%%%%%ASYMPOTIC EXPANSION%%%%%%%%%%%%%%%%%%%%%%%%%%%%%%%%%%%%%%%%%
%%%%%%%%%%%%%%%%%%%%%%%%%%%%%%%%%%%%%%%%%%%%%%%%%%%%%%%%%%%

\chapter{Asymptotic expansion of the Wigner eigenfunctions}\label{ch7}
%The phase-space eigenfunctions  are the distribution functions corresponding to the density matrix elements $W^{\hb}_{nm}=W_{v_{m}^{\hb}}[v_{n}^{\hb}](x,p)$ where $v_{n}^{\hb}$, $v_{m}^{\hb}$ are the eigenfunctions of the Schr\"odinger operator $\widehat{H}^{\hb}$. 
%The eigenfunctions in configuration space have approximated using the WKB method, as $\hb\rightarrow 0$. 
%We are interested for the asymptotic approximation of eigenfunctions for the harmonic oscillator between the turning points of Schr\"odinger equation, as $n\rightarrow \infty$ and $\hb\rightarrow 0$.

In this chapter we construct  Airy-type asymptotic expansions of the Wigner transform of the two-phase WKB eigenfunctions $(\ref{wkbhosc})$  of the harmonic oscillator. For this purpose we extend the {\it uniformization procedure} developed in \cite{G, GM} for eliminating turning point singularities of the WKB solution of a simplified scattering problem for the semiclassical Airy equation. 

However, there are certain fundamental differences between the two problems. These differences arise from the fact that the spectrum of the semiclassical Airy function is continuous, while the spectrum of the harmonic oscillator is discrete, a situation which reflects to the geometry of the Lagrangian manifolds (actually curves). The Lagrangian curve of the semiclassical Airy equation is an open curve extending to infinity (parabola), while the Lagrangian curves of the  Schr\"odinger eigenequations are closed (circles). 

Although the local behaviour at the turning points is the same in both cases, namely that of a fold singularity, it turns out that in the case of the harmonic oscillator the presence of the second turning point affects the interaction of the upper and lower branches of the Lagrangian curves and the stationary points of the Wigner integrals, and it generates a complicated asymptotic structure related to the asymptotic behaviour of Laguerre polynomials.

We recall  that the eigenfunction series solution $(\ref{wigner_exp})$ of the initial value problem $(\ref{wignereq_moy_1})$-$(\ref{wignereq_in_1})$ for the Wigner equation, reads as follows
\begin{eqnarray*}\label{wf_exp}
W^{\hb}[u^\epsilon](x,p,t)=\sum_{n=0}^{\infty}\sum_{m=0}^{\infty} c_{nm}^{\hb}(0)\, e^{-\frac{i}{\hb}(E_{n}^{\hb}-E_{m}^{\hb})t}\, W_{nm}^{\hb}(x,p) \ .
\end{eqnarray*}
The coefficients 
\begin{eqnarray*}
c_{nm}^{\hb}(0)=
(W_{0}^{\hb}[u^\hb],W_{nm}^{\hb})_{L^2(R_{xp}^2)} \  
\end{eqnarray*}
are the projections of the initial Wigner function 
\begin{eqnarray}
W_{0}^{\hb}[u^\hb](x,p):=W^{\hb}[u^{\hb}_{0}](x,p)=(\pi\hb)^{-1}\int_{R}^{} e^{-\frac{i}{2\epsilon} p\sigma} u^\hb_{0}\left(x+\sigma\right)\overline{u^\hb_{0}}\left(x-\sigma\right) \, d\sigma \ ,
\end{eqnarray}
 $u^\hb_{0}$ being the initial wavefunction  (see eq. $(\ref{initialdata5}))$,
onto the Wigner eigenfunctions 
\begin{eqnarray}\label{wigner_nm}
W^{\hb}_{nm}(x,p)=W^{\epsilon}_{u^\hb_{m}}[u^\hb_{n}](x,p)=(\pi\hb)^{-1}\int_{R}^{} e^{-\frac{i}{2\epsilon}p\sigma} v^\hb_{n}\left(x+\sigma\right)\overline{v^\hb_{m}}\left(x-\sigma\right) \, d\sigma 
\end{eqnarray}
(cross Wigner transform of the eigenfunctions $u^\hb_{n}$; see eqs $(\ref{statschr})$, $(\ref{eigenf_harm})$) . 

As we have explained in Section \ref{sec52}, an asymptotic approximation of $W^{\hb}[u^\epsilon](x,p,t)$ can be constructed through the approximation of  the Wigner functions $W_{nm}^{\hb}(x,p)$  for small parameter $\epsilon$. 

The proposed approximations $\mathcal{W}_{nm}^{\hb}$ of  $W_{nm}^{\hb}$ are based on  the integrals 
$(\ref{wigner_nm})$,  where instead of the exact eigenfunctions $v^\hb_{n}$ , we use their WKB approximations 
$\psi^{\epsilon}_{n}$  (eq $(\ref{wkbhosc})$), that is,
\begin{eqnarray}\label{wigner_approx_nm}
\mathcal{W}^{\hb}_{nm}(x,p):=W^{\epsilon}_{\psi_m^\epsilon}[\psi_n^\epsilon]=(\pi\hb)^{-1}\int_{R}^{} e^{-\frac{i}{2\epsilon}p\sigma} \psi^{\epsilon}_{n}\left(x+\sigma\right)\overline{\psi^{\epsilon}_{m}}\left(x-\sigma\right) \, d\sigma \ ,
\end{eqnarray}
and then, we apply the above mentioned uniformization procedure on the integrals $(\ref{wigner_approx_nm})$.

\section{Wigner transform of the WKB eigenfunctions}\label{7.1}

The diagonal Wigner eigenfunctions $(\ref{wigner_approx_nm})$,  $m=n$, are given by

\begin{eqnarray} \label{eigenf_int}
{\mathcal{W}}^\epsilon_{n}(x,p)\equiv \mathcal{W}^{\hb}_{nn}(x,p):=W^{\epsilon}[\psi^{\epsilon}_{n}](x,p)=\frac{1}{\pi\epsilon}\sum_{\ell=1}^{4}\int_{R}{}
D_{\ell,n}^{\hb}(\sigma;x)\ e^{\frac{i}{\epsilon}F_{\ell,n}^{\hb}(\sigma;x,p)}\, d\sigma=
\sum_{\ell=1}^{4}\mathcal{W}^\epsilon_{\ell,n}(x,p)\nonumber\\
& &
\end{eqnarray}
where
\begin{equation}\label{wigner_int_n}
\mathcal {W}^\epsilon_{\ell,n}(x,p):=\int_{R}^{} D_{\ell,n}^{\hb}(\sigma;x)\,
e^{\frac{i}{\epsilon}F_{\ell,n}^{\hb}(\sigma;x,p)}\, d\sigma \quad , \quad \ell=1,..,4 \ .
\end{equation}
The amplitudes and phases of the above four Wigner integrals $\mathcal {W}^\epsilon_{\ell,n}$ are
given by
\begin{eqnarray*}
D_{1,n}^{\hb}(\sigma;x)&=&{A_{n}^{\hb}}^{+}(x+\sigma)\overline{{A_{n}^\hb}^{+}}(x-\sigma) \ , \\
D_{2,n}^{\hb}(\sigma;x)&=&{A_{n}^\hb}^{-}(x+\sigma)\overline{{A_{n}^{\hb}}^{-}}(x-\sigma)  \ , \\
D_{3,n}^{\hb}(\sigma;x)&=&{A_{n}^{\hb}}^{+}(x+\sigma)\overline{{A_{n}^\hb}^{-}}(x-\sigma)\ , \\
D_{4,n}^{\hb}(\sigma;x)&=&{A_{n}^{\hb}}^{-}(x+\sigma)\overline{{A_{n}^\hb}^{+}}(x-\sigma) \ ,
\end{eqnarray*}
and
 \begin{eqnarray*}
F_{1,n}^{\hb}(\sigma;x,p)&=&{S_{n}^\hb}^{+}(x+\sigma)-{S_{n}^\hb}^{+}(x-\sigma)-2p\sigma \ , \\
F_{2,n}^{\hb}(\sigma;x,p)&=&{S_{n}^\hb}^{-}(x+\sigma)-{S_{n}^\hb}^{-}(x-\sigma)-2p\sigma \ ,  \\
F_{3,n}^{\hb}(\sigma;x,p)&=&{S_{n}^\hb}^{+}(x+\sigma)-{S_{n}^{\hb}}^{-}(x-\sigma)-2p\sigma \ ,   \\
F_{4,n}^{\hb}(\sigma;x,p)&=&{S_{n}^\hb}^{-}(x+\sigma)-{S_{n}^\hb}^{+}(x-\sigma)-2p\sigma   \ ,
\end{eqnarray*}
 where the amplitudes ${A_{n}^{\hb}}^{\pm}(x)$ and phases ${S_{n}^{\hb}}^{\pm}(x)$ are given by
 
\begin{eqnarray}
{A_{n}^{\hb}}^{\pm}(x)=A_{n}^{\hb}(x) e^{\pm i\pi /4} \ ,\ \ \
A_{n}^{\hb}(x):=\frac{1}{2}\left(\frac{2}{\pi}\right)^{1/2}\left(2E_{n}^{\hb}-x^2 \right)^{-1/4}
\end{eqnarray}
and
\begin{eqnarray}\label{ansn}
{S_{n}^{\hb}}^{\pm}(x)= \pm {S_{n}^{\hb}}(x) \ , \ \ \ {S_{n}^{\hb}}(x)=\int_{\sqrt{2E_{n}^{\hb}}}^{x}\sqrt{2E_{n}^{\hb}-t^2}dt \ ,
\end{eqnarray}
(see eqs. $(\ref{wkbampl_1})$-$(\ref{wkbampl_2})$  and  $(\ref{wkbph_1})$-$(\ref{wkbph_2})$, respectively).
 
Then,  we can express all $F_{\ell,n}^{\hb}$ only  in terms of ${S_{n}^{\hb}}$,
\begin{eqnarray}
F_{1,n}^{\hb}(\sigma;x,p)&=&{S_{n}^\hb}(x+\sigma)-{S_{n}^\hb}(x-\sigma)-2p\sigma \ ,\label{phase_1n}\\
F_{2,n}^{\hb}(\sigma;x,p)&=&-\left({S_{n}^\hb}(x+\sigma)-{S_{n}^\hb}(x-\sigma)+2p\sigma \right) \ ,  \label{phase_2n}\\
F_{3,n}^{\hb}(\sigma;x,p)&=&{S_{n}^\hb}(x+\sigma)+{S_{n}^{\hb}}(x-\sigma)-2p\sigma \ , \label{phase_3n} \\
F_{4,n}^{\hb}(\sigma;x,p)&=&-\left({S_{n}^\hb}(x+\sigma)+{S_{n}^\hb}(x-\sigma)+2p\sigma \right)  \label{phase_4n} \ ,
\end{eqnarray}
and we see that they satisfy the following relations 
\begin{eqnarray}
F_{2,n}^{\hb}(\sigma;x,p)&=&-F_{1,n}^{\hb}(\sigma;x,-p) \ , \label{diagsym_12} \\
F_{4,n}^{\hb}(\sigma;x,p)&=&-F_{3,n}^{\hb}(\sigma;x,-p) \ .   \label{diagsym_34}
\end{eqnarray}

Also we can express the amplitudes only in terms of $A_{n}^{\hb}$, and they read as follows
\begin{eqnarray}
D_{1,n}^{\hb}(\sigma;x)&=&{A_{n}^{\hb}}(x+\sigma)\overline{{A_{n}^\hb}}(x-\sigma) \ ,\nonumber \\
D_{2,n}^{\hb}(\sigma;x)&=&{A_{n}^\hb}(x+\sigma)\overline{{A_{n}^{\hb}}}(x-\sigma)  \ ,\nonumber \\
D_{3,n}^{\hb}(\sigma;x)&=&i{A_{n}^{\hb}}(x+\sigma)\overline{{A_{n}^\hb}}(x-\sigma)\ , \nonumber \\
D_{4,n}^{\hb}(\sigma;x)&=&-i{A_{n}^{\hb}}(x+\sigma)\overline{{A_{n}^\hb}}(x-\sigma) \ ,
\end{eqnarray}
Similarly, the  off-diagonal terms $(\ref{wigner_approx_nm}),$ $ n \neq m$, are given by

\begin{equation}\label{wigner_int_nm}
{\mathcal{W}}_{nm}^\epsilon(x,p):=W^{\epsilon}_{\psi_m^\epsilon}[\psi_n^\epsilon](x,p)=
\sum_{\ell=1}^{4}\mathcal{W}^\epsilon_{\ell,nm}(x,p)
\end{equation}
where
\begin{equation}
\mathcal{W}^\epsilon_{\ell,nm}(x,p):=\int_{R}^{} D_{\ell,nm}^{\hb}(\sigma;x)\,
e^{\frac{i}{\epsilon}F_{\ell,nm}^{\hb}(\sigma;x,p)}\, d\sigma \, , \quad \ell=1,\ldots,4 \, , \quad n,m=0,1,\ldots \, .
\end{equation}
The amplitudes and the phases of the Wigner integrals $\mathcal{W}^\epsilon_{\ell,nm}$ are given by
\begin{eqnarray*}
D_{1,nm}^{\hb}(\sigma;x)&=&{A_{n}^\hb}^{+}(x+\sigma)\overline{{A_{m}^\hb}^{+}}(x-\sigma) \ ,  \\
D_{2,nm}^{\hb}(\sigma;x)&=&{A_{n}^\hb}^{-}(x+\sigma)\overline{{A_{m}^\hb}^{-}}(x-\sigma)  \ , 
\\
D_{3,nm}^{\hb}(\sigma;x)&=&{A_{n}^\hb}^{+}(x+\sigma)\overline{{A_{m}^\hb}^{-}}(x-\sigma) \ , 
 \\
D_{4,nm}^{\hb}(\sigma;x)&=&{A_{n}^\hb}^{-}(x+\sigma)\overline{{A_{m}^\hb}^{+}}(x-\sigma) \ ,
\end{eqnarray*}
and
 \begin{eqnarray*}
F_{1,nm}^{\hb}(\sigma;x,p)&=&{S_{n}^\hb}^{+}(x+\sigma)-{S_{m}^\hb}^{+}(x-\sigma)-2p\sigma \ , \\
F_{2,nm}^{\hb}(\sigma;x,p)&=&{S_{n}^\hb}^{-}(x+\sigma)-{S_{m}^\hb}^{-}(x-\sigma)-2p\sigma \ ,  \\
F_{3,nm}^{\hb}(\sigma;x,p)&=&{S_{n}^\hb}^{+}(x+\sigma)-{S_{m}^\hb}^{-}(x-\sigma)-2p\sigma \ ,   \\
F_{4,nm}^{\hb}(\sigma;x,p)&=&{S_{n}^\hb}^{-}(x+\sigma)-{S_{m}^\hb}^{+}(x-\sigma)-2p\sigma  \  ,
\end{eqnarray*}
where the amplitudes ${A_{n}^{\hb}}^{\pm}$ and phases ${S_{n}^{\hb}}^{\pm}$ for $n$ and $m$ are given by  $(\ref{wkbampl_1})$-$(\ref{wkbampl_2})$  and  $(\ref{wkbph_1})$-$(\ref{wkbph_2})$, respectively.

As we did for the diagonal terms, we can express $F_{\ell,nm}^{\hb}$ only in terms of $S_{n}^{\hb}$ and $S_{m}^{\hb}$, 

\begin{eqnarray}
F_{1,nm}^{\hb}(\sigma;x,p)&=&{S_{n}^\hb}(x+\sigma)-{S_{m}^\hb}(x-\sigma)-2p\sigma \ , \label{phase_1nm}\\
F_{2,nm}^{\hb}(\sigma;x,p)&=&-\left({S_{n}^\hb}(x+\sigma)-{S_{m}^\hb}(x-\sigma)+2p\sigma \right) \ , \label{phase_2nm}\\
F_{3,nm}^{\hb}(\sigma;x,p)&=&{S_{n}^\hb}(x+\sigma)+{S_{m}^\hb}(x-\sigma)-2p\sigma \  , \label{phase_3nm} \\
F_{4,nm}^{\hb}(\sigma;x,p)&=&-\left({S_{n}^\hb}(x+\sigma)+{S_{m}^\hb}(x-\sigma)+2p\sigma \right) \ ,\label{phase_4nm}
\end{eqnarray}
and we easily see that they satisfy the relations 
\begin{eqnarray}
F_{2,nm}^{\hb}(\sigma;x,p)&=&-F_{1,nm}^{\hb}(\sigma;x,-p) \ , \label{offdiagsym_12} \\
F_{4,nm}^{\hb}(\sigma;x,p)&=&-F_{3,nm}^{\hb}(\sigma;x,-p) \ .   \label{offdiagsym_34}
\end{eqnarray}
Similarly the amplitudes are expressed only on terms of $A_{n}^\hb$,
\begin{eqnarray}
D_{1,nm}^{\hb}(\sigma;x)&=&{A_{n}^\hb}(x+\sigma)\overline{{A_{m}^\hb}}(x-\sigma) \ , \nonumber \\
D_{2,nm}^{\hb}(\sigma;x)&=&{A_{n}^\hb}(x+\sigma)\overline{{A_{m}^\hb}}(x-\sigma)  \ , \nonumber \\
D_{3,nm}^{\hb}(\sigma;x)&=&i{A_{n}^\hb}(x+\sigma)\overline{{A_{m}^\hb}}(x-\sigma) \ , \nonumber \\
D_{4,nm}^{\hb}(\sigma;x)&=&-i{A_{n}^\hb}(x+\sigma)\overline{{A_{m}^\hb}}(x-\sigma)  \ .
\end{eqnarray}

Throughout the  asymptotic calculations of the Wigner integrals $\mathcal {W}^\epsilon_{\ell,n}$ , $\mathcal {W}^\epsilon_{\ell,nm}$ we must consider that  $\hb$ is small and  $n$ large enough, so that $E^{\hb}_{n}$ is constant, independent of $n$ and $\hb$, because the eigenvalues $E^{\hb}_{n}$ entering  the  phase ${S_{n}^{\hb}}$  are obtained from the Bohr-Sommerfeld quantization rule $(\ref{bohrsom})$. More precisely, for the case of the harmonic oscillator, this means that $E^{\hb}_{n}=(n+1/2)\,\hb\approx n\hb=\mathrm{constant}$. For this reason, and for simplifying the notation as much as possible, in the course of the asymptotic calculations,
the amplitudes  and phases of the Wigner integrals are dimmed to be independent of  $\epsilon$, and for this reason, we omit the superscript $\hb$, and we write $E_{n}=E_{n}^{\hb}$, and 
$$A_{n}^{\pm}(x)={A_{n}^{\hb}}^{\pm}(x) \ , \ \ \ A_{n}(x)=A_{n}^{\hb}(x) \ ,$$
$$ S_{n}^{\pm}(x)={S_{n}^{\hb}}^{\pm}(x) \ , \ \ \ S_{n}(x)=S_{n}^{\hb}(x) \ ,$$
$$D_{\ell,n}(\sigma;x)= D_{\ell,n}^{\hb}(\sigma;x) \ , \ \ \ D_{\ell,nm}(\sigma;x)= D_{\ell,nm}^{\hb}(\sigma;x)  \ ,$$
$$ F_{\ell,n}(\sigma;x,p)= F_{\ell,n}^{\hb}(\sigma;x,p) \ , \ \ \ F_{\ell,nm}(\sigma;x,p)= F_{\ell,nm}^{\hb}(\sigma;x,p) \ ,$$
for $\ell=1,\ldots,4$ and $n,m =0,1,\ldots$ (recall the definitions of   amplitudes and phases that are given by $(\ref{wkbampl_1})$-$(\ref{wkbampl_2})$, $(\ref{wkbph_1})$-$(\ref{wkbph_2})$ and $(\ref{ansn})$).

In the sequel we compute  asymptotic
expansions of the Wigner integrals  $\mathcal{W}^\epsilon_{\ell,nm}$,  by applying either the standard or  the uniform  stationary-phase formulae (in this case we use Berry's semiclassical Wigner function $(\ref{sclwigairy})$) accordingly to the nature of the stationary points in various regions of the phase space. More precisely, it turns out that we must  use the semiclassical Wigner function in the cases $\ell=1,2$ and standard stationary-phase approximation in the cases $\ell=3,4$. In Sections \ref{sec72} and \ref{sec73}
we collect the derived approximation formulae. Some details of the long and cumbersome asymptotic calculations are presented  in Sections \ref{sec74} and \ref{sec75}.

%%%%%%%%%%%%%%%%%%%%%%%%%%%%%%%%%%%%%%%%%%%%%%%%%%%%%%%%%%%%%%%%%%%%%%%%%%%%%%%%%%%%%
\section{Asymptotics of the diagonal eigenfunctions}\label{sec72}

For the computation of the asymptotics of the Wigner integrals $\mathcal{W}^{\epsilon}_{\ell,n}$ , $\ell=1,2$ (see (\ref{wigner_int_n})), we apply Berry's semiclassical formula $(\ref{sclwigairy})$ (i.e. the uniform stationary phase approximation of Appendix  \ref{app_a}), because  the symmetric phases $F_{\ell,n}$  given by $(\ref{phase_1n})$-$(\ref{phase_2n})$,  have a double stationary point $\si=0$, when $(x,p)$ lies on  the upper or on the lower branch of  Lagrangian curve (see Fig. \ref{stat_points_n})
\begin{eqnarray}\label{curve_n}
\Lambda_{n}:=\lbrace (x,p)\in R^2: x^2 + p^2= 2E_{n}\rbrace \ .
\end{eqnarray}
Note that $\Lambda_{n}$ is the Lagrangian eigencurve $H(x,p)= \frac12(x^2+p^2)=E_{n}$ of the harmonic oscillator.

The phases $F_{\ell,n}$ ,  $\ell=1,2$, have a couple of symmetric stationary points 
when $(x,p)$ lies in the meniscus
$$\Sigma_{n}=\Sigma_{n}^{+}\cup\Sigma_{n}^{-}$$
defined by the intersection of the interior $p^2+x^2\leq2E_{n}$ of  the Lagrangian curve  $\Lambda_{n}$ and the exterior of its dual curve $\Lambda^{*}_{n}$ .
This dual curve is defined by
\begin{eqnarray}\label{union}
\Lambda^{*}_{n}:=\Lambda^{*}_{1,n}\cup \Lambda^{*}_{2,n} \ ,
\end{eqnarray}
where
\begin{eqnarray}\label{dual_n1}
\Lambda^{*}_{1,n}:=\lbrace (x,p)\in R^2:p^2+(x-\sqrt{E_{n}/2})^2= E_{n}/2\rbrace \ ,
\end{eqnarray}
and 
\begin{eqnarray}\label{dual_n2}
\Lambda^{*}_{2,n}:=\lbrace (x,p)\in R^2:p^2+(x+\sqrt{E_{n}/2})^2= E_{n}/2 \rbrace \ .
\end{eqnarray}
The simple stationary points coalesce to the double point $\si=0$, for $(x,p) \in \Lambda_{n}$ .

\begin{figure}[h]
\centering
\includegraphics[width=0.5\textwidth]{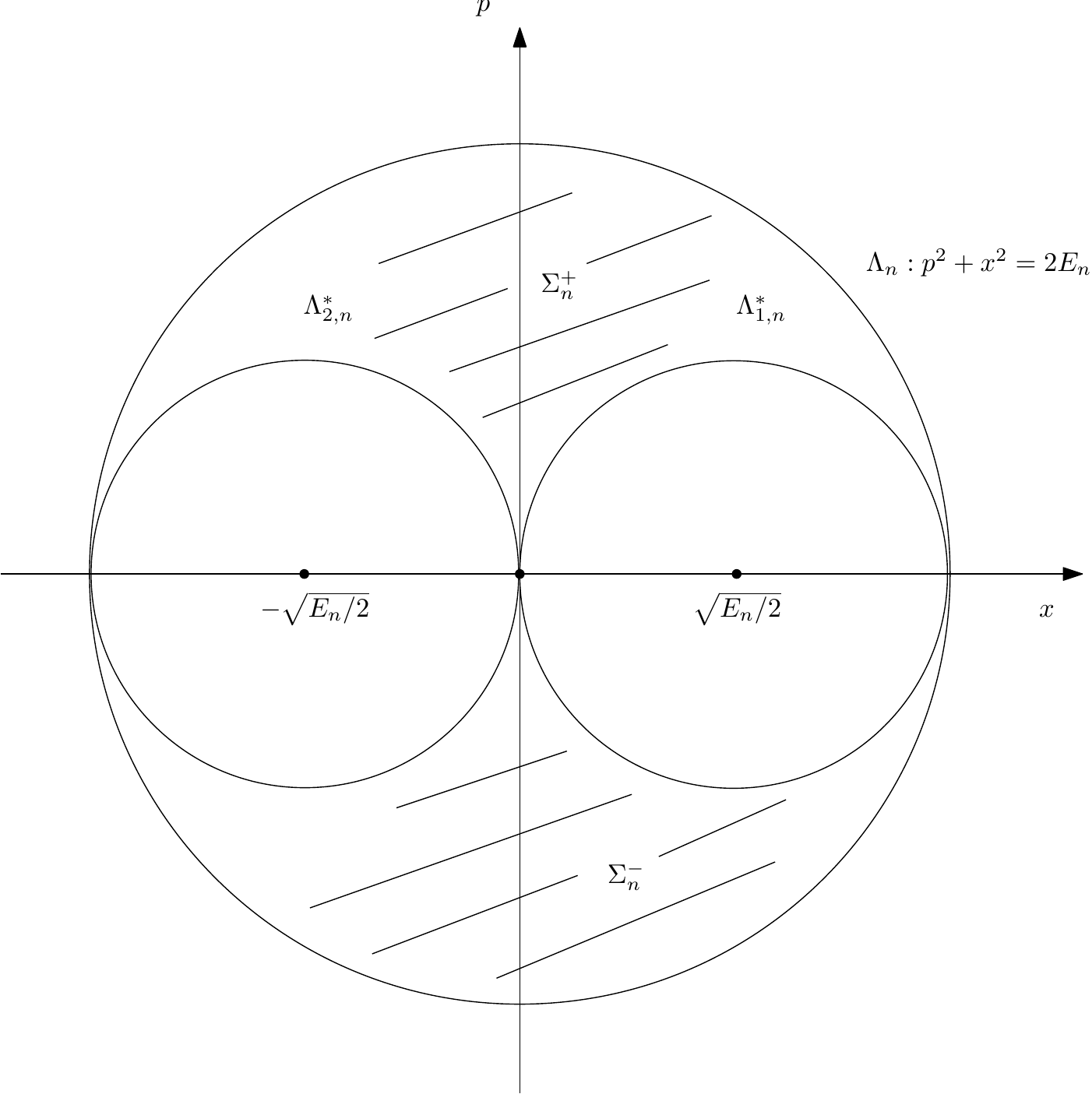}
\caption{{\it Lagrangian eigencurve $\Lambda_n$ }}
\label{stat_points_n}
\end{figure}

We then derive the  approximation
\begin{eqnarray*}
\mathcal{W}_{1,n}^{\epsilon}(x,p)\approx {\pi}^{-1}\, \hb^{-2/3}{\left(2E_{n}\right)}^{-1/3}\,
Ai\left(\frac{p^2+x^2-2E_n}{\epsilon^{2/3}{(2E_{n})}^{1/3}}\right)\quad ,
\end{eqnarray*}
near the branch $p=\sqrt{2E_{n}-x^2}$ and the approximation
\begin{eqnarray*}
\mathcal{W}_{2,n}^{\epsilon}(x,p)\approx{\pi}^{-1}\, \hb^{-2/3}{\left(2E_{n}\right)}^{-1/3}\,
Ai\left(\frac{p^2+x^2-2E_n}{\epsilon^{2/3}{(2E_{n})}^{1/3}}\right)\quad ,
\end{eqnarray*}
near the branch $p=-\sqrt{2E_{n}-x^2}$ .

It is important to emphasize that although the above two formulas are formally the same, they are valid in different regions and they have been derived by applying Berry's  semiclassical formula $(\ref{sclwigairy})$ near two different branches of the Lagrangian curve. We thus lead to define
\begin{eqnarray}\label{new_notation}
\widetilde{\mathcal{W}^{\hb}_{n}}(x,p):= \pi^{-1}\, \hb^{-2/3}{\left(2E_{n}\right)}^{-1/3}\,
Ai\left(\frac{p^2+x^2-2E_n}{\epsilon^{2/3}{(2E_{n})}^{1/3}}\right)\quad .
\end{eqnarray}

For the computation of the asymptotics of $\mathcal{W}_{\ell,n}^{\epsilon}(x,p)$ , $\ell=3,4$ (see (\ref{wigner_int_n})), we apply the standard stationary phase formula, because the stationary points of $F_{\ell,n}$ ,  $\ell=3,4$ , are simple, when $(x,p)$ lies in the interior of the dual curve $\Lambda^{*}_{n}$ of $\Lambda_{n}$ .

We then derive the approximations

\begin{itemize}
\item for $0\leqslant x<\sqrt{2E_{n}}$ and $p>0$ ,
\begin{eqnarray*}
\mathcal{W}_{3,n}^{\epsilon}(x,p)&\approx&\frac{1}{2\pi^{3/2}\sqrt{\epsilon}}\, 
e^{\frac{i}{\epsilon}F_{3,n}(-\sigma_{0})}e^{i\pi/4}(p^2+x^2)^{-1/4}(2E_{n}-p^2-x^2)^{-1/4}\quad ,\nonumber\\
\mathcal{W}_{4,n}^{\epsilon}(x,p)&\approx &\frac{1}{2\pi^{3/2}\sqrt{\epsilon}}\,
e^{-\frac{i}{\epsilon}F_{3,n}(-\sigma_{0})}e^{-i\pi/4}(p^2+x^2)^{-1/4}(2E_{n}-p^2-x^2)^{-1/4}\nonumber
\end{eqnarray*}
\item for $0\leqslant x<\sqrt{2E_{n}}$ and $p<0$ ,
\begin{eqnarray*}
\mathcal{W}_{3,n}^{\epsilon}(x,p)&\approx&\frac{1}{2\pi^{3/2}\sqrt{\epsilon}}\,
e^{\frac{i}{\epsilon}F_{3,n}(+\sigma_{0})}e^{i\pi/4}(p^2+x^2)^{-1/4}(2E_{n}-p^2-x^2)^{-1/4}\quad ,\nonumber\\
\mathcal{W}_{4,n}^{\epsilon}(x,p)&\approx&\frac{1}{2\pi^{3/2}\sqrt{\epsilon}}\,
e^{-\frac{i}{\epsilon}F_{3,n}(+\sigma_{0})}e^{-i\pi/4}(p^2+x^2)^{-1/4}(2E_{n}-p^2-x^2)^{-1/4}\nonumber
\end{eqnarray*}
\item for  $-\sqrt{2E_{n}}<x\leqslant 0$ and $p>0$ ,
\begin{eqnarray*}
\mathcal{W}_{3,n}^{\epsilon}(x,p)&\approx&\frac{1}{2\pi^{3/2}\sqrt{\epsilon}}\,
e^{\frac{i}{\epsilon}F_{3,n}(+\sigma_{0})}e^{i3\pi/4}(p^2+x^2)^{-1/4}(2E_{n}-p^2-x^2)^{-1/4}\quad , \nonumber\\
\mathcal{W}_{4,n}^{\epsilon}(x,p)&\approx&\frac{1}{2\pi^{3/2}\sqrt{\epsilon}}\,
e^{-\frac{i}{\epsilon}F_{3,n}(+\sigma_{0})}e^{-i3\pi/4}(p^2+x^2)^{-1/4}(2E_{n}-p^2-x^2)^{-1/4}\nonumber
\end{eqnarray*}
\item for $-\sqrt{2E_{n}}<x\leqslant 0$ and $p<0$ ,
\begin{eqnarray*}
\mathcal{W}_{3,n}^{\epsilon}(x,p)&\approx&\frac{1}{2\pi^{3/2}\sqrt{\epsilon}}\,
e^{\frac{i}{\epsilon}F_{3,n}(-\sigma_{0})}e^{i3\pi/4}(p^2+x^2)^{-1/4}(2E_{n}-p^2-x^2)^{-1/4} \ ,\nonumber\\
\mathcal{W}_{4,n}^{\epsilon}(x,p)&\approx&\frac{1}{2\pi^{3/2}\sqrt{\epsilon}}\,
e^{-\frac{i}{\epsilon}F_{3,n}(-\sigma_{0})}e^{-i3\pi/4}(p^2+x^2)^{-1/4}(2E_{n}-p^2-x^2)^{-1/4}  \nonumber
\end{eqnarray*}
\end{itemize}
where $F_{3,n}(-\sigma_{0})$ denotes  the value of 
$$F_{3,n}(\si;x,p)=\int_{\sqrt{2E_{n}}}^{x+\si}\sqrt{2E_{n}-t^2}dt+\int_{\sqrt{2E_{n}}}^{x-\si}\sqrt{2E_{n}-t^2}dt-2p\si$$
at the stationary point $\sigma=-\sigma_0$ , with
\begin{eqnarray*}
\si_{0}(x,p):=\frac{|p|}{\sqrt{p^2+x^2}}\sqrt{|2E_{n}-p^2-x^2|} \ ,
\end{eqnarray*}
(see eq.  $(\ref{sigma_0})$ below).

Outside the curve $\Lambda_{n}$ , we derive that
\begin{eqnarray*}
\mathcal{W}_{3,n}^{\epsilon}(x,p)=\mathcal{W}_{4,n}^{\epsilon}(x,p) = 0 \ .
\end{eqnarray*}

The asymptotic contributions to the Wigner integral  $\mathcal{W}^{\hb}_{n}(x,p)$ when $(x,p)$ lies in various regions of phase space,  are summarized  in the Table \ref{table1}. For fixed $x$ with $|x|<\sqrt{2E_{n}}$, we have 
\begin{center}
\begin{tabular}[!hbp]{|c|c|}\hline
region & main contribution to $W^{\hb}_{n}(x,p)$ \\
\hline
$p>\sqrt{2E_{n}-x^2}$ & $\mathcal{W}^{\hb}_{1,n}\approx\widetilde{\mathcal{W}^{\hb}_{n}}$ \\
\hline
$p\approx\sqrt{2E_{n}-x^2}$ & $\mathcal{W}^{\hb}_{1,n}\approx\widetilde{\mathcal{W}^{\hb}_{n}}$ \\
\hline
inside $\Lambda^{*}_{n}$ & $\mathcal{W}^{\hb}_{3,n}+\mathcal{W}^{\hb}_{4,n} \approx\widetilde{\mathcal{W}^{\hb}_{n}}   $\\
\hline
$p\approx-\sqrt{2E_{n}-x^2}$ & $\mathcal{W}^{\hb}_{2,n}\approx\widetilde{\mathcal{W}^{\hb}_{n}}$ \\
\hline
$p>-\sqrt{2E_{n}-x^2}$ & $\mathcal{W}^{\hb}_{2,n}\approx\widetilde{\mathcal{W}^{\hb}_{n}}$ \\
\hline
\end{tabular}
\end{center}
\begin{table}[h]
\caption{\it{ The main contribution to ${\mathcal{W}}^\epsilon_{n}$} }
\label{table1}
\end{table}
where  $\Lambda^{*}_{n}$ is the dual curve (see definition (\ref{union}) and Figure \ref{stat_points_n}), and $\widetilde{\mathcal{W}^{\hb}_{n}}$ is given by (\ref{new_notation}). 

By rather involved transformations, related to  the asymptotics of Laguerre polynomials (which are the exact Wigner eigenfunctions; se Section \ref{761}), it turns out that the phases $F_{3,n}(\pm \sigma_{0})$ are related to the phase of the Airy function in the approximation $(\ref {new_notation})$  of  
$\mathcal{W}_{\ell,n}^{\epsilon} \ , \ell=1,2$. Then, by standard asymptotics of the Airy function, we can see that inside the manifold $\Lambda^{*}_{n}$ the contribution  $(\mathcal{W}^{\hb}_{3,n}+\mathcal{W}^{\hb}_{4,n})$ to  $\mathcal{W}^{\hb}_{n}$ matches with the asymptotics of $(\ref {new_notation})$, and therefore $(\mathcal{W}^{\hb}_{3,n}+\mathcal{W}^{\hb}_{4,n})\approx \widetilde{\mathcal{W}^{\hb}_{n}}$ .

Thus, the leading approximation of  ${\mathcal{W}}^\epsilon_{n}$ is  given by
\begin{eqnarray}\label{approx_w_n}
{\mathcal{W}}^\epsilon_{n}(x,p) \approx \widetilde{\mathcal{W}^{\hb}_{n}}(x,p):={\pi}^{-1}\, \hb^{-2/3}{\left(2E_{n}\right)}^{-1/3}\,
Ai\left(\frac{p^2+x^2-2E_n}{\epsilon^{2/3}{(2E_{n})}^{1/3}}\right)\quad .
\end{eqnarray}

\section{Asymptotics of the off-diagonal eigenfunctions}\label{sec73}

For the construction of the asymptotics of the  Wigner integrals $\mathcal{W}^\epsilon_{\ell,nm}$, $\ell=1,2$ (eq. (\ref{wigner_int_nm})) we apply the uniform stationary phase approximation (\ref{ap4}) of Appendix  \ref{app_a}. Berry's semiclassical Wigner function cannot be used directly in this case, because the double stationary points of $F_{\ell,nm}(\sigma;x,p)$ ,  $\ell=1,2$, for $(x,p)$ on certain Lagrangian manifolds are not zero contrary to what happens when $n=m$. The details of the construction will be presented in Section \ref{sec75}.

We suppose, without loss of generality, that $n>m$, thus $E_n>E_m$ by $(\ref{energy_harm})$. 
Now, in the study of the stationary points of the Wigner integrals  $\mathcal{W}_{\ell,nm}^{\epsilon}$ arise the couple of Lagrangian curves $\Lambda _{1,nm}$ , $\Lambda _{2,nm}$ , where
\begin{eqnarray}
\Lambda _{1,nm}&=&\lbrace (x,p)\in R^2:p^2+x^2=R_{nm}^2\rbrace\quad ,\label{man_nm_a}\\
\Lambda _{2,nm}&=&\lbrace (x,p)\in R^2:p^2+x^2=\rho_{nm}^2\rbrace\quad ,\label{man_nm_b}
\end{eqnarray}
where
\begin{eqnarray}
R_{nm}&:=&\frac{1}{2}(\sqrt{2E_{n}}+\sqrt{2E_{m}})\quad ,\label{z1}\\
\rho_{nm}&:=&\frac{1}{2}(\sqrt{2E_{n}}-\sqrt{2E_{m}})\ . \label{z2} 
\end{eqnarray}
Somehow, these curves may by thought of as the analogue of $\Lambda_n$ which arouse in the diagonal case, since formally $\rho_{nm}=0$ and $R_{nm}=R_{n}$ for $n=m$.

In the construction of the approximation we assume that $n,m$ are large and $\epsilon$ small, so that $n\hb,m\hb=\mathrm{constant}$ and $n-m=\mathrm{constant}>0$. Then, the eigenvalues $E_{n}=E^{\hb}_{n}=(n+1/2)\hb\approx n\hb$ and $E_{m}=E^{\hb}_{m}=(m+1/2)\hb\approx m\hb$ are treated as constants.

The phases $F_{\ell,nm}$ , $\ell=1,2$, have a couple of  stationary points $\sigma=\si_{1,2}$
when $(x,p)$ lies in the meniscus 
$$\Sigma_{nm}=\Sigma_{nm}^{+}\cup\Sigma_{nm}^{-}$$
defined by the intersection of the ring $\rho^{2}_{nm} \le x^2+p^2 \leq R^{2}_{nm}$ (the area between the Lagrangian curves  
$\Lambda _{1,nm}$, $\Lambda _{2,nm}$), and the exterior of its dual curve $\Lambda^{*}_{nm}$ .

%%%%%%%%%%%%%%%%%%%%%%%%%%%%%%%%%%%%%%%%
This dual curve is defined by
\begin{eqnarray}\label{union_nm}
\Lambda^{*}_{nm}:=\Lambda^{*}_{1,nm}\cup \Lambda^{*}_{2,nm} \ ,
\end{eqnarray}
where
\begin{itemize}
\item{ if $\sigma=\si_{1,2} \geq 0$ (see Figure \ref{spnm_1a})

\begin{eqnarray}\label{dual_n2}
\Lambda^{*}_{1,nm}:=\lbrace (x,p)\in R^2:   p^2+{(x-\sqrt{E_n/2})}^2 =E_{m}/2      \rbrace \ .
\end{eqnarray}
}
and
\begin{eqnarray}\label{dual_n1}
\Lambda^{*}_{2,nm}:=\lbrace (x,p)\in R^2:    p^2+{(x+\sqrt{E_m/2})}^2 = E_{n}/2   \rbrace \ ,
\end{eqnarray}
\newpage
\begin{figure}[h]
\centering
\includegraphics[width=0.5 \textwidth]{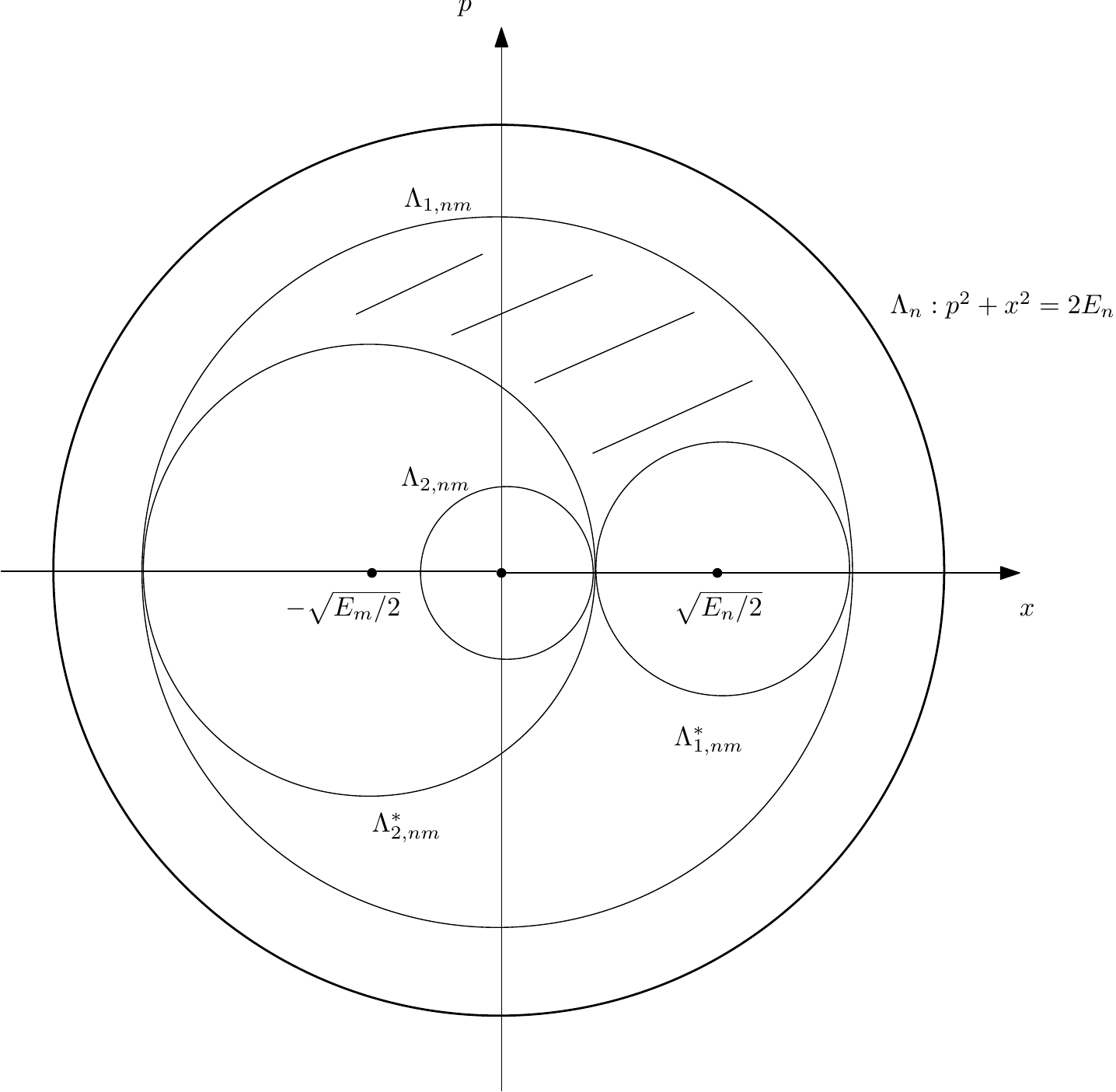}
\caption{{\it Area of existence of stationary points of $F_{1,nm} \   (\sigma=\si_{1,2} >0)$}}
\label{spnm_1a}
\end{figure}

\item{if $\sigma=\si_{1,2} \leq 0$ (see Figure \ref{spnm_1b})

\begin{eqnarray}\label{dual_n1}
\Lambda^{*}_{1,nm}:=\lbrace (x,p)\in R^2:    p^2+{(x-\sqrt{E_m/2})}^2 = E_{n}/2   \rbrace \ ,
\end{eqnarray}
and 
\begin{eqnarray}\label{dual_n2}
\Lambda^{*}_{2,nm}:=\lbrace (x,p)\in R^2:   p^2+{(x+\sqrt{E_n/2})}^2 = E_{m}/2   \rbrace \ .
\end{eqnarray}
\newpage
\begin{figure}[h]
\centering
\includegraphics[width=0.5 \textwidth]{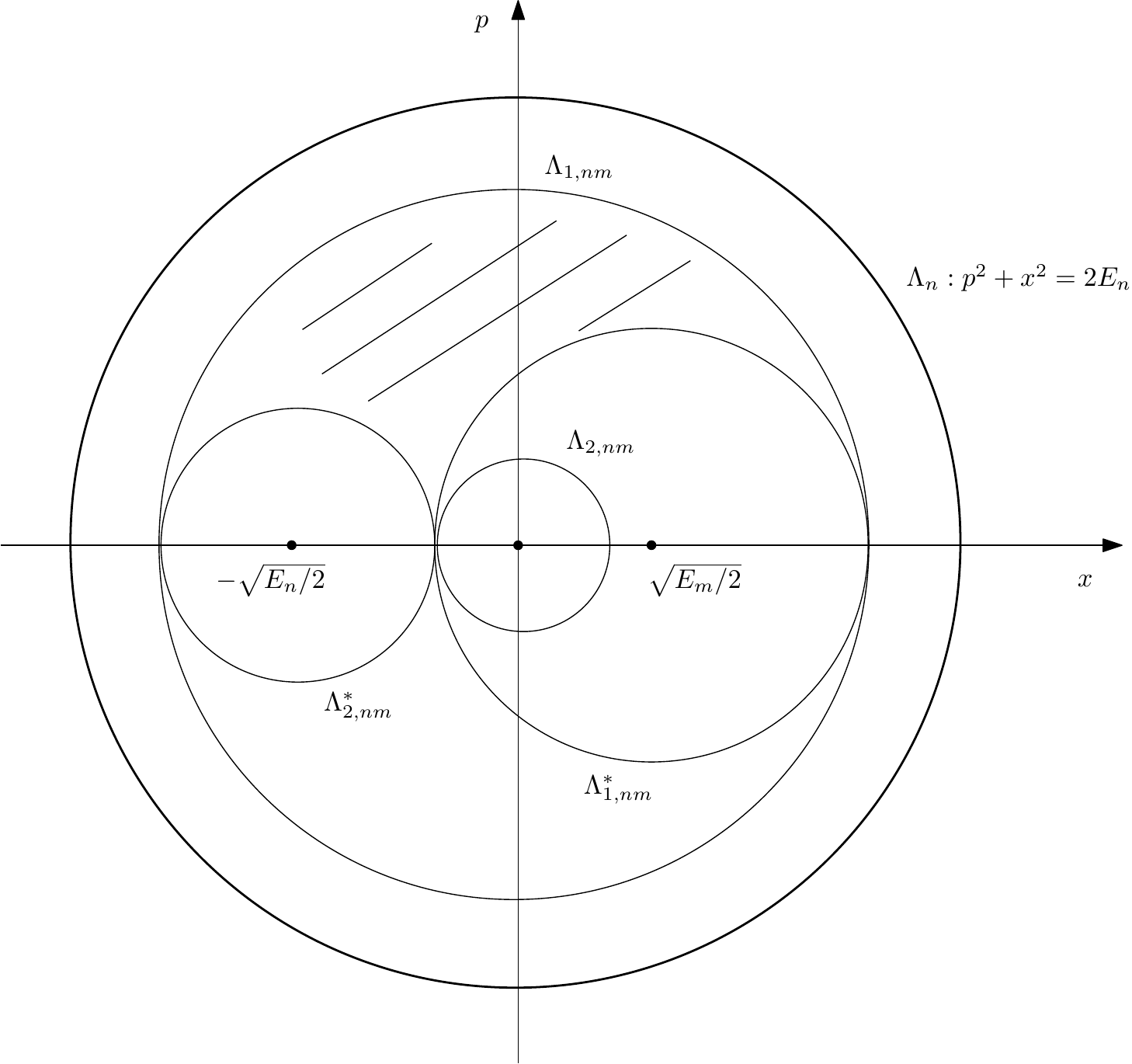}
\caption{{\it \it Area of existence of stationary points of $F_{1,nm} \  (\sigma=\si_{1,2} <0)$}}
\label{spnm_1b}
\end{figure}
}

\end{itemize}

%%%%%%%%%%%%%%%%%%%%%%%%%%%%%%%%%%%%%%%%%%

It turns out  that simple stationary points exist only for $(x,p)$ in the shaded part of the ring $\rho_{nm}^2<p^2+x^2<R_{nm}^2$ , and there are no real stationary points for $(x,p)$ lying on the interior of $\Lambda _{2,nm}$ . 
The simple stationary points coalesce to the double point $\si \ne 0$ , for $(x,p) \in \Lambda _{1,nm}\cup \Lambda _{2,nm}$ .

Then, applying the uniform stationary phase formula  to $\mathcal{W}_{1,nm}^{\epsilon}$ , for small $\hb$, we derive  the approximation 
\begin{eqnarray*}
\mathcal{W}_{1,nm}^{\epsilon}(x,p)\approx
{\pi}^{-1}e^{-i(n-m)\phi}{\epsilon}^{-2/3}R_{nm}^{-4/3}(R_{nm}^2-\rho_{nm}^2)^{1/3}
Ai\left[\frac{p^2+x^2-R_{nm}^2}{\epsilon^{2/3}{R_{nm}^{4/3}(R_{nm}^2-\rho_{nm}^2)}^{-1/3}}\right]\, ,
\end{eqnarray*}
for $(x,p)$ near the branch $p=\sqrt{R^{2}_{nm}-x^2}>0$ . We have put $\phi:=\arctan(p/x)$ , and recall that $R_{nm}$ , $\rho_{nm}$ are given by $(\ref{z1})$, $(\ref{z2})$. 
 
For the second integral $\mathcal{W}^{\epsilon}_{2,nm}$ , in a similar way, we obtain the approximation
\begin{eqnarray*}
\mathcal{W}_{2,nm}^{\epsilon}(x,p)\approx {\pi}^{-1}
e^{-i(n-m)\phi}{\epsilon}^{-2/3}R_{nm}^{-4/3}(R_{nm}^2-\rho_{nm}^2)^{1/3}
Ai\left[\frac{p^2+x^2-R_{nm}^2}{\epsilon^{2/3}{R_{nm}^{4/3}(R_{nm}^2-\rho_{nm}^2)}^{-1/3}}\right]\, ,
\end{eqnarray*}
near the branch $p=-\sqrt{R^{2}_{nm}-x^2}<0$ .
 
It must be emphasized that although the above two formulas are formally the same, they are valid for different values of $p$. Thus we are lead to define
\begin{eqnarray}\label{new_notation_nm}
\widetilde{\mathcal{W}_{nm}^{\epsilon}}(x,p):= {\pi}^{-1}
e^{-i(n-m)\phi}{\epsilon}^{-2/3}R_{nm}^{-4/3}(R_{nm}^2-\rho_{nm}^2)^{1/3}
Ai\left[\frac{p^2+x^2-R_{nm}^2}{\epsilon^{2/3}{R_{nm}^{4/3}(R_{nm}^2-\rho_{nm}^2)}^{-1/3}}\right] \ . \nonumber\\
\end{eqnarray}

The  phases $F_{3,nm}(\sigma,x,p)$ and $F_{4,nm}(\sigma,x,p)$ (rel. (\ref{phase_3nm}) and  (\ref{phase_4nm}), respectively)  have simple real stationary points in the shaded region of Figures $(\ref{stationaryp_34nma})$ , 
$(\ref{stationaryp_34nmb})$
 and a pair of complex stationary points outside this area. 
%\newpage
\begin{figure}[h]
\centering
\includegraphics[width=0.5 \textwidth]{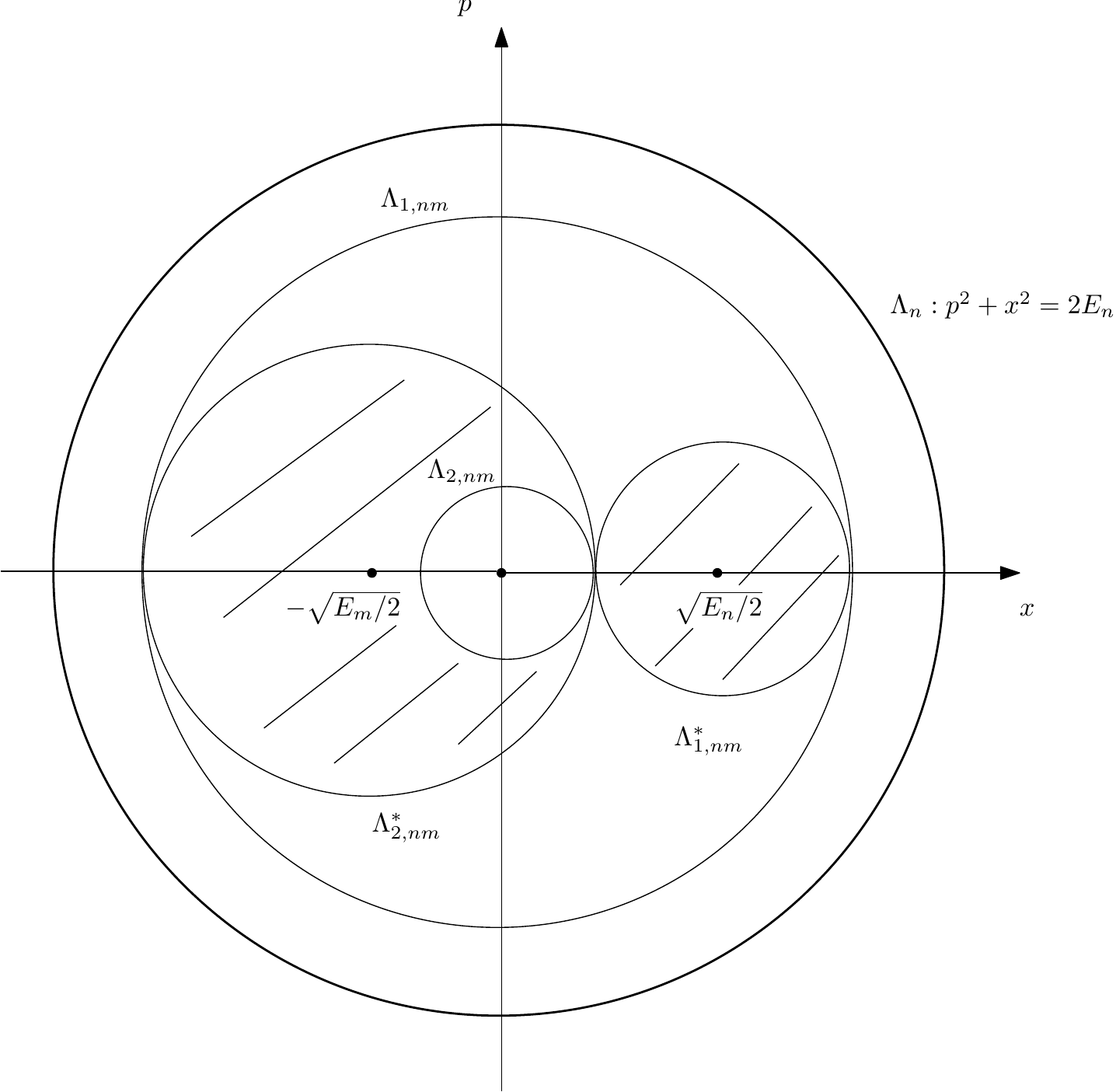}
\caption{{\it Area of existence of stationary points of $F_{\ell,nm}$, $\ell=3,4$  $(\sigma=\si_{1,2} >0)$}}
\label{stationaryp_34nma}
\end{figure}
\newpage
\begin{figure}[h]
\centering
\includegraphics[width=0.7 \textwidth]{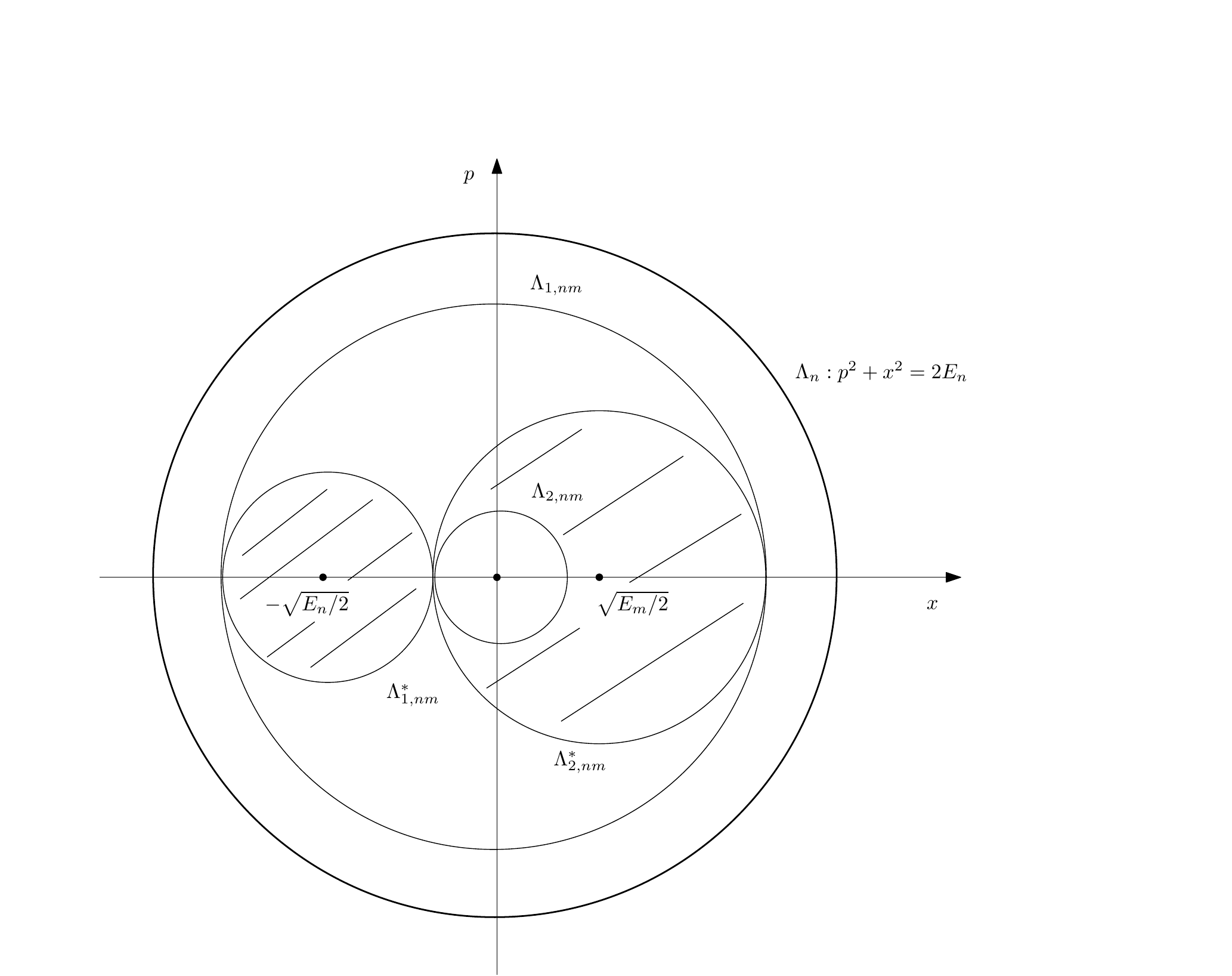}
\caption{{\it Area of existence of stationary points of $F_{\ell,nm}$, $\ell=3,4$  $(\sigma=\si_{1,2} <0)$}}
\label{stationaryp_34nmb}
\end{figure}
By standard stationary phase method, and using involved transformations  which are quite analogous to those applied to the approximation of $\mathcal{W}_{\ell,n}^{\epsilon}$ , $\ell=3,4$,  we obtain that $\mathcal{W}_{\ell,nm}^{\epsilon}$ , $\ell=3,4$ admit of asymptotic approximations that match with Airy approximation $(\ref{new_notation_nm})$.

The asymptotic contribution to the Wigner integral  $\mathcal{W}^{\hb}_{nm}(x,p)$ when $(x,p)$ lies in various regions of phase space,  are summarized  in the Table \ref{table2}. For fixed $x$ with $\rho_{nm}<|x|<R_{nm}$, we have 
\begin{center}
\begin{tabular}[!hbp]{|c|c|}\hline
region & main contribution to $\mathcal{W}^{\hb}_{nm}(x,p)$ \\
\hline
$p>\sqrt{R^{2}_{nm}-x^2}$ & $\mathcal{W}^{\hb}_{1,nm}\approx\widetilde{\mathcal{W}_{nm}^{\epsilon}}$ \\
\hline
$p\approx\sqrt{R^{2}_{nm}-x^2}$ & $\mathcal{W}^{\hb}_{1,nm}\approx\widetilde{\mathcal{W}_{nm}^{\epsilon}}$ \\
\hline
inside $\Lambda^{*}_{nm}$ & $\mathcal{W}^{\hb}_{3,nm}+\mathcal{W}^{\hb}_{4,nm}\approx\widetilde{\mathcal{W}_{nm}^{\epsilon}}$\\
\hline
$p\approx-\sqrt{R^{2}_{nm}-x^2}$ & $\mathcal{W}^{\hb}_{2,nm}\approx\widetilde{\mathcal{W}_{nm}^{\epsilon}}$ \\
\hline
$p>-\sqrt{R^{2}_{nm}-x^2}$ & $\mathcal{W}^{\hb}_{2,nm}\approx\widetilde{\mathcal{W}_{nm}^{\epsilon}}$ \\
\hline
\end{tabular}
\end{center}
\begin{table}[h!]
\caption{\it{ The main contribution to $\mathcal{W}^{\hb}_{nm}(x,p)$} }
\label{table2}
\end{table}
\newpage
Therefore,  the leading approximation of $W^{\hb}_{nm}(x,p)$ is 
\begin{eqnarray}\label{approx_w_nm}
\mathcal{W}_{nm}^{\epsilon}(x,p)&\approx& \widetilde{\mathcal{W}_{nm}^{\epsilon}}(x,p) \nonumber\\
&:=& {\pi}^{-1}
e^{-i(n-m)\phi}{\epsilon}^{-2/3}R_{nm}^{-4/3}(R_{nm}^2-\rho_{nm}^2)^{1/3}
Ai\left(\frac{p^2+x^2-R_{nm}^2}{\epsilon^{2/3}{R_{nm}^{4/3}(R_{nm}^2-\rho_{nm}^2)}^{-1/3}}\right)\ , \nonumber \\
\end{eqnarray}
where $R_{nm}$, $\rho_{nm}$ are given by $(\ref{z1})$, $(\ref{z2})$.
%%%%%%%%%%%%%%%%%%%%%%%%%%%%%%%%%%%%%%%%%%%%%%%%%%%%%%%%%%%%%%%%%%%%%%%%%%%%%%%%%%%%% 
%%%%%%%%%%%%%%%%%%%%%%%%%%%%%%%%%%%%%%%%%%%%%%%%%%%%%%%%%%%%%%%%%%%%%%%%%%%%%%%%%%%%%%%
\section{Derivation of the asymptotics: diagonal case }\label{sec74}

\subsubsection*{Stationary points of the Wigner phases $F_{\ell,n}(\sigma;x,p) \ ,  \ \  \ell=1,2$ .}
For a fixed point $(x,p)$ in phase space, we consider the stationay points of $F_{\ell,n}(\sigma;x,p)$  (eq. $(\ref{phase_1n})$) w.r.t. $\si$, only  in the region
\begin{equation}\label{illum}
|x\pm\si|<\sqrt{2E_{n}}  \ ,
 \end{equation}
because outside of this region the phase $S_{n}$ becomes imaginary, and the contribution to the integral is exponentially small. 

The stationary points of $F_{1,n}$ are roots of the equation
\begin{equation}\label{symeq}
\partial_{\si}{F_{1,n}}(\sigma;x,p)={S_{n}}^{\prime}(x+\sigma)+{S_{n}}^{\prime}(x-\sigma)-2p=0
\ ,
\end{equation}
that is
\begin{equation}\label{stateqa}
\sqrt{2E_{n}-(x+\sigma)^2}+\sqrt{2E_{n}-(x-\sigma)^2}=2p \ .
\end{equation}

For $p<0$ there are no roots since $\sqrt{2E_{n}-(x\pm\sigma)^2}>0$ , 
while for $p>0$ we see that, due to the symmetry, the  possible roots of $(\ref{stateqa})$ must appear in symmetric pairs. 
By repeated elimination of the square roots in $(\ref{stateqa})$, under the constraint
\begin{eqnarray}\label{cond}
\sqrt{2E_{n}-(x+\sigma)^2}\, \, \sqrt{2E_{n}-(x-\sigma)^2}=2p^{2}+x^{2}-2E_n+\sigma ^2 \ge 0 \ ,
\end{eqnarray}
the solution of $(\ref{stateqa})$ provides the stationary points of $F_{1,n}$ ,
\begin{equation}\label{critpoints}
\sigma(x,p) =\pm \si_{0}(x,p) \ , 
\end{equation}
where
\begin{equation}\label{sigma_0}
 \si_{0}(x,p):=\frac{|p|}{\sqrt{p^2+x^2}}\sqrt{|2E_{n}-p^2-x^2|} \ .
\end{equation}
These stationary points may coalesce to a double point $\sigma(x,p)=0$ for $(x,p) \in \Lambda_{n}$ . 

For the constraints  $(\ref{illum})$ and $(\ref{cond})$ to be satisfied by the stationary points 
$(\ref{critpoints})$, it turns out that $(x,p)$ must lie in the meniscus $\Sigma_{n}=\Sigma_{n}^{+}\cup\Sigma_{n}^{-}$ defined by the intersection of the interior $p^2+x^2\leq2E_{n}$ of  the Lagrangian curve 
$\Lambda_{n}$ and the exterior of its dual curve $\Lambda^{*}_{n}$.

Now by
\begin{eqnarray*}
\partial_{\si\si}{F_{1,n}}(\sigma;x,p)=\frac{-2\si(p^2+x^2)}{p\sqrt{2E_{n}-{(x+\si)}^2}
\sqrt{2E_{n}-{(x-\si)}^2}}
\end{eqnarray*}
we have
\begin{eqnarray*}
\partial_{\si\si}{F_{1,n}}(\sigma=\pm
\sigma_{0};x,p)\not=0 \ , 
\end{eqnarray*}
and 
\begin{eqnarray*}
\partial_{\si\si}{F_{1,n}}(\sigma=0 ;x,p)=0 \ , \ \  \partial_{\si\si\si}{F_{1,n}}(\sigma=0 ;x,p)\not=0 \ .
\end{eqnarray*}
Thus, the stationary points $\pm \sigma_{0}(x,p)$ given by $(\ref{sigma_0})$ are simple, and
the point $\sigma(x,p)=0$  formatted by the coalescence of 
$\pm\sigma_{0}(x,p)$ , $(x,p) \in \Lambda_{n}$  is double.

Therefore,  in the upper meniscus $\Sigma_{n}^{+}$ , the phase $F_{1,n}$  has two real stationary points, 
$$\si(x,p)=\pm\si_{0}(x,p)=\pm\frac{p}{\sqrt{p^2+x^2}}\sqrt{2E_{n}-p^2-x^2}
\quad \mathrm{for}\ \ \ p>0 \ , $$
which coalesce to $\sigma(x,p)=0$ on the upper branch $p=+\sqrt{2E_{n}-x^2}$ of  $\Lambda_{n}$ . This result  would also come out  by Berry's chord construction (see Subsection  \ref{section422}) near this branch of $\Lambda_{n}$ .
\newpage
\begin{figure}[h]
\centering
\includegraphics[width=0.5\textwidth]{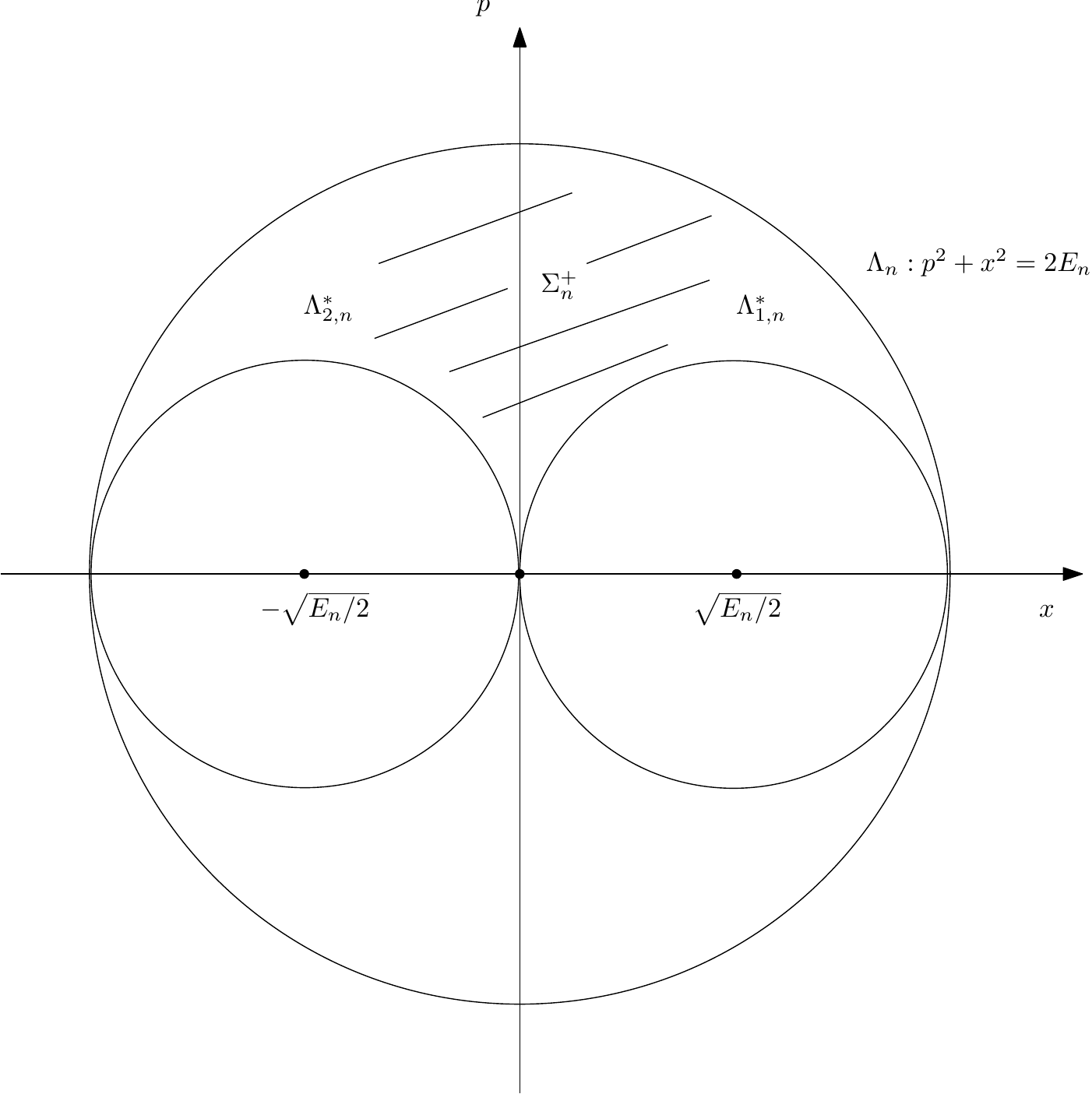}
\caption{{\it Region of critical points of $F_{1,n}$  }}
\end{figure}

In the exterior $p^2+x^2\geq2E_{n}$ , $p>0$  of $\Lambda_{n}$ , obviously the phase $F_{1,n}$ has two simple imaginary stationary points
$$\sigma(x,p)=\pm i \sigma_{0}(x,p)=\pm i\frac{p}{\sqrt{p^2+x^2}}\sqrt{p^2+x^2-2E_{n}}\quad \mathrm{for} \  p>0$$
 which also coalesce to $\sigma(x,p)=0$ on the upper branch of $p^2+x^2=2E_{n}$ . Such points have  exponentially small contributions to the corresponding Wigner integral.
 
Because of the relation $(\ref{diagsym_12} )$, a quite similar analysis can be done for the stationary points of $F_{2,n}$ , and we conclude that in the lower meniscus $\Sigma_{n}^{-}$ , the phase $F_{2,n}$  has two real stationary points, 
$$\si(x,p)=\pm\si_{0}(x,p)=\pm\frac{p}{\sqrt{p^2+x^2}}\sqrt{2E_{n}-p^2-x^2}
\quad \mathrm{for}\ \ \ p<0 \ , $$
which coalesce to $\sigma(x,p)=0$ on the lower branch $p=-\sqrt{2E_{n}-x^2}$ of  $\Lambda_{n}$ . This result  would again come out  by Berry's chord construction (see Subsection  \ref{section422}) near the lower branch of $\Lambda_{n}$ . Also, in the region $p^2+x^2\geq2E_{n}$ ,  $p< 0$ there are two simple imaginary stationary points $\pm i \sigma_{0}$ of the phase $F_{2,n}$ , which again coalesce to the double point $\sigma(x,p)=0$ on the lower branch of the Lagrangian manifold $p^2+x^2=2E_{n}$ .

%%%%%%%%%%%%%%%%%%%%%%%%%%%%%%%%%%%%%%%%%%%%%%%%%%
\subsubsection*{Asymptotic expansion of the integrals $\mathcal{W}_{\ell,n}^{\epsilon} \ , \ell=1,2$ .}

From the above analysis of the stationary points, it follows that we can use the semiclassical Wigner function $(\ref{sclwigairy})$ in order to approximate the Wigner functions $\mathcal{W}_{1,n}^{\epsilon}$ (resp. $\mathcal{W}_{2,n}^{\epsilon}$), near the upper (resp. lower) branch $p=+\sqrt{2E_{n}-x^2}$ 
(resp. $p=-\sqrt{2E_{n}-x^2}$) of the Lagrangian curve $\Lambda_{n}$. The parameter $\alpha$ must be accordingly selected to be  $\alpha=p-\sqrt{2E_{n}-x^2}$ (resp. $\alpha=p+\sqrt{2E_{n}-x^2}$).
In both cases we have 
$$\xi\approx2(-\alpha){\left[\frac{p^2+x^2}{p(2E_{n}-x^2)}\right]}^{-1/3} \ ,$$ 
with the respective $\alpha$ for each branch.
Also by $(\ref{delta})$ we get
$$A_0\approx\frac{1}{4}{(p^2+x^2)}^{-1/3} \ ,  \ \ \ B_0=0 \ .$$

Thus, $(\ref{sclwigairy})$ leads to the following approximation formulae 
\begin{eqnarray}\label{1n_appr}
\mathcal{W}_{1,n}^{\epsilon}(x,p)&\approx&
{1\over \pi}\, \hb^{-2/3}{\left(2E_{n}\right)}^{-1/3}
Ai\left(\frac{2\epsilon^{-2/3}}{\left(2E_{n}\right)^{1/3}\left(2E_{n}-x^2\right)^{-1/2}}\left(p-\sqrt{2E_{n}-x^2}\right)\right)\nonumber \\ 
&\approx&{1\over \pi}\, \hb^{-2/3} {\left(2E_{n}\right)}^{-1/3}\, Ai\left(\frac{ p^2+x^2-2E_n}{\hb^{2/3}{(2E_{n})}^{1/3}}\right) \ ,
\end{eqnarray}
near the branch $p=\sqrt{2E_{n}-x^2}$ , and 
\begin{eqnarray}\label{2n_appr}
\mathcal{W}_{2,n}^{\epsilon}(x,p)\approx
{1\over \pi}\, \hb^{-2/3} {\left(2E_{n}\right)}^{-1/3}\, Ai\left(\frac{ p^2+x^2-2E_n}{\hb^{2/3}{(2E_{n})}^{1/3}}\right)  \ ,
\end{eqnarray}
near the branch $p=-\sqrt{2E_{n}-x^2}$ .
Someone should keep in mind that the approximations hold  for small $\epsilon$ and large enough $n$.

Outside $\Lambda_n$ , where the stationary points are imaginary and again coalesce to a double point on $\Lambda_n$ , we can formally repeat the asymptotic calculations with the uniform stationary phase formula, in the same way as we derived the semiclassical Wigner function $(\ref{sclwigairy})$. This procedure somehow corresponds to the analytic continuation of the asymptotic formulae in the complex space but it is not rigorous since the Airy function becomes exponentially small and the algebraic remainders cease to carry any approximation information. Nevertheless, the formal asymptotic calculation leads to the same approximations $(\ref{1n_appr})$, $(\ref{2n_appr})$.

%%%%%%%%%%%%%%%%%%%%%%%%%%%%%%%%%%%%%%%%%%%%%%%%%%%%%%%%%%%%%%
\subsubsection*{Stationary points of the Wigner phases $F_{\ell,n}$ , $\ell=3,4$ .}

The stationary points of $F_{3,n}$ are roots of the equation
\begin{equation}\label{asymeq}
\partial_{\si}{F_{3,n}}(\sigma;x,p)={S_{n}}^{\prime}(x+\sigma)-{S_{n}}^{\prime}(x-\sigma)-2p=0
\ ,
\end{equation}
that is
\begin{equation}\label{astateqa}
\sqrt{2E_{n}-(x+\sigma)^2}-\sqrt{2E_{n}-(x-\sigma)^2}=2p \ .
\end{equation}

Proceeding analogously to the calculation of stationary points of $F_{1,n}$ , we obtain that for any fixed $(x,p)$ interior point of  $\Lambda^{*}_{n}$ (see $(\ref{union})$),  the stationary point is $\sigma=-\sigma_0$ for $(x\geq 0,p> 0)$ or $(x\leq 0,p< 0)$, while it  is $\sigma=+\sigma_0$  if $(x\geq 0,p<0)$ or $(x\leq 0,p> 0)$, where 
$\sigma_0=\si_0(x,p)$ given by $(\ref{sigma_0})$.

Outside of  the Lagrangian curve $\Lambda_{n}$ , for $p^2+x^2>2E_n$ ,  the stationary points are imaginary
$\sigma(x,p)=\pm i\sigma_0$ . 
%\newpage
\begin{figure}[h]
\centering
\includegraphics[width=0.5 \textwidth]{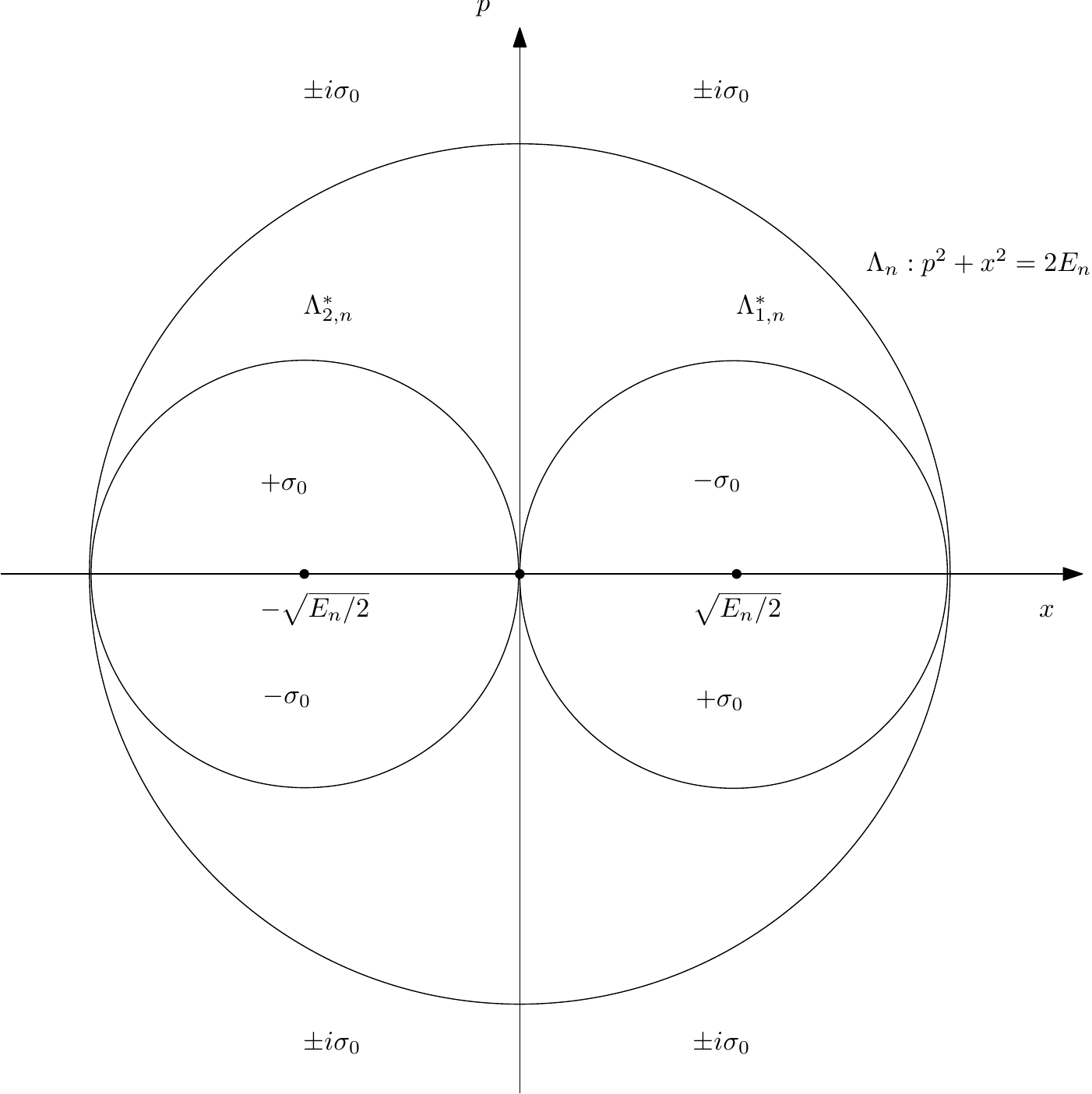}
\caption{{\it Stationary points of $F_{3,n}$}}
\label{statpoints3n}
\end{figure}

In this case, all the stationary points are simple, since by
\begin{eqnarray*}
\partial_{\si\si}{F_{3,n}}(\sigma;x,p)=\frac{2\sigma (p^2+x^2)}{p\sqrt{2E_{n}-(x+\sigma)^2}\sqrt{2E_{n}-(x-\sigma)^2}}\quad ,\quad \mathrm{for} \quad p\neq 0
\end{eqnarray*}
we have 
$$
\partial_{\si\si}{F_{3,n}}(\sigma=\pm
\sigma_0;x,p)\not=0\quad , \quad
\partial_{\si\si}{F_{3,n}}(\sigma=\pm i \sigma_0;x,p)\not=0 \quad .
$$
Note that $\partial_{\si\si}{F_{3,n}}(\sigma=\pm \sigma_0;x,p)$ becomes
infinite on $\Lambda^{*}_{1,n}$ and $\Lambda^{*}_{2,n}$ . Finally, it is important to observe that in this case there are no
stationary points between $\Lambda_n$ and $\Lambda^{*}_{n}=\Lambda^{*}_{1,n}\cup\Lambda^{*}_{2,n}$.
%\newpage
\begin{figure}[h]
\centering
\includegraphics[width=0.5 \textwidth]{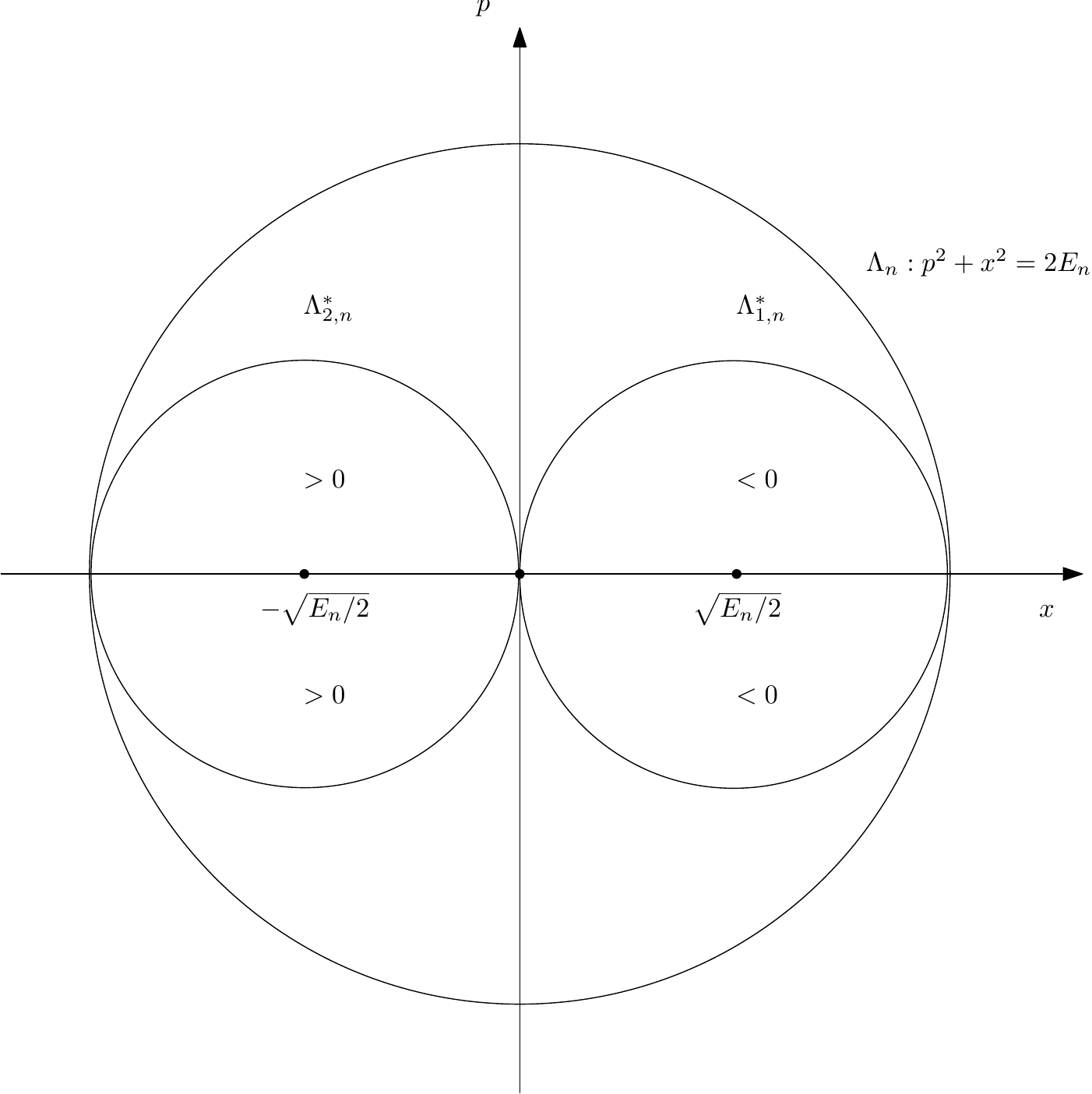}
\caption{{\it The sign of  $\partial_{\si\si}F_{3,n}$}}
\end{figure}

Because of the relation $(\ref{diagsym_12} )$, a quite similar analysis can be done for the stationary points of $F_{4,n}$ , and we conclude that  for $(x\geq 0,p> 0)$ or $(x\leq 0,p< 0)$, the stationary
point is $\sigma=+\sigma_0$ , while for
$(x\geq 0,p< 0)$ or $(x\leq 0,p>0)$  it is $\sigma=-\sigma_0$. 
As before, outside of  the Lagrangian curve $\Lambda_{n}$ , for $p^2+x^2>2E_n$ ,  the stationary points are imaginary $\sigma(x,p)=\pm i\sigma_0$ , and all stationary points $\pm \sigma_0$ and $\pm i\sigma_0$ are simple.
\newpage
\begin{figure}[h]
\centering
\includegraphics[width=0.5 \textwidth]{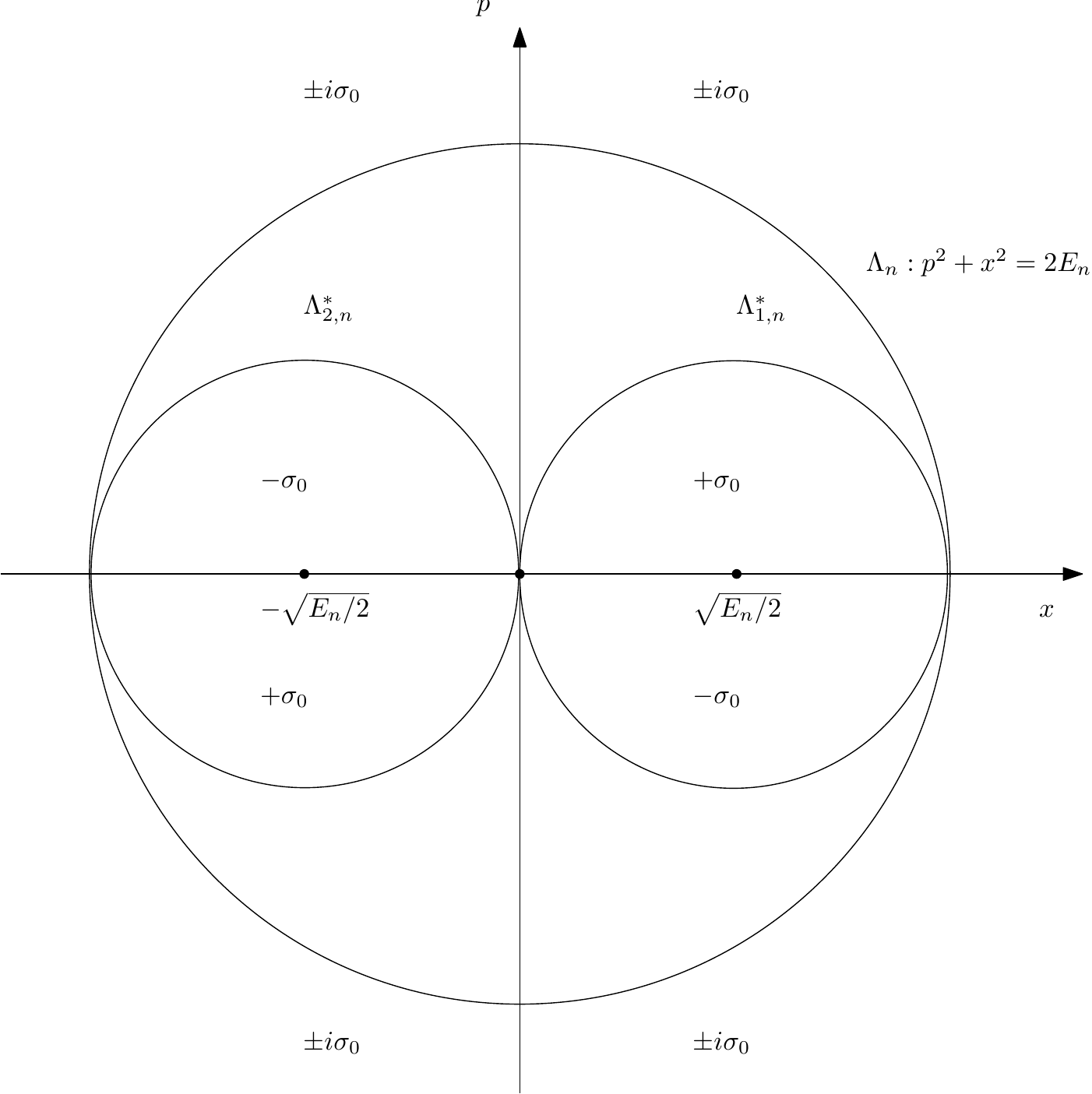}
\caption{{\it Stationary points of $F_{4,n}$}}
\label{statpoints4n}
\end{figure}

\subsubsection*{Asymptotic expansion of the integrals $\mathcal{W}_{\ell,n}^{\epsilon} \ , \ell=3,4$ .}

Therefore, we can apply the standard stationary phase formula inside $\Lambda^{*}_{n}$ , and we get the following approximations

\begin{itemize}
\item for $(x\geq0,p>0)$ ,
\begin{eqnarray}
\mathcal{W}_{3,n}^{\epsilon}(x,p)&\approx& 
\frac{1}{2\pi^{3/2}\sqrt{\epsilon}}\,
e^{\frac{i}{\epsilon}F_{3,n}(-\sigma_{0})}e^{i\pi/4}(p^2+x^2)^{-1/4}(2E_{n}-p^2-x^2)^{-1/4}\quad ,\nonumber\\
\mathcal{W}_{4,n}^{\epsilon}(x,p)&\approx&\frac{1}{2\pi^{3/2}\sqrt{\epsilon}}\,
e^{-\frac{i}{\epsilon}F_{3,n}(-\sigma_{0})}e^{-i\pi/4}(p^2+x^2)^{-1/4}(2E_{n}-p^2-x^2)^{-1/4}\quad , \nonumber\\
&&\,
\end{eqnarray}
\item for $(x\geq0,p<0)$ ,
\begin{eqnarray}
\mathcal{W}_{3,n}^{\epsilon}(x,p)&\approx&\frac{1}{2\pi^{3/2}\sqrt{\epsilon}}\,
e^{\frac{i}{\epsilon}F_{3,n}(+\sigma_{0})}\, e^{i\pi/4}(p^2+x^2)^{-1/4}(2E_{n}-p^2-x^2)^{-1/4}\quad , \nonumber\\
\mathcal{W}_{4,n}^{\epsilon}(x,p)&\approx&\frac{1}{2\pi^{3/2}\sqrt{\epsilon}}\,
e^{-\frac{i}{\epsilon}F_{3,n}(+\sigma_{0})}e^{-i\pi/4}(p^2+x^2)^{-1/4}(2E_{n}-p^2-x^2)^{-1/4}\quad , \nonumber\\
&&\,
\end{eqnarray}
\item for $(x\leq 0,p>0)$ ,
\begin{eqnarray}
\mathcal{W}_{3,n}^{\epsilon}(x,p)&\approx&\frac{1}{2\pi^{3/2}\sqrt{\epsilon}}\,
e^{\frac{i}{\epsilon}F_{3,n}(+\sigma_{0})}e^{i3\pi/4}(p^2+x^2)^{-1/4}(2E_{n}-p^2-x^2)^{-1/4}\quad , \nonumber\\
\mathcal{W}_{4,n}^{\epsilon}(x,p)&\approx&\frac{1}{2\pi^{3/2}\sqrt{\epsilon}}\,
e^{-\frac{i}{\epsilon}F_{3,n}(+\sigma_{0})}e^{-i3\pi/4}(p^2+x^2)^{-1/4}(2E_{n}-p^2-x^2)^{-1/4}\quad , \nonumber\\
&&\,
\end{eqnarray}
\item for $(x\leq0,p<0)$ ,
\begin{eqnarray}
\mathcal{W}_{3,n}^{\epsilon}(x,p)&\approx&\frac{1}{2\pi^{3/2}\sqrt{\epsilon}}\,
e^{\frac{i}{\epsilon}F_{3,n}(-\sigma_{0})}e^{i3\pi/4}(p^2+x^2)^{-1/4}(2E_{n}-p^2-x^2)^{-1/4}\quad , \nonumber\\
\mathcal{W}_{4,n}^{\epsilon}(x,p)&\approx&\frac{1}{2\pi^{3/2}\sqrt{\epsilon}}\,
e^{-\frac{i}{\epsilon}F_{3,n}(-\sigma_{0})}e^{-i3\pi/4}(p^2+x^2)^{-1/4}(2E_{n}-p^2-x^2)^{-1/4}\quad , \nonumber\\
&&\,
\end{eqnarray}
\end{itemize}
We observe that the approximation of $\mathcal{W}_{4,n}^{\epsilon}(x,p)$ is the complex conjugate of the approximation of $\mathcal{W}^{\epsilon}_{3,n}(x,p)$ .

Outside $\Lambda_n$ we can formally apply either the standard stationary phase formula with imaginary stationary points (this is essentially the so called complex stationary phase formula), or the steepest descents formula (see \cite{BH}, rel.(7.2.10) p.265) to conclude that
\begin{eqnarray}
\mathcal{W}_{3,n}^{\epsilon}(x,p)=0 \ , \ \ \  \mathcal{W}_{4,n}^{\epsilon}(x,p)=0  \ .
\end{eqnarray}

\subsubsection*{}

For any fixed $x$, with $|x|<\sqrt{2E_{n}}$, we summarize the main asymptotic contribution to the phase space eigenfunction  $\mathcal{W}^{\hb}_{n}(x,p)$ in the following table
\begin{center}
\begin{tabular}[!hbp]{|c|c|}\hline
region & main contribution to $\mathcal{W}^{\hb}_{n}(x,p)$ \\
\hline
$p>\sqrt{2E_{n}-x^2}$ & $\mathcal{W}^{\hb}_{1,n}$ \\
\hline
$p\approx\sqrt{2E_{n}-x^2}$ & $\mathcal{W}^{\hb}_{1,n}$ \\
\hline
inside $\Lambda^{*}_{n}$ & $\mathcal{W}^{\hb}_{3,n}+\mathcal{W}^{\hb}_{4,n}$\\
\hline
$p\approx-\sqrt{2E_{n}-x^2}$ & $\mathcal{W}^{\hb}_{2,n}$ \\
\hline
$p>-\sqrt{2E_{n}-x^2}$ & $\mathcal{W}^{\hb}_{2,n}$ \\
\hline
\end{tabular}
\end{center}
%\begin{table}[h]
%\caption{\it{ The main contribution to $\mathcal{W}^{\hb}_{n}(x,p)$} }
%\end{table}
Recall that $\Lambda^{*}_{n}$ is given by (\ref{union}), (see also  Figure \ref{stat_points_n}). 
The integrals ${\mathcal W}^\epsilon_{1,n}$ and ${\mathcal W}^\epsilon_{2,n}$ have formally the same asymptotic formula (see $(\ref{1n_appr})$, $(\ref{2n_appr})$). 

Moreover, by rather involved transformations (which are suggested by the asymptotics of Laguerre polynomials which are the exact Wigner eigenfunctions, Section \ref{761}), the phase $F_{3,n}(-\sigma_{0})$ can be related to the phase of the Airy function in the approximations  $(\ref{1n_appr})$, $(\ref{2n_appr})$ of  $\mathcal{W}_{\ell,n}^{\epsilon} \ , \ell=1,2$. In this way we can see that inside the manifold $\Lambda^{*}_{n}$ the contribution  $(\mathcal{W}^{\hb}_{3,n}+\mathcal{W}^{\hb}_{4,n})$ to  $\mathcal{W}^{\hb}_{n}$ has an asymptotic formula which is given by the same Airy function. Also, in Section \ref{76} we will compute the  asymptotics of the exact Wigner eigenfunctions of the harmonic oscillator in terms of Laguerre polynomials (see $(\ref{eigenf_lag})$) and we will explain how these eigenfunctions can be approximated  by an  Airy function concentrated near the Lagrangian curve  $\Lambda_{n}$, for large $n$ and small $\hb$ (see  eq.  $(\ref{ex_eigenf_as_n})$ below).

 Therefore, we  conclude that $\mathcal{W}^{\hb}_{n}(x,p)$ is approximated by the formula
\begin{eqnarray}
\mathcal{W}_{n}^{\epsilon}(x,p)\approx {\pi}^{-1}\, \hb^{-2/3}{\left(2E_{n}\right)}^{-1/3}\,
Ai\left(\frac{p^2+x^2-2E_n}{\epsilon^{2/3}{(2E_{n})}^{1/3}}\right)\quad ,
\end{eqnarray}
 for small $\hb$ , large $n$, and $E_{n}=E^{\hb}_{n}=(n+1/2)\hb$ .

%%%%%%%%%%%%%%%%%%%%%%%%%%%%%%%%%%%%%%%%%%%%%%%%%%%%%%%%%%%%%%%%%%%%%%%%%%%%%%%%%%%%%%

\section{Derivation of the asymptotics: off-diagonal case}\label{sec75}

\subsubsection{Stationary points of the Wigner phase $F_{\ell,nm}(\sigma;x,p) \ ,  \ \ell=1,2$. } 

For a fixed point $(x,p)$ in the interior $\Lambda_{1,nm}$ (see $(\ref{man_nm_a}))$, we consider the stationary points of $F_{\ell,nm}(\sigma;x,p)$  (eq. $(\ref{phase_1n})$) w.r.t. $\si$, only  in the region
\begin{equation}\label{illummn}
|{x\pm\sigma}|<\sqrt{2E_n} \ , \ \ |{x\pm\sigma}|<\sqrt{2E_m} 
\end{equation}
because outside of this region the phase $S_{n} \ , S_{m}$ become imaginary, and their contribution to the integral is exponentially small. 

The stationary points of $F_{1,nm}(\sigma;x,p)$ (see $(\ref{phase_1nm})$) are  roots of the equation
\begin{eqnarray}
\partial_{\si} {F_{1,nm}}(\sigma;x,p)=S'_{n}(x+\si)-S'_{m}(x-\si)-2p=0 \ ,
\end{eqnarray}
that is
\begin{eqnarray}\label{stpointsnm1}
\sqrt{ 2E_{n}-(x+\sigma)^2 }+\sqrt{ 2E_{m}-(x-\sigma)^2 }-2p=0 \ .
\end{eqnarray}
For $p<0$ there are no roots since $\sqrt{2E_{n}-(x\pm\sigma)^2}>0$. 

For $p>0$, by repeated elimination of the square roots in $(\ref{stpointsnm1})$, under the constraint
\begin{eqnarray}\label{condnm1}
\sqrt{2E_{n}-(x+\sigma)^2}\, \, \sqrt{2E_{n}-(x-\sigma)^2}=2p^{2}+x^{2}- (E_n+ E_m)+ \sigma ^2 \ge 0 \ ,
\end{eqnarray}
we get the quadratic equation
\begin{eqnarray}\label{quadraticeq}
(p^2+x^2)\, \si^2-2x e_{nm}\, \si+p^2(p^2+x^2-2E_{nm})+{e}^{2}_{nm}=0
\end{eqnarray}
where
\begin{eqnarray}\label{Epsilon}
E_{nm}:=\frac{1}{2}\left(E_{n}+E_{m}\right)\quad \mathrm{and}\quad e_{nm}:=\frac{1}{2}\left(E_{n}-E_{m}\right) \ .
\end{eqnarray}
The determinant of the quadratic equation is
\begin{eqnarray*}
\Delta=-4p^2\left[(p^2+x^2)(p^2+x^2-2E_{nm})+{e}^{2}_{nm}\right] \ .
\end{eqnarray*}

It follows that there are  two real roots, when $p^2+x^2-2E_{nm}<0$ with $(p^2+x^2)(p^2+x^2-2E_{nm})+{e}^{2}_{nm}<0$ and $p>0$. This means that the stationary points are real at $(x,p)$ lying in the ring $\rho_{nm}^2<p^2+x^2<R_{nm}^2$
where $R_{nm}:=(\sqrt{2E_{n}}+\sqrt{2E_{m}})/2$ and $\rho_{nm}:=(\sqrt{2E_{n}}-\sqrt{2E_{m}})/2$ . 

Solving $(\ref{quadraticeq})$ w.r.t. $\sigma$, for $(x,p)\neq (0,0)$, we get the stationary points  
\begin{eqnarray}\label{statpointsnm}
\si_{1,2}(x,p)=\frac{x e_{nm}}{p^2+x^2}\pm\frac{|p|}{p^2+x^2}\sqrt{(p^2+x^2)(2E_{nm}-p^2-x^2)-{e}^{2}_{nm}} \ .
\end{eqnarray}

The constraints $(\ref{illummn})$, $(\ref{condnm1})$, that is $|{x+\sigma}|<\sqrt{2E_n}$ , $|{x-\sigma}|<\sqrt{2E_m}$ and $\sigma^2\geq 2E_{nm}-2p^2-x^2$, impose the following restrictions. First, $\sigma=\si_{1,2} \geq 0$
if $p^2+{(x-\sqrt{E_n/2})}^2\geq E_{m}/2$ and $p^2+{(x+\sqrt{E_m/2})}^2\geq E_{n}/2$ , and second, $\sigma=\si_{1,2} \leq 0$, if $p^2+{(x-\sqrt{E_m/2})}^2\geq E_{n}/2$ and $p^2+{(x+\sqrt{E_n/2})}^2\geq E_{m}/2$. These regions, for $p>0$, as the shaded regions in Figures \ref{spnm_1a}, \ref{spnm_1b}. It follows that stationary points exist only in the shaded part of the ring $\rho_{nm}^2<p^2+x^2<R_{nm}^2$, and in particular there are no stationary points for $(x,p)$ lying on the inner boundary $x^2+p^2= \rho^{2}_{nm}$. Thus, although the discriminant $\Delta$ vanishes on this circle, we get double stationary points only on the circle $x^2+p^2= R^{2}_{nm}$.

Since $\partial_{\si\si}{F_{1,nm}}(\si=\si_{1,2}(x,p))\neq 0 \ ,$ the stationary points $(\ref{statpointsnm})$ are simple, and coalesce to the double stationary point 
\begin{eqnarray}\label{si0bar}
\bar{\si}_0(x,p):=\frac{xe_{nm}}{p^2+x^2} \ ,
\end{eqnarray}
 as $(x,p)$ approaches the upper branch of the curve $\Lambda _{1,nm}=\{p^2+x^2=R_{nm}^2\}$ where $\Delta$ vanishes.
 
Because of the relation $(\ref{offdiagsym_12})$, a quite similar analysis can be done for the stationary points of $F_{2,nm}$. In this case we conclude that for $p<0$ there are two simple stationary points are given by $(\ref{statpointsnm})$. These are a couple of simple stationary points that again coalesce to the double point $\bar{\si}_0$ as  $(x,p)$ approach the lower branch of the Lagrangian curve $\Lambda _{1,nm}$.
 
Note that in this case, the situation is quite different than in the case of $F_{\ell,n}$ where the double stationary point is zero. This difference significantly affects the asymptotics of the Wigner integrals $\mathcal{W}_{\ell,nm}^{\epsilon} \ , \ell=1,2$, since $F_{\ell,nm}(\bar{\si}_0(x,p)) \neq 0$ contributes an extra oscillatory term, which, although weakly vanishes in the high frequency limit,  is crucial in deriving the correct behaviour of the classical limit of the Wigner eigenfunctions.

\subsubsection*{Asymptotic expansion of the integrals $\mathcal{W}_{\ell,nm}^{\epsilon} \ , \ell=1,2$ .}

The double stationary point is not zero, $\bar{\si}_0 \neq 0$ (see eq.$(\ref{si0bar})$). For this reason, we cannot rely directly on the semiclassical Wigner function $(\ref{sclwigairy})$ as we did in the corresponding cases ($\ell=1,2)$ for the diagonal ($n=m$) integrals where $\si_{0}=0$ . Instead, we apply the  uniform stationary formula (\ref{ap4}), by computing the various quantities $\xi$ , $A_0$ , $B_0$  and
$F_{1,nm}(\bar{\si}_0,\alpha,x)$, etc., that enters  the  formula with their definitions (\ref{ap13}), (\ref{A0}), (\ref{B0}).

The double point  appears at the Lagrangian curve  $\Lambda _{1,nm}=\{x^2+p^2=R_{nm}^2\}$, where $R_{nm}$ given by $(\ref{z1})$. Accordingly the parameter $\alpha$ is selected to be  
$$\alpha=\alpha(x,p)=p-\sqrt{R_{nm}^2-x^2}\leq0 \ .$$

First of all, we write the stationary points in terms of  of $\alpha$ ,
\begin{eqnarray}
\sigma_{1,2}(\alpha)=\bar{\si}_{0}\pm \mu\sqrt{-\alpha}\label{si_12}
\end{eqnarray}
where $\mu:={p}(p^2+x^2)^{-1} \sqrt{p^2+x^2-\rho_{nm}^2}\sqrt{p+\sqrt{R_{nm}^2-x^2}}$ (obviously the simple stationary  points  $\si_{1,2}$ coalesce to the double s point $\bar{\sigma}_0=\bar{\si}_0(x,p)$, as 
$\alpha \rightarrow 0$). Moreover, we have 

\begin{eqnarray*}
\partial_{\si\si}F_{1,nm}(\si;x,p)=\frac{2xe_{nm}-2\sigma (p^2+x^2)}{p\sqrt{2E_{n}-(x+\sigma)^2}
\sqrt{2E_{m}-(x-\sigma)^2}}  \ , 
\end{eqnarray*}
\begin{eqnarray*}
\partial_{\si\alpha}{F_{1,nm}}(\si,x,p)=-2
\end{eqnarray*}
and 
\begin{eqnarray*}
\partial_{\si\si\si}{F_{1,nm}}(\si=\bar\si_0;x,p)=\frac{-2(p^2+x^2)}{p\sqrt{2E_{n}-(x+\bar\sigma_0)^2}
\sqrt{2E_{m}-(x-\bar\sigma_0)^2}} \neq 0 \ .
\end{eqnarray*}

We approximate  $\xi$ by (\ref{ap13}), 
\begin{eqnarray*}
\xi\approx2(-\alpha){\left[\frac{p^2+x^2}{p\sqrt{2E_{n}-(x+\bar\sigma_{0})^2} \, 
\sqrt{2E_{m}-(x-\bar\sigma_{0})^2}}\right]}^{-1/3}\quad .
\end{eqnarray*}

and then, by (\ref{A0}), (\ref{B0}), we get the amplitudes
\begin{eqnarray*}
A_{0}\approx\frac{1}{4}R_{nm}^{-1/3}\, (R_{nm}^2-\rho_{nm}^2)^{-1/6}\ , \ \ \ B_{0}\approx 0 \ ,
\end{eqnarray*}
where $R_{nm}$ and $\rho_{nm}$ are given by $(\ref{z1})$ and $(\ref{z2})$, respectively.

Finally, we must compute the phase $F_{1,nm}(\bar{\si}_0;x,p)$ at the double stationary point. For this, we recall that by 
$(\ref{phase_1nm})$ and $(\ref{ansn})$ we have 
 \begin{eqnarray*}
F_{1,nm}^{\hb}(\sigma;x,p)&=&{S_{n}^\hb}(x+\sigma)-{S_{m}^\hb}(x-\sigma)-2p\sigma \\
&=&\int_{\sqrt{2E_{n}^{\hb}}}^{x+\sigma}\sqrt{2E_{n}^{\hb}-t^2}dt -\int_{\sqrt{2E_{m}^{\hb}}}^{x-\sigma}\sqrt{2E_{m}^{\hb}-t^2}dt \ ,
\end{eqnarray*}
Using formula (\ref{wkbphase}) for the integrals, and putting $\sigma=\bar{\si}_0$, we get 

\begin{eqnarray*}
F_{1,nm}(\si=\bar{\si}_0;x,p)&=&
\frac{x+\bar{\si}_0}{ 2}\sqrt{2E_{n}-(x+\bar{\si}_0)^2}-
\frac{x-\bar{\si}_0}{ 2}\sqrt{2E_{m}-(x-\bar{\si}_0)^2} \nonumber \\
&&+ E_{n}\arcsin \left(\frac{x+\bar{\si}_0}{\sqrt{2E_{n}}}\right)-E_{m}\arcsin \left(\frac{x-\bar{\si}_0}{\sqrt{2E_{m}}}\right)  \nonumber \\
&&-\pi e_{nm}-2p\bar{\si}_0\, .
\end{eqnarray*}

The geometrical interpretation of the phase $F_{1,nm}(\si=\bar{\sigma}_0;x,p)$ is obtained by the help of Figure \ref{geom_1nm}.

%\newpage
\begin{figure}[h]
\centering
\includegraphics[width=0.5 \textwidth]{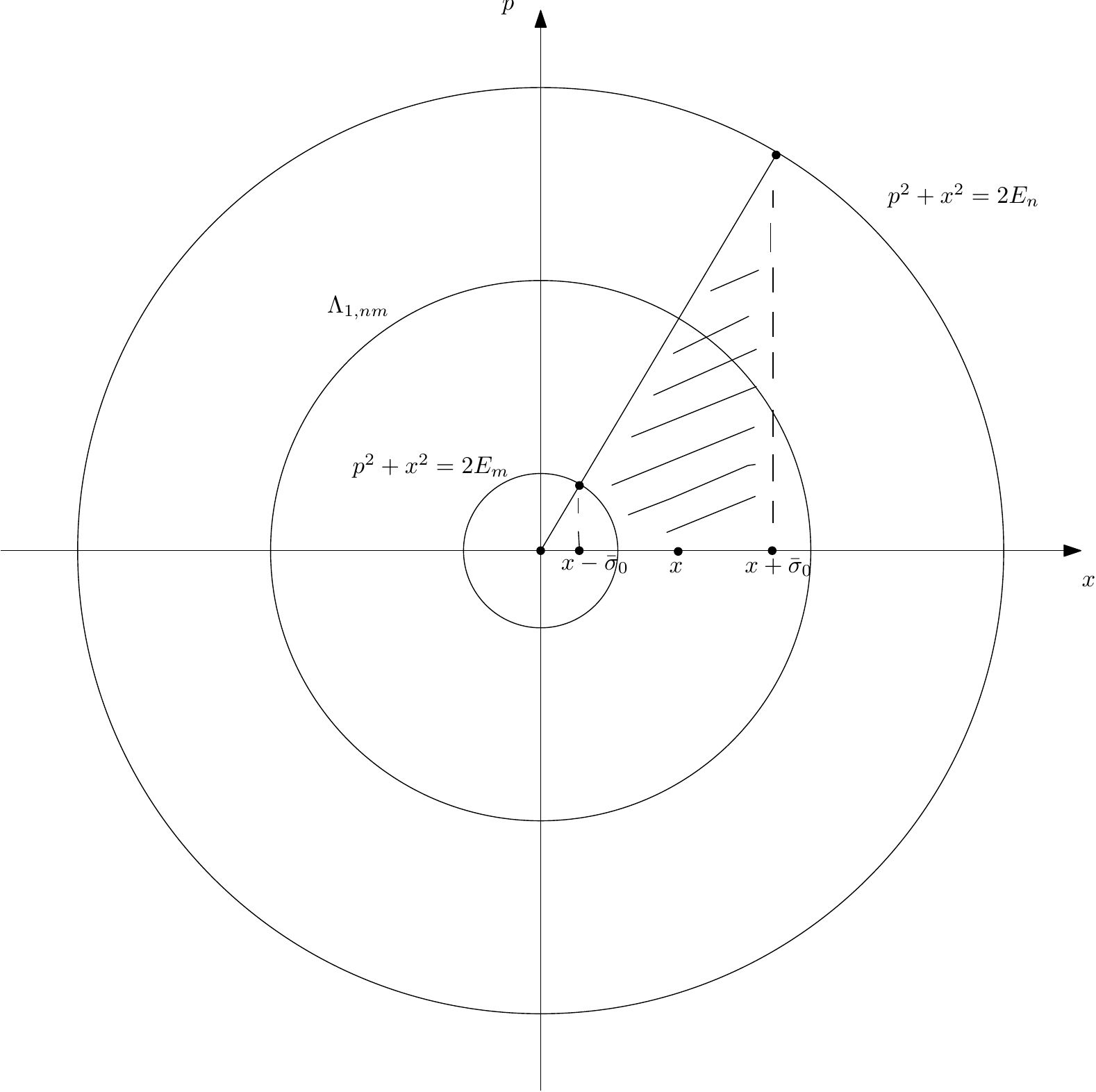}
\caption{{\it Geometrical interpretation of $F_{1,nm}(\si=\bar{\sigma}_0;x,p)$}}
\label{geom_1nm}
\end{figure}
The first term in the hrs. of the last formula is equal to the area of the small triangle with base $(0,x-\bar{\si}_0)$ . The second one, is equal to the area of the big triangle, with base $(0, x+\bar{\si}_0)$. Their difference is obviously equal to the shaded area of trapezoid,  which is equal to $2\pi\bar{\si}_0$ . Thus,
\begin{eqnarray*}
F_{1,nm}(\si=\bar{\si}_0;x,p)=
-\pi e_{nm}+E_{n}\arcsin \left(\frac{x+\bar{\si}_0}{\sqrt{2E_{n}}}\right)-E_{m}\arcsin \left(\frac{x-\bar{\si}_0}{\tpm}\right)
\end{eqnarray*}
Now we have 
\begin{eqnarray*}
&&E_{n}\arcsin \left(\frac{x+\bar{\si}_0}{\sqrt{2E_{n}}}\right)-E_{m}\arcsin \left(\frac{x-\bar{\si}_0}{\tpm}\right)-\pi e_{nm}=\\
&&E_{n} \arcsin(\cos\phi)-E_{m}\arcsin(\cos\phi)-\pi e_{nm}=\\
&&E_{n}\left(\frac{\pi}{2}-\phi\right)-E_{m}\left(\frac{\pi}{2}-\phi\right)
-\pi e_{nm}= -2\phi e_{nm} \ ,
\end{eqnarray*} 
where $\phi:=\arctan(p/x)$, $p> 0$. Recall that $e_{nm}=\frac{1}{2}\left(E_{n}-E_{m}\right)$ (eq. $(\ref{Epsilon})$).

Then, (\ref{ap4}) leads to the following approximation
formula of $\mathcal{W}_{1,nm}^{\epsilon}$ , for small  $\epsilon$  and large $n,m$
\begin{eqnarray}
 \mathcal{W}_{1,nm}^{\epsilon}(x,p)&\approx& {\pi}^{-1}\, e^{-\frac{i}{\epsilon} (E_{n}-E_{m})\phi}{\epsilon}^{-2/3}R_{nm}^{-4/3}(R_{nm}^2-\rho_{nm}^2)^{1/3}\nonumber \\
&&\cdot Ai\left(\frac{2 \left(p-\sqrt{R_{nm}^2-x^2}\right)}{\hb^{2/3}{R_{nm}^{4/3}(R_{nm}^2-x^2)^{-1/2}(R_{nm}^2-\rho_{nm}^2)}^{-1/3}}\right)\nonumber\\
&\approx & {\pi}^{-1}\,e^{-\frac{i}{\epsilon} (E_{n}-E_{m})\phi}{\epsilon}^{-2/3}R_{nm}^{-4/3}(R_{nm}^2-\rho_{nm}^2)^{1/3}
Ai\left(\frac{p^2+x^2-R_{nm}^2}{\epsilon^{2/3}{R_{nm}^{4/3}(R_{nm}^2-\rho_{nm}^2)}^{-1/3}}\right) \ , \nonumber \\\label{1nm_appr}
\end{eqnarray}
near the branch $p=\sqrt{R_{nm}^2-x^2}$ , where $R_{nm}$ , $\rho_{nm}$ given by $(\ref{z1})$ and  $(\ref{z2})$, respectively. 

By an analogous calculation of 
$\mathcal{W}^{\epsilon}_{2,nm}$ , $(\ref{wigner_int_nm})$, with $\alpha(x,p):=p+\sqrt{R_{nm}^2-x^2}\geq0$ to localise near the lower branch of Lagrangian curve  $\Lambda _{1,nm}$ ,  we obtain the  approximation  
\begin{eqnarray}
\mathcal{W}_{2,nm}^{\epsilon}(x,p)\approx
{\pi}^{-1}\,e^{-\frac{i}{\epsilon} (E_{n}-E_{m})\phi}{\epsilon}^{-4/3}R_{nm}^{-1/3}(R_{nm}^2-\rho_{nm}^2)^{1/3}
Ai\left(\frac{p^2+x^2-R_{nm}^2}{\epsilon^{2/3}{R_{nm}^{4/3}(R_{nm}^2-\rho_{nm}^2)}^{-1/3}}\right)\, ,\nonumber\\
&&\label{2nm_appr} \ ,
\end{eqnarray}
where $\phi=\arctan(p/x)$ ,  $p<0$ .

Outside $\Lambda_{1,nm}$ , the corresponding stationary points are complex and again coalesce to the double point $\bar{\si}_0=x e_{nm}/R_{nm}^2$ on $\Lambda_{1,nm}$ . We can {\it formally} repeat the asymptotic calculations with the uniform stationary phase formula to derive $(\ref{1nm_appr})$, $(\ref{2nm_appr})$.
As we have already explained in the derivation  of the approximations the diagonal terms ($n=m$), this procedure somehow corresponds to the analytic continuation of the asymptotic formulae in the complex space but it is not rigorous since the Airy function becomes exponentially small and the algebraic remainders cease to carry any approximation information. Nevertheless, the formal asymptotic calculation leads to the same approximations.

%%%%%%%%%%%%%%%%%%%%%%%%%%%%%%%%%%%%%%%%%%%%%%%%%%%%%%%%%%%%%%%%%%%%%%%%%%%%
%%%%%%%%%%%%%%%%%%%%%%%%%%%%%%%%%%%%%%%%%%%%%%%%%%%%%%%%%%%%%%%%%%%%%%%%%%%%

\subsubsection*{Stationary points of the Wigner phase $F_{\ell,nm}(\sigma,x,p) \ , \ell=3,4$ .}

The  phases $F_{3,nm}(\sigma,x,p)$ and $F_{4,nm}(\sigma,x,p)$ (rel. (\ref{phase_3nm}) and  (\ref{phase_4nm}), respectively)  have simple real stationary points inside $\Lambda_{1,nm}$ and a pair of complex stationary points outside $\Lambda_{1,nm}$. The calculation is quite analogous to the calculation of the stationary points of $F_{\ell,n}$ , $\ell=3,4$ and the regions of existence have also similar geometry (cf. Figures \ref{statpoints3n}, \ref{statpoints4n}).

By standard stationary phase method and considerations, which are based on transformations suggested by the asymptotics of Laguerre polynomials, and which are analogous to those applied to the approximation of $\mathcal{W}_{\ell,n}^{\epsilon} $ ,  $\ell=3,4$,  we obtain that $\mathcal{W}_{\ell,nm}^{\epsilon}$ ,  $\ell=3,4$ admit of asymptotic approximations that match with Airy approximations of $\mathcal{W}_{\ell,nm}^{\epsilon}$ ,  $\ell=1,2$. 

Therefore, we  conclude that
$\mathcal{W}^{\hb}_{nm}(x,p)$ is approximated by the formula
\begin{eqnarray}\label{eigen_exp_nm}
\mathcal{W}_{nm}^{\epsilon}(x,p)\approx {\pi}^{-1}
e^{-\frac{i}{\epsilon} (E_{n}-E_{m})\phi}{\epsilon}^{-2/3}R_{nm}^{-4/3}(R_{nm}^2-\rho_{nm}^2)^{1/3}
Ai\left(\frac{p^2+x^2-R_{nm}^2}{\epsilon^{2/3}{R_{nm}^{4/3}(R_{nm}^2-\rho_{nm}^2)}^{-1/3}}\right) \ ,\nonumber\\
\end{eqnarray}
for small $\hb$, and large $n,m$ such that  $n\hb,m\hb$ are constants, and $n-m$ is constant.
Recall again the definitions $R_{nm}=(\sqrt{2E_{n}}+\sqrt{2E_{m}})/2$ , $\rho_{nm}=(\sqrt{2E_{n}}-\sqrt{2E_{m}})/2$ ,  with $E_{n}=E^{\hb}_{n}=(n+1/2)\hb$, $E_{m}=E^{\hb}_{m}=(m+1/2)\hb$ and $\phi:=\arctan(p/x)$ .
It is important here to note that $\mathcal{W}_{nm}^{\epsilon}$ is significant only if $n \approx m$ due to the fastly oscillating term $e^{-\frac{i}{\epsilon} (E_{n}-E_{m})\phi}$ arising from angular anisotropy of the non-diagonal eigenfunctions.

In the following Section  \ref{76}, we will compare the above constructed  approximations  with the exact Wigner eigenfunctions. This is possible in the particular case of the harmonic oscillator, because the Wigner eigenfunctions are expressed in terms of Laguerre polynomials, which in  turn admit uniform asymptotic approximations in terms of Airy functions.

%%%%%%%%%%%%%%%%%%%%%%%%%%%%%%%%%%%%%%%%%%%%%%%%%%%%%%%%%%%%%%%%%%%%%%%%%%%%%%%%%%%%%%%%%%%%%%%%%%%%%%%%%%%%%%%%%%%%%%%%%%%%%%%%
\section{Comparison of  exact and approximate Wigner eigenfunctions}\label{76}

We have seen in Section \ref{314} that the exact Schr\"{o}dinger eigenfunctions $v_{m}^{\hb}$ of the harmonic oscillator are the Hermite functions $(\ref{eigenf_harm})$.  As we have shown  in Chapter \ref{ch5}, the corresponding exact Wigner eigenfunctions $\Phi_{nm}^{\hb}$  are obtained by the Wigner transform of the Schr\"{o}dinger eigenfunctions, i.e. $\Phi_{nm}^{\hb}:=W^{\hb}_{v_{m}^{\hb}}[v_{n}^{\hb}]$.
M.Bartlett and J.Moyal \cite{BM} have calculated the Wigner transform of Hermite functions in terms of Laguerre polynomials. By their result we get  

\begin{eqnarray}\label{eigenf_lag}
\Phi_{nm}^{\hb}(x,p)= 
\left\{
	\begin{array}{lll}
		\frac{(-1)^m}{\pi\hb}\, 2^{(n-m)/2}\sqrt{\frac{m!}{n!}}\left(\frac{x-i p}{\sqrt{\hb}}\right)^{n-m}e^{-(p^2+x^2)/\hb} L_{m}^{(n-m)}\left[\frac{2}{\hb}(p^2+x^2)\right]\quad , & \mbox{if } n\geq m\\
		&\nonumber \\
		\frac{(-1)^n}{\pi\hb}\, 2^{(m-n)/2}\sqrt{\frac{n!}{m!}}\left(\frac{x-i p}{\sqrt{\hb}}\right)^{m-n}e^{-(p^2+x^2)/\hb} L_{n}^{(m-n)}\left[\frac{2}{\hb}(p^2+x^2)\right]\quad , & \mbox{if } n\leq m 
	\end{array}
\right.\\
\end{eqnarray}

In the sequel, we use the fact that the Laguerre polynomials in $(\ref{eigenf_lag})$ can be approximated for large $n$ and small $\hb$ by using the asymptotic formula (\ref{lag_asymp}) in Appendix \ref{app_e}, to show that the approximations $(\ref{approx_w_n})$ and $(\ref{approx_w_nm})$ of $\mathcal{W}_{n}^{\epsilon}$ and 
$\mathcal{W}_{nm}^{\epsilon}$, respectively, are indeed approximations of the exact Wigner eigenfunctions $\Phi_{nm}^{\hb}$. 
%{\color{red} Moreover,  this comparison clarifies the fact that .....}

\subsection{Comparison  of the diagonal eigenfunctions}\label{761}

In order to use the asymptotic formula  (\ref{lag_asymp}) for the diagonal eigenfunctions $(\ref{eigenf_lag})$ for ($n=m$)
\begin{eqnarray*}
\Phi_{n}^{\hb}(x,p)\equiv \Phi_{nn}^{\hb}(x,p)=\frac{(-1)^n}{\pi\hb}e^{-(p^2+x^2)/\hb} L_{n}^{(0)}\left[\frac{2}{\hb}(p^2+x^2)\right] \ ,
\end{eqnarray*}
we put  $a=0$, $\nu=4(n+1/2)=4E^{\hb}_{n}/\hb$  and 
$t=(p^{2}+x^2)/2E^{\hb}_{n}$ . We are looking for the region $0<p^{2}+x^2\leq 2E^{\hb}_{n}$ , thus $0<t\leq 1$, and more precisely for $(x,p)$ close to the Lagrangian curve $\Lambda_{n}:=\lbrace (x,p)\in R^2: x^2 + p^2= 2E_{n}\rbrace$ , that is $t \approx 1$. In this regime, by the use of appropriate Taylor expansions we approximate the quantities in (\ref{lag_asymp}) by $[B(t)]^2\approx -(1-t)/2^{2/3}$ , $\alpha_0(t)\approx 2^{1/3}$ and $\beta_{1}(t)\approx 0$  .

Then, we find that the leading term of the asymptotic expansion of  $\Phi_{n}^{\hb}(x,p)$ is given by
 \begin{eqnarray}\label{ex_eigenf_as_n}
\Phi_{n}^{\hb}(x,p)\approx {\pi}^{-1}{\hb^{-2/3}}(2E_{n}^{\hb})^{-1/3}Ai\left(\frac{p^{2}+x^2-2E^{\hb}_{n}}{\hb^{2/3}\left(2E^{\hb}_{n}\right)^{1/3}}\right)\quad ,
\end{eqnarray}
for small $\hb$ and large $n$, such that $n\hb=\mathrm{constant}$.
We observe that {\it this approximation coincides with the approximation $(\ref{approx_w_n})$ of ${\mathcal{W}}^\epsilon_{n}$} .

\subsection{Comparison of the off-diagonal eigenfunctions}

For the non-diagonal eigenfunctions with $n>m$, in order to use the asymptotic formula  (\ref{lag_asymp})  for the approximation of the Laguerre polynomial in $(\ref{eigenf_lag})$, we put
$a=n-m$ , $\nu=2(E^{\hb}_{n}+E^{\hb}_{m})/\hb$ , $t=(p^2+x^2)/(E^{\hb}_{n}+E^{\hb}_{m})$ , 
We are looking for points $(x,p)$ close to
$\Lambda _{1,nm}=\lbrace (x,p)\in R^2:p^2+x^2=R_{nm}^2\rbrace$ , and for $n\approx m$. Thus, by  $R_{nm}=(\sqrt{2E^{\hb}_{n}}+\sqrt{2E^{\hb}_{m}})/2$ and $\rho_{nm}=(\sqrt{2E^{\hb}_{n}}-\sqrt{2E^{\hb}_{m}})/2$ , we have   
 $t \approx 1$, and therefore we use the same approximations of $B(t) \ , \alpha_0(t) \ , \beta_{1}(t)$ as above.  Also we have $x-ip\approx e^{-i\phi}R_{nm}\approx e^{-i\phi} \sqrt{E^{\hb}_{n}+E^{\hb}_{m}}$ , where $\phi = Arg(x+ip)=\arctan(p/x)$.
 Thus, $(\ref{eigenf_lag})$ leads to the approximation
\begin{eqnarray*}
\Phi^{\hb}_{nm}(x,p)\approx \frac{\hb^{-2/3}}{\pi}e^{-i(n-m)\phi}{(E^{\hb}_{n}+E^{\hb}_{m})}^{-1/3}Ai\left(\frac{p^2+x^2-R_{nm}^2}{\hb^{2/3} R_{nm}^{2/3}}\right) \ .
\end{eqnarray*}
The last formula, by  $E^{\hb}_{n} \approx n\hb = \mathrm{const.} \ , E^{\hb}_{m} \approx m\hb= \mathrm{const.}$, for small $\hb$ and large $n, m$ such that $n\approx m$, becomes
\begin{eqnarray}\label{exeigen_as_nm}
\Phi^{\hb}_{nm}(x,p)\approx\frac{1}{\pi\hb}e^{-i(n-m)\phi}(n+m)^{-1/3}Ai\left(\frac{p^2+x^2-R^2_{nm}}{\hb^{2/3}R^{2/3}_{nm}}\right)\ .
\end{eqnarray} 

Recall now the approximation $(\ref{approx_w_nm})$ of $\mathcal{W}_{nm}^{\epsilon}$ , which reads as 
\begin{eqnarray*}
\mathcal{W}_{nm}^{\hb}(x,p)\approx
\frac{1}{\pi}\, e^{-i(n-m)\phi}\, {\epsilon}^{-2/3}R_{nm}^{-4/3}(R_{nm}^2-\rho_{nm}^2)^{1/3}
Ai\left(\frac{p^2+x^2-R_{nm}^2}{\epsilon^{2/3}{R_{nm}^{4/3}(R_{nm}^2-\rho_{nm}^2)}^{-1/3}}\right)\ .
\end{eqnarray*}
For small $\hb$ and large $n, m$ such that $n\approx m$, we have
$${\epsilon}^{-2/3}R_{nm}^{-4/3}(R_{nm}^2-\rho_{nm}^2)^{1/3}\approx {\epsilon}^{-1}(n+m)^{-1/3} \ .$$
Thus, we observe that {\it the approximation $(\ref{exeigen_as_nm})$ coincides with the approximation 
$(\ref{approx_w_nm})$ of $\mathcal{W}^\epsilon_{nm}$ .}

\section{Classical limit of the Wigner eigenfunctions}\label{limitwigner}

As we have already noted in Section \ref{7.1}, it is not possible to evaluate the classical limit $\hb\rightarrow 0$ of the phase space eigenfunctions  $W^{\hb}_{nm}$ for any  $n,m$ independent of each other, and indpendent of $\hb$, a fact which is connected with the Bohr-Sommerefeld quantization condition. 
So there is a fundamental obstruction in obtaining an eigenfunction series solution of  Liouville  equation, by simply taking  the limit $\epsilon \to 0$ of the  eigenfunction series solution of the Wigner equation. 

What can be evaluated is the classical limit of the Wigner eigenfunctions $W_{nm}^{\hb}$. By taking the {\it classical limit} we mean that $\hb\rightarrow 0$, and  $n,m\rightarrow\infty$ so that $n\hb,m\hb=\mathrm{constants}$ and $n-m=\mathrm{constant}$. We refer  to these weak limits of $W^{\hb}_{nm}$ as the {\it Liouville-Wigner eigenfunctions}, and we denote them by  $W^0_{nm}$.

For integrable Hamiltonians, M.V.Berry \& N.L.Balazs, \cite{Ber, BeBa}, have  computed the classical limit of  eigenfunctions $W_{nm}^{\hb}$ in the case  $n=m$, assuming that $n$ is large enough so that $E_{n}^{\hb}\rightarrow E^{0}_{n}$ (constant) as $\hb\rightarrow 0$. This limit which reads as $W^0_{n}(I,\theta)=\delta(H(I)-E^{0}_{n})$ , where $H(I)$ is the Hamiltonian expressed in action-angle variables $(I,\theta)$ . 

In the ``simplest" solvable case of the harmonic oscillator,  N.Ripamonti  \cite{Rip}, and   A.Truman \& H.Z.Zhao  \cite{TZ}, have given independent proofs for the classical limit of the Wigner eigenfunctions  $W_{nm}^{\hb}$ for all $n,m=0,1,\dots$. Their derivation is based on the asymptotics of Laguerre polynomials (which could be derived for  
(\ref{lag_asymp}) in Appendix \ref{app_e}).

Moreover,  in \cite{WB} it has been stated that the classical limit of  eigenfunctions, in terms of action-angle variables $(I,\theta)$ and for $n\neq m$ , reads as
\begin{eqnarray}\label{limit_nm}
W^0_{nm}(I,\theta)=e^{-i\frac{E^{0}_{nm}}{\gamma_{nm}}\theta}\delta\left(H(I)-\frac{1}{2}(E^{0}_{n} +E^{0}_{m})\right) \quad ,
\end{eqnarray}
where $(E_{n}^{\hb}-
E_{m}^{\hb})/\hb\rightarrow E^{0}_{nm}$ (constant) as $\hb\rightarrow 0$ , $\gamma_{nm}=H'(H^{-1}((E^{0}_{n}+E^{0}_{m})/2)$ and $\theta=\arctan(p/x)$. 

The formula  $(\ref{limit_nm})$ has been derived in \cite{Kal} from the eigenvalue equations  $(\ref{eig1})$ and $(\ref{eig2})$, respectively. For the sake of completeness and easier reference, we repeat the computation here. 

\subsection{The Liouville-Wigner eigenfunctions}

Recall the eigenvalue equations 
$\mathcal{M}^{\hb}$ ,  
\begin{equation}\label{eig1_new}
\mathcal{L}^{\hb}\wh_{nm}(x,p)=\frac{i}{\hb}(E^{\hb}_n-E^{\hb}_m
)\wh_{nm}(x,p)\ ,
\end{equation}
and
\begin{equation}\label{eig2_new}
\mathcal{M}^{\hb}\wh_{nm}(x,p)=\frac12(E^{\hb}_n+E^{\hb}_m)\wh_{nm}(x,p)\ .
\end{equation}
The operaors $\mathcal{L}^{\hb} \ ,\mathcal{M}^{\hb}$ are given by
\begin{eqnarray*}
\mathcal{L}^{\hb}\, \bullet&:=&-\frac{2}{\hb}\, H(x,p)\sin\left[\frac{\hb}{2}\left({\overleftarrow{\partial_
x}}\overrightarrow{\partial_p}-\overleftarrow{\partial_p}\overrightarrow{\partial_
x}\right)\right]\, \bullet \quad ,\\
\mathcal{M}^{\hb}\, \bullet&:=& H(x,p)\cos\left[\frac{\hb}{2}\left({\overleftarrow{\partial_
x}}\overrightarrow{\partial_p}-\overleftarrow{\partial_p}\overrightarrow{\partial_
x}\right)\right]\, \bullet \quad ,
\end{eqnarray*}
with $H(x,p)=p^2/2+V(x)$. For smooth potential $V (x)$ they are written also in the form
\begin{eqnarray*}
\mathcal{L}^{\hb}&=&p\partial_{x}-V'(x)\partial_{p}-\sum_{j=1}^{\infty}\hb^{2j}\left(\frac{i}{2}\right)^{2j} \frac{1}{(2j+1)!} V^{(2j+1)}(x)\, \partial^{2j+1}_p\quad ,\\
\mathcal{M}^{\hb}&=&-\frac{\hb^2}{8}\Delta_{xp}+H(x,p)+\sum_{j=1}^{\infty}\hb^{2j}\left(\frac{i}{2}\right)^{2j}\frac{1}{(2j)!}V^{(2j)}(x)\, \partial^{2j}_{p}+\frac{\hb^2}{8}(\partial_{p})^2\quad .
\end{eqnarray*}

By taking formally the classical limit $\hb\rightarrow 0$, and assuming that $W_{nm}^{\hb}\rightarrow W^{0}_{nm}$ ,
$(E_{n}^{\hb}-E_{m}^{\hb})/\hb \rightarrow E^{0}_{nm}$ , $E_{n}^{\hb}\rightarrow E^{0}_{n}$ and $E_{m}^{\hb}\rightarrow E^{0}_{m}$  as $\hb\rightarrow 0$ and $n\rightarrow \infty$
and $E^{0}_{nm}$ , $E^{0}_{n}$ , $E^{0}_{m}$ are constant, 
the equations  $(\ref{eig1_new}) \ ,(\ref{eig2_new})$, reduce to 
\begin{eqnarray}\label{eql}
\left(p\partial_{x}-V'(x)\partial_{p}\right)W_{nm}^{0}(x,p)&=&i E^{0}_{nm} W_{nm}^{0}(x,p) \ ,
\end{eqnarray}
and 
\begin{eqnarray}\label{eqm}
H(x,p)W_{nm}^{0}(x,p)&=&\frac{1}{2}(E^{0}_{n}+E^{0}_{m})W_{nm}^{0}(x,p) \ ,
\end{eqnarray}
respectively.

Introducing the action-angle variables $(I,\theta)$ (see, e.g., Goldstein \cite{GOLD}) , the Hamiltonian becomes $H(x,p)=H(I)$ and the equations $(\ref{eql})$ and $(\ref{eqm})$ are transformed to 
\begin{eqnarray}\label{eql_newvar}
-H'(I)\ \partial_{\theta}W_{nm}^{0}(I,\theta)=i E^{0}_{nm}W_{nm}^{0}(I,\theta) \ ,
\end{eqnarray}
and 
\begin{eqnarray}\label{eqm_newvar}
H(I)W_{nm}^{0}(I,\theta)=\frac{1}{2}(E^{0}_{n}+E^{0}_{m})W_{nm}^{0}(I,\theta) \ .
\end{eqnarray}

The equation $(\ref{eqm_newvar})$ implies that $W_{nm}^{0}(I,\theta)$ is a distribution of the form 
$$
W_{nm}^{0}(I,\theta)=g(\theta)\ \delta\left(H(I)-\frac{1}{2}(E^{0}_{n}+E^{0}_{m})\right) \ .
$$
Substituting the above representation into $(\ref{eql_newvar})$ we get
$$
-H'(I)\ \partial_{\theta}g(\theta)\ \delta\left(H(I)-\frac{1}{2}(E^{0}_{n}+E^{0}_{m})\right)=i E^{0}_{nm}\  g(\theta)\ \delta\left(H(I)-\frac{1}{2}(E^{0}_{n}+E^{0}_{m})\right) \ ,
$$
which is written in the form
$$-\gamma_{nm}\partial_{\theta}g(\theta)=iE^{0}_{nm}g(\theta) \ ,$$ 
with $\gamma_{nm}=H'\left(H^{-1}\left(\frac{1}{2}(E^{0}_{n}+E^{0}_{m})\right)\right)$.  
Thus
$$g(\theta)=e^{-i\frac{E^{0}_{nm}}{\gamma_{nm}}\theta} \ .$$ 
Therefore, the solution of the "classical"  eigenvalue equations $(\ref{eql}) \ , (\ref{eqm})$ is 
$$
W_{nm}^{0}(I,\theta)=e^{-i\frac{E^{0}_{nm}}{\gamma_{nm}}\theta}\  \delta\left(H(I)-\frac{1}{2}(E^{0}_{n}+E^{0}_{m})\right)\quad ,
$$
that is the formula $(\ref{limit_nm})$.

\subsection{Liouville-Wigner eigenfuctions of the harmonic oscillator}

In the case of the harmonic oscillator, where $H(x,p)=(p^2 + x^2)/2$, the action-angle variables are essentially the polar coordinates    $\left(I=(p^2 + x^2)/2 \ , \theta=\arctan(p/x)\right)$ in $xp-$plane,  thus $H(I)=I$. 

The eigenvalue of the harmonic oscillator are $E^{\hb}_{n}=(n+1/2)\hb$. Then, $(E_{n}^{\hb}-E_{m}^{\hb})/\hb \rightarrow E^{0}_{nm}=n-m=\mathrm{const.}$, and the Liouville-Wigner eigenfunctions are
\begin{eqnarray}\label{limit_harmonic}
W_{nm}^{0}(x,p)=e^{-i(n-m)\theta}\  \delta\left(p^2+x^2-(E^{0}_{n}+E^{0}_{m})\right)\ .
\end{eqnarray}

In particular, for $n=m$, we get the diagonal Liouville-Wigner eigenfunctions
\begin{eqnarray}\label{limit_harmonic_d}
W_{n}^{0}(x,p)=\delta\left(p^2+x^2-2E^{0}_{n}\right) \ .
\end{eqnarray}

It is important to observe that the classical limits of  the approximate Wigner eigenfunctions ${\mathcal{W}}^\epsilon_{n}$ (eq. $(\ref{approx_w_n})$) and $\mathcal{W}_{nm}^{\epsilon}$ (eq. $(\ref{approx_w_nm})$), are given (modulo the non essential factor ${\pi}^{-1}$) by $W_{nm}^{0}$ and $W_{n}^{0}$, respectively.  This follows using the formula
$$\frac{1}{\epsilon}Ai\left(\frac{x}{\epsilon}\right) \rightarrow \delta(x)\ ,\  \mathrm{in} \ \ \mathcal{D'}   \ \mathrm{as} \ \  \epsilon \to 0 \ , $$
to take the limit $\epsilon \to 0$, of  
\begin{eqnarray*}
{\mathcal{W}}^\epsilon_{n}(x,p) \approx {\pi}^{-1}\, \hb^{-2/3}{\left(2E_{n}\right)}^{-1/3}\,
Ai\left(\frac{p^2+x^2-2E_n}{\epsilon^{2/3}{(2E_{n})}^{1/3}}\right) 
\end{eqnarray*}
and
\begin{eqnarray*}
\mathcal{W}_{nm}^{\epsilon}(x,p)\approx {\pi}^{-1}
e^{-i(n-m)\phi}{\epsilon}^{-2/3}R_{nm}^{-4/3}(R_{nm}^2-\rho_{nm}^2)^{1/3}
Ai\left(\frac{p^2+x^2-R_{nm}^2}{\epsilon^{2/3}{R_{nm}^{4/3}(R_{nm}^2-\rho_{nm}^2)}^{-1/3}}\right) \ ,
\end{eqnarray*}
respectively, by holding $E_{n} \ ,R_{nm} \ ,\rho_{nm}$ constant, according to the definition of the classical limit.

\section{Remarks on the asymptotics of Wigner eigenfunctions}

In this chapter, we have confirmed that  the approximate Wigner eigenfunctions ${\mathcal{W}}^\epsilon_{n}$ (eq. $(\ref{approx_w_n})$) and $\mathcal{W}_{nm}^{\epsilon}$ (eq. $(\ref{approx_w_nm})$, have the following important features.

First, they are approximations of the exact Wigner eigenfunctions (Section $\ref{76}$). Second, they converge weakly to the Liouville-Wigner eigenfunctions in the classical limit (Section $\ref{limitwigner}$).

These two facts provide significant evidence that   {\it the uniformization procedure}, which was defined at the beginning of this chapter, as {\it  the construction of  uniform stationary phase asymptotics of the Wigner transform of WKB eigenfunctions}, leads to the correct {\it approximate Wigner eigenfunctions}.
 
%%%%%%%%%%%%%%%%%%%%%%%%%%%%%%%%%%%%%%%%%%%%%%%%%%%%%
%%%%%%%%%%%%%%%%%%%%%%%%%%%%%%%%%%%%%%%%%%%%%%%%%%%%%%

%%%%%%%%%%%%%%%%%%%%%%%%%%%%%%%%%%%%%%%%%%%%%%%%%%%%%
%%%%%%%%%%%%%%%%%%%%%%%%%%%%%%%%%%%%%%%%%%%%%%%%%%%%%%
\chapter{Approximation of the  solution  of the Wigner equation}\label{ch8}
%\section{Asymptotics of the coefficients}
%\section{The semiclassical eigenfunction series expansion}
%\section{Nonlinear decomposition relations for the WKB solution of time-dependent Schr\"odinger equation}

 The eigenfunction series solution  $W^{\epsilon}[u^\epsilon](x,p,t)$ (eq. $(\ref{wigner_exp})$) of the initial value problem $(\ref{wignereq_moy_1})$-$(\ref{wignereq_in_1})$ for the Wigner equation was constructed in Chapter \ref{ch5}. This is given by 
 \begin{eqnarray}\label{wf_exp_new}
W^{\hb}[u^\epsilon](x,p,t)= W^{\hb}[u^\epsilon]_{coh}(x,p) + W^{\hb}[u^\epsilon]_{incoh}(x,p,t)
\end{eqnarray}
where
\begin{equation}\label{wf_exp_coh}
W^{\hb}[u^\epsilon]_{coh}(x,p):=\sum_{n=0}^{\infty}c_{nn}^{\hb}(0)\,W_{n}^{\hb}(x,p) \ ,
\end{equation}
and
\begin{equation}\label{wf_exp_incoh}
W^{\hb}[u^\epsilon]_{incoh}(x,p,t):=\sum _{n=0}^{\infty}\sum_{m=0, m\ne n}^{\infty} c_{nm}^{\hb}(0)\, e^{-\frac{i}{\hb}(E_{n}^{\hb}-E_{m}^{\hb})t}\, W_{nm}^{\hb}(x,p) \ .
\end{equation}

The time-independent single series $(\ref{wf_exp_coh})$ is the {\it coherent part} of the solution of the Wigner equation, and it survives for all time. The time-dependent double series $(\ref{wf_exp_incoh})$ is the {\it  incoherent part} of the solution, which  is weakly vanishing for large time\footnote{This property  is referred as {\it decoherence}  in the theory of open quantum systems(see, e.g., \cite{BP}, Ch. 4).}, due to fast oscillations of the term  $e^{-\frac{i}{\hb}(E_{n}^{\hb}-E_{m}^{\hb})t}$ . This means that the solution of the Wigner equation, for large time, converges to the stationary solution $W^{\hb}[u^\epsilon]_{coh}$.
Note also that the incoherent part has zero net contribution to the wave energy, that is $\int\int_{R^{2}_{xp} }W^{\hb}[u^\epsilon]_{incoh}(x,p,t)dxdp = 0$ . The role of the off-diagonal Wigner eigenfunctions $W_{nm}^{\hb}$ is the exchange of energy between the modes of the solution.
This mechanism seems to be important in understanding quantum effects, but its complete investigation is not possible for the moment and it requires further deep understanding of the spectral theory for the Wigner equation. For this reason, in the sequel we will deal mainly with some aspects of the approximation of the coherent part.

The expansion coefficients 
\begin{eqnarray}\label{coeff_in}
c_{nm}^{\hb}(0)=2\pi\hb
\left(W_{}^{\hb}[u_{0}^\hb],W_{nm}^{\hb}\right)_{L^2(R_{xp}^2)} \ ,
\end{eqnarray}
are the phase space projections of the initial Wigner function $W_{}^{\hb}[u^\hb](x,p)$ , given by $(\ref{wignereq_in})$, onto the Wigner eigenfunctions $W^{\epsilon}_{nm}(x,p):=W^{\epsilon}_{v_{m}^{\hb}}[v_{n}^{\hb}](x,p)$ . Recall that these eigenfunctions are the cross Wigner transforms  $(\ref{cross_wigner_v})$
of the Schr\"odinger eigenfunctions $v_{n}^{\hb}(x)$ and $v_{m}^{\hb}(x)$ for $n,m=0,1,\ldots$ . 
Note also that by using the definition of Wigner eigenfunctions, the coefficients can be also expressed through the  Schr\"odinger eigenfunctions and the initial wavefunction by the formula
\begin{eqnarray}\label{coeff_config}
c_{nm}^{\hb}(0)=
(u_{0}^{\hb},u_{n}^{\hb})_{L^2(R_x)}\overline{(u_{0}^{\hb},u_{m}^{\hb})}_{L^2(R_x)} \ .
\end{eqnarray}

\section{Approximate solution of Wigner equation}

As we have already explained  in Section \ref{sec53},  we want to construct an asymptotic approximation of $W^{\hb}[u^\epsilon](x,p,t)$ starting from  the eigenfunction series $(\ref{wf_exp_new})$.  
For this purpose we substitute the approximate  Wigner eigenfunctions (see eqs. $(\ref{approx_w_n})$,  $(\ref{approx_w_nm})$)
\begin{eqnarray}\label{eigen_exp_n8}
\widetilde{\mathcal{W}^{\hb}_{n}}(x,p)= {\pi}^{-1}\, \hb^{-2/3}{\left(2E^{\hb}_{n}\right)}^{-1/3}\,
Ai\left(\frac{p^2+x^2-2E^{\hb}_n}{\epsilon^{2/3}{(2E^{\hb}_{n})}^{1/3}}\right)\ ,
\end{eqnarray}
and
\begin{eqnarray}\label{eigen_exp_nm8}
\widetilde{\mathcal{W}^{\hb}_{nm}}(x,p)={\pi}^{-1}
{\epsilon}^{-2/3}R_{nm}^{-4/3}(R_{nm}^2-\rho_{nm}^2)^{1/3}e^{-\frac{i}{\epsilon} (E^{\hb}_{n}-E^{\hb}_{m})\phi (p)}
Ai\left(\frac{p^2+x^2-R_{nm}^2}{\epsilon^{2/3}{R_{nm}^{4/3}(R_{nm}^2-\rho_{nm}^2)}^{-1/3}}\right) \ ,\nonumber\\
\end{eqnarray}
in place of the exact eigenfunctions $W_{nm}^{\hb}(x,p)$ into  $(\ref{wf_exp_new})$.
Recall that $R_{nm}\equiv R^{\hb}_{nm}=(\sqrt{2E^{\hb}_{n}}+\sqrt{2E^{\hb}_{m}})/2$ , $\rho_{nm}\equiv\rho^{\hb}_{nm}=(\sqrt{2E^{\hb}_{n}}-\sqrt{2E^{\hb}_{m}})/2$, with $E^{\hb}_{n}=(n+1/2)\hb$, and  $\phi (p)=\arctan(p/x)$ .

Also, we approximate the coefficients $c_{nm}^{\hb}(0)$ by
\begin{eqnarray}
\widetilde{c_{nm}^{\hb}}(0)= 2\pi\hb
\left({\widetilde{\mathcal{W}}}_{0}^\epsilon, \widetilde{\mathcal{W}^{\hb}_{nm}}\right)_{L^2(R_{xp}^2)} \ ,
\end{eqnarray}
where ${\widetilde{\mathcal{W}}}_{0}^\epsilon$ is the semiclassical Wigner function 
\begin{eqnarray}\label{sclwignereq_in}
{\widetilde{\mathcal{W}}}_{0}^\epsilon(x,p)&=& \frac{2^{2/3}}{\epsilon^{2/3}}\left(\frac{2}{\mid
S_{0}'''(x)\mid} \right)^{1/3} [A_{0}(x)]^{2} \nonumber \\
&&\cdot Ai\left(-\frac{2^{2/3}}{\epsilon^{2/3}}\left(\frac{2}{
S_{0}'''(x)} \right)^{1/3}\left(p-S_{0}'(x)\right)\right)     \  ,
\end{eqnarray}
corresponding to the WKB initial wavefunction  
$$u_{0}^{\epsilon}(x)=A_{0}(x)\, e^{\frac{i}{\epsilon}S_{0}(x)} \ ,$$ 
with  $A_{0}(x)\in C_{0}^{\infty}(R_x)$ and $S_{0}(x) \in C^{\infty}(R_x)$ (see eq. $(\ref{initialdata})$). 

Thus, we claim that the approximation

 \begin{eqnarray}\label{wf_exp_approx}
\widetilde{W^{\hb}[u^\epsilon]}(x,p,t)= \widetilde{W^{\hb}[u^\epsilon]}_{coh}(x,p) + \widetilde{W^{\hb}[u^\epsilon]}_{incoh}(x,p,t)
\end{eqnarray}
where
\begin{eqnarray}\label{wf_exp_coh_approx}
\widetilde{W^{\hb}[u^\epsilon]}_{coh}(x,p):=\sum_{n=0}^{\infty}c_{nn}^{\hb}(0)\,\widetilde{W_{n}^{\hb}}(x,p) 
\end{eqnarray}
and
\begin{eqnarray}\label{wf_exp_incoh_approx}
 \widetilde{W^{\hb}[u^\epsilon]}_{incoh}(x,p,t):=\sum _{n=0}^{\infty}\sum_{m=0, m\ne n}^{\infty} c_{nm}^{\hb}(0)\, e^{-\frac{i}{\hb}(E_{n}^{\hb}-E_{m}^{\hb})t}\, \widetilde{W_{nm}^{\hb}}(x,p) \ ,
\end{eqnarray}
as an approximate solution of the Wigner equation. 

In fact, the $L_2$-asymptotic nearness of $W^{\hb}[u^\epsilon]$ and $\widetilde{W^{\hb}[u^\epsilon]}$ is derived by employing standard WKB estimates of the Schr\"odinger eigenfunctions \cite{Fed}. A long straightforward calculation leads to
\begin{eqnarray}\label{wf_exp_near}
\| W^{\hb}[u^\epsilon] - \widetilde{W^{\hb}[u^\epsilon]} \|_{L^{2}(R^{2}_{xp})} =o (\epsilon) \ ,  \ \ \mathrm{as} \ \ \epsilon \to 0 \ .
\end{eqnarray}
The exact form of the estimate $o (\epsilon)$ depends on the initial phase $S_{0}(x)$, through the dependence of $\widetilde{ c_{nm}^{\hb}}(0)$ on the initial phase as it will be shown in next section.

Note also that by integration in polar coordinates $r=(x^2 +p^2)^{1/2} \ , \  \ \phi=\arctan(p/x)$, we get
$$\int\int_{R^{2}_{xp} } \widetilde{W_{nm}^{\hb}[u^\epsilon]}_{incoh}(x,p,t)dxdp = 0 \ ,$$
because $\int_{0}^{2\pi}e^{-\frac{i}{\epsilon} (E^{\hb}_{n}-E^{\hb}_{m})\phi}d\phi =0$. In general,  for anharmonic oscillators
we have the approximation
$$\int\int_{R^{2}_{xp} } \widetilde{W^{\hb}[u^\epsilon]}_{incoh}(x,p,t)dxdp = O(\epsilon) \ .$$

\section{Approximation of the expansion coefficients}\label{sec82}

We consider approximations of the coefficients of the coherent part $(\ref{wf_exp_coh})$, which is the principal asymptotic part of the solution of the Wigner equation. Quite similar results can be drawn for the  coefficients  of the incoherent part 
$(\ref{wf_exp_incoh})$, due to the analogous  structure of the argument of the Airy functions in $\widetilde{\mathcal{W}^{\hb}_{n}}(x,p)$ and $\widetilde{\mathcal{W}^{\hb}_{nm}}(x,p)$ as we easily observe from equations $(\ref{eigen_exp_n8})$, $(\ref{eigen_exp_nm8})$. The approximations of the coefficients depend crucially  on the phase $S_0(x)$ of the initial wavefunction. For this reason we present here two particular examples.

\subsection{Example 1: $S_{0}'''(x)=0$.}

The simplest initial phase having this property is the quadratic phase $S_{0}(x)=\pm x^2/2$.
In this case, the semiclassical Wigner function  $(\ref{sclwignereq_in})$ reduces to
$${\widetilde{\mathcal{W}}}_{0}^\epsilon(x,p)=A^{2}_{0}(x)\delta(p\mp x) \ , \ \  A_{0}(x)\in C_{0}^{\infty}(R_x) \ . $$

It is very interesting to observe that the different signs in the argument of the Dirac function do not affect the diagonal coefficients $\widetilde{ c_{nn}^{\hb}}(0)$ as  it follows from  $(\ref{coeff_in})$ and $(\ref{eigen_exp_n8})$, because the argument of the Airy function depends on $p^2$. On the other hand, it follows from $(\ref{coeff_in})$ and $(\ref{eigen_exp_nm8})$, that the off-diagonal coefficients $\widetilde{ c_{nm}^{\hb}}(0)$ are affected by the different signs through the exponential term $e^{-\frac{i}{\epsilon} (E^{\hb}_{n}-E^{\hb}_{m})\phi (p)}$ which for $p=\pm x$ gives $e^{\mp\frac{i}{\epsilon} (E^{\hb}_{n}-E^{\hb}_{m})\pi/4}$. This means that only the incoherent part is sensitive to the sign of the phase. Therefore, the concentration effects and the formation of focal points which are predicted by geometrical optics in the case of $(-)$ sign in the initial phase, is expected to be a phenomenon associated with the incoherent part of the Wigner function. In fact, the dependence on $\epsilon$ of the coefficients of the coherent part of the Wigner function implies that the coherent amplitude is bounded for $\epsilon \ll 1$.

In fact, by using  $(\ref{coeff_in})$, $(\ref{eigen_exp_n8})$ and 
the formula (\cite{VS}, eq. (3.93), p. 54)
\begin{eqnarray}\label{aisq_int}
\int_{-\infty}^{+\infty}Ai(z^2-y)dz= 2^{2/3}\pi Ai^{2}\left(  \frac{-y}{2^{2/3}}\right) \ ,
\end{eqnarray}
to do the $p$-integration, we obtain
\begin{equation}\label{coef_cnn_example}
\widetilde{ c_{nn}^{\hb}}(0)=2\pi\hb
\left<{\widetilde{\mathcal{W}}}_{0}^\epsilon, \widetilde{\mathcal{W}^{\hb}_{nn}}\right>_{L^2(R_{xp}^2)} \approx \epsilon \frac{1}{2\sqrt{E^{\hb}_{n}}}\left(A^{2}_{0}(\sqrt{E^{\hb}_n})  + A^{2}_{0}(-\sqrt{E^{\hb}_n}) \right) \ .
\end{equation}

\begin{itemize}
\item
In the special case $A_{0}(x)\equiv 1$, $u_{0}^{\epsilon} \notin L^{2}(R_x)$.
Then, all integrals can be calculated  analytically by using the formula $(\ref{aisq_int})$,
the coefficient $\widetilde{ c_{nn}^{\hb}}(0)$ can be calculated explicitly, and we get
\begin{equation}
\widetilde{ c_{nn}^{\hb}}(0)=2^{7/6}\pi \epsilon^{2/3}{(2E^{\hb}_{n})}^{-1/6} Ai^{2}\left(\frac{-(2E^{\hb}_{n})^{2/3}}{ 2^{2/3}\epsilon^{2/3}} \right) \ .
\end{equation}

We must recall that the eigenvalues $E^{\hb}_{n}$ have been derived by the Bohr-Sommerfeld rule for large $n$ and small $\epsilon$, so that $n\epsilon=$ const. However, if we proceed {\it formally} \footnote{this is the way that semiclassical series are used in physical applications.} by considering that $\epsilon \ll 1$ and $n$  fixed, we get that $\widetilde{ c_{nn}^{\hb}}(0)=O\left(\epsilon^{1/2}\right)$ . Moreover, when $A_{0}(x)$ has compact support, it turns out that only finitely many terms have significant contribution to the coherent part  $(\ref{wf_exp_coh_approx})$. Similar approximation holds for $\widetilde{ c_{nm}^{\hb}}(0)$ .

\item
$A_{0}(x)= e^{{-x^2}/2}$, $u_{0}^{\epsilon} \in L^{2}(R_x)$.
From $(\ref{coeff_in})$ we have that
\begin{eqnarray}
c_{nn}^{\hb}(0)=2\hb^{1/3}\, {(2E_{n}^{\hb})}^{-1/3}\int_{R_x}^{}e^{-x^2}Ai\left[\frac{2x^2-2E_{n}^{\hb}}{\hb^{2/3}(2E_{n}^{\hb})^{1/3}}\right]\, dx
\end{eqnarray}
and using the asymptotic decomposition formula $(\ref{airydec_pm})$ we obtain,
\begin{eqnarray}\label{sumgaussian}
c_{nn}^{\hb}(0)\approx \frac {1}{2a}\left[\int_{R_x}^{}e^{-x^2}Ai\left(\frac{x+a}{b}\right)\, dx+
\int_{R_x}^{} e^{-x^2}Ai\left(\frac{x-a}{b}\right)\, dx \right]
\end{eqnarray}
where $a:=\sqrt{E_{n}^{\hb}}$ and $b:=2^{-1}\hb^{2/3}{(2E_{n}^{\hb})}^{1/3}$. 

The integrals into $(\ref{sumgaussian})$ are the Airy transform of a Gaussian function. For their computation we use the formula (eq. (4.31), p. 78 , \cite{VS}),
\begin{eqnarray*}
\phi_{\alpha}(x)=\frac{1}{|\alpha|}\int_{R}^{}e^{-y^2}Ai\left(\frac{x-y}{\alpha}\right)\, dy=\frac{\sqrt{\pi}}{|\alpha|}e^{(x+\frac{1}{24\alpha^3})/4\alpha^3}Ai\left(\frac{x}{\alpha}+\frac{1}{16\alpha^4}\right)
\end{eqnarray*}
$\alpha\in R$, 
so finally we have an approximate coefficient of the form
\begin{eqnarray}
c_{nn}^{\hb}(0)\approx \frac{\sqrt{\pi}e^{\frac{1}{96 b^6}}}{2ab}\left[e^{a/4b^3} Ai\left(\frac{a}{b}+\frac{1}{16b^4}\right)+e^{-a/4b^3} Ai\left(\frac{-a}{b}+\frac{1}{16b^4}\right)\right]
\end{eqnarray}
with $a:=\sqrt{E_{n}^{\hb}}$ and $b:=2^{-1}\hb^{2/3}{(2E_{n}^{\hb})}^{1/3}$.

\end{itemize}

\subsection{Example 2: $S_{0}'''(x) \neq 0$.}

The simplest initial phase having this property is $S_{0}(x)=-x^3/3$ . 

%By standard geometrical optics' calculations, it can be shown that the evolution of this phase generates a fold caustic (cf. \cite{FM1}, Sec. 4.1).

From $(\ref{coeff_in})$ we have that
\begin{eqnarray}\label{approx_cnm}
c_{nn}^{\hb}(0)&=&2\pi\hb\left<W_{0}^{\hb}[u_{0}^\hb],W_{n}^{\hb}\right>_{L^2(R_{xp}^2)} \nonumber \\
&\approx & 2\pi\hb\left<{\widetilde{W}}_{0}^\epsilon(x,p), \widetilde{\mathcal{W}^{\hb}_{n}}(x,p)\right> \nonumber \\
&=&\frac{4\pi}{\hb^{1/3}{(2E^{\hb}_n)^{1/3}}}\int_{R_{x}}\frac{A_{0}^{2}(x)}{\mid
S_{0}'''(x)\mid^{1/3}} Q^{\hb}_{n}(x)dx  \ ,
\end{eqnarray}
where
\begin{eqnarray}\label{int_qn}
Q^{\hb}_{n}(x):=\int_{R_{p}} Ai\left(-\frac{2^{2/3}}{\epsilon^{2/3}}\left(\frac{2}{
S_{0}'''(x)} \right)^{1/3}(p-S_{0}'(x))\right) Ai\left(\frac{p^2+x^2-2E^{\hb}_n}{\epsilon^{2/3}{(2E^{\hb}_{n})}^{1/3}}\right)dp \ . \nonumber
\\
\end{eqnarray}
The integrals $(\ref{int_qn})$ cannot be calculated analytically. The obstruction of the calculation seems to be the quadratic term $p^2$ in the argument of the second Airy function. An approximate calculation of the integrals $Q^{\hb}_{n}(x)$ is done by using a decomposition of   the Airy function with quadratic argument into Airy functions with linear argument w.r.t. $p$ . Such an asymptotic decomposition formula is derived in the next subsection.

\subsubsection{An asymptotic decomposition of Airy function.}
We will derive the approximate decomposition formula
\begin{eqnarray}\label{airydec}
\frac{1}{\epsilon}Ai\left( \frac{x^2 -\alpha^2}{\epsilon}  \right)  \asymp \frac{1}{2\alpha}\left[\frac{1}{\epsilon}Ai\left( \frac{x +\alpha}{\epsilon}  \right) +\frac{1}{\epsilon}Ai\left( \frac{x -\alpha}{\epsilon}  \right) \right] \ , \ \ 
\mathrm{as} \ \  \epsilon \to 0 \ .
\end{eqnarray}
Here the symbol $\asymp$ is used to denote that eq. $(\ref{airydec})$ holds as a distributional asymptotic approximation \footnote{Distributional asymptotic approximations are constructed, in general, by expressing stationary phase or Laplace asymptotic expansions, in terms of generalized functions, and using weak limits to interpret the vanishing remainders. A clear exposition of such ideas and techniques is given in the book by R.Estrada \& R.P.Kanwal \cite{EK}.}. This means that the principal (Airy) part of the approximation is derived by appropriate use stationary phase method, but the remainder of the approximation formula  contains derivatives of Dirac function modulated by a certain rapidly oscillating exponential function. This remainder, considered as a distribution in $\mathcal{D}'(R)$,  tends to zero as $\epsilon \to 0 \ $.

\begin{remark}
For any $\alpha>0$, the following identity
\begin{eqnarray}\label{deltadec}
\delta\left(x^2 -\alpha^2 \right) = \frac{1}{2\alpha}\left[ \delta\left(x +\alpha\right) + 
\delta\left(x -\alpha\right) \right] \ ,
\end{eqnarray}
holds in $\mathcal{D}'(R)$.
We observe that as $\epsilon \rightarrow 0$, equation $(\ref{airydec})$ implies formally equation
$(\ref{deltadec})$, thanks to the $\mathcal{D'}$- limit
\begin{equation}\label{airy-delta}
\frac{1}{\epsilon}Ai\left(\frac{x}{\epsilon}\right) \rightarrow \delta(x) \ , \ \  \mathrm{as} \ \  \epsilon \to 0 \ . 
\end{equation}
Therefore, we can think of $(\ref{airydec})$ as a \emph{dispersive regularization} of $(\ref{deltadec})$, which of course is not unique. Note that regularizations of $(\ref{deltadec})$ in terms of other special functions are also possible. Different regularizations are necessary to copy with hyperbolic, parabolic or mixed mechanisms, according to the nature of the problem under consideration.
\end{remark}

We give a brief derivation of equation $(\ref{airydec})$. Let $\phi \in C^{\infty}_{0}(R_x)$ , and consider the integral
\begin{eqnarray*}
I^{\epsilon}(\alpha)= \int_{-\infty}^{+\infty} Ai\left( \frac{x^2 -\alpha^2}{\epsilon}  \right)\phi(x)\, dx  \ .
\end{eqnarray*}
By the integral representation of the Airy function
\begin{eqnarray*}
Ai(x)=\frac{1}{2\pi}\int_{-\infty}^{+\infty}e^{i\left(z^3/3 + xz\right)}\, dz \ ,
\end{eqnarray*}
we rewrite $I^{\epsilon}(\alpha)$ as a double integral
\begin{eqnarray*}
I^{\epsilon}(\alpha)&=&\frac{1}{2\pi}\int_{-\infty}^{+\infty}\int_{-\infty}^{+\infty}e^{i\left( \sigma^3/3 +\sigma \frac{x^2 -\alpha^2}{\epsilon}\right)}\phi(x)\, d\sigma dx  \nonumber \\
&=& \frac{1}{2\pi \epsilon^{1/2}}\int_{-\infty}^{+\infty}\int_{-\infty}^{+\infty}e^{i\epsilon^{-3/2}\left( \xi^3/3 +\xi (x^2 -\alpha^2)\right)}\phi(x)\, d\xi dx \ .
\end{eqnarray*}
The last integral is approximated by the stationary-phase formula for double integrals.
The phase $\Phi(\xi, x)= \xi^3/3 +\xi (x^2 -\alpha^2)$ has four stationary points $(x=0 \ , \xi=\pm\alpha)$ and $(\xi=0 \ , x=\pm\alpha)$. These points are non-degenerate since $\Phi_{xx}\Phi_{\xi \xi}-\Phi^{2}_{x\xi}
=-4\alpha^2 \neq 0$. The eigenvalues of the Jacobian determinant $\det\left(\frac{\partial \Phi(\xi ,x)}{\partial(\xi ,x)}\right)$ are $\pm 2\alpha$, thus $\delta=0$. Then applying  the stationary phase formula (\ref{doublestat}) for each stationary point and summing up the contributions, we derive the approximation
\begin{eqnarray}
I^{\epsilon}(\alpha)=
\epsilon \left[\frac{1}{2\alpha}\left( \phi(\alpha) +\phi(-\alpha) \right)+2 \cos\left(\epsilon^{-3/2} \frac23 \alpha^3\right) \phi(0) \right]+O(\epsilon^{3/2})
\end{eqnarray}
Note that, the remainder term $O(\epsilon^{3/2})$ contains derivatives  $\phi^{(\ell)}(\pm \alpha)$ , $\ell =1,2, \dots$ of the test function, and also oscillatory terms $ \cos\left(\epsilon^{-3/2} \frac23 \alpha^3\right)  \phi^{(\ell)}(0)$ ,  $\ell =1,2, \dots$  ,  which are multiplied by powers of 
$\epsilon$ smaller than $\epsilon^{3/2}$ .

Now,  we interpret the approximation of $I^{\epsilon}(\alpha)$ as follows

\begin{eqnarray*}
\left<\frac{1}{\epsilon}Ai\left( \frac{x^2 -\alpha^2}{\epsilon}  \right), \phi(x)\right>_{\cal{D}'}&=&
\left< \frac{1}{2\alpha}\left( \delta\left(x +\alpha\right) + 
\delta\left(x -\alpha\right) \right), \phi(x)\right>_{\cal{D}'} \nonumber\\
&&+2 \cos\left(\epsilon^{-3/2} \frac23 \alpha^3\right)\left<\delta(x), \phi(x)\right>_{\cal{D}'} 
+{\widetilde O}(\epsilon^{3/2}) \ .
\end{eqnarray*}
Here the symbol $\left< . , . \right>_{\cal{D}'}$ denotes the duality of $\cal{D}$ and 
$\cal{D}'$, and the symbol ${\widetilde O}$ in the remainder terms means that the remainder  contains terms of the form $\left<\delta^{(\ell)}(x\pm \alpha), \phi(x)\right>_{\cal{D}'}$ and 
$\left<\cos\left(\epsilon^{-3/2} \frac23 \alpha^3\right)\delta^{(\ell)}(x), \phi(x)\right>_{\cal{D}'}$ .

Thus, we obtain
\begin{eqnarray}\label{airydec_1}
\frac{1}{\epsilon}Ai\left( \frac{x^2 -\alpha^2}{\epsilon}  \right) &=& \frac{1}{2\alpha}\left[ \delta\left(x +\alpha\right) +  \delta\left(x -\alpha\right) \right] \nonumber \\
&&+2 \cos\left(\epsilon^{-3/2} \frac23 \alpha^3\right)\delta(x) +O_{\cal{D}'}(\epsilon^{3/2}) \ .
\end{eqnarray}
The symbol $O_{\cal{D}'}$ in the remainder terms means that the remainder contains terms of the form $\delta^{(\ell)}(x\pm \alpha)$  and $\cos\left(\epsilon^{-3/2} \frac23 \alpha^3\right)\delta^{(\ell)}(x)$ multiplied by powers of 
$\epsilon$ smaller than $\epsilon^{3/2}$.

Similarly by considering the integrals
\begin{eqnarray*}
I_{\pm}^{\epsilon}(\alpha)= \int_{-\infty}^{+\infty} Ai\left( \frac{x \pm \alpha}{\epsilon}  \right)\phi(x)\, dx  \ , 
\end{eqnarray*}
and proceeding along similar lines as for $I^{\epsilon}(\alpha)$, we obtain
\begin{eqnarray}\label{airydec_pm}
\frac{1}{\epsilon}Ai\left( \frac{x \pm \alpha}{\epsilon}  \right) =  \delta\left(x \pm \alpha\right)
+O_{\cal{D}'}(\epsilon^{3/2})
\end{eqnarray}
Here, the symbol $O_{\cal{D}'}$ in the remainder terms means that the remainder contains the derivatives $\delta^{(\ell)}(x\pm \alpha)$ multiplied by terms smaller than $\epsilon^{3/2}$.

Therefore, by combining the equations $(\ref{airydec_1})$ and $(\ref{airydec_pm})$, we derive the approximation formula

\begin{eqnarray}\label{airydec_osc}
\frac{1}{\epsilon}Ai\left( \frac{x^2 -\alpha^2}{\epsilon}  \right) &=&
\frac{1}{2\alpha}\left(\frac{1}{\epsilon}Ai\left( \frac{x + \alpha}{\epsilon}  \right) + \frac{1}{\epsilon}Ai\left( \frac{x - \alpha}{\epsilon}  \right)\right) \nonumber\\
\nonumber\\
&&+2 \cos\left(\epsilon^{-3/2} \frac23 \alpha^3\right)\delta(x) +O_{\cal{D}'}(\epsilon^{3/2}) \ .
\end{eqnarray}
Here, the symbol $O_{\cal{D}'}$ in the remainder terms means that the remainder contains terms of the form $\delta^{(\ell)}(x\pm \alpha)$  and $\cos\left(\epsilon^{-3/2} \frac23 \alpha^3\right)\delta^{(\ell)}(x)$ multiplied by powers of  $\epsilon$ smaller than $\epsilon^{3/2}$.

Formally, equation $(\ref{airydec})$ follows from $(\ref{airydec_osc})$ by rejecting the remainder term, and also the highly oscillatory term $2 \cos\left(\epsilon^{-3/2} \frac23 \alpha^3\right)\delta(x)$ since it is weakly negligible.

\subsubsection{Approximation of the integrals $Q^{\epsilon}_{n}$}

By using the formula $(\ref{airydec})$, we approximate the integral $Q^{\epsilon}_{n}$ (eq. 
$(\ref{int_qn})$) by
\begin{equation}
Q^{\hb}_{n}(x) \approx \frac{1}{2\sqrt{2E^{\hb}_n -x^2}}\left[ Q^{\hb}_{n+}(x)+ Q^{\hb }_{n-}(x)\right]
\end{equation}
where
\begin{eqnarray}\label{int_qnpm}
Q^{\hb}_{n\pm}(x):=\int_{R_{p}} Ai\left(-\frac{2^{2/3}}{\epsilon^{2/3}}\left(\frac{2}{
S_{0}'''(x)} \right)^{1/3}(p-S_{0}'(x))\right) Ai\left( \frac{p \pm \sqrt{2E^{\hb}_n -x^2}}{\epsilon^{2/3}{(2E^{\hb}_{n})}^{1/6}}\right)dp \ . \nonumber \\
\nonumber\\
\end{eqnarray}

The integrals $Q^{\hb}_{n\pm}$ are calculated by using the formula (\cite{VS}, eq. (3.108), p. 57)
%\begin{eqnarray}
\[\frac{1}{\mid \alpha \beta\mid}\int_{-\infty}^{\infty}Ai\left(\frac{z+a}{\alpha}\right)Ai\left(\frac{z+b}{\beta}\right)dz =
\left\{\begin{array}{lr}
\delta(b-a) \ \ \hfill  \mathrm{if} \ \   \beta=\alpha \\
\frac{1}{|\beta^3 -\alpha^3|^{1/3}}Ai\left( \frac{b-a}{ (\beta^3 -\alpha^3)^{1/3}}\right) \ \ \ \ \mathrm{if} \ \ \beta \neq \alpha \ .
\end{array} \right.
\]
%\end{eqnarray}
The integration leads to
\begin{eqnarray}\label{qnpm}
Q^{\hb}_{n\pm}(x)= \epsilon^{2/3}\frac{(2E^{\hb}_n)^{1/6}  \mid S_{0}'''(x) \mid^{1/3}}
 {\mid (2E^{\hb}_n)^{1/2}  +\frac{1}{8} S_{0}'''(x) \mid^{1/3}}\,
Ai\left( \frac{1}{\epsilon^{2/3}} \frac{ S_{0}'(x)\pm  \sqrt{2E^{\hb}_n -x^2}}{\mid (2E^{\hb}_n)^{1/2}  +\frac{1}{8} S_{0}'''(x)  \mid^{1/3}}  \right) \ .
\end{eqnarray}

Let $x_{\ell}=x_{\ell}^{\pm}(E^{\hb}_n)$ be the roots of 
\begin{equation}\label{n_roots}
S_{0}'(x)\mp  \sqrt{2E^{\hb}_n -x^2}=0 \ ,
\end{equation}
and we assume for simplicity that these roots are simple.

Then, by equations $(\ref{approx_cnm})$ and $(\ref{qnpm})$ we obtain the leading term in the approximation of the coefficients
\begin{eqnarray}\label{approx_cnm_1}
\widetilde{c_{nn}^{\hb}}(0)&\approx&\frac{4\pi}{\hb^{1/3}{(2E^{\hb}_n)^{1/3}}}\int_{R_{x}}\frac{A_{0}^{2}(x)}{\mid
S_{0}'''(x)\mid^{1/3}} Q^{\hb}_{n}(x)dx  \nonumber \\
&\approx &  \epsilon (2\pi) (2E^{\hb}_n)^{-1/6}
\sum_{\ell} \Biggl( A_{0}^{2}(x_{\ell}^{+})\mid S_{0}''(x_{\ell}^{+})\sqrt{2 E^{\hb}_n -(x_{\ell}^{+})^2} 
-x_{\ell}^{+} \mid ^{-1}\nonumber\\
&&+A_{0}^{2}(x_{\ell}^{-})
\mid S_{0}''(x_{\ell}^{-})\sqrt{2 E^{\hb}_n -(x_{\ell}^{-})^2} 
+x_{\ell}^{-} \mid ^{-1} \Biggr)
\ .
\end{eqnarray}
Similar approximation can be derived for $\widetilde{c_{nm}^{\hb}}(0)$ .

We observe that these approximation formulas are meaningful, even when $x_{\ell}$ coincide with the turning points.  This observation reveals the usefulness of the proposed  {\it uniformization procedure} for the construction of the approximate Wigner eigenfunctions. In fact, such approximations of the coefficients cannot be derived  directly in the configuration space by using $(\ref{coeff_config})$,  because the Schr\"odinger eigenfunctions are weakly singular at the turning points.

\section{Approximation of the wavefunction amplitude}

The above presented approximate solution of the Wigner equation, implies an approximation of the amplitude of the wavefunction, and, more precisely, a decomposition  into a coherent and an incoherent component of the wave intensity \footnote {This decomposition provides, in principle, a way to study large-time asymptotics of the transport and the Hamilton-Jacobi equation.}. 

Let $u^\epsilon(x,t)=\alpha^\epsilon (x,t) e^{i\phi^\epsilon (x,t)}$ be the polar decomposition of the wavefunction. 
By integrating  $(\ref{wf_exp_new})$ w.r.t. momentum, for some fixed $(x,t)$, we  get the following approximate decomposition of the wave intensity $\eta^{\epsilon} (x,t)= \mid\alpha^\epsilon (x,t)\mid^{2}$
\begin{equation}\label{approx_ampl}
\eta^{\epsilon} (x,t)= \eta^{\epsilon}_{coh} (x,t) + \eta^{\epsilon}_{incoh} (x,t) \ ,
\end{equation}
where
\begin{equation}\label{approx_coh_ampl}
\eta^{\epsilon}_{coh} (x,t) =\int_{R_p}W^{\hb}[u^\epsilon]_{coh}(x,p)dp \approx
\sum_{n=0}^{\infty}\widetilde{ c_{nn}^{\hb}}(0) \int_{R_p} \widetilde{W_{nn}^{\hb}}(x,p) dp  
\end{equation}
and 
\begin{equation}\label{approx_incoh_ampl}
\eta^{\epsilon}_{incoh} (x,t)=\int_{R_p}W^{\hb}[u^\epsilon]_{coh}(x,p,t)dp \approx
\sum_{n=0}^{\infty}\sum_{m=0 \ , m\ne n}^{\infty}\widetilde{ c_{nm}^{\hb}}(0)\, e^{-\frac{i}{\hb}(E_{n}^{\hb}-E_{m}^{\hb})t}\, \int_{R_p} \widetilde{W_{nm}^{\hb}}(x,p) dp\ .
\end{equation}
Note that $ \eta^{\epsilon}_{coh} (x,t) $ is a positive quantity, while $\eta^{\epsilon}_{incoh} (x,t)$, in general, oscillates and changes sign as time evolves. Since the exponential terms $e^{-\frac{i}{\hb}(E_{n}^{\hb}-E_{m}^{\hb})t}$ tend weakly to zero as $t$ increases, we expect that time averages of the incoherent part  converge to their stationary limits for large time.

The integrals $\int_{R_p} \widetilde{W_{nn}^{\hb}}(x,p) dp$  in equation $(\ref{approx_coh_ampl})$, are calculated by using the formula $(\ref{aisq_int})$.
Thus, we obtain the following approximation of the {\it coherent component of the intensity}
\begin{eqnarray}\label{coh_ampl_approx}
\eta^{\epsilon}_{coh} (x)&\approx & \sum_{n=0}^{\infty}\widetilde{ c_{nn}^{\hb}}(0) \int_{R_p} \widetilde{W_{nn}^{\hb}}(x,p) dp\nonumber\\
&=&\sum_{n=0}^{\infty}\widetilde{ c_{nn}^{\hb}}(0)\, 2^{2/3}\epsilon^{-1/3}(2E_{n}^{\hb})^{-1/6}Ai^{2}\left(\frac{-(2E^{\hb}_n -x^2)}{2^{2/3}\epsilon^{2/3}(2E_{n}^{\hb})^{1/3}}\right) \ .
\end{eqnarray}
Obviously the R.H.S of the above equation is positive, since $c_{nn}^{\hb}(0) > 0$ by their definition (cf. eq.
$(\ref{coeff_config})$).

For {\it the incoherent component of the intensity} we have 
\begin{eqnarray}\label{incoh_ampl_approx}
&&\eta^{\epsilon}_{incoh} (x,t) \approx\nonumber\\
&&
\sum_{n=0}^{\infty}\sum_{m=0 \ , m\ne n}^{\infty}\widetilde{ c_{nm}^{\hb}}(0)\, e^{-\frac{i}{\hb}(E_{n}^{\hb}-E_{m}^{\hb})t}\,{\pi}^{-1}{\epsilon}^{-2/3}R_{nm}^{-4/3}(R_{nm}^2-\rho_{nm}^2)^{1/3} \nonumber\\
&&\cdot \int_{-\infty}^{\infty}
e^{-\frac{i}{\epsilon} (E^{\hb}_{n}-E^{\hb}_{m})\arctan(p/x)}
Ai\left(\frac{x^2+p^2-R_{nm}^2}{\epsilon^{2/3}{R_{nm}^{4/3}(R_{nm}^2-\rho_{nm}^2)}^{-1/3}}\right) dp \ .
\end{eqnarray}
The integrals in $(\ref{incoh_ampl_approx})$ cannot calculated analytically at a general position $x$. Nevertheless, It can be shown, by using the Riemann-Lebesgue lemma  and the exponential decay of the Airy function for large positive argument, that they are convergent. At the special position $x=0$, the exponential term  
disappears. Then we can calculate the integrals by using the formula $(\ref{aisq_int})$, and we get 
\begin{eqnarray}\label{incoh_ampl_approx0}
&&\eta^{\epsilon}_{incoh} (x=0,t) \approx \nonumber %\\&&
\sum_{n=0}^{\infty}\sum_{m=0 \ , m\ne n}^{\infty}\widetilde{ c_{nm}^{\hb}}(0)\, e^{-\frac{i}{\hb}(E_{n}^{\hb}-E_{m}^{\hb})t}\nonumber\\
&&\cdot 2^{2/3}
{\epsilon}^{-1/3}R_{nm}^{-2/3}(R_{nm}^2-\rho_{nm}^2)^{1/6} 
Ai^{2}\left(\frac{R_{nm}^{2}}{{\epsilon}^{2/3}R_{nm}^{4/3}(R_{nm}^2-\rho_{nm}^2)^{-1/3}}\right) \ .
\end{eqnarray}

At this point we can make the following important remark. In geometrical optics, focal points and caustic formation appear along space-time curves.
Therefore, from the fact that the coherent part of the amplitude is time-independent, we expect  that such effects must be associated with the incoherent part. This implies, in turn,  that possible amplification of the wave intensity, as $\epsilon$ diminishes, is encoded in the off-diagonal coefficients.

The investigation of how the singularities are built up as the time evolves, is an {\bf open problem} which requires, first, the calculation or the approximation of the integrals in 
$(\ref{incoh_ampl_approx})$, and second, the systematic study of the trigonometric series, starting with $(\ref{incoh_ampl_approx0})$ which is simpler, but still very complicated due to the Airy functions in the coefficients.

\section{Some remarks on the use of the approximation}

The significant terms in the expansion $(\ref{wf_exp_approx})$ are those with $n \ ,m \ge  [\frac{p^2 + x^2}{2\epsilon}]$, for any fixed $p \ , x$, since then the Airy functions are oscillatory  for $\epsilon \ll 1$ (otheriwse, they are exponentially small). Thus, the main asymptotic contribution to the Wigner function comes from the high order eigenvalues. This fact  is consistent with the fact that the approximate Wigner eigenfunctions where constructed by using WKB method for constructing the Schr\"odinger eigenfunctions and the Bohr-Sommerfeld rule for the eigenvalues.
Moreover, the formulas for the approximate coefficients $\widetilde{ c_{nm}^{\hb}}(0)$ that we derived in Section \ref{sec82} (see, for example, eq. $(\ref{coef_cnn_example})$), show that at least for the considered special cases of the initial phases, it must be $n \ , m \le [\frac{d^2}{\epsilon}]$, where $supp A_{0}(x) =(-d, d)$. Therefore, for each fixed $x \ ,p$ and $\epsilon$, at most,  finitely many terms have significant contribution in the approximation of the Wigner function. This situation makes probably this approximation useful in testing numerical approximations of the solution of the Wigner equation in the semiclassical regime.

On the other hand, $(\ref{coh_ampl_approx})$ shows that significant contributions to the coherent component of the intensity comes from terms with $n  \le  [\frac{x^2}{2\epsilon}]$, for any fixed $x$ and $\epsilon \ll 1$. Thus,  low lying eigenvalues may also contribute in the approximation of the amplitude. In the case of the harmonic oscillator we consider in this work, the eiegenvalues provided by the Bohr-Sommerfeld rule are exact. But in the general case of potential wells, corrections to the low lying eigenvalues must be taken into account by using, for example, the approximations constructed in \cite{Si} by employing harmonic approximations near the bottom of the well (see also \cite{Kal, KalMak} for the role of low lying eigenvalues in certain approximation of the Wigner function).

%%%%%%%%%%%%%%%%%%%%%%%%%%%%%%%%%%%%%%%%%%%%%%%%%%%%%
%%%%%%%%%%%%%%%%%%%%%%%%%%%%%%%%%%%%%%%%%%%%%%%%%%%%%%

%%%%%%%%%%%%%%%%%%%%%%%%%%%%%%%%%%%%%%%%%%%%%%%%%%%%%%%%%%
%%%%%%%%%%%%%%%%%%%%%%%%%%%%%%%%%%%%%%%%%%%%%%%%%%%%%%%%%%
%%%%%%%%%%%%%%%%%%%%%%%%%%%%%%%%%%%%%%%%%%%%%%%%%%%%%%%%%%
%%%%%%%%%%%%%%%%%%%%%%%%%%%%%%%%%%%%%%%%%%%%%%%%%%%%%%%%%%
%%%%%%%%%%%%%%%%%%%%%%%%%%%%%%%%%%%%%
\chapter{Epilogue}\label{ch9}

We close the present work with a discussion of  the main results and  the basic underlying ideas which have guided our approach to the solution of the Wigner equation, and with the statement of a naturally arising and very  interesting  problem which awaits for future investigation.

\section{Discussion of the results }

\subsubsection{Wigner equation and the deformation point of view}

The Wigner equation [eq. (\ref{wignereq_moy})]   is a linear evolution equation in phase space. It is of first order with respect to time, but  it is  non local in phase space variables.  The primitive and most general derivation of this equation comes from the Weyl representation [eq. (\ref{wel_op})] of the Liouville-von Neumann equation [eq. (\ref{vonneumann})] for the evolution of quantum density operator, which is equivalent to the Schr\"odinger equation only for pure states. Non locality arises from the Moyal product [eq. (\ref{starproduct_1}); symbolically eq. (\ref{starexp})]   which couples the Hamiltonian with the Wigner function. This non-commutative product, which is the Weyl image of operator composition, encodes the  important features of quantum interferences, and it is  interpreted  in a physically plausible and mathematically elegant way in the framework of deformation quantization of classical Hamiltonian mechanics. 

It is instructive to recall that for smooth potentials the Wigner equation can be reformulated as a {\it mixed equation} which {\it combines a classical transport Hamiltonian operator and a dispersive operator of  infinite order with respect to the momentum variable}. This form of the equation is a manifestation of the deformation  structure of quantum mechanics \footnote{The same structure is inherent also in classical wave theories, where we look for classical waves as deformations of their corresponding geometrical ray skeletons.}, and it has inspired our {\it uniformization procedure} for deriving asymptotic approximations of the Wigner functions corresponding to stationary multiphase WKB functions. 

The WKB functions [Chapt. \ref{chapter3}] are  essentially classical objects which are constructed as solutions of the Hamilton-Jacobi and transport equations of classical mechanics.
From the view point of {\it phase-space geometry}, WKB solutions are localized on Lagrangian manifolds (actually curves in two-dimensional phase space), in the sense that their limit Wigner distributions are supported on these manifolds  \footnote{Recall that the Lagrangian curves are defined in phase space by the condition $H(x,p)=E=\mathrm{const.}$, where $H$ is the classical Hamiltonian function and $E$ the energy of the waves ($=$ eigen-energy for eigenvalue problems or total energy in scattering problems).}. The uniformization procedure somehow smooths out the Wigner distribution and reveals the correct  scales of the semiclassical parameter, by taking  account of the singularities (caustics) of the Lagrangian curve. This is realized by a {\it semiclassical} (uniform Airy asymptotic) {\it approximation} of the Wigner transform of the WKB function.
From the {\it energetic point of view} the uniformization  restores the dispersion of energy from the Lagrangian curve into the surrounding phase space, a phenomenon which cannot be captured by geometrical optics. In this way, the integration of Wigner functions with respect to the momentum leads to finite wave amplitudes even on the caustics, thus uniformizing the WKB solutions at the singularities.

This approach was applied to construct {\it semiclassical approximations of the Wigner eigenfunctions}, which are the basic ingredients in the construction of the approximate solution of the Wigner equation from its eigenfunction series.

\subsubsection*{The eigenfunction series for the Wigner function}

The linearity of the Wigner equation allows us to apply {\it directly in phase space} the method of separation of variables and to construct an eigenfunction series of the Wigner function by separating the time from the phase space variables [eq. (\ref{wigner_exp})]. The separation procedure leads to  {\it  a pair of phase-space eigenvalue equations}, one corresponding to the Moyal bracket [eq. (\ref{moyalbr})] and the other to the Baker bracket [eq. (\ref{bakerbr})]. 

The emergence of these phase space eigenvalue equations made clear that the natural way to derive  the Cauchy problem for the Wigner equation is to start from the Liouville-von Neumann equation, and exploit the Weyl representation and the properties of the Moyal product. In contrary, when one starts from configuration space and transforms the Cauchy problem for the Schr\"odinger equation by introducing the Wigner transform of the wavefunction in an operational way [Sec. \ref{433}], it is not possible to derive the second eigenvalue equations in a natural way. 

However,  the eigenfunction expansion of the Wigner function can be also  derived  by applying the Wigner transform onto the eigenfunction series expansion of the wavefunction in configuration space.
It is probably this way that J.E.Moyal \cite{Mo} deviced this expansion in his statistical treatment of quantum mechanics through phase space techniques. 
This independent can be thought of as a {\it confirmation of the validity of the phase space direct derivation}. In this approach, the necessity of the second equation  becomes a posteriori clear, when one wants to understand the completeness of the eigenfunctions of the first equation\footnote{In \cite{Kal} this step was made possible only after  observing the relevance of the Wigner eigenfunctions with the so called Laguerre expansions studied by Thangavelu \cite{Th}.}. 

The study of the eigenvalues equations was based on known fundamental results for the spectra of quantum Liouvillian [Sec. \ref{432}]. For the eigenvalues of the first (Moyal bracket) equation we used the results of  H.Spohn \cite{SP}  and I.Antoniou et al. \cite{ASS}, while for the eigenvalues of second (Baker bracket) equation we used the results derived in the thesis by E.Kalligiannaki \cite{Kal} (see also \cite{KalMak}),
for the special case {\it when the Hamiltonian operator has discrete spectrum}. Then, the {\it representation of the Wigner eigenfunctions as Wigner transforms of the Schr\"odinger eigenfunctions} was proved by using the recent spectral results by M.A.de Gosson \& F.Luef \cite{GL} for the $\star$-genvalue equation.

\subsubsection*{Approximation of the Wigner eigenfunctions}

The Wigner eigenfunctions cannot, in general, be calculated in closed form, although in the particular example of the harmonic oscillator $H(x,p)=(x^2 + p^2)/2$ , which we used as model  in our study, the eigenfunctions can be expressed in terms of the Laguerre polynomials. In general, the eigenvalues $E_{n}^{\epsilon}$ of the Hamiltonian operator are approximated by the Bohr-Sommerefeld rule [eq. (\ref{bohrsom})]. In the simple model  of harmonic oscillator this rule provides the exact eigenvalues $E_{n}^{\epsilon}=( n+1/2)\epsilon$ .

Our purpose was to  extend the uniformization  procedure that we proposed  in \cite{G, GM}  for the semiclassical Airy equation (a very simple scattering problem with continuous spectrum and parabolic Lagrangian curve possessing the standard fold singularity), to the case of the discrete spectrum. In particular we studied the approximation of the off-diagonal Wigner eigenfunctions which take account of the interference between the energy levels of the oscillator and span  the incoherent  component of the Wigner function which dies for long times.

We constructed approximations of the Wigner eigenfunctions starting from the Wigner transforms of the WKB approximations of the Schr\"odinger eigenfunctions by applying Airy-type uniform approximations of the related Wigner integrals. It turned out that the diagonal eigenfunctions are semiclassically concentrated as Airy functions on the Lagrangian eigencurves $H(x,p)=E_{n}^{\epsilon}$ [Sec. \ref{sec72}, eq. (\ref{approx_w_n})].  In the classical limit $\epsilon \to 0$ they converge to Dirac (Wigner) distributions on these curves. On the other hand the off-diagonal eigenfunctions are semiclassically concentrated on the intermediate curve $H(x,p)=\frac{1}{4}(\sqrt{2E_{n}^{\epsilon}}+\sqrt{2E_{m}^{\epsilon}})^{2}$  but, they are modulated by a rapidly oscillating factor $e^{-\frac{i}{\epsilon}\left(E_{n}^{\epsilon}-E_{m}^{\epsilon}\right)\phi}=e^{-i(n-m)\phi}$ , where $\phi= \arctan(p/x)$ is the angular direction in phase space \footnote{In general the modulating factor for a broad class of potential wells is of the form $e^{-i\left((n-m)+o(\epsilon)\right)\phi}$ .} [Sec. \ref{sec73}, eq. (\ref{approx_w_nm})].

\subsubsection*{The approximation of the Wigner function \& the wave amplitude}

An approximation of the solution of the Cauchy problem for the Wigner equation  [eqs. (\ref{wignereq_moy_1}), (\ref{wignereq_in_1})] was constructed by approximating the eigenfunctions in the eigenfunction series [eq. (\ref{wigner_exp})] by the approximate Wigner eigenfunctions [eqs. (\ref{approx_w_n}), (\ref{approx_w_nm})], and the initial datum [eq. (\ref{wignereq_in_1})] by Berry's semiclassical Wigner function [eq. (\ref{sclwignereq_in})] (see also Sec. \ref{section422})]. 

The coefficients of the expansion of the coherent part (and in special case also of the incoherent part) of the solution were evaluated either analytically for quadratic initial phases, or approximately by combining a novel asymptotic decomposition of the Airy function [eq. (\ref{airydec})] with analytic integration, for more general phases. 

The related calculations show dependence of the coefficients on the semiclassical parameter $\epsilon$, is crucially dependent on the initial phase. The derived approximate Wigner function is well-behaved everywhere in the phase space. Moreover, its integration with respect to the momentum results in a wave amplitude which is meaningful even on the turning points  of the WKB approximations of the Schr\"odinger eigenfunctions. Therefore, it turned out that the approximate Wigner function derived by the uniformization procedure is  a promising to smooth out the caustic singularities in time-dependent problems. 

We must emphasize that our detailed calculations performed for the special case of the harmonic oscillator, in which case we are able to check many of our approximations through existing analytical solutions. However,  the whole procedure can be applied for a broad class of potential wells which behave like the harmonic oscillator near the bottom of the well  \footnote{This is possible by using the approximations of the Schr\"odinger eigenfunctions which have been constructed in \cite{Si} by employing harmonic approximations near the bottom of the well. See also \cite{Kal, KalMak} for the use of these approximations in the perturbative construction  of the time-dependent semiclassical Wigner function.}.

\section{An interesting open problem}

We close the final discussion by noting that it would be very interesting and useful for understanding the approximate solutions of the Wigner equation, to investigate the interrelation of Berry's semiclassical Wigner function with Airy series approximation. 

Let us consider the time-dependent WKB function [eq. (\ref{wkb})]
\begin{eqnarray*}
u^{\epsilon}(x,t)\approx \psi^{\epsilon}(x,t)=A(x,t)\, e^{\frac{i}{\epsilon} S(x,t)}\quad ,
\end{eqnarray*}
which is an approximate (geometrical optics) solution of the Cauchy problem
\begin{eqnarray*}
i\epsilon{\partial_t}u^{\epsilon}(x,t)&=&\left[-\frac{\epsilon ^2}{2}\partial_{xx}+V(x)\right]u^{\epsilon}(x,t)\quad , \quad x\in R_{x} \quad ,\, \, \, t\in [0,T) \ , \\
u^{\epsilon}(x,t=0)&=&A_{0}(x)\, e^{\frac{i}{\epsilon}S_{0}(x)} \ . 
\end{eqnarray*}
In \cite{FM1} it has been shown that Berry's semiclassical Wigner function 
\begin{eqnarray*}
{\widetilde{\mathcal{W}}}^\epsilon (x,p,t)= \frac{2}{\epsilon^\frac23}\, \frac{ A^2 (x,t) }{\mid S_{xxx} (x,t)\mid ^\frac13}
Ai\left(-\frac{2}{\epsilon^\frac23} \,\frac{ p-S_x (x,t) }{(S_{xxx} (x,t))^\frac13}\right) 
\end{eqnarray*}
is an approximate solution of the Wigner equation [eq. (\ref{wignereq_moy})]. 
The phase $S(x,t)$ and  amplitude $A(x,t)$ are solutions of the eikonal and transport equations
\begin{eqnarray*}
&&\partial_{t}S(x,t) + \frac{1}{2}\left({\partial_x}S(x,t)\right)^{2} +V(x) =0 \quad , \quad S(x, t=0)= S_{0}(x) \ ,\\
&&{\partial_t}A(x,t) +{\partial_x}A(x,t)\,\partial_{x} S(x,t)+\frac{1}{2}A(x,t)\,\partial_{xx}S(x,t)  =0
\quad , \quad A(x,t=0)=A_{0}(x) \ .
\end{eqnarray*}

In the presnt work we have shown that the solution $W^{\hb}[u^\epsilon](x,p,t)$ of the Wigner equation admits of the approximation
[eqs. (\ref{eigen_exp_n8}), (\ref{eigen_exp_nm8}), (\ref {wf_exp_approx}), (\ref{wf_exp_coh_approx}),(\ref{wf_exp_incoh_approx})]
 \begin{eqnarray*}
\widetilde{W^{\hb}[u^\epsilon]}(x,p,t)&=& \widetilde{W^{\hb}[u^\epsilon]}_{coh}(x,p) + \widetilde{W^{\hb}[u^\epsilon]}_{incoh}(x,p,t)\\
\\
&=&\sum_{n=0}^{\infty}c_{nn}^{\hb}(0)\,{\pi}^{-1}\, \hb^{-2/3}{\left(2E^{\hb}_{n}\right)}^{-1/3}\,
Ai\left(\frac{p^2+x^2-2E^{\hb}_n}{\epsilon^{2/3}{(2E^{\hb}_{n})}^{1/3}}\right)\\
&&+ \sum _{n=0}^{\infty}\sum_{m=0, m\ne n}^{\infty} c_{nm}^{\hb}(0)\, e^{-\frac{i}{\hb}(E_{n}^{\hb}-E_{m}^{\hb})t}\, {\pi}^{-1}
{\epsilon}^{-2/3}R_{nm}^{-4/3}(R_{nm}^2-\rho_{nm}^2)^{1/3}\\
&&\cdot \, e^{-\frac{i}{\epsilon} (E^{\hb}_{n}-E^{\hb}_{m})\phi}
Ai\left(\frac{p^2+x^2-R_{nm}^2}{\epsilon^{2/3}{R_{nm}^{4/3}(R_{nm}^2-\rho_{nm}^2)}^{-1/3}}\right) 
\end{eqnarray*}
where
\begin{eqnarray*}
c_{nm}^{\hb}(0)=
(u_{0}^{\hb},u_{n}^{\hb})_{L^2(R_x)}\overline{(u_{0}^{\hb},u_{m}^{\hb})}_{L^2(R_x)} \quad ,
\end{eqnarray*}
\begin{eqnarray*}
R_{nm}\equiv R^{\hb}_{nm}=\frac{1}{2}(\sqrt{2E^{\hb}_{n}}+\sqrt{2E^{\hb}_{m}})\quad , \ \ 
\rho_{nm}\equiv \rho^{\hb}_{nm}=\frac{1}{2}(\sqrt{2E^{\hb}_{n}}-\sqrt{2E^{\hb}_{m}})\quad ,
\end{eqnarray*}
with $E^{\hb}_{n}=(n+1/2)\hb$ . 

Thus, it is natural to  investigate {\it how  someone could pass from the one approximation to that other}. The answer to this question seems to be rather difficult, and it requires the summation of the involved series of Airy functions. For the harmonic oscillator \footnote{Note that even in the configuration space, the passage from the eigenfunction series expansion of the solution of the Schr\"odinger equation to the corresponding WKB solution for small $\epsilon$ is open problem.}, a possible way would be to start with the Wigner transform of the integral representation of the Schr\"odinger wavefunction through Mehler's formula for the Schr\"odinger propagator, and apply uniform asymptotics on this Wigner transform.

%%%%%%%%%%%%
%%%%%%%%%%%%%%%%%%%%%%%%%%%%%%%%%%%%%%%%%%%%%%%%%%%%%%%%%%
%%%%%%%%%%%%%%%%%%%%%%%%%%%%%%%%%%%%

\appendix
\chapter{Some spectral results for the Moyal $\star$-product }\label{appendix_a}

\section{$\star$-properties of the Wigner eigenfunctions}

Suppose that $v^{\hb}_n(x)$ is an eigenfunction of the Schr\"odinger operator $\widehat{H}^{\epsilon}$ , and $E^{\hb}_{n}$ the corresponding eigenvalue, $\widehat{H}^{\epsilon}v^{\epsilon}_{n}(x)=E^{\epsilon}_{n}v^{\epsilon}_{n}(x)$ . Let the Wigner (eigen)function $W^{\hb}_{n}(x,p):=W^{\hb}[v^{\hb}_n](x,p) \ ,$ for $n=0,1,2,\ldots \ ,$
which is given by the semiclassical Wigner transform of $v^{\hb}_n$ ,
\begin{eqnarray}\label{diag_phsp_eiegenf}
W^{\hb}_{n}(x,p):=W^{\hb}[v^{\hb}_n](x,p)=(2\pi\hb)^{-1}\int_{R}^{} e^{-\frac{i}{\epsilon}y p} v^{\hb}_n\left(x+\frac{y}{2}\right)\overline{v^{\hb}_n}\left(x-\frac{y}{2}\right) \, dy \, ,
\end{eqnarray}
$\overline{v^{\hb}_n}$  being the complex conjugate of $v^{\hb}_n$. This function satisfies the eigenvalue equation 
\begin{eqnarray}\label{genvalue1}
H(x,p)\star_{\M} W_{n}^{\epsilon}(x,p)=E^{\epsilon}_{n}W_{n}^{\epsilon}(x,p)\quad \left(=W_{n}^{\epsilon}(x,p)\star_{\M} H(x,p)\right) \ ,
\end{eqnarray}
where $\star_{\M}$ is the Moyal star product,
\begin{eqnarray*}
\star_{\M}:=
\exp\left[\frac{i\epsilon}{2}\left(
\overleftarrow{\partial _x}
\overrightarrow{\partial _p}-
\overleftarrow{\partial _p}
\overrightarrow{\partial _x}\right)\right]\ .
\end{eqnarray*}
Recall that, for any smooth functions $f_{1}(x,p) \ , f_{2}(x,p)$, we have
\begin{eqnarray}
(f_{1}\star_{\M} f_{2})(x,p)&=&
f_{1}(x,p)\exp\left[\frac{i\epsilon}{2}\left(
\overleftarrow{\partial _x}
\overrightarrow{\partial _p}-
\overleftarrow{\partial _p}
\overrightarrow{\partial _x}\right)\right]
f_{2}(x,p)\nonumber\\
&=&
f_{1}\left(x+\frac{i\epsilon}{2}\overrightarrow{\partial_ p},p-\frac{i\epsilon}{2}\overrightarrow{\partial _x}\right)f_{2}(x,p)\nonumber\\
&=&f_{1}(x,p)f_{2}\left(x-\frac{i\epsilon}{2}\overleftarrow{\partial_ p},p+\frac{i\epsilon}{2}\overleftarrow{\partial_ x}\right)\label{starformula_2new}\\
&=& f_{1}\left(x+\frac{i\epsilon}{2}\overrightarrow{\partial_ p},p\right)f_{2}\left(x-\frac{i\epsilon}{2}\overleftarrow{\partial _p},p\right)\nonumber\\
&=&f_{1}\left(x,p-\frac{i\epsilon}{2}\overrightarrow{\partial _x}\right)f_{2}\left(x,p+\frac{i\epsilon}{2}\overleftarrow{\partial_ x}\right)\label{starformula1new} \, ,
\end{eqnarray}
for smooth functions $f_{1}$ , $f_{2}$ .

The eigenvalue equation (\ref{genvalue1}) has been proved in \cite{CFZ1} by using the definition (\ref{diag_phsp_eiegenf}) of the phase space eigenfunction, and the representation (\ref{starformula1new}) of the Moyal product, as follows.  Let  $H(x,p)=\frac{p^2}{2\bf m}+V(x)$, be the Hamiltonian function (Weyl symbol) of the Schr\"odinger operator. Then, we have
\begin{eqnarray*}
&&H(x,p)\star_{\M}W_{n}^{\epsilon}(x,p)=\\
&& H\left(x,p-\frac{i\epsilon}{2}{\overrightarrow{\partial_x}}\right)
W_{n}^{\epsilon}\left(x,p+\frac{i\epsilon}{2}{\overleftarrow{\partial_x}}\right)=\\
&&\frac{1}{2\pi}\int_{R}^{}
\left[\frac{1}{2\bf m}{\left(p-\frac{i\epsilon}{2}{\overrightarrow{\partial}_x}\right)}^{2}+V(x)\right]e^{-iy\left(p+\frac{i\epsilon}{2}{\overleftarrow{\partial_x}}\right)}v_{n}^{\hb}\left(x+\frac{\epsilon y}{2}\right)\overline{v_{n}^{\hb}}\left(x-\frac{\epsilon y}{2}\right)\, dy =\\
&&\frac{1}{2\pi}\int_{R}^{}
\left[\frac{1}{2\bf m}{\left(p-\frac{i\epsilon}{2}{\overrightarrow{\partial_x}}\right)}^{2}+V\left(x+\frac{\epsilon y}{2}\right)\right]e^{-iyp}v_{n}^{\hb}\left(x+\frac{\epsilon y}{2}\right)\overline{v_{n}^{\hb}}\left(x-\frac{\epsilon y}{2}\right)\, dy=\\
&&\frac{1}{2\pi}\int_{R}^{}
\left[\frac{1}{2\bf m}{\left(i{\overrightarrow{\partial_y}}+\frac{i\epsilon}{2}{\overrightarrow{\partial_x}}\right)}^{2}+V\left(x+\frac{\epsilon y}{2}\right)\right]e^{-iyp}v_{n}^{\hb}\left(x+\frac{\epsilon y}{2}\right)\overline{v_{n}^{\hb}}\left(x-\frac{\epsilon y}{2}\right)\, dy=\\
&&\frac{1}{2\pi}\int_{R}^{}e^{-iyp}\overline{v_{n}^{\hb}}\left(x-\frac{\epsilon y}{2}\right)E^{\epsilon}_{n}v_{n}^{\hb}\left(x+\frac{\epsilon y}{2}\right)\, dy=\\
&&\frac{1}{2\pi}\int_{R}^{}e^{-iyp}
H\left[x+\frac{\epsilon y}{2},-\left({\overrightarrow{\partial_y}}+\frac{i\epsilon}{2}{\overrightarrow{\partial_x}}\right)\right]
v_{n}^{\hb}\left(x+\frac{\epsilon y}{2}\right)\overline{v_{n}^{\hb}}\left(x-\frac{\epsilon y}{2}\right)\, dy=\\
&&E^{\epsilon}_{n}W_{n}^{\epsilon}(x,p) \ .
\end{eqnarray*}

We define the deformed (quantized) Hamiltonian $\mathbb{H}^\hb:=H\left(x+\frac{i\epsilon}{2}\overrightarrow{\partial_ p},p-\frac{i\epsilon}{2}\overrightarrow{\partial _x}\right)$.  Then, equation (\ref{genvalue1}) is written as
\begin{eqnarray}\label{genvaluetilde}
\mathbb{H}^\hb W_{n}^{\epsilon}(x,p)=E^{\epsilon}_{n}W_{n}^{\epsilon}(x,p) \ .
\end{eqnarray}

We now consider a pair of eigenfuctions $v^{\hb}_n(x) \ ,v^{\hb}_m(x)$   of the Schr\"odinger operator $\widehat{H}^{\hb}$, with eigenvalues $E^{\hb}_n$ and $E^{\hb}_m$, respectively. The non-diagonal Wigner functions
\begin{eqnarray}\label{nondiag_phsp_eiegenf}
%\label{cross_wig}
W_{nm}^{\hb}(x,p):= W_{v^{\hb}_m}^{\hb}[v^{\hb}_n](x,p)=(2\pi\hb)^{-1}\int_{R}^{} e^{-\frac{i}{\epsilon}p y} v^{\hb}_{n}\left(x+\frac{y}{2}\right)\overline{v^{\hb}_m}\left(x-\frac{y}{2}\right) \, dy \, ,
\end{eqnarray}
satisfy the system of eigenvalue equations
\begin{eqnarray}
H(x,p)\star_{\M} W_{nm}^{\epsilon}(x,p)=\mathbb{H}^\hb W_{nm}^{\epsilon}(x,p)&=&E^{\epsilon}_{n}W_{nm}^{\epsilon}(x,p)\ , \label{genvaluetilde_2}\\
W_{nm}^{\epsilon}(x,p)\star_{\M} H(x,p) &=&E^{\epsilon}_{m}W_{nm}^{\epsilon}(x,p)\label{genvaluetilde_3}
\end{eqnarray}
for all $n,m=0,1,2,\ldots$ .

Now by using the representation $(\ref{starformula_2new})$ of Moyal product in equation (\ref{genvaluetilde_3}) we obtain the eigenvalue equation  
\begin{eqnarray}\label{genvaluetildecon}
\overline{\mathbb{H}^\hb} W_{nm}^{\epsilon}(x,p)=E^{\epsilon}_{m}W_{nm}^{\epsilon}(x,p) \ ,
\end{eqnarray}
for the operator $\overline{\mathbb{H}^\hb}=H\left(x-\frac{i\epsilon}{2}\overrightarrow{\partial_ p},p+\frac{i\epsilon}{2}\overrightarrow{\partial_ x}\right)$.

%\subsection*{$\star$-Genfunctions and $\star$-Genvalues}
%The eigenvalue problem leads to $\star$-orthogonality and spectral projection properties of stationary Wigner functions. A standard reference is the work of Bayen, Flato, Fronsdal, Lichnerowicz and Sternheimer \cite{BaFFLS}, Go.Torres Vega and J.H. Frederick \cite{TVF1} ???,Mlodawski???, also a series of papers by Zachos +++ (and the references cited there). Furthermore, a new approach to the $\star$-genvalue equation was developed by M.Gosson and F.Luef \cite{GL}. 
%In this paper, they showed that the $\star$-genvalue problem can be completely solved in terms of the eigenvalue problem $\widehat{H}\psi=E\psi$, where $\widehat{H}$ is the Weyl operator of symbol $H$. Indeed, the solutions of these equations can be obtained from each other using the cross Wigner transform.

By equation  (\ref{genvalue1}), the  Wigner functions $W_{n}^{\epsilon}(x,p)$ and $W_{m}^{\epsilon}(x,p)$  have the properties
\begin{eqnarray*}
W_{n}^{\epsilon}\star_{\M}H\star_{\M} W_{m}^{\epsilon}&=&E_{n}W_{n}^{\epsilon}\star_{\M}W_{m}^{\epsilon}=E_{m}W_{n}^{\epsilon}\star_{\M}W_{m}^{\epsilon} \ .
\end{eqnarray*} 
If  $E_n\neq E_m$, the last equation is satisfied only by  $W_{n}^{\epsilon}\star_{\M}W_{m}^{\epsilon}=0$. This  can be also checked directly by the definition of the Moyal product in the form (\ref{starformula1new}).
Then, it follows that the phase space traces satisfy
\begin{eqnarray*}
\int_{}^{}W_{n}^{\epsilon}\star_{\M}W_{m}^{\epsilon}\, dpdx=
\int_{}^{}W_{n}^{\epsilon}W_{m}^{\epsilon}\, dpdx=
\int_{}^{}W_{m}^{\epsilon}\star_{\M}W_{n}^{\epsilon}\, dpdx=0 \ ,
\end{eqnarray*}
which implies that all overlapping Wigner functions cannot be everywhere positive.

Moreover, by (\ref{starformula1new}) it follows that $W^{\epsilon}_{n} \ , W_{m}^{\epsilon}$ satisfy Baker's projection identity \cite{Ba}
$$W^{\epsilon}_n\star_{\M}W^{\epsilon}_n=\frac{1}{\epsilon}W^{\epsilon}_n \ ,$$ 
and
$$W_{n}^{\epsilon}\star_{\M}W_{m}^{\epsilon}=\frac{1}{\epsilon}\delta_{nm}W_{n}^{\epsilon} \ .$$

%%%%%%%%%%%%%%%%%%%%%%%%%%%%%%%%%%%%%%%%%%%%%%%%%%%%%%%%%%%%%%%%%%%%%%%%
%%%%%%%%%%%%%%%%%%%%%%%%%%%%%%%%%%%%%%%%%%%%%%%%%%%%%%%%%%%%%%%%%%%%%%%%
\section{Spectral properties of the Hamiltonian $\mathbb{H}^\hb$}

In this section we present briefly some recent spectral results obtained 
 by de Gosson \& Luef \cite{GL} for the deformed Hamiltonian $\mathbb{H}^\hb$. Their  results exploits the connection between the Schr\"{o}dinger eigenvalue  equation and the time-independent Wigner equation  ($\star$-genvalue equation)  in phase space, and it given in the Theorem \ref{mainresult} which stated below.
 
Let $W_{\psi}\phi$ be the operator defined by
$$W_{\psi}\phi (x,p)=(2\pi\hb)^{-1}\int_{R}^{} e^{-\frac{i}{\epsilon}p y} \psi\left(x+\frac{y}{2}\right)\overline\phi\left(x-\frac{y}{2}\right) \, dy \ \  .
$$
The adjoint of $W_{\phi}$ is given by 
$$W^{*}_{\phi}\Psi(x,p)=\frac{2}{\pi\epsilon}\int_{R}^{}\int_{R}^{}e^{\frac{2i}{\epsilon}p(x-y)}\phi(2y-x)\Psi(y,p) \, dp dy \ \ .$$

Then, the following theorem holds.

\begin{theorem}\label{mainresult}
The following properties are true:
\\
i. The eigenvalues of the operators $\widehat{H}^\hb$ and $\mathbb{H}^\hb=H\star\cdot$ are the same.\\
ii. Let $\psi$ be an eigenfunction of $\widehat{H}^\hb$ , $\widehat{H}^\hb\psi=\lambda\psi$. Then, for every $\phi$, the function $\Psi=W_{\phi}{\psi}$ is an eigenfunction of 
$\mathbb{H}^\hb$ corresponding to the same eigenvalue , $\mathbb{H}^\hb \Psi=\lambda\Psi$.
\\
iii. Conversely, if $\Psi$ is an eigenfunction of $\mathbb{H}^\hb$ then $\psi=\wphi^{\star}\Psi$ is an eigenfunction of $\hh^\hb$ corresponding to the same eigenvalue.
\end{theorem}
The proof of thetheorem is based on the following lemma.
\begin{lemma}
Let $\{\phi_j\}_j$ be an arbitrary orthonormal basis of $L^2(R)$ the vectors $\Phi_{jk}=W_{\phi_j}\phi_{k}$ form an orthonormal basis of $L^2({R}^{2})$.
\end{lemma}

The Theorem \ref{mainresult} is quite general. There is no assumption on the multiplicity of $\star$-genvalues,and it is not assumed that $\widehat{H}$ is essentially self-adjoint. If such extra conditions are assumed, then the proof of the theorem implies the following corollary.

\begin{corollary}\label{corollary}
Suppose that $\hh$ is an essentially self-adjoint operator on $L^2(R)$ and that each of the eigenvalues $\lambda_{0},\lambda_{1},\ldots,\lambda_{j},\ldots$ , has multiplicity one.
\\
Let 
$\psi_{0},\psi_{1},\ldots,\psi_{j},\ldots$,  be a corresponding sequence of orthonormal eigenfunctions. Also $\Psi_{j}$ is an eigenfunction of $\mathbb{H}^\hb$ corresponding
 to the eigenvalue $\lambda_{j}$. There exists a sequence $(\alpha_{jk})_k$ of the complex numbers such that 
\begin{eqnarray*}
\Psi_{j}=\sum_{\ell}^{}\alpha_{j\ell}\Psi_{j,\ell} 
\end{eqnarray*}
with
\begin{eqnarray*}
\Psi_{j,\ell}=W_{\psi_{j}}\psi_{j}\in\mathcal{H}_j\cap\mathcal{H}_\ell \ .
\end{eqnarray*}
\end{corollary}

%%%%%%%%%%%%%%%%%%%%%%%%%%%%%%%%%%%%%%%%%%%%%%%%%%%%%%%%%%%%%%%%%%%%%%%%%%%%%%%%%%%
%%%%%%%%%%%%%%%%%%%%%%%%%%%%%%%%%%%%%%%%%%%%%%%%%%%%%%%%%%%%%%%%%%%%%%%%%%%%%%%%%%%
%%%%%%%%%%%%%%%%%%%%%%%%%%%%%%%%%%%%%%%%%%%%%%%%%%%%%%%%%%%%%%%%%%%%%%%%%%%%%%%%%%%%%%%%%%%%%%%%%%%%%%%%%%%%%%%%%%%%%%%%%%%%%%%%%%%%%%%%%%%%%%%%%%%%%%%%%%%%%%%%%%%%%%
%%%%%%%%%%%%%%%%%%%%%%%%%%%%%%%%%%%%%%%%%%%%%%%%%%%%%%%%%%%%%%%%%%%%%%%%%%%%%%%%%%%
\chapter{Stationary phase asymptotics}\label{app_a}
\section{The standard formula for a simple stationary point}

We consider the integral
\begin{equation}\label{integral}
I(\lambda)=\int_{a}^{b}f(t)\, e^{i\lambda\phi(t)} \,  dt
\end{equation}
and suppose that $f\in C[a,b]$ while $\phi\in C^2[a,b]$, real-valued function.
Suppose also that $t=c$ is the only point in $[a,b]$ where $\phi'(t)$
vanishes, and $\phi''(c)\not =0$. We rewrite $I(\lambda)$ as
\begin{eqnarray*}
I(\lambda)=e^{i\lambda\phi(c)}\int_{a}^{b}f(t)\, e^{i\lambda(\phi(t)-\phi(c))}dt \ .
\end{eqnarray*}
The main contribution to the integral (\ref{integral}) is expected to
come from a small neighbourhood of $c$ where the oscillations of the exponential term dissappear, that is
\begin{eqnarray*}
\int_{a}^{b}f(t)\, e^{i\lambda(\phi(t)-\phi(c))} dt \approx \int_{c-r}^{c+r}f(c)\, e^{i\lambda[\phi(c)+\frac{(t-c)^2}{2}\phi''(c)]}\,  dt
\end{eqnarray*}
where $r$ is small but finite. To evaluate this integral, we make the change of variable
\begin{eqnarray*}
\mu\tau^2=(t-c)^2 \ \frac{\phi''(c)}{2} \ \lambda,\quad  \textrm{or} \quad
\tau=(t-c)\sqrt{\frac{|\phi''(c)|\lambda}{2}}
\end{eqnarray*}
where $\mu=sgn\phi''(c)$. Then, the integral becomes
\begin{eqnarray*}
f(c)e^{i\lambda\phi(c)}\sqrt{\frac{2}{|\phi''(c)|\lambda}}\int_{-r\sqrt{\lambda|\phi''(c)|/2}}^{r\sqrt{\lambda|\phi''(c)|/2}}e^{i\mu\tau^2}
d\tau
\end{eqnarray*}
As $\lambda\rightarrow \infty$ the last integral reduces to $\int_{-\infty}^{\infty}
e^{i\mu\tau^2} d\tau$, which can be evaluated
exactly
\begin{eqnarray*}
\int_{-\infty}^{\infty} e^{i\mu\tau^2} d\tau=2\int_{0}^{\infty} e^{i\mu\tau^2}
d\tau=\sqrt{\pi}e^{\frac{i\pi\mu}{4}}
\end{eqnarray*}
by using the formula
\begin{eqnarray*}
\int_{0}^{\infty}t^{\gamma}\, e^{i\nu t^{p}}dt
={\left(\frac{1}{|\nu |}\right)}^{\frac{\gamma+1}{p}}\ \frac{\Gamma
(\frac{\gamma+1}{p})}{p}\ e^{i\frac{\pi}{2p}(\gamma+1)sgn\nu } \ , \  \ \gamma>-1 \ , p >0  \ , \nu \in R \ ,
\end{eqnarray*}
with $\gamma=0$, $p=2$, and $\nu =\mu$ (recall that $\Gamma(1/2)=\sqrt{\pi}$ ).

Thus, we obtain the approximation formula(\cite{BH}, Ch. 6, or \cite{Bor}, Ch. 2)

\begin{equation}\label{statphform}
I(\lambda)\approx e^{i\lambda\phi(c)+i\mu
\pi/4}f(c)\left[\frac{2\pi}{\lambda|\phi''(c)|}\right]^{1/2} \ , \ \ \ \lambda\rightarrow \infty \ ,
\end{equation}
with $\mu=\sgn\phi''(c)$. The remainder of the approximation is of the order $O(\lambda^{-3/2})$.

%%%%%%%%%%%%%%%%%%%%%%%%%%%%%%%%%%%%%%%%%

\section{The uniform formula for two coalescing stationary points}

We consider the integral
$$I(\lambda ,\alpha)=\int_{-\infty}^{\infty} e^{i\lambda \phi(x,\alpha)}f(x)dx, $$
where $\alpha>0$ and $\lambda \rightarrow \infty$. We assume
that $f$ is smooth, and that the phase function $\phi\in
C^{\infty}$ has two stationary points, $x_1(\alpha)$ and
$x_2(\alpha),$ which approach the same limit $x_0$ when
$\alpha\rightarrow 0.$ Let $\partial_{xx}\phi(x_1,\alpha)<0$ and
$\partial_{xx}\phi(x_2,\alpha)>0$.

The standard stationary-phase formula $(\ref{statphform})$ applied to $I(\lambda,\alpha)$ gives the approximation
\begin{eqnarray}\label{ap1}
I(\lambda ,\alpha)\approx \left(\frac{2\pi}{\lambda
}\right)^{1/2}\sum_{l=1,2}\frac{f(x_l(\alpha))e^{i\lambda
\phi(x_{l}( \alpha ),\alpha)}}{\mid \partial_{xx}\phi(x_l(\alpha))\mid
^{1/2}} e^{i\frac{\pi}{4}\delta_l} \ . \ \ \delta_l=sgn\partial_{xx}\phi(x_l(\alpha)) \ .
\end{eqnarray}
Here $\delta_1=-1,\delta_2=1$, since we assumed $\partial_{xx}\phi(x_1,\alpha)<0$ and
$\partial_{xx}\phi(x_2,\alpha)>0$. This approximation fails when $\alpha =0$ since $\partial_{xx}\phi(x_1,\alpha=0)=\partial_{xx}\phi(x_2,\alpha=0)=0$. Hence, we need to develop an approximation formula which holds uniformly wrt. $\alpha$.

We further assume also that $\phi(x,\alpha)$ is analytic for small $(x-x_0)$
and small $\alpha>0,$ and that we have
\begin{eqnarray}\label{ap20}
\partial_{xxx}\phi\not=0,\ \partial_{x}\phi=\partial_{xx}\phi=0,\
\partial_{x\alpha}\phi\not=0
\end{eqnarray}
at $(x=x_{0},\alpha=0)$.

Under these conditions a theorem by Chester, Friedman and Ursell
\cite{CFU} (see also \cite{Bor}, Ch. 2) imples that there exists a change of
variable $x=x(\tau),$ analytic and invertible for small $(x-x_0)$
and small $\alpha>0,$ depending parametrically on $\alpha,$ such
that

\begin{eqnarray}\label{ap14}
\phi(x,\alpha)=\phi_0(\alpha)+\frac{\tau^3}{3}-\xi(\alpha)\ \tau
\end{eqnarray}
where $\phi_0(\alpha)$ and $\xi(\alpha)$ are analytic functions of
$\alpha.$

Then, we write the integral in the form
\begin{eqnarray}\label{ap2}
I(\lambda ,\alpha)=e^{i\lambda
\phi_0}\int_{-\infty}^{\infty}e^{i\lambda
(\tau^3/3-\tau\xi(\alpha))}f(x(\tau))\frac{dx(\tau)}{d\tau}\ d\tau \, .
\end{eqnarray}
where, by a version of Malgrange's preparation theorem (that is a kind of "division" theorem), we have the representation
\begin{eqnarray}\label{ap3}
f(x(\tau))\frac{dx(\tau)}{d\tau}\
d\tau=A_{0}(\alpha)+B_{0}(\alpha)\tau+h(\tau)(\tau^2-\xi)
\end{eqnarray}
where $h(\tau)$ is smooth function.

Substituting $\mathrm{(\ref{ap3})}$ into $\mathrm{(\ref{ap2})}$, and by expressing analytically the first two terms in the integral in terms of the Airy function, we get
\begin{eqnarray}
I(\lambda ,\alpha)= e^{i\lambda \phi_0(\alpha)}\left[2\pi
A_{0}(\alpha)\lambda ^{-1/3}Ai(-\lambda ^{2/3}\xi)-2\pi i
B_{0}(\alpha)\lambda ^{-2/3} Ai'(-\lambda ^{2/3}\xi)+C(\lambda
,\xi)\right] \ ,
\end{eqnarray}
where $$C(\lambda,\xi)=\frac{i}{\lambda}\int_{-\infty}^{\infty}h'(\tau)e^{i\lambda
(\tau^3/3-\tau\xi)}d\tau=O(\lambda ^{-4/3}) \ .$$
The integral  $C$ can be integrated by parts, as many time as we wish, and we obtain
\begin{eqnarray}\label{ap4}
I(\lambda ,\alpha)=e^{i\lambda \phi_0(\alpha)}\left[2\pi A \lambda
^{-1/3}Ai(-\lambda ^{2/3}\xi) -2\pi i B\lambda ^{-2/3}Ai'(-\lambda
^{2/3}\xi)\right]
\end{eqnarray}
where
\begin{eqnarray}
A=\sum_{n=0}^{\infty}A_n(\alpha)\left(\frac{i}{\lambda }\right)^n \ , \ \
B=\sum_{n=0}^{\infty}B_n(\alpha)\left(\frac{i}{\lambda }\right)^n \ .
\end{eqnarray}

In order to compute $\phi_{0}(\alpha),\xi(\alpha)$ and the leading
coefficients $A_{0}(\alpha),B_{0}(\alpha)$ we apply the so called  {\it principle of
asymptotic matching}.

We fix $\alpha>0$ and we consider $\lambda \rightarrow \infty.$
Then the asymptotics of $Ai,Ai'$ read as follows
\begin{eqnarray}\label{ap5}
Ai(-\lambda ^{2/3}\xi)\approx \frac{1}{2\sqrt{\pi}}\lambda
^{-1/6}\xi^{-1/4}\left[e^{2i\lambda \xi^{3/2}/3-i \pi/4}+
e^{-2i\lambda \xi^{3/2}/3+i \pi/4 }\right]
\end{eqnarray}

\begin{eqnarray}\label{ap6}
Ai'(-\lambda ^{2/3}\xi)\approx \frac{-1}{2\sqrt{\pi}}\lambda
^{1/6}\xi^{1/4}\left[e^{2i\lambda \xi^{3/2}/3+i \pi/4}+
e^{2i\lambda \xi^{3/2}/3-i \pi/4 }\right]
\end{eqnarray}

Substituting $\mathrm{(\ref{ap5})}$ and $\mathrm{(\ref{ap6})}$
into $\mathrm{(\ref{ap4})},$  we get the expansion
\begin{eqnarray}\label{ap7}
I(\lambda ,\alpha)&\approx &\left(\frac{\pi}{i\lambda
}\right)^{1/2}(A_{0}\xi^{-1/4}-B_{0}\xi^{1/4})\ e^{i\lambda
(\phi_{0}+2\xi^{3/2}/3)}\nonumber\\
&=&\left(\frac{\pi}{i\lambda
}\right)^{1/2}(A_{0}\xi^{-1/4}+B_{0}\xi^{1/4})\ e^{i\lambda
(\phi_{0}-2\xi^{3/2}/3)}+O(\lambda^ {-3/2})
\end{eqnarray}

{\it The principle of asymptotic matching requires that the expansion
$\mathrm{(\ref{ap7})}$ must coincide with the non-uniform
expansion $\mathrm{(\ref{ap1})}$.}

Comparing $\mathrm{(\ref{ap7})}$ with  $\mathrm{(\ref{ap1})}$, and
taking into account that $\phi(x_1,\alpha)>\phi(x_2,\alpha)$ we
obtain
\begin{eqnarray}
\phi_0+\frac{2}{3}\ \xi^{3/2}=\phi(x_1,\alpha)
\end{eqnarray}

\begin{eqnarray}
\phi_0-\frac{2}{3}\ \xi^{3/2}=\phi(x_2,\alpha)
\end{eqnarray}
from the phases, and
\begin{eqnarray}\label{ap14}
A_{0}\xi^{-1/4}+B_{0}\xi^{1/4}=\sqrt{2}\ \frac{f(x_2)}{
(\partial_{xx}\phi(x_2,\alpha))^{1/2}}
\end{eqnarray}

\begin{eqnarray}\label{ap15}
A_{0}\xi^{-1/4}-B_{0}\xi^{1/4}=\sqrt{2}\ \frac{f(x_1)}{\mid
\partial_{xx}\phi(x_1,\alpha)\mid ^{1/2}} \ ,
\end{eqnarray}
from the amplitudes.

Therefore, we obtain

\begin{eqnarray}\label{ap23}
\phi_0(\alpha)=\frac{1}{2}\left(\phi(x_1(\alpha),\alpha)+\phi(x_2(\alpha),\alpha)\right) \ ,
\end{eqnarray}

\begin{eqnarray}\label{ap13}
\xi(\alpha)=\left[\
\frac{3}{4}\left(\phi(x_1(\alpha),\alpha)-\phi(x_2(\alpha),\alpha)\right)\
\right]^{2/3} \ ,
\end{eqnarray}
and
\begin{eqnarray}\label{A0}
A_0=2^{-1/2}\xi^{1/4}\left[\frac{f(x_2)}{\sqrt{\phi_{xx}(x_2,\alpha)}}+\frac{f(x_1)}{\sqrt{|\phi_{xx}(x_1,\alpha)|}}\right] \ ,
\end{eqnarray}
\begin{eqnarray}\label{B0}
B_0=-2^{-1/2}\xi^{-1/4}\left[\frac{f(x_1)}{\sqrt{|\phi_{xx}(x_1,\alpha)|}}-\frac{f(x_2)}{\sqrt{\phi_{xx}(x_2,\alpha)}}\right] \ .
\end{eqnarray}

In the applications, and in order to do efficiently tha analytical calculations, it is desirable to use simpler approximations of the quantities $\phi_0(\alpha) \ , \xi(\alpha)$, etc.,
as $\alpha\rightarrow 0^+$. In the sequel we construct such approximations.

Without loss of generality, we set $x_{0}=0$ (which amounts for changing the variable
$x$ to $x'=x-x_0$), and we expand $\phi(x,\alpha)$ near $(x=0,\alpha=0)$ up to terms of order $O(\alpha^2)$,
\begin{eqnarray}\label{ap14}
\phi(x,\alpha)&=&\phi(0,0)+\partial_{x}\phi(0,0)\ x+\partial_{\alpha}\phi(0,0)\ \alpha\nonumber\\
&&+\frac{1}{2}\ \partial_{xx}\phi(0,0)\ x^2+\partial_{x\alpha}\phi(0,0)\ \alpha
x+\frac{1}{2}\
\partial_{\alpha\alpha}\phi(0,0)\ \alpha^2\nonumber\\
&&+\frac{1}{6}\ \partial_{xxx}\phi(0,0)\ x^3+\frac{1}{2}\
\partial_{xx\alpha}\phi(0,0)\ x^2\alpha
+\frac{1}{2}\ \partial_{x\alpha\alpha}\phi(0,0)\ x\alpha^2\nonumber\\
&&+\frac{1}{6}\ \partial_{\alpha\alpha\alpha}\phi(0,0)\
\alpha^3+({4^{th}-order \
terms})\nonumber\\
&=&\phi+\partial_{\alpha}\phi\ \alpha+\partial_{x\alpha}\phi\ \alpha
x+\frac{1}{2}\
\partial_{\alpha\alpha}\phi\ \alpha^2+\frac{1}{6}\ \partial_{xxx}\phi\ x^3\nonumber\\
&&+\frac{1}{2}\ \partial_{xx\alpha}\phi\ x^2\alpha+\frac{1}{2}\
\partial_{x\alpha\alpha}\phi\ x \alpha^2+\frac{1}{6}\
\partial_{\alpha\alpha\alpha}\phi\ \alpha^2 \ .
\end{eqnarray}
(From here on we omit the argument $(0,0)$ in the coefficients. In the general case the coefficients are computed at the point $(x_0, 0)$).

By differentiating the last equation wrt. $x$, we have
\begin{eqnarray}\label{ap8}
\partial_{x}\phi(x,\alpha)=\partial_{x\alpha}\phi\ \alpha+\frac{1}{2}\ \partial_{xxx}\phi\
x^2+\partial_{xx\alpha}\phi\ x\alpha+\frac{1}{2}\ \partial_{x\alpha\alpha}\phi\ \alpha^2
\end{eqnarray}
and
\begin{eqnarray}\label{ap9}
\partial_{xx}\phi(x,\alpha)=\partial_{xxx}\phi\ x+\partial_{xx\alpha}\phi\ \alpha
\end{eqnarray}

We approximate eq. $\mathrm{(\ref{ap8})}$ up to   $O(\alpha^2)$, and we compute the approximate stationary points $x_1(\alpha)$, $x_2(\alpha)$ by solving the equation

\begin{eqnarray}
\partial_{x}\phi(x,\alpha)=\frac{1}{2}\ \partial_{xxx}\phi(0,0)\
x^2+\partial_{xx\alpha}\phi(0,0)\ x\alpha+\partial_{x\alpha}\phi(0,0)\ \alpha=0
\end{eqnarray}
The roots of the last equation are
\begin{eqnarray*}
x_{1,2}(\alpha)&=&\frac{-\alpha\ \partial_{xx\alpha}\phi\pm
((\partial_{xx\alpha}\phi) ^2\
\alpha^2-2\partial_{xxx}\phi\ \partial_{x\alpha}\phi\ \alpha)^{1/2}}{\partial_{xxx}\phi}\\
&\approx &\pm (-2\partial_{xxx}\phi\ \partial_{x\alpha}\phi\
\alpha)^{1/2}(\partial_{xxx}\phi)^{-1}
\end{eqnarray*}
since $\sqrt{\alpha}>>\alpha>\alpha^2$ as $\alpha\rightarrow 0^+$ .
The assumptions
$\partial_{xx}\phi(x_1(\alpha),\alpha)<0$ , $\partial_{xx}\phi(x_2(\alpha),\alpha)>0$ and
$\mathrm{(\ref{ap9})}$, for $\alpha\rightarrow 0^+,$ imply that
\begin{eqnarray}\label{ap10}
x_1(\alpha)\approx -(-2\partial_{xxx}\phi\ \partial_{x\alpha}\phi\
\alpha)^{1/2}(\partial_{xxx}\phi)^{-1}
\end{eqnarray}
\begin{eqnarray}\label{ap11}
x_2(\alpha)\approx +(-2\partial_{xxx}\phi\ \partial_{x\alpha}\phi\
\alpha)^{1/2}(\partial_{xxx}\phi)^{-1}
\end{eqnarray}
Then, by $(\ref{ap23})$,
\begin{eqnarray}\label{app10}
\phi_{0}(\alpha)&\approx& \phi(0,0) \ .
\end{eqnarray}

We also need to approximate the phase difference
\begin{eqnarray}\label{ap12}
\delta\phi=\phi(x_1(\alpha),\alpha)-\phi(x_2(\alpha),\alpha)\approx
\frac{1}{6}\ \partial_{xxx}\phi\ ({x_1}^3-{x_2}^3)+\partial_{x\alpha}\phi\
\alpha(x_1-x_2)+\ldots \nonumber\\
\end{eqnarray}
From  $\mathrm{(\ref{ap10})},$  $\mathrm{(\ref{ap11})}$,
we have
$${x_1}^3-{x_2}^3=-2(\partial_{xxx}\phi)^{-3}(-2\partial_{xxx}\phi\partial_{x\alpha}\phi)^{3/2} \alpha^{3/2}  \ ,$$
$$x_1-x_2=-2(\partial_{xxx}\phi)^{-1}(-2\partial_{xxx}\phi\ \partial_{x\alpha}\phi)^{1/2}\alpha^{1/2} \ ,$$
and therefore and $\mathrm{(\ref{ap12})}$,
\begin{eqnarray}
\delta\phi&=&-2\alpha^{3/2}(\partial_{xxx}\phi)^{-1}(-2\partial_{xxx}\phi\
\partial_{x\alpha}\phi)^{1/2}\nonumber\\
&& \cdot \left[\ \frac{1}{6}(\partial_{xxx}\phi)^{-1}(-2\partial_{xxx}\phi\
\partial_{x\alpha}\phi)+\partial_{x\alpha}\phi\ \right]\nonumber\\
&=&-\frac{2}{3}\ \alpha^{3/2}(-2\partial_{xxx}\phi\
\partial_{x\alpha}\phi)^{3/2}(\partial_{xxx}\phi)^{-2}
\end{eqnarray}

Then, using $\mathrm{(\ref{ap13})}$ we obtain
\begin{eqnarray}\label{289}
\xi(\alpha)&\approx& -\partial_{x\alpha}\phi\ \left(\frac{\partial_{xxx}\phi}{2}\right)^{-1/3}\alpha \ .
\end{eqnarray}

Finally by using $\partial_{xx}\phi(x,\alpha)=\partial_{xxx}\phi(0,0)\
x+\partial_{xx\alpha}\phi(0,0)\ \alpha,$ we have
\begin{eqnarray*}
\partial_{xx}\phi(x_1(\alpha),\alpha)&=&-\partial_{xxx}\phi\ (-2\partial_{xxx}\phi\
\partial_{x\alpha}\phi\
\alpha)^{1/2}(\partial_{xxx}\phi)^{-1}+\partial_{xx\alpha}\phi\ \alpha \nonumber\\
&\approx &-(-2\partial_{xxx}\phi\ \partial_{x\alpha}\phi\ \alpha)^{1/2}
\end{eqnarray*}
which gives
\begin{eqnarray}\label{30}
\mid \partial_{xx}\phi(x_1(\alpha),\alpha)\mid \ \approx (-2\partial_{xxx}\phi\
\partial_{x\alpha}\phi\ \alpha)^{1/2}\, ,
\end{eqnarray}
\begin{eqnarray}\label{31}
\mid \partial_{xx}\phi(x_2(\alpha),\alpha)\mid \ \approx (-2\partial_{xxx}\phi\
\partial_{x\alpha}\phi\ \alpha)^{1/2}
\end{eqnarray}
and thus we have tha approximation
\begin{eqnarray}\label{32}
\xi^{1/4}\approx \left[-\partial_{x\alpha}\phi\
\left(\frac{\partial_{xxx}\phi}{2}\right)^{-1/3}\alpha\right]^{1/4} \ .
\end{eqnarray}

\section{The stationary-phase formula for double integrals}

Let the fucntions $f(x,y)$ and $\phi(x,y)$ in the integral
$$I(\lambda)=\int\int_{\Omega} f(x,y)g(x,y)e^{i\lambda \phi(x,y)} dxdy$$
be infinitely differentiable, $\phi(x,y)$ has the stationary point $x=x_0 \ , y=y_0$ with the non degenerate  matrix of second derivatives
\[ A=\left ( \begin{array}{ccc}
\phi_{xx} & \phi_{xy}  \\
\phi_{yx} & \phi_{yy}  \end{array} \right)\ , \]
and let $g(x,y)$ be a cutting function that singles out a sufficiently small vicinity of the stationary point $x=x_0 \ , y=y_0$. then, as $\lambda \to \infty$, the integral $I(\lambda)$ has the asymptotic expansion  (\cite{Bor}, Ch. 4)
\begin{equation}\label{doublestat}
I(\lambda)\approx \alpha_0  \lambda^{-1} e^{i\lambda\phi(x_0, y_0)} + O(\lambda^{-2}) \ ,
\end{equation}
where
$$\alpha_0 = \frac{2\pi f(x_0, y_0)e^{i\delta\pi/2}}
{\sqrt{\mid \phi_{xx}\phi_{yy}-\phi_{xy}^{2}\mid}} \ ,$$
and $\delta=1$, if both eigenvalues of the matrix $A$ are positive, $\delta=0$, if they have opposite signs, and $\delta=-1$, if both are negative.
%%%%%%%%%%%%%%%%%%%%%%%%%%%%%%%%%%%%%%%%%%%%%%%%%%%%%%%%%%%%%%%%%%%%%%%%%%%%%%%%%%%%%%%%%%%%%%%%%%%%%%%%%%%%%%%%%%%%%%%%%%%%%%%%%%%%%%%%%%%%%%%%%%%%%%%%%%%%%%%%%%
%%%%%%%%%%%%%%%%%%%%%%%%%%%%%%%%%%%%%

\chapter{The harmonic oscillator}
\section{The WKB phase of the harmonic oscillator}\label{app_b}
The WKB  solution for the harmonic oscillator in the interval $|x|<\sqrt{2E_{n}^{\hb}}\, $ between the turning points   (oscillatory solution), is given by
\begin{eqnarray}\label{psin}
\psi^{\epsilon}_{n}(x)=\left(2E_{n}^{\hb}-x^2\right)^{-1/4}\cos\left(\frac{1}{\epsilon}\int_{\sqrt{2E_{n}^{\hb}}}^{x}\sqrt{2E_{n}^{\hb}-t^2}\, dt+\frac{\pi}{4}\right)
\end{eqnarray}
wher $E_{n}^{\hb}=\left(n+\frac{1}{2}\right)\epsilon$ is the corresponding eigenvalue (\cite{Fed}, Ch. 3).

The phase of integral $(\ref{psin})$ can be calculated analytically. By the change of variable $s=t/\sqrt{2E_{n}^{\hb}}$, we have
\begin{eqnarray}\label{defint}
\int_{\sqrt{2E_{n}^{\hb}}}^{x}\sqrt{2E_{n}^{\hb}-t^2}\, dt= 2E_{n}^{\hb}\int_{1}^{x/\sqrt{2E_{n}^{\hb}}}\sqrt{1-s^2}\, ds \, .
\end{eqnarray}
The integral in the last relation is calculated by the formulae ( \cite{GR}, eqs.  (2.261),(2.262))
\begin{eqnarray*}%\label{int}
\int_{}^{}\sqrt{R}\, ds=\frac{(2cs+b)\sqrt{R}}{4c}+\frac{\Delta}{8c}\int_{}^{}\frac{ds}{\sqrt{R}} \ ,
\end{eqnarray*}
\begin{eqnarray*}
\int_{}^{}\frac{ds}{\sqrt{R}}=\frac{-1}{\sqrt{-c}}\arcsin\left({\frac{2cs+b}{\sqrt{-\Delta}}}\right) \ ,
\end{eqnarray*}
where $R=a+bs+cs^2$ and $\Delta=4ac-b^2$.

In our case $a=1$, $b=0$, $c=-1$ and $\Delta=-4<0$ and we obtain
\begin{eqnarray*}
2E_{n}^{\hb}\int_{1}^{x/\sqrt{2E_{n}^{\hb}}}\sqrt{1-s^2}\, ds&=&E_{n}^{\hb}\left[s\sqrt{1-s^2}+\arcsin(s)\right]_{s=1}^{s=x/\sqrt{2E_{n}^{\hb}}}\nonumber\\
&=&E_{n}^{\hb}\left[\frac{x\sqrt{2E_{n}^{\hb}-x^2}}{2E_{n}^{\hb}}+\arcsin\left(\frac{x}{\sqrt{2E_{n}^{\hb}}}\right)-\frac{\pi}{2}\right]\, .
\end{eqnarray*}
Thus, the phase of (\ref{psin}) becomes
\begin{eqnarray}\label{wkbphase}
\int_{\sqrt{2E_{n}^{\hb}}}^{x}\sqrt{2E_{n}^{\hb}-t^2}\, dt=
\frac{x}{2}\sqrt{2E_{n}^{\hb}-x^2}+
{E_{n}^{\hb}}\arcsin\left(\frac{x}{\sqrt{2E_{n}^{\hb}}}\right)-{E_{n}^{\hb}}\frac{\pi}{2}  \ .
\end{eqnarray}

\section{Comparison of exact and WKB eigenfunctions}\label{comparison_sol}
Consider the eigenvalue equation
\begin{eqnarray}\label{harmonicosc}
\widehat{H}^{\epsilon}v^{\epsilon}(x)=\left[-\frac{\epsilon^2}{2}\frac{d^2}{dx^2}+\frac{x^2}{2}\right]\vh(x)=E_{n}^{\hb} \vh(x) \ ,
\end{eqnarray}
for the harmonic oscillator.
The eigenfunctions of $\widehat{H}^{\hb}$ are known and they are given by
\begin{eqnarray}\label{hermitesol_new}
\vh_{n}(x)=\epsilon^{-1/4}\, v_{n}\left(\frac{x}{\sqrt{\epsilon}}\right)\, , \quad \mathrm{for}\quad n=0,1,\ldots
\end{eqnarray}
with
\begin{eqnarray}\label{hermitefun_new}
v_{n}(x)={(2^nn!\sqrt{\pi})}^{-1/2}e^{-x^2/2}\, H_{n}(x)
\end{eqnarray}
where the functions $H_n$ are the Hermite polynomials
$H_n(x):=(-1)^n e^{x^2}\frac{d^n}{dx^n}\left(e^{-x^2}\right)$ and $\vh_{n}(x)$ are normalized to $1$. For the basic properties of Hermite polynomials and functions see, e.g., the book by Wong \cite{WoM}.

Thus, the eigenfunction given by (\ref{hermitesol_new}), can be written as
\begin{eqnarray*}
%\label{eigenf_harm}
\vh_{n}(x)=\frac{e^{-x^2/2\epsilon}}{(\pi\epsilon)^{1/4}\sqrt{2^{n}n!}}\, H_{n}\left(\frac{x}{\sqrt{\epsilon}}\right)\, ,\quad \mathrm{for}\quad n=0,1,\ldots \, .
\end{eqnarray*}

Dominici \cite{Do} has obtained an asymptotic
formula for the Hermite polynomials $H_{n}(x)$ in the oscillatory range $|x|<\sqrt{2n}$, which has the form  (with $x=\sqrt{2n}\, sin(\theta)$, $|\theta|<\pi/2$)
\begin{eqnarray*}
H_{n}(\sqrt{2n}\sin\theta)\approx
\sqrt{\frac{2}{\cos\theta}}\, e^{\frac{n}{2}ln(2n)}e^{-\frac{n}{2}\cos(2\theta)}\cos\left[\frac{n}{2}\sin(2\theta)+\left(n+\frac{1}{2}\right)\theta-\frac{n\pi}{2}\right]\, .
\end{eqnarray*}
\\
If we set $\xi=x/\sqrt{\hb}$ then $\sin(2\theta)=2\sin\theta\cos\theta={\xi\sqrt{2n-\xi^2}}/{n}$, $\cos(2\theta)=\cos^2(\theta)-\sin^2(\theta)=1-\xi^2/n$, and
\begin{eqnarray*}
H_{n}(\xi)\approx \sqrt{2} {\left(\frac{2n}{2n-\xi^2}\right)}^{1/4}{(2n)}^{n/2} e^{\frac{\xi^{2}-n}{2}} \cos \left[\frac{\xi}{2}\sqrt{2n-\xi^2}+\left(n+\frac{1}{2}\right)\arcsin\left(\frac{\xi}{\sqrt{2n}}\right)-\frac{n\pi}{2}\right] \ .
\end{eqnarray*}
Thus the eiegnfunction $v_n(x)$ has the approximation
\begin{eqnarray*}
v_{n}(\xi)\approx{(2^{n}n!\sqrt{\pi})}^{-1/2}c_n(\xi)\cos \left[\frac{\xi}{2}\sqrt{2n-\xi^2}+\left(n+\frac{1}{2}\right)\arcsin\frac{\xi}{\sqrt{2n}}-\frac{n\pi}{2}\right] \ ,
\end{eqnarray*}
where
\begin{eqnarray*}
c_{n}(\xi):=&\sqrt{2}{\left(\frac{2}{\pi}\right)}^{1/4}\frac{n^{1/4}n^{n/2}e^{-n/2}}{\sqrt{n!}}{(2n-\xi^2)}^{-1/4}
\end{eqnarray*}
By Striling's formula \cite{Leb}, $n!\approx(2\pi n)^{1/2}(n/e)^n=\sqrt{2\pi}n^{n+\frac{1}{2}}e^{-n} $ for $n>>1$, we obtain
\begin{eqnarray}
v_{n}(\xi)\approx \left(\frac{2}{\pi}\right)^{1/2}{(2n-\xi^2)}^{-1/4}
\cos \left[\frac{\xi}{2}\sqrt{2n-\xi^2}+\left(n+\frac{1}{2}\right)\arcsin\frac{\xi}{\sqrt{2n}}-\frac{n\pi}{2}\right]\, ,\nonumber\\
\label{hermsolut}
\end{eqnarray}
when $\xi^2<2n+1\approx 2n$ for large $n$, $\xi=x/\sqrt{\hb}$.

Finally, substituting (\ref{hermsolut}) into (\ref{hermitesol_new}), we obtain the following asymptotic approximation
\begin{eqnarray}\label{hermsolut_n}
v^{\epsilon}_{n}(x)&\approx& \tilde{v}_{n}^{\hb}(x)\nonumber\\
&=&
\left(\frac{2}{\pi}\right)^{1/2}{\left(2n\hb-x^{2}\right)}^{-1/4}
\cos \left[\frac{x}{2\hb}\sqrt{2n\hb-x^2}+\left(n+\frac{1}{2}\right)\arcsin\left(\frac{x}{\sqrt{2n\hb}}\right)-\frac{n\pi}{2}\right]\nonumber
\end{eqnarray}
for $|x|<\sqrt{2n}$, with large $n$ and small $\hb$ .

We now suppose that $\psi^{\epsilon}_{n}(x)$ is a WKB-approximate eigenfunction of  $\widehat{H}^{\hb}$ in the  oscillatory region  $|x|<\sqrt{2E_{n}^{\hb}}$ , by Fedoriuk in \cite{Fed} we obtain
\begin{eqnarray}\label{psin2}
\psi^{\epsilon}_{n}(x)=\left(2E_{n}^{\hb}-x^2\right)^{-1/4}\cos\left(\frac{1}{\epsilon}\int_{\sqrt{2E_{n}^{\hb}}}^{x}\sqrt{2E_{n}^{\hb}-t^2}\, dt+\frac{\pi}{4}\right)
\end{eqnarray}
with corresponding eigenvalue  $E_{n}^{\hb}=\left(n+1/2\right)\epsilon$ . Using the change of variable
$\xi=x/\sqrt{\epsilon}$  the amplitude can be written as,
\begin{eqnarray}\label{ampl_appr}
\left(2E_{n}^{\hb}-x^2\right)^{-1/4}=\left[(2n+1)\epsilon-\xi^2\epsilon\right]^{-1/4}=\epsilon^{-1/4}\left[(2n+1)-\xi^2\right]^{-1/4}
\end{eqnarray}
and $(2E_{n}^{\hb}-x^2)^{-1/4}\approx \epsilon^{-1/4} (2n-\xi^2)^{-1/4}$ for large $n$ and small $\hb$.

Moreover, by (\ref{wkbphase}), the WKB-phase is given by
\begin{eqnarray*}
\frac{1}{\epsilon}\int_{\sqrt{2E_{n}^{\hb}}}^{x}\sqrt{2E_{n}^{\hb}-t^2}\, dt+\frac{\pi}{4}=
\frac{1}{\hb}\left[\frac{x}{2}\sqrt{2E_{n}^{\hb}-x^2}+
{E_{n}^{\hb}}\arcsin\left(\frac{x}{\sqrt{2E_{n}^{\hb}}}\right)-{E_{n}^{\hb}}\frac{\pi}{2}\right]+{\pi\over 4} \ . \nonumber\\
&&
\end{eqnarray*}
Then, for large $n$ and small $\hb$ , and using the auxiliary variable  $\xi=x/\sqrt{\hb}$, the phase becomes
\begin{eqnarray}\label{phase_appr}
&&\frac{1}{\hb}\int_{\sqrt{2E_{n}^{\hb}}}^{x}\sqrt{2E_{n}^{\hb}-t^2}\, dt+{\pi\over 4}
\approx \nonumber\\
&& \frac{1}{\hb}\left[\frac{x\sqrt{2n \epsilon-x^2}}{2}+\epsilon\left(n+\frac{1}{2}\right)\arcsin\left(\frac{x}{\sqrt{2n\epsilon}}\right)-\epsilon\left(n+\frac{1}{2}\right)\frac{\pi}{2}\right]+{\pi\over 4}=\nonumber\\
&&\frac{\xi }{2}\sqrt{2n-\xi^2}+\left(n+\frac{1}{2}\right)\arcsin\left(\frac{\xi}{\sqrt{2n}}\right)- n\frac{\pi}{2} \ .
 \end{eqnarray}

Thus, inserting into (\ref{psin2}) the approximate amplitude  (\ref{ampl_appr}) and approximate phase(\ref{phase_appr}), respectively, the approximation of the asymptotic solution in the oscillatory range $|x|<\sqrt{2n}$, is given by
\begin{eqnarray*}
\psi^{\epsilon}_{n}(x)\approx\tilde{\psi}^{\epsilon}_{n}(x)=\left({2n\hb-x^2}\right)^{-1/4}\cos\left[\frac{x}{2\hb}\sqrt{2n\hb-x^2}+\left(n+\frac{1}{2}\right)\arcsin\left(\frac{x}{\sqrt{2n\hb}}\right)-\frac{n\pi}{2}\right]\quad \quad
\end{eqnarray*}
for large $n$ and small $\hb$.

Comparing the above formula with (\ref{hermsolut_n}), we conclude that
$$v^{\epsilon}_{n}(x)\approx\tilde{v}_{n}^{\hb}(x)=\left(\frac{2}{\pi}\right)^{1/2} \tilde{\psi}^{\epsilon}_{n}(x)\approx \left(\frac{2}{\pi}\right)^{1/2} \psi^{\epsilon}_{n}(x) \, ,$$
where $v^{\epsilon}_{n}(x)$ are the exact normalized eigenfunctions of the harmonic oscillator. Thus for small $\hb$ , the factor $\left({2}/{\pi}\right)^{1/2}$ normalizes the WKB-eigenfunctions.

%%%%%%%%%%%%%%%%%%%%%%%%%%%%%%%%%%%%%%%%%%%%%%%%%%%%%
%%%%%%%%%%%%%%%%%%%%%%%%%%%%%%%%%%%%%%%%%%%%%%%%%%%%%%%%

\chapter{Airy asymptotics of Laguerre polynomials}\label{app_e}
The classical Laguerre orthonormal polynomial of degree $n$ and order $a$, for $n=0,1,2,\ldots$ and $a>-1$ fixed , is defined by 
\begin{eqnarray}\label{laguerrep}
L_{n}^{(a)}(x)=\sum_{m=0}^{n}{{n+a}\choose{n-m}}\frac{(-x)^m}{m!}
\end{eqnarray}
where ${{z}\choose{w}}$ are the binomial coefficients (see e.g. \cite{Sz, Olv}) .

Employing contour  integral representations  of $L_{n}^{(a)}(x)$, Wong \cite{WoR} has derived uniform Airy approximations for large $n$ and fixed  $x, a$. This expansion reads as follows.

\begin{proposition}
Let 
\begin{eqnarray}\label{Betat}
B(t)=
\left\{
	\begin{array}{ll}
	i\left[\frac{3}{2}\beta(t)\right]^{1/3}\, , & \mbox{if }\, 0<t\leq 1\\
	\left[\frac{3}{2}\gamma(t)\right]^{1/3}\, , & \mbox{if }\, t>1 \, ,
	\end{array}
\right.
\end{eqnarray}
%%%%%%%%%%%%%%%%%%%%%%%%%%%%%%%%%%%%%%%%%%%%%%%%%%
\begin{eqnarray}\label{betat}
\beta(t)={1\over 2}\left[\arccos\sqrt{t} \,-\sqrt{t-t^2}\,\right]\, ,\quad \mathrm{if} \,\,\,  0<t\leq 1
\end{eqnarray}
%%%%%%%%%%%%%%%%%%%%%%%%%%%%%%%%%%%%%%%%%%%%%%%%%%%%%%
and
\begin{eqnarray}
\gamma(t)={1\over 2}\left[\sqrt{t^{2}-t}-\mathrm{arccosh}\sqrt{t}\,\right]\, ,\quad \mathrm{if} \,\,\,  t>1 \, .
\end{eqnarray}
%%%%%%%%%%%%%%%%%%%%%%%%%%%%%%%%%%%%%%%%%%%%%%%%%%%%%%
Then for $p=0,1,2,\ldots$, and parameter $\nu$ defined by
\begin{eqnarray}\label{param_nu}
\nu=4n+2a+2\, , 
\end{eqnarray}
the Airy-type approximation of Laguerre polynomials is 
$$
L_{n}^{(\alpha)}(\nu t)=(-1)^{n}2^{-a}e^{\nu t/2} 
$$
\begin{equation}\label{lag_asymp}
\cdot\left\{ Ai\left[\nu^{2/3}\left(B(t)\right)^2\right]\sum_{k=0}^{[(p-1)/2]}\alpha_{2k}\nu^{-2k-1/3}
- Ai'\left[\nu^{2/3}\left(B(t)\right)^2\right]\sum_{k=0}^{[(p-1)/2]}\beta_{2k+1}\nu^{-2k-5/3}+\varepsilon_{p}\right \}
\end{equation}
\\
for $\nu>>1$, where $Ai(\cdot)$ is the Airy function and $Ai'(\cdot)$ its derivative, and $\varepsilon_{p}$ is a remainder  $\varepsilon_{p}=O(\nu^{-(p+2/3)})$ if $p=\mathrm{odd}$, and $\varepsilon_{p}=O(\nu^{-(p+1/3)})$ if $p=\mathrm{even}$.

The first coefficients $\alpha_0$, $\beta_0$ in the asymptotic series are given by
\begin{eqnarray*}
\alpha_{0}(t)=
\left\{
	\begin{array}{lll}
	t^{(1-a)/2}\frac{\sqrt{2B(t)}}{(t-1)^{1/4} t^{3/4}}
	\, , & \mbox{if }\, t>1\\
	& \\
	t^{(1-a)/2}\frac{\sqrt{2|B(t)|}}{(1-t)^{1/4} t^{3/4}}
	\, , & \mbox{if }\, 0<t\leq 1
	\end{array}
\right.
\end{eqnarray*}
%%%%%%%%%%%%%%%%%%%%%%%%%%%%%%%%%%%%%%%%%%%%%%%%%%%%%%%%%
\begin{eqnarray*}
\beta_{1}(t)=\frac{\alpha_0}{2B(t)}\left[\frac{5}{24}(B(t))^{-3}-{3\over 4}\, \frac{\sqrt{t-1}}{\sqrt{t}}+{1\over 2}\, \frac{\sqrt{t}}{\sqrt{t-1}}-\frac{5}{12}\, \frac{t^{3/2}}{{(t-1)}^{3/2}}-\left(a^{2}-1\right)\, \frac{\sqrt{t-1}}{\sqrt{t}}\right]
\end{eqnarray*}
if $t>1$ . A corresponding formula exists for $0<t<1$ and obtained from the last formula for $\beta_1$ by replacing $\sqrt{t-1}$ by $i\sqrt{1-t}$ .
\end{proposition}

\renewcommand{\bibname}{References}
%\bibliography{mybib}

\begin{thebibliography}{ZZ999999}

%%%%%%%%%%%%%%%%%%%%%%%%%%%%%%%%%%%%%A%%%%%%%%%%%%%%%%%%%%%%%%%%%%%%%%%%%%%%%%%%%%%%%%%%%%%%%%%%%%%%%%%%%%%%%%%%%%%%%%%%%%%%

%\bibitem[AK]{AK}
%G.Avila and J.Keller,
%\emph{The high-frequency asymptotic field of a point source in an
%inhomogeneous medium},
%Comm. Pure Appl. Math.  \textbf{XVI}, 1963

\bibitem[AN]{AN}
A. Arnold \&  F. Nier, \emph{Numerical analysis of the deterministic particle method applied to the Wigner equation}, Math. Comp. , Vol. {\bf 58}, No 198, 645-669, 1992


\bibitem[ARN1]{ARN1}
V.I. Arnold, \emph{Mathematical methods of classical mechanics}, Springer-Verlag, New York, 1989

%\bibitem[AS]{AS}
%M.Abramowitz, I.A.Stegun,  \emph{ Handbook of Mathematical Functions with %Formulas, Graphs, and Mathematical Tables} Dover, New York, 1972

%\bibitem[AsK]{AsK}
%A.A. Asatryan and Yu.A. Kravtsov, \emph{Longitudinal
%caustic scale and boundaries of applicability of uniform
%Airy-asymptotics}, Wave Motion \textbf{19} (1994), 1-10

\bibitem[ASS]{ASS}
I. Antoniou, S.A. Shkarin \& Z. Suchanecki, \emph{The spectrum of the Liouville-von Neumann
operator in the Hilbert-Schmidt space}, J. Math. Phys., {\bf 40(9)}, 459-469, 1989 


\bibitem[Ath1]{Ath1} A. Athanassoulis, \emph{Smoothed Wigner transforms in the numerical simulation of semiclassical (high-frequency) wave propagation}, 2007, \url{http://arxiv.org/pdf/0704.3404v1.pdf}


\bibitem[Ath2]{Ath2} A. Athanassoulis, 
\emph{The Smoothed Wigner Transform Method: Precise Coarse-Scale Simulation of Wave Propagation}, AIP Conf. Proc. 936, 2007, 58



%\bibitem[AVH]{AVH}
 %V.I.Arnold, A.N.Varchenko and S.M.Husein-Zade,
%\emph{Singularities of Differentiable Maps}, Vol. 1,
 %Birkh\"{a}user Verlag, Basel, 1985
 
%%%%%%%%%%%%%%%%%%%%%%%%%%%%%%B%%%%%%%%%%%%%%%%%%%%%%%%%%%%%%%%%%%%%%%%%%%%
%%%%%%%%%%%%%%%%%%%%%%%%%%%%%%%%%%%%%%%%%%%%%%%%%%%%%%%%%%%%%%%%%%%%%%%%%%%%

\bibitem[B]{B}
L.E. Ballentine, \emph{Quantum mechanics: a modern development}, World Scientific, 1998

\bibitem[Ba]{Ba}
G.A. Baker, Jr., \emph{Formulation of Quantum Mechanics based on the
quasi-probability
 distribution induced on phase-space},  Phys. Rev. \textbf{109(6)}, 2198-2206, 1958

%\bibitem[BaB]{BaB}
%V.B. Babich \& V.S. Buldyrev {\it Short-wavelength diffraction theory. Asymptotic methods}, Springer-Verlag, Berlin, 1991

\bibitem[BaBMo]{BaBMo} 
C. Bardos \& L. Boutet de Monvel, \emph{From the atomic hypothesis to microlocal analysis}, Unesco EOLSS Encyclopedia of Mathematical Science, 2005

\bibitem[BaFFLS1]{BaFFLS1} F. Bayen, M. Flato, C. Fronsdal, A. Lichnerowicz \& D. Sternheimer,  \emph{Deformation Theory and Quantization I. Deformations of Symplectic Structures}, Annals of Physics \textbf{111}, 61-110, 1978 

\bibitem[BaFFLS2]{BaFFLS2} F. Bayen, M. Flato, C. Fronsdal, A. Lichnerowicz \& D. Sternheimer,  \emph{Deformation Theory and Quantization II. Physical Applications}, Annals of Physics \textbf{110}, 111-151, 1978 

\bibitem[BaKi]{BaKi}
V.M. Babich \& N.Y. Kirpichnikova,
\emph{The Boundary-Layer Method in Diffraction Problems},
 Springer-Verlag, Berlin-Heidelberg, 1979
 
  
\bibitem[BB]{BB}
V.B. Babich \& V.S. Buldyrev,
\emph{ Short-Wavelength Diffraction Theory. Asymptotic Methods },
Springer-Verlag, Berlin-Heidelberg, 1991

%\bibitem[BCKP]{BCKP}
%J.D.Benamou, F.Castella, T.Katsaounis \& B.Perthame, \emph{
%High frequency limit of the Helmholtz equations}, Rev. Mat.
%Iberoamericana \textbf{18}, 2002 

\bibitem[BeBa]{BeBa}
M.V. Berry \& N.L. Balazs, \emph{Evolution of semiclassical states in phase space}, J.
Phys. A, {\bf 12}, 625-642, 1979

\bibitem[Ben1]{Ben1}J.D. Benamou, \emph{Big ray tracing: Multivalued travel time
field computation using viscosity solutions of the eikonal
equation}, J. Comp. Phys. \textbf{128} 1996

\bibitem[Ben]{Ben}
J.D. Benamou, \emph{Direct computation of multivalued phase space solutions for Hamilton-Jacobi equations}, Comm. Pure Appl. Math., {\bf 52}, 1443-1475, 1999

\bibitem[Ber]{Ber} M.V. Berry,
\emph{Semi-classical mechanics in phase space: A study of Wigner's function},
Phil. Trans. of the Royal Society of London,
\textbf{287(1343)}, 237-273, 1977

\bibitem[BeSh]{BeSh} F.A. Berezin \& M.A. Shubin, \emph{Symbols of operators and quantization},  Colloquia Math. Soc. Janos Bolyai, Tihany (Hungary), 1970

\bibitem[BH]{BH}
 N. Bleistein \& R. Handelsman,
\emph{Asymptotic Expansions of Integrals}, Dover Publications
Inc., New York, 1986

%\bibitem[Bl]{Bl}
 %N.Bleistein,
%\emph{Mathematical methods for wave phenomena}, Academic Press
%Inc., California, 1984

\bibitem[BLP]{BLP}
A. Bensoussan, J.L. Lions \& G. Papanicolaou, \emph{Asymptotic
Analysis for Periodic Structures}, North-Holland, Amsterdam, New
York, Oxford, 1978

\bibitem[BM]{BM}
M. Bartlett \& J. Moyal, \emph{The Exact Transition Probabilities of Quantum Mechanical
Oscillators Calculated by the Phase-Space Method},  Math. Proc. Camb.
Phil. Soc. {\bf 45}, 545-553, 1949

\bibitem[Bor]{Bor}
V.A. Borovikov, \emph{Uniform stationary phase method}, The Institution of
Electrical Engineers, London, 1994

\bibitem[BP]{BP}
H-P. Breuer \& F. Petruccione, \emph{The Theory of Open Quantum Systems}
Oxford University Press Inc., New York, 2002

\bibitem[BR]{BR} A. Bouzouina \& D. Robert, 
\emph{Uniform semiclassical estimates for the propagation of quantum observables}, Duke Math. J. {\bf III(2)}, 223-252, 2002

\bibitem[BS]{BS} 
F.A. Berezin \& M.A. Shubin, \emph{ The Schr\"odinger Equation}, Kluwer Academic Publishers, Dordrecht/Boston/London, 1991 (Translated from Russian) 

\bibitem[BuKe]{BuKe}
R.N. Buchal \& J.B. Keller,
\emph{Boundary layer problems in diffraction theory},
Comm. Pure Appl. Math. \textbf{13}, 85-144, 1960

%%%%%%%%%%%%%%%%%%%%%%%%%%%%%%%%%%C%%%%%%%%%%%%%%%%%%%%%%%%%%%%%%%%%%%%%%%%%%
%%%%%%%%%%%%%%%%%%%%%%%%%%%%%%%%%%%%%%%%%%%%%%%%%%%%%%%%%%%%%%%%%%%%%%%%%%%%%

%\bibitem[Ca]{Ca}
%F.Castella, \emph{The radiation condition at infinity for the
%high-frequancy Helmholtz equation with source term: a wave packet
%approach}, J. Func. Anal. \textbf{223} (2005), 204-257

\bibitem[CFU]{CFU}
C. Chester, B. Friedman \& F. Ursell,
\emph{An extension of the method of steepest descent},
Proc. Camb. Philos. Soc. \textbf{53}, 599-611, 1957

\bibitem[CFZ1]{CFZ1}
T. Curtright, D. Fairlie \& C. Zachos,
\emph{Features of time-independent Wigner functions}, 
The American Physical Society, Physical Review D \textbf{58} 025002-(1-14), 1998

\bibitem[CFZ2]{CFZ2}
C.K. Zachos, D.B. Fairlie \& T.L. Curtright, \emph{ Quantum Mechanics in Phase
Space, An Overview with Selected Papers}, World Seientific, Singapore, 2005

%\bibitem[CH]{CH}
%R. Courant \& D. Hilbert, \emph{Methods of Mathematical Physics}, Wiley, New York, 1962

%\bibitem[CH1]{CH1} C.H.Chapman and R.Drummond, \emph{Body-wave
%seismograms in inhomogeneous media using Maslov asymptotic
%theory}, Bull. Seism. Soc. Amer. \textbf{72(6)}, S277-S317, 1982

%\bibitem[CH2]{CH2}C.H.Chapman, \emph{Ray theory and its extensions: WKBJ
%and Maslov seismograms}, J. Geophys. \textbf{58}, 27-43, 1985

%\bibitem[CK]{CK}
%D.Colton \& R.Kress,
%\emph{Inverse Acoustic and Electromagnetic Scattering},
%Springer-Verlag, New York, 1992

%\bibitem[CLOT]{CLOT}
%S.J.Chapman, J.M.H.Lawry, J.R.Ockendon \& R.H.Tew, \emph{On the theory of complex rays}, SIAM
%Rev. {\textbf{41(3)}}, 417-509, 1999

\bibitem[CMP]{CMP}
V.C\u{e}rven\`{y}, I.A.Molotkov \& I. P\u{s}en\u{c}ik,
\emph{Ray Method in Seismology},
Univerzita Karlova, Praha, 1977

%%%%%%%%%%%%%%%%%%%%%%%%%%%%%%%%%%D%%%%%%%%%%%%%%%%%%%%%%%%%%%%%%%%%%%%%%%%%
%%%%%%%%%%%%%%%%%%%%%%%%%%%%%%%%%%%%%%%%%%%%%%%%%%%%%%%%%%%%%%%%%%%%%%%%%%

\bibitem[D]{D}
P.A.M. Dirac, \emph{A new notation for quantum mechanics}, Mathematical Proceedings of the Cambridge Philosophical Society, Vol. {\bf 35}, Issue 03, 416-418, 1939

\bibitem[Do]{Do}
D. Dominici, \emph {Asymptotic analysis of the Hermite polynomials from their differential-difference equation}, J. Difference Equ. Appl. {\textbf{13(12)}}, 1115-1128, 2007

\bibitem[Dui1]{Dui1}
J.J. Duistermaat,
\emph{Oscillatory integrals, Lagrangian immersions and unfolding of singularities},
Comm. Pure Appl. Math \textbf{ XXVII}, 207-281, 1974

\bibitem[Dui2]{Dui2}
J.J. Duistermaat, 
\emph{The light in the neighborhood of a caustic}, Seminaire N. Bourbaki--1976/77, \textbf{exp. no. 490}, 19-29, 1976

\bibitem[Dui3]{Dui3}
J.J. Duistermaat, 
\emph{Fourier Integral Operators}, Progress in Mathematics 130,
Birkhauser, Boston, 1996

%%%%%%%%%%%%%%%%%%%%%%%%%%%%%E%%%%%%%%%%%%%%%%%%%%%%%%%%%%%%%%%%%%%%%%%%%%%%%%%
%%%%%%%%%%%%%%%%%%%%%%%%%%%%%%%%%%%%%%%%%%%%%%%%%%%%%%%%%%%%%%%%%%%%%%%%%%%%%%%
\bibitem[EK]{EK}
R. Estrada \& R.P. Kanwal, \emph{A distributional approach to asymptotics. Theory and applications}, Birkh\"auser, Boston, 2002

\bibitem[ER]{ER}
B. Engquist \& O. Runborg, \emph{Multi-phase computations in geometrical optics}, J. Comput. Appl. Math., {\bf 74}, 175-192, 1996


\bibitem[EmRo1]{EmRo1}
H. Emamirad \& P. Rogeon, \emph{Existence of wave operators for the Wigner equation in $L^{2,p}$ spaces}, C. R. Math. Acad. Sci. Paris, {\bf{334(9)}}, 811-816, 2002

\bibitem[EmRo2]{EmRo2}
H. Emamirad \& P. Rogeon, \emph{Scattering theory for the Wigner equation}, Math.
Methods Appl. Sci., {\bf{28(8)}}, 947-960, 2005


\bibitem[EW]{EW}
E. Wigner, \emph{On the quantum correction for thermodynamic equilibrium}, Wiley, Physical Review, \textbf{40}, 749-759, 1932

%%%%%%%%%%%%%%%%%%%%%%%%%%%%%%%%F%%%%%%%%%%%%%%%%%%%%%%%%%%%%%%%%%%%%%%%%%%%%%%%
%%%%%%%%%%%%%%%%%%%%%%%%%%%%%%%%%%%%%%%%%%%%%%%%%%%%%%%%%%%%%%%%%%%%%%%%%%%%%%%%

\bibitem[F]{F}
 D.B. Fairlie, \emph{ The formulation of quantum mechanics in terms of
phase space functions }, Proc. Camb. Phil. Soc., {\bf 60}, 581-586, 1964

\bibitem[Fed]{Fed}
M. Fedoriouk,
\emph{M$\acute{e}$thodes asymptotiques pour les équations différentielles ordinaires linéaires}, traduction francaise, Edition Mir, 1987

\bibitem[Fedos]{Fedos}
B. Fedosov, \emph{Deformation quantisation and index theory}, Academic Verlag, Berlin, 1996

\bibitem[FEO]{FEO}E. Fatemi, B. Enquist \& S. Osher,
\emph{Numerical solution of the high frequency asymptotic
expansion for the scalar wave equation}, J. Comput. Phys.
\textbf{120(1)}, 145-155, 1995

\bibitem[Ferg]{Ferg} 
E. Fergadakis, \emph{Numerical Experiments with the Particle Method for the Wigner equation in high-frequency paraxial propagation}, MSc thesis, University of Crete, 2004, \url{http://www.math.uoc.gr:1080/erevna/diplomatikes/Fregadakis_MDE.pdf}

\bibitem[Flat]{Flat}
S.M. Flatt$\grave{e}$, \emph{The Schr\"odinger equation in classical physics}, Amer. J. Phys., {\bf 54}, 1088-1092, 1986

\bibitem[FM1]{FM1}
S. Filippas \& G.N. Makrakis,
\emph{Semiclassical Wigner function and geometrical optics},
Multiscale Model. Simul., \textbf{1(4)}, 674-710,  2003

\bibitem[FM2]{FM2}
S. Filippas \& G.N. Makrakis,
\emph{On the evolution of the semi-classical Wigner function in higher dimensions},
Euro. Jnl of Applied Mathematics, \textbf{(17)}, 33-62, 2003

\bibitem[FMan]{FMan}
D.B. Fairlie \& C.A. Manogue,
\emph{The formulation of quantum mechanics in terms f phase space functions- the third equation}, J. Phys. A: Math. Gen. {\bf{24}}, 3807-3815, 1991

\bibitem[Foc]{Foc}
V.A. Fock, \emph{Fundamentals of quantum mechanics}, MIR Publishers, Moscow, 1982

\bibitem[Foc1]{Foc1}
V.A. Fock, \emph{Theory of radio-wave propagation in an inhomogeneous atmosphere for a raised
source}, Bull. Acad. Sci. USSR Ser. Phys., {\bf 14}, 70-94, 1950 (in Russian)

\bibitem[Foc2]{Foc2}
V.A. Fock, \emph{Theory of radiowave propagation in an inhomogeneous  atmosphere for a raised source}, Electromagnetic  Diffraction and Propagation Problems,
Pergamon Press, Oxford, ch. 14, 276-307, 1965 

\bibitem[FOL]{FOL}
G.B. Folland, \emph{Harmonic Analysis in Phase Space}, Princeton Univ. Press, Princeton, 1989

%%%%%%%%%%%%%%%%%%%%%%%%%%%%%%%%G%%%%%%%%%%%%%%%%%%%%%%%%%%%%%%%%%%%%%%%%
%%%%%%%%%%%%%%%%%%%%%%%%%%%%%%%%%%%%%%%%%%%%%%%%%%%%%%%%%%%%%%%%%%%%%%%%
\bibitem[G]{G} K.-S. Giannopoulou, \emph{Asymptotic approximation of the Wigner function in two-phase geometric optics}, MSc thesis, University of Crete, 2009, \url{http://www.math.uoc.gr:1080/erevna/diplomatikes/Giannopoulou_K_S_MDE.pdf}


\bibitem[G1]{G1}
M.A. de Gosson,
\emph{Symplectic Geometry and Quantum Mechanics}, Birkhauser Basel, series ``Operator Theory:Advances and Applications"(subseries:"Advances in Partial Differential Equations"), vol.{\bf 166}, 2006

%\bibitem[G2]{G2}
%M.A.de Gosson,
%\emph{Symplectic Methods in Harmonic Analysis and in Mathematical Physics}, Springer, Basel, 2011.

\bibitem[GG]{GG}
M.A. de Gosson \& S.M. de Gosson,
\emph{Weak values of a quantum
observable and the cross-Wigner distribution},
2013,  \url{http://arxiv.org/pdf/1109.3665v1.pdf}

\bibitem[GL]{GL}
M.A. de Gosson \& F. Luef,
\emph{A New Approach to the $\star$-Genvalue Equation},
Lett Math Phys, \textbf{85}, 173-183, 2008

\bibitem[GM]{GM}
K.-S. Giannopoulou \& G.N. Makrakis, \emph{Uniformization of WKB functions by Wigner transform}, Applicable Analysis, Vol. {\bf 93}, No. 3, 624-645, 2014 

\bibitem[GMMP]{GMMP}
P. Gerard, P.A. Markowich, N.J. Mauser \& F. Poupaud,
\emph{Homogenization limits and Wigner transforms},
Comm. Pure Appl. Math. \textbf{50}, 323-380, 1997


\bibitem[GOLD]{GOLD} H. Goldstein, \emph{Classical Mechanics}, Addison Wesley, 2th edition, 1980

\bibitem[Gr]{Gr}
D.J. Griffths
\emph{Introduction to Quantum Mechanics}, Prentice Hall, Inc., 1995

\bibitem[GR]{GR} 
I.S. Gradshteyn \& I. M. Ryzhik,
\emph{Table of Integrals, Series, and Products}, Academic Press, New York, 7th edition, 2007

\bibitem[Gro]{Gro}
H.J. Groenewold, \emph{On the Principles of elementary quantum mechanics}, Physica, {\bf{12}}, 405-460, 1946

\bibitem[GS]{GS}
V. Guillemin \& S. Sternberg, \emph{Geometric Asymptotics},
American Mathematical Society, Providence, R.I., 1977

%\bibitem[GSc]{GSc}V. Guillemin and D. Schaeffer, \emph{Remarks on a paper
%of D. Ludwig}, Bull. Am. Math. Soc. \textbf{ 79(2)} (1973),
%382-385

%%%%%%%%%%%%%%%%%%%%%%%%%%%%%%%%H%%%%%%%%%%%%%%%%%%%%%%%%%%%%%%%%%%%%%%%
%%%%%%%%%%%%%%%%%%%%%%%%%%%%%%%%%%%%%%%%%%%%%%%%%%%%%%%%%%%%%%%%%%%%%%%%%

%\bibitem[Ha]{Ha}
%P. Hartman, \emph{Ordinary differential equations},Classics in
%Applied Mathematics, 38, SIAM,Philadelphia, 2002

\bibitem[He]{He}
E.J. Heller, \emph{Wigner phase space method: Analysis for semiclassical applications}, J. Chem. Phys., {\bf 65(4)}, 1289-1298, 1976

\bibitem[HMS1]{HMS1} 
M. Hug, C. Menke \& W.P. Schleich, \emph{Modified spectral method in phase space: calculation of the Wigner function. I. Fundamentals}, Phys. Rev. A {\bf(3)}, {\bf 57(5)}, 3188-3205, 1998

\bibitem[HMS2]{HMS2} M. Hug, C. Menke \& W.P. Schleich, 
\emph{Modified spectral method in phase space: calculation of the Wigner function. II. Generalizations}, Phys. Rev. A {\bf (3)}, {\bf 57(5)}, 3206-3224, 1998


\bibitem[Ho]{Ho}
L. H\"{o}rmander,
\emph{The Analysis of Linear Partial Differential Operators I},
Springer-Verlag, New York, 1983

\bibitem[HO]{HO}
L. H\"{o}rmander,
\emph{The Weyl calculus of pseudo-differential operators},
  Comm. Pure Appl. Math., \textbf{XXXII }, 359-443, 1979



\bibitem[HS]{HS} P.D. Hislop \& I.M. Sigal, 
\emph{Introduction to Spectral Theory, With Applications to Schr\"odinger Operators}, Springer-Verlag, New York, 1996

%%%%%%%%%%%%%%%%%%%%%%%%%%%%%J%%%%%%%%%%%%%%%%%%%%%%%%%%%%%%%%%%%%%%%%%%%%%
%%%%%%%%%%%%%%%%%%%%%%%%%%%%%%%%%%%%%%%%%%%%%%%%%%%%%%%%%%%%%%%%%%%%%%%%%%%

\bibitem[JL]{JL}S. Jin \& X. Li, \emph{Multi-Phase Computations of the Semiclassical
limit of the
 Schr\"odinger equation and related problems:Whitham vs Wigner}, Physica D 
\textbf{182}, 46-85, 2003

\bibitem[Jo]{Jo}
F. John, \emph{Partial differential equations}, Springer-Verlag, New York, 1980

%%%%%%%%%%%%%%%%%%%%%%%%%%%K%%%%%%%%%%%%%%%%%%%%%%%%%%%%%%%%%%%%%%%%%%%%%%%%
%%%%%%%%%%%%%%%%%%%%%%%%%%%%%%%%%%%%%%%%%%%%%%%%%%%%%%%%%%%%%%%%%%%%%%%%%%%

\bibitem[Kal]{Kal}
E.K. Kalligiannaki, \emph{Asymptotic Solutions of the Wigner Equation and High Frequency Waves Propagation near caustics}, PhD thesis, University of Crete, 2007 (in Greek), \url{http://web-server.math.uoc.gr:1080/erevna/didaktorikes/Kalligiannaki_PhD.pdf}

\bibitem[Kal1]{Kal1} 
E. Kalligiannaki, \emph{Particle Method for the Wigner equation in high frequency paraxial propagation}, MSc thesis, University of Crete, 2002, \url{http://elocus.lib.uoc.gr/php/pdf_pager.php?filename=/var/www/dlib-portal//dlib/2/b/d/attached-metadata-dlib-2002kalligiannaki/2002kalligiannaki.pdf&lang=el&pageno=1&pagestart=1&width=&height=&maxpage=
}

\bibitem[KalMak]{KalMak}
E.K. Kalligiannaki \& G.N. Makrakis, \emph{Perturbation solutions of the semiclassical Wigner equation}, \url{http://arxiv.org/abs/1402.6194}

%\bibitem[Kel]{Kel}J.B. Keller, \emph{Corrected Bohr-Sommerfeld quantum conditions
%for non-separable systems}, Ann. Phys. \textbf{4} (1958), 180-188

%\bibitem[KKM]{KKM}T. Katsaounis, G.T. Kossioris and G.N. Makrakis,
%\emph{Computation of high frequency fields near caustics}, Math.
%Meth. Model. Appl. Sci. \textbf{11(2)} (2001), 1-30

\bibitem[KO]{KO}
Yu.A. Kravtsov \& Yu.I. Orlov,
\emph{Caustics, Catastrophes and Wave Fields},
Springer Series on Wave Phenomena 15,
Springer-Verlag, Berlin, 1999

%\bibitem[KO1]{KO1}Yu.A.Kravtsov and Yu.I.Orlov, \emph{Geometrical Optics
%of Inhomogeneous Media}, Springer Series on Wave Phenomena 6,
%Springer-Verlag, Berlin, 1990

\bibitem[KP]{KP}
J.G. Kr\"uger \& A. Poffyn,
\emph{Quantum mechanics in phase space II. Eigenfunctions of the Liouville operator}, Physica {\textbf{87A}}, 132-144, 1977

\bibitem[Kra]{Kra}
Yu.A. Kravtsov, \emph{Two new asymptotic methods in the theory of
wave propagation in inhomogeneous media(review)}, Sov. Phys.
Acoust. \textbf{14(1)}, 1-17, 1968


%\bibitem[Ku]{Ku}
%V.V.Kucherenko,
%\emph{Quasiclassical asymptotics of a point-source function for
%the stationary Schr\"odinger equation},
%Theoret. Math. Phys. (English Translation) \textbf{ 1(3)}, 294-310, 1969

%%%%%%%%%%%%%%%%%%%%%%%%%%%%%%%%%L%%%%%%%%%%%%%%%%%%%%%%%%%%%%%%%%%%%%%%
%%%%%%%%%%%%%%%%%%%%%%%%%%%%%%%%%%%%%%%%%%%%%%%%%%%%%%%%%%%%%%%%%%%%%%%%%

\bibitem[L]{L}
H.W. Lee, \emph{ Theory and applications of quantum phase-space distribution functions}, Physics Reports, {\textbf{259}}, 147-211, 1995


%\bibitem[Le]{Le}R.M. Lewis,  \emph{Asymptotic theory of wave propagation},
%Arch. Rat. Mech. Anal., \textbf{20}, 191- 250, 1965

\bibitem[Leb]{Leb}
N.N. Lebedev, \emph{Special Functions and Their Applications},
Dover Publications, Inc., New York, 1972

\bibitem[LF]{LF}
M. Leontovich \& V. Fock, \emph{Solution of the problem of electromagnetic wave propagation along the Earth's surface by the method of parabolic equation}, J. Phus. USSR, Vol. {\bf 10}, 13-23, 1946

%\bibitem[Lit]{Lit}
%R.G. Littlejohn, \emph{The Van Vleck formula, Maslov theory and phase space
%geometry}, J. Statist. Phys. \textbf{68(1/2)}, 7-50, 1992

\bibitem[LP]{LP}
P.L. Lions \& T. Paul,
\emph{Sur les measures de Wigner},
Rev. Math. Iberoamericana \textbf{9}, 563-618, 1993

\bibitem[Lu]{Lu}
D. Ludwig, \emph{Uniform asymptotic expansions at a caustic},
Comm. Pure Appl. Math. \textbf{XIX }, 215-250, 1966

%%%%%%%%%%%%%%%%%%%%%%%%%%%%%%%%%%M%%%%%%%%%%%%%%%%%%%%%%%%%%%%%%%%%%%%
%%%%%%%%%%%%%%%%%%%%%%%%%%%%%%%%%%%%%%%%%%%%%%%%%%%%%%%%%%%%%%%%%%%%%%%%

\bibitem[MA]{MA} P. Markowich,
 \emph {On the Equivalence of the {S}chr\"odinger and the Quantum Liouville Equations}, Math. Methods in the Applied Sci. \textbf{11}, 4106-4118, 1999

%\bibitem[Ma1]{Ma1} V.P.Maslov, \emph{Operational Methods}, Mir
%Publishers, Moscow, 1979

%\bibitem[Ma2]{Ma2}V.P. Maslov, \emph{Theory of Perturbations and
%Asymptotic Methods}, Dunod, Paris, 1972

\bibitem[Mar]{Mar}
A. Martinez, \emph{An Introduction to Semiclassical and Microlocal Analysis}, Lecture Notes, Italy, 2001

\bibitem[MCD]{MCD} S. McDonald, 
\emph{Phase-space representations of wave equations with applications to the eikonal approximation for short-wavelength waves}, Phys. Rep.,{\textbf{158(6)}} ,  337-416, 1988

\bibitem[MF]{MF}
V.P. Maslov \& V.M. Fedoriuk,
\emph{   Semi-classical
approximations in quantum mechanics },
D. Reidel, Dordrecht, 1981

%\bibitem[MN]{MN}V.P. Maslov and Nazaikinskii, \emph{Asymptotics of
%Operator and Pseudodifferential Equations}

\bibitem[Mo]{Mo}
J.E. Moyal,
\emph{Quantum mechanics as a statistical theory},
Proc. Camb. Phil. Soc.,
\textbf {45}, 99-124, 1949




\bibitem[MSS]{MSS} A. Mishchenko, V. Shatalov, \& B. Sternin,\emph{
Lagrangian Manifolds and the Maslov Operator}, Springer-Verlag,
Berlin-Heidelberg, 1990





%%%%%%%%%%%%%%%%%%%%%%%%%%%%%%%%%%%%N%%%%%%%%%%%%%%%%%%%%%%%%%%%%%%%%%%%%%%%
%%%%%%%%%%%%%%%%%%%%%%%%%%%%%%%%%%%%%%%%%%%%%%%%%%%%%%%%%%%%%%%%%%%%%%%%%%%%

\bibitem[N]{N}
F. Narkowich, 
\emph{On the quantum Liouville equation}, Physica, {\bf 134A}, 193-208, 1985

\bibitem[NEUM]{NEUM}
J.von Neumann, \emph{Mathematical Foundations of Quantum Mechanics},
Princeton University Press, Princeton, 1955

\bibitem[NSS]{NSS}
V.E. Nazaikinskii, B.W. Schulze \& B.Yu. Sternin \emph{Quantization Methods in differential equations}, Taylor \& Francis, London \& New York, 2002

%%%%%%%%%%%%%%%%%%%%%%%%%%%%%%%%O%%%%%%%%%%%%%%%%%%%%%%%%%%%%%%%%%%%%%%%%%
%%%%%%%%%%%%%%%%%%%%%%%%%%%%%%%%%%%%%%%%%%%%%%%%%%%%%%%%%%%%%%%%%%%%%%%%%%

\bibitem[Olv]{Olv}
F.W. Olver, \emph{Asymptotics and Special Functions}, Acad.Press, New York, 1974


%%%%%%%%%%%%%%%%%%%%%%%%%%%%%%P%%%%%%%%%%%%%%%%%%%%%%%%%%%%%%%%%%%%%%%%%%%%
%%%%%%%%%%%%%%%%%%%%%%%%%%%%%%%%%%%%%%%%%%%%%%%%%%%%%%%%%%%%%%%%%%%%%%%%%%%%

\bibitem[P]{P}
M. Pulvirenti, \emph{Semiclassical expansion of Wigner functions}, J. Math. Phys. {\bf 47(5)}, 052103-(1-12), 2006


\bibitem[PR]{PR}
G. Papanikolaou \& L. Ryzhik,
\emph{Waves and Transport, Hyperbolic Equations and Frequency Interactions}, (Eds L.
Caffarelli and E. Weinan), IAS/Park City Mathematical
Series, AMS, 1999

%\bibitem[PV]{PV}
%B.Perthame \& L.Vega,
%\emph{Sommerfeld condition for a Liouville equation and concentration of trajectories},
%Bull. Braz. Math. Soc., New Series \textbf{34(1)},43-57, 2003

%%%%%%%%%%%%%%%%%%%%%%%%%%%%%%%%%%R%%%%%%%%%%%%%%%%%%%%%%%%%%%%%%%%%%%%%%%%%%
%%%%%%%%%%%%%%%%%%%%%%%%%%%%%%%%%%%%%%%%%%%%%%%%%%%%%%%%%%%%%%%%%%%%%%%%%%%%%%
\bibitem[Raz]{Raz} 
M. Razavy, \emph{Quantum theory of tunneling}, World Scientific, Singapore, 2003

\bibitem[Rip]{Rip}
N. Ripamonti, \emph{Classical limit of the harmonic oscillator Wigner functions in the Bargmann representation}, J.Phys. A: Math. Gen., \textbf{29}, 5137-5151, 1996

\bibitem[Rob]{Rob} D. Robert,  
\emph{Autour de l' Approximation Semi-Classique}, Birkh\"auser, Boston, 1987


\bibitem[Ru1]{Ru1}
O. Runborg, \emph{ Multiscale and Multiphase Methods for Wave
Propagation}, Doctoral Dissertation, Dept. Num. Anal. Comp. Sci.,
Roy. Inst. Techn., Stockholm, 1998, \url{https://www.nada.kth.se/~olofr/Publications/}

\bibitem[Ru2]{Ru2}
O. Runborg, \emph{Some new results in multiphase geometrical
optics},J. Math. Model. Num. Anal. \textbf{ 34(6)}, 1203-1231, 2000

%%%%%%%%%%%%%%%%%%%%%%%%%%%%%S%%%%%%%%%%%%%%%%%%%%%%%%%%%%%%%%%%%%%%%%%%
%%%%%%%%%%%%%%%%%%%%%%%%%%%%%%%%%%%%%%%%%%%%%%%%%%%%%%%%%%%%%%%%%%%%%%%

\bibitem[S]{S}
R. Shankar, \emph{Principles of Quantum Mechanics}, Second Edition, Kluwer Academic/Plenum Publishers, 1994

\bibitem[Schr]{Schr}
E. Schr\"odinger, \emph{An Undulatory Theory of the Mechanics of Atoms and Molecules}, Physical Review {\bf 28 (6)}, 1049-1070, 1926

\bibitem[SF]{SF}
H. Sato \& M.C. Fehler, \emph{Seismic Wave Propagation and Scattering in the Heterogeneous Earth}, Springer-Verlag, New York, 1998


%\bibitem[SK]{SK} B.D., Seckler \& J.B. Keller,
%\emph{Geometrical theory of diffraction in inhomogeneous media}, J. Acoust. Soc. Amer. \textbf{31} 1959, 192-205

\bibitem[Si]{Si}
B. Simon, \emph{Semiclassical analysis of low lying eigenvalues, I. Non-degenerate minima: Asymptotic expansions}, Ann. Inst. H. Poincare, {\bf 38(3)}, 295-307, 1983

\bibitem[SMM]{SMM}
C. Sparber, P.A. Markowich \& N.J. Mauser,
\emph{Wigner functions versus WKB-methods in multivalued geometrical optics},
Asymptot. Anal. \textbf{33(2)}, 153-187, 2003

\bibitem[SP]{SP}
H. Spohn, \emph{The spectrum of the Liouville-von Neumann operator}, J. Math. Phys., {\bf{17}}, 57-60, 1976

%\bibitem[Sta]{Sta}
%O.N. Stavroudis, \emph{ The Optics of Rays, Wavefronts and
%Caustics}, Academic Press, New York, 1972

\bibitem[Ste]{Ste}
H. Steinr\"uck, 
\emph{Asymptotic Analysis of the Quantum Liouville Equation},
Mathematical Methods in the Applied Sciences, Vol. {\bf 13}, 143-157, 1990

\bibitem[Sz]{Sz}
G. Szeg\"o, 
\emph{Orthogonal Polynomials}, Amer. Math. Soc., Colloq. Publ., Vol. {\bf 23}, New York, 1958

%%%%%%%%%%%%%%%%%%%%%%%%%%%%%%T%%%%%%%%%%%%%%%%%%%%%%%%%%%%%%%%%%%%%%%%%%%%
%%%%%%%%%%%%%%%%%%%%%%%%%%%%%%%%%%%%%%%%%%%%%%%%%%%%%%%%%%%%%%%%%%%%%%%%%%%%%

\bibitem[Tak]{Tak}
L.A. Takhtajan, \emph{Quantum mechanics for mathematicians}, Graduate Studies in Mathematics Vol. {\bf 95}, Amer. Math. Soc. 2008

\bibitem[Tap1]{Tap1}
F.D. Tappert, \emph{Diffractive ray tracing of laser beams}, J. Opt. Soc. Amer., {\bf 66}, 1368-1373, 1976

\bibitem[Tap2]{Tap2}
F.D. Tappert, \emph{The parabolic approximation method, in Wave Propagation and Underwater
Acoustics}, J.B. Keller and J.S. Papadakis, eds., Lecture Notes in Phys. 70,
Springer-Verlag, Berlin, 224-287, 1977


\bibitem[Tat1]{Tat1}
V.I. Tatarskii,
\emph{The Effects of the Turbulent Atmosphere on Wave Propagation} Israel
Program for Scientific Translation, Jerusalem, 1971

%\bibitem[Tat2]{Tat2}
%V.I.Tatarskii,
%\emph{The Wigner representation of quantum mechanics},
%Soviet Phys. Uspekhi \textbf{26}, 311-327, 1984

\bibitem[Tayl]{Tayl}
M. Taylor, \emph{Pseudodifferential operators}, Princeton Univ. Press, 1981

\bibitem[TC]{TC}
I. Tolstoy \& C.S. Clay, \emph{Ocean Acoustics. Theory
and Experiment in Underwater Sound}, American Institute of
Physics, New York, 1966

\bibitem[Th]{Th} 
S. Thangavelu, \emph{Lectures on Hermitte and Laguerre expansions },
Princeton University Press, 1993

\bibitem[Tr]{Tr}
F. Treves, \emph{Introduction to Pseudodifferential and Fourier Integral Operators} Vols 1 \& 2, Plenum Press, New York, 1980

%\bibitem[TVF1]{TVF1} 
%Go.Torres-Vega and J.H.Frederick,
%\emph{A quantum mechanical representation in phase space}, J.Chem. Phys. \textbf{98} (4), 1993

\bibitem[TZ]{TZ}
A. Truman \& H.Z. Zhao, \emph{Semi-classical limit of wave functions}, Proc. Am. Math. Soc., \textbf{128(3)}, 1003-1009, 2000

%%%%%%%%%%%%%%%%%%%%%%%%%%%%%%%%%%%V%%%%%%%%%%%%%%%%%%%%%%%%%%%%%%%%%%%%%%%
%%%%%%%%%%%%%%%%%%%%%%%%%%%%%%%%%%%%%%%%%%%%%%%%%%%%%%%%%%%%%%%%%%%%%%%%%%%

%\bibitem[Va]{Va}
%B.R.Vainberg,
%\emph{Quasiclassical approximation in stationary scattering problems},
%Func. Anal. Appl. \textbf{11}, 247-257, 1977

\bibitem[Va1]{Va1}
B.R. Vainberg, \emph{Asymptotic Methods in Equations of
Mathematical Physics}, Gordon and Breach, New York, 1989

\bibitem[VS]{VS} 
O. Vallee \& M. Soares, \emph{Airy functions and applications to
physics}, Imperial College Press, London, 2004

%%%%%%%%%%%%%%%%%%%%%%%%%%%%%%W%%%%%%%%%%%%%%%%%%%%%%%%%%%%%%%%%%%%%%%%%
%%%%%%%%%%%%%%%%%%%%%%%%%%%%%%%%%%%%%%%%%%%%%%%%%%%%%%%%%%%%%%%%%%%%%%%%%%%%

\bibitem[WB]{WB}
J. Wilkie \& P. Brumer, \emph{Quantum classical correspondence via Liouville dynamics: I. Integrable systems and the chaotic spectral decomposition }, Phys. Rev. A, \textbf{55(9)}, 27-42, 1997

%\bibitem[Wed]{Wed}
%R.Weder,
%\emph{Spectral and Scattering Theory for Wave Propagation in Perturbed Stratified
%Media},
%Springer-Verlag, New York, 1991


\bibitem[Wig]{Wig}
E. Wigner, \emph{ On the quantum correction for thermodynamic equilibrium}, Physical Review , \textbf{40}, 749-759, 1932


\bibitem[WoM]{WoM}
M.W. Wong, \emph{Weyl Transforms}, Springer-Verlag, New York, 1998

\bibitem[WoR]{WoR}
R. Wong,\emph{ Asymptotic Approximations of Integrals}, Classics in Applied Mathematics, Vol. {\bf 34}, (SIAM), Philadelphia, 2001


%%%%%%%%%%%%%%%%%%%%%%%%%%%%%%Z%%%%%%%%%%%%%%%%%%%%%%%%%%%%%%%%%%%%%%%%%%%
%%%%%%%%%%%%%%%%%%%%%%%%%%%%%%%%%%%%%%%%%%%%%%%%%%%%%%%%%%%%%%%%%%%%%%%%%

\bibitem[Y]{Y} 
K. Yosida, \emph{Lectures on differential and integral equation}, Interscience Publishers, New York, 1960

\bibitem[Z]{Z}
E. Zeidler, \emph{Applied Functional Analysis. Applications to Mathematical Physics}, Applied Mathematical Sciences 108, Springer-Verlag, New York, 1995

%\bibitem[Za]{Za}
%C.Zachos,
%\emph{Deformation quantization: Quantum mechanics lives and works in phase-spase},
%Int. J. Mod.Phys. A,
%\textbf{17(3)},297-316, 2002

\bibitem[Za1]{Za1}
C. Zachos,
\emph{A survey of star product geometry}, 2000, \url{ http://arxiv.org/pdf/hep-th/0008010.pdf}

\bibitem[Za2]{Za2}
C. Zachos, \emph{ Crib Notes on Campbell-Baker-Hausdorff expansions}, unpublished, 1999 , \url{www.hep.anl.gov/czachos/index.html}

\bibitem[Zau]{Zau}
E. Zauderer, \emph{Partial differential equations of applied mathematics}, Wiley, New York, 1983

\bibitem[Zw]{Zw}
M. Zworski, \emph{Semiclassical analysis}, Grad. Studies Math., vol.138, AMS, Providence, 2012

%\bibitem[ZH]{ZH}
%\emph{Wigner Measure and Semiclassical Limits of Nonlinear
%Schrodinger Equations}, Courant Lecture Notes \textbf{17}, AMS, New
%York, 2008

%\bibitem[B2]{B2}
%V.M. Babich, \emph{The short wave asymptotic form of the
%solutionfor the problem of a point source in an inhomogeneous medium}, USSR Comp.
%Math.  Math. Phys. \textbf{5} (1965), 247-251

%\bibitem[HUD]{HUD}
%R.L. Hudson, \emph{When is the Wigner quasi-probability density non-negative?}, Reports on Mathematical Physics, Vol. {\textbf{6}} (1974), 249-252


\end{thebibliography}
%\begin{thebibliography}{plain}

\end{document}